# Study of Vortex Dynamics and Phase Transitions in Superconducting Thin Films

*A thesis*

*Submitted to the*

*Tata Institute of Fundamental Research, Mumbai*

*for the degree of Doctor of Philosophy*

*in Physics*

by

**Indranil Roy**

Department of Condensed Matter Physics and Materials Science

Tata Institute of Fundamental Research

Mumbai, India

15th January, 2020

Final version is submitted on 20th March, 2020

> "আমি তাই করি ভাই যখন চাহে এ মন যা,
> করি শত্রুর সাথে গলাগলি, ধরি মৃত্যুর সাথে পাঞ্জা,
> আমি উন্মাদ, আমি ঝঞ্ঝা।
> আমি মহামারী, আমি ভীতি এ ধরিত্রীর;
> আমি শাসন-ত্রাসন, সংহার আমি উষ্ণ চির অধীর!
> বল বীর-
> আমি চির উন্নত-শির!"

-কাজী নজরুল ইসলাম, 'বিদ্রোহী'

# DECLARATION

This thesis is a presentation of my original research work. Wherever contributions of others are involved, every effort is made to indicate this clearly, with due reference to the literature, and acknowledgement of collaborative research and discussions.

The work was done under the guidance of Professor Pratap Raychaudhuri, at the Tata Institute of Fundamental Research, Mumbai.

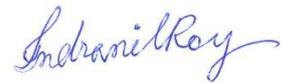

**Indranil Roy**

In my capacity as supervisor of the candidate's thesis, I certify that the above statements are true to the best of my knowledge.

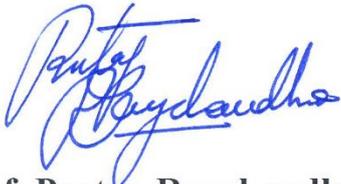

**Prof. Pratap Raychaudhuri**

Date: 13/01/2020

# Table of Contents

# PREAMBLE

The work reported in my doctoral thesis is an experimental study of vortex dynamics and phase transitions in thin films of type II superconductors using scanning tunnelling spectroscopy, low frequency ac susceptibility measurements and complimentary transport measurements. The thesis is organized in the following fashion:

In *Chapter I*, an introduction is given to basics of superconductivity and various aspects of vortex lattice in a type II superconductor.

In *Chapter II*, the instrumentation and analysis techniques are discussed. Details of real space imaging of vortex lattice using scanning tunnelling spectroscopy (STS) and analysis of the vortex lattice and movements of the vortices are explained. It also contains low frequency ac susceptibility measurement technique using two coil mutual inductance technique. The details of preparation and characterisation of NbN and *a*-MoGe thin films are also described.

In *Chapter III*, I have discussed the effect of periodic pinning centres on the geometry and dynamics of vortex lattice in NbN thin films. Consequent study of dynamic transition of vortex Mott-like to vortex metal-like state is described.

In *Chapter IV*, effect of strong disorder on vortex lattice in NbN thin films is studied. Here we show magnetic field induced granularity gives rise to pseudogap phase which is utilized to explain superconductor to insulator-like transition in stronger disorder.

*Chapter V* contains the study of 2-dimensional vortex lattice melting in *a*-MoGe thin films. We discuss the observation of hexatic vortex fluid phase and the BKTHNY two-step melting in this scenario. Effect of sample thickness on this phases is also described.

In *Chapter VI*, we study the possibility of quantum fluctuation of vortices in weakly pinned *a*-MoGe thin films. Effect of pinning on this fluctuation and a possibility of a quantum to thermal crossover is also explored.

Finally, in *Chapter VII*, we conclude about the findings and set a stage for future directions from both experimental and theoretical aspects.

*Appendix A* contains a short study exploring superconductivity in $Nb_xBi_2Se_3$.

# STATEMENT OF JOINT WORK

The experiments reported in this thesis are carried out in the Department of Condensed Matter Physics and Materials Science at the Tata Institute of Fundamental Research under the supervision of Prof. Pratap Raychaudhuri. The major portions of the results presented in this thesis have already been published in refereed journals.

All the STM/S experiments and analysis of data from *a*-MoGe thin films discussed in this thesis were performed by me. The deposition and mutual inductance measurements on NbN films on AAM (anodic alumina membrane) and subsequent analysis are done by me. The details of collaborative work areas follows:

Deposition and two coil measurements of NbN films on AAM are done in collaboration with Prashant Chauhan and John Jesudasan.

NbN thin films for the study of strong disorder were deposited by Dr. Harkirat Singh. The transport data on these films and some of the STM/S data are taken by Dr. Rini Ganguly.

*a*-MoGe thin films for STM/S studies were grown in collaboration with Surajit Dutta, John Jesudasan and Dr. Aditya N. Roy Choudhury. MoGe pellet for sample deposition in Pulsed Laser Deposition (PLD) were grown by Dr. Aditya N. Roy Choudhury and Dr. Harkirat Singh in collaboration with Dr. Rajib Mondal in the laboratory of Prof. Arumugam Thamizhavel. The a-MoGe thin films for transport were grown and consequent transport measurements on these samples were done by Surajit Dutta. Mutual inductance measurements on these samples were performed by Soumyajit Mandal and Somak Basistha.

NbSe$_2$ single crystals for the standardization of techniques were grown by Vivas Bagwe in the laboratory of Prof. Arumugam Thamizhavel.

Some of the basic codes for vortex image analysis were written by Dr. Somesh Chandra Ganguli. Surajit Dutta, Soumyajit Mandal and Somak Basistha helped in numerous occasions in STM/S data acquisition.

# ACKNOWLDGEMENT

I want to begin with expressing my sincere gratitude to my thesis advisor *Prof. Pratap Raychaudhuri* for providing the continuous guidance and supervision during my PhD. His contagious enthusiasm for physics has encouraged and motivated to pursue this exciting journey. Writing some popular science articles in Bengali with him were also very intriguing.

I thank my past and present colleagues *Dr. Somesh Chandra Ganguli, Dr. Rini Ganguly, Dr. Harkirat Singh, John Jesudasan, Vivas Bagwe, Surajit Dutta, Soumyajit Mandal, Somak Basistha, Dr. Aditya Narayan Roy Choudhury, Rishabh Duhan, Srijita Das* and project student *Prashant Chauhan* for their collaboration, support and company.

I would like to thank *Prof. Sangita Bose, Prof. Rajdeep Sensarma, Prof. Amit Ghoshal, Dr. Anurag Banerjee, Prof. Lara Benfatto, Dr. Ilaria Maccari, Prof. T V Ramakrishnan, Prof. Chandan Dasgupta, Prof. Mikhail Feigel'man, Prof. Nandini Trivedi, Prof. Dragana Popovic, Prof. Steve Kivelson, Prof. Valerii Vinokur, Prof. Mukul Lad, Prof. Christophe Brun, Prof. Sandip Ghosh, Prof. Vikram Tripathi, Prof. Kedar Damle, Prof. Mandar Desmukh, Prof. Jacob Linder, Morten Amundsen* for collaborating, discussing and critiquing our works.

I thank *Subash Pai, Supriya, Pranay* and *Anas* from Excel Instruments for their prompt technical support. I also thank *Bagyashri Chalke and Rudhir Bapat* for sample characterization using EDX and SEM and *Atul Raut* for technical support. Most importantly I want to thank the *Low Temperature Facility* team of TIFR: *Srinivasan, Vijay Arolkar, Bosco, Arvind, Jaison, Purao* and *Nikhil* for continuous supply of liquid helium and nitrogen, without which these experiments were hardly possible.

This journey has its roots with teachers from my school, *Subhabrata Chaudhuri* and *Indrajit Chatterjee* who has always motivated me to pursue my career. In my college days *Prof. Indranath Chaudhuri, Prof. Suparna Roy Chowdhury, Prof. Tanaya Bhattacharya, Prof. Partha Ghosh* have ignited love for physics. In TIFR, coursework with *Prof. Vaibhav Prabhudesai, Prof. Sreerup Raychaudhuri, Prof. Amol Dighe, Prof. R Vijayraghavan* were very much enjoyable.

Coming to the life outside academic bounds, the journey have been so smooth only because a friend like *Sudip Chakraborty* was always there from my college days. Talking with him in countless topics for countless hours has always cheered me up. Along with him, *Soham Bhattacharya* and *Dr. Nairit Sur* have also provided support and company for numerous cooking and food-ventures. *Suman Chatterjee, Dibyendu Bala, Sumeru Hazra, Arkadipta Sarkar, Sazedur Laskar* have been the perfect supportive classmates. Time spent with seniors like *Dr. Manibrata Sen, Dr. Chandroday Chattopadhyay, Dr. Touseek Samui, Dr. Shamik Bhattacharya, Dr. Soureek Mitra, Dr. Debjyoti Bardhan, Dr. Shubhadeep Biswas, Srimanta Bandopadhyay,* and juniors like *Sandip Mandal, Pratap Chandra Adak, Rabisankar Samanta, Anupam Roy* and many others have given me the energy to overcome the daily stress of academic life. At the end of five and half years, I can safely claim to have visited more than seventy percent eateries in Colaba and Fort area in Mumbai.

I have obtained a continuous encouragement from many family members, *dida*, *dadu*, *mama*, *thakuma*, *thakurda* and people from whom my parents and I have got enormous support, *Tarunkaku, Prabirkaku, Shambhu-jethu* and *jethima* and *family*.

And last but not the least this part would remain incomplete if I don't mention my family, baba and maa for continuously believing in me and supporting me. *Srijoyee* has been a source of support in my highs and lows with whom I have shared and debated over numerous topics. These three persons have given me a boost like no one else.

# List of Publications

**Publications relevant to the thesis**

1. Dynamic transition from Mott-like to metal-like state of the vortex lattice in a superconducting film with a periodic array of holes; **Indranil Roy**, Prashant Chauhan, Harkirat Singh, Sanjeev Kumar, John Jesudasan, Pradnya Parab, Rajdeep Sensarma, Sangita Bose, Pratap Raychaudhuri; *Physical Review B, 95 (054513), 2017.*

2. Magnetic field induced emergent inhomogeneity in a superconducting film with weak and homogeneous disorder; Rini Ganguly and **Indranil Roy**, Anurag Banerjee, Harkirat Singh, Amit Ghosal, Pratap Raychaudhuri; *Physical Review B, 96 (054509), 2017.*

3. Melting of the Vortex Lattice through Intermediate Hexatic Fluid in an $a-$MoGe Thin Film; **Indranil Roy**, Surajit Dutta, Aditya N. Roy Choudhury, Somak Basistha, Ilaria Maccari, Soumyajit Mandal, John Jesudasan, Vivas Bagwe, Claudio Castellani, Lara Benfatto, Pratap Raychaudhuri; *Physical Review Letters, 122 (047001), 2019.*

4. Robust pseudogap across the magnetic field driven superconductor to insulator-like transition in strongly disordered NbN films; **Indranil Roy**, Rini Ganguly, Harkirat Singh, Pratap Raychaudhuri; *European Physical Journal B, 92 (49), 2019.*

**Other Publications**

1. Disorder-induced two-step melting of vortex matter in Co-intercalated $NbSe_2$ single crystals; Somesh Chandra Ganguli, Harkirat Singh, **Indranil Roy**, Vivas Bagwe, Dibyendu Bala, Arumugam Thamizhavel, and Pratap Raychaudhuri; *Physical Review B, 93 (144503), 2016.*

2. Collective flux pinning in hexatic vortex fluid in a-MoGe thin film; Surajit Dutta, **Indranil Roy**, Somak Basistha, Soumyajit Mandal, John Jesudasan, Vivas Bagwe and Pratap Raychaudhuri; *Journal of Physics: Condensed Matter, 31 (075601), 2020.*

3. Extreme sensitivity of the vortex state in a-MoGe films to radio-frequency electromagnetic perturbation; Surajit Dutta, **Indranil Roy**, Soumyajit Mandal, John Jesudasan, Vivas Bagwe and Pratap Raychaudhuri; *Physical Review B, 100 (214518), 2019.*

# List of symbols and abbreviations

## Symbols

| | |
|---|---|
| $\Phi_0$ | Magnetic flux quantum |
| $e$ | Electronic charge |
| $h$ | Planck's constant |
| $a_0$ | Vortex lattice constant |
| $E_c$ | Condensation energy |
| $E_F$ | Fermi energy |
| $\omega_D$ | Debye cut-off frequency |
| $N(0)$ | Density of states at Fermi energy |
| $\Delta$ | Superconducting energy gap |
| $H_{c1}$ | Lower critical field |
| $H_{c2}$ | Upper critical field |
| $J_s$ | Supercurrent density |
| $J_S$ | Superfluid stiffness |
| $\lambda_L$ | London penetration depth |
| $\lambda_C$ | Campbell penetration depth |
| $\xi_{GL}$ | Ginzburg Landau coherence length |
| $J_c$ | Critical current density |
| $T_c$ | Superconducting transition temperature |
| $G_{\vec{k}}(\vec{r})$ | Positional correlation function |
| $G_6(\vec{r})$ | Orientational correlation function |

# Abbreviations

| | |
|---|---|
| 2D | 2 dimensional |
| 3D | 3 dimensional |
| BKT | Berezenskii Kosterlitz Thouless |
| BKTHNY | Berezenskii Kosterlitz Thouless Halperin Nelson Young |
| VL | Vortex lattice |
| STM | Scanning tunnelling microscope/ microscopy |
| STS | Scanning tunnelling spectroscopy |
| FC | Field cooled |
| ZFC | Zero field cooled |
| FT | Fourier transform |
| FFT | Fast Fourier transform |
| SIT | Superconductor to Insulator Transition |
| ZBC | Zero bias conductance |
| CPH | Coherence peak height |
| HVF | Hexatic Vortex Fluid |
| IVL | Isotropic Vortex Liquid |

# Chapter I: Introduction

Since its inception more than a century ago in 1911 by Heike Kamerlingh Onnes[1], superconductivity has garnered enormous attention from experimentalists and theorists alike. Although the perspective of applicability in real life scenarios has been for a long time, the driving force behind the vast research in this topic, a fundamental perspective is the other side of the same coin which has helped the topic to gather its present impetus and flourishment. However a fundamental understanding of this phenomenon had been elusive until in 1950s Bardeen, Cooper and Schrieffer (BCS) presented their microscopic explanation.[2] Seven years before BCS, Ginzburg and Landau[3] introduced a complex pseudo-wavefunction $\psi$ as an order parameter within Landau's general theory of second order phase transitions. This situation was again overturned and the subject rejuvenated with the discovery of a new class of high-temperature (high-$T_c$) superconductors by Bednorz and Muller.[4] Basic microscopic mechanism in these new superconductors remains an open question till date, however general phenomenology of them obeys those of the classic superconductors. Hence it is a common strategy to address several high-$T_c$ phenomena like existence of a pseudogap phase[5], existence of quantum fluctuations of vortices[6], phase fluctuations in superconductors with small superfluid density[7] in specially designed classic superconductors and try to find the parallel between these two cases. Classic superconductors can also act as a platform to test several other statistical mechanics problems, which we will discuss in this thesis. It is interesting to note that recent advancement towards room temperature superconductivity has been established with observation of significant drop in resistivity in hydrogen rich compounds (for example, $H_2S$,[8] $LaH_{10\pm x}$[9] etc.) under several hundreds of Giga-Pascal pressure. This phenomenon is also being explored at present. However we shall deal with thin films of classic superconductors and related mechanisms, which can enrich our arsenal to deal with the fascinating and versatile field of superconductivity. We start with reviewing the basic observed electrodynamic phenomena.

## 1.1  Basics of superconductivity

The two hallmarks of superconductivity are its zero resistance state and its perfect diamagnetism. Both of these phenomenon have been proven experimentally over the years since its discovery.

### 1.1.1 Zero resistance state

Electrical resistance of superconducting materials drops below measurable limit below a certain temperature called critical temperature, $T_c$ which is characteristic of the material. This property has been successfully demonstrated (and is being demonstrated in countless superconducting magnets in their persistent mode) with persistent current in superconducting loops, which is expected to die out in no less than $10^{10^{10}}$ years at the least. This perfect conductivity is one of the basic prerequisite for most potential applications such as transmission of high current and high field magnets.

### 1.1.2 Perfect diamagnetism

The next hallmark was discovered by Meissner and Oschsenfeld[10] in 1933, to be perfect diamagnetism. Although a perfect diamagnetism can be explained by perfect conductivity alone, superconductors are special in the sense that they expel magnetic field as they are cooled below $T_c$ with the magnetic field on. In contrast a perfect conductor would trap the flux inside if they are cooled in field. This perfect diamagnetism, often called as *Meissner effect*, also shows that superconductivity gets destroyed above a critical field, $H_c$ which also is a material parameter. However it is important to note that diamagnetism is perfect only for bulk samples since the field does penetrate the superconductor to a finite distance, $\lambda_L$.

### 1.1.3 London equations

These two basic electrodynamic properties were described well by F. and H. London[11] in 1935, who proposed two equations to govern the microscopic electric and magnetic fields:

$$\vec{E} = \frac{\partial}{\partial t}(\Lambda \vec{J_s}) \tag{1.1}$$

$$\vec{h} = \vec{\nabla} \times \vec{A} = -\vec{\nabla} \times (\Lambda \vec{J_s}) \tag{1.2}$$

Here, $\Lambda = \frac{m}{n_s e^2}$ is a phenomenological parameter, which connects to the London penetration depth as,

$$\lambda_L = \sqrt{\frac{m}{\mu_0 n_s e^2}} = \sqrt{\frac{\Lambda}{\mu_0}} \tag{1.3}$$

London penetration depth determines the lengthscale to which magnetic field can penetrate the superconductor, expressed as: $|\vec{B}| = |\vec{B_0}|e^{-x/\lambda_L}$. The number density of superconducting electrons, $n_s$ is expected to increase below $T_c$, from zero at $T_c$.

## 1.2 Basics of Bardeen-Cooper-Schrieffer (*BCS*) theory

The giant leap in the understanding of superconductivity was the establishment of an energy gap $\Delta$ of the order of $k_B T_c$ between the ground state and the quasiparticle excitations of the system. The history making theory developed by BCS[2] showed that "*even a weak attractive interaction between electrons, such as that caused in second order by electron-phonon interaction, which causes an instability of the ordinary Fermi-sea ground state of the electron gas with respect to the formation of bound pairs of electrons occupying states with equal and opposite momentum and spin. These so-called Cooper pairs have a spatial extension of order $\xi_0$ and, crudely speaking, comprise the superconducting charge carriers anticipated in the phenomenological theories.*" (Tinkham[12]) We explore in the next subsections, the basics of the theory.

### 1.2.1 BCS Hamiltonian

Starting from Schrodinger equation, $H\psi_n = -\frac{\hbar^2}{2m}\nabla^2\psi_n + v(x)\psi_n = E_n\psi_n$, taking $\psi(x) = \sum_k b_k \frac{e^{i\vec{k}.\vec{x}}}{\sqrt{\Omega}}$, in second quantization form, the kinetic energy part of the Hamiltonian comes as:

$$H_k = \int \frac{d^3x}{\Omega} \sum_{k,k'} b_{k'}^+ e^{-i\vec{k}.\vec{x}} \left(-\frac{\hbar^2}{2m}\nabla^2\right) b_k e^{i\vec{k}.\vec{x}} = \sum_k \frac{\hbar^2 k^2}{2m} b_k^+ b_k \qquad (1.4)$$

If the interaction potential is given as, $v(x - x') = \sum_q e^{i\vec{q}.(\vec{x}-\vec{x}')}$, in second quantization language it becomes,

$$H' = \sum_{k_1,k_2,q} b_{k_1+q}^+ b_{k_2-q}^+ b_{k_2} b_{k_1} v(q) \qquad (1.5)$$

Together this becomes:

$$H^s = \sum_k \frac{\hbar^2 k^2}{2m} b_k^+ b_k + \sum_{k_1,k_2,q} b_{k_1+q}^+ b_{k_2-q}^+ b_{k_2} b_{k_1} v(q) \qquad (1.6)$$

This Hamiltonian, when expressed in the basis of BCS wavefunction becomes the BCS Hamiltonian, $H_{BCS}$.

### 1.2.2 BCS wavefunction

BCS took their wavefunction as a product of different momenta components instead of a linear superposition of them. If a single pair of electrons with opposite momentum and spin can be expressed as, $|\psi\rangle = \sum_k g_k c_{k\uparrow}^+ c_{-k\downarrow}^+ |0\rangle$, where $c_{k\uparrow}^+$ is the creation operator of an electron with momentum $\boldsymbol{k}$ and spin up. Similarly, $c_{k\uparrow}$ is the annihilation operator of the electron with momentum $\boldsymbol{k}$ and spin up. The operators obey characteristic anticommutation relations of fermionic operators: $\{c_{k\sigma}, c^+_{k'\sigma'}\} = \delta_{kk'}\delta_{\sigma\sigma'}$ and $\{c_{k\sigma}, c_{k'\sigma'}\} = \{c^+_{k\sigma}, c^+_{k'\sigma'}\} = 0$. The BCS wavefunction is given by,

$$|\psi_{BCS}\rangle = \prod_k (u_k + v_k e^{i\phi} c^+_{k\uparrow} c^+_{-k\downarrow})|0\rangle \tag{1.7}$$

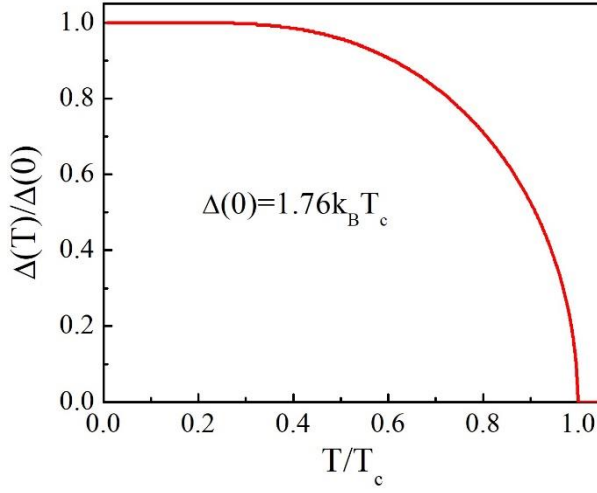

*Figure 1.1:* BCS universal gap variation. NbN and *a*-MoGe thin films in their clean limit follow this trend of $\Delta(T)/\Delta(0)$ vs. $T/T_c$.

This choice of wavefunction instead of a wavefunction written in terms of Slater determinants of momentum eigenfunctions, is valid only because there are a large number of particles involved. In this mean field approach, probability of having a Cooper pair is $|v_k|^2$ and not having is $|u_k|^2$, with $|u_k|^2 + |v_k|^2 = 1$. Using variational method it can be shown that the phase $\phi$ is same for all Cooper pairs.

Working in a grand canonical ensemble, the mean number of particles is to be regulated. This is done by inclusion of a chemical potential term in the effective Hamiltonian, which in the basis of BCS wavefunction looks like:

$$\bar{H}_{BCS} = \langle \psi_{BCS}|H_{BCS}|\psi_{BCS}\rangle = \left\langle 0 \left| \sum_{k,\sigma}\left(\frac{\hbar^2 k^2}{2m}-\mu\right)c^+_{k\sigma}c_{k\sigma} + \sum_{k,l} V_{kl} c^+_{k\uparrow}c^+_{-k\downarrow}c_{-l\downarrow}c_{l\uparrow} \right| 0\right\rangle \tag{1.8}$$

Next variational method is applied to find out $u_k$ and $v_k$, with approximations such as: Pairing amplitude, $\Delta_k = -\sum_l V_{kl} u_l v_l$ and $\xi_k = \sqrt{E_k^2 - \Delta_k^2}$; And $V_{kl} = -V$ for $|\xi_k| < \hbar\omega_D$ and zero otherwise; $\Delta_k = \Delta$ for $|\xi_k| < \hbar\omega_D$ and zero otherwise. ($\omega_D$ is the Debye frequency and $E_k$ is the energy of an elementary excitation of momentum $\hbar\vec{k}$.)

### 1.2.3 BCS gap variation

Bogoliubov[13] and Valatin[14] transformation of the BCS Hamiltonian and subsequent mean field approximations gives a beautiful and sophisticated approach to deal with excited states. $\Delta_k$ plays the role of an energy gap, temperature dependence of which ($\Delta(T)$) is obtained from numerically solving,

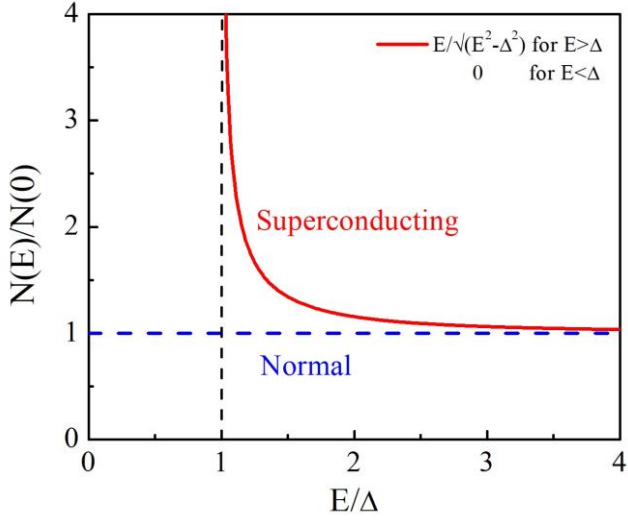

*Figure 1.2:* Red line represents BCS density of states (DOS), while blue dotted line represent DOS of a normal metal. At $E = \Delta$ (shown by the vertical dotted line) the superconducting DOS shows a divergence. Both of the DOS are normalized with respect to the DOS of normal metal at $E = E_F = 0$. For $\frac{E}{\Delta} \gg 1$, both the DOS match with each other.

$$\frac{1}{N(0)V} = \int_0^{\hbar\omega_D} \frac{\tanh\frac{1}{2}\beta(\xi^2 + \Delta^2)^{\frac{1}{2}}}{(\xi^2 + \Delta^2)^{\frac{1}{2}}} d\xi \quad (1.9)$$

This variation of gap gives a universal curve in weak-coupling limit shown in *Figure 1.1*. This in turn can be reduced to a thumb rule for relating $T_c$ with $\Delta(0)$, using,

$$\Delta(0) = 1.76 \times k_B T_c \quad (1.10)$$

### 1.2.4 BCS density of states (DOS)

The quasi-particle excitation can be simply described as fermions created by the Bogoliubon creation operator which are in one-to-one correspondence with $c_k^+$ of the normal metal, which prompts us that one can obtain the superconducting density of states $N_s(E)$ by equating,

$$N_s(E)dE = N_n(\xi)d\xi \quad (1.11)$$

Since we are dealing with energies $\xi$ only a few millielectronvolts from Fermi energy, we take $N_n(\xi) = N(0)$ a constant. This leads to,

$$\frac{N_s(E)}{N(0)} = \frac{d\xi}{dE} = \begin{cases} \frac{E}{(E^2 - \Delta^2)^{\frac{1}{2}}} & (E > \Delta) \\ 0 & (E < \Delta) \end{cases} \quad (1.12)$$

A divergent density of states is observed just above $E = \Delta$, as is shown in *Figure 1.2*.

### 1.2.5 Tunnelling: The semiconductor model

The most detailed experimental verification of the superconducting DOS is obtained from electron tunnelling measurements, pioneered by Giaever[15] in 1960. This is based on the idea of quantum mechanical tunnelling of electrons between two conductors separated by a thin insulating barrier. To understand tunnelling in cases where superconductors are involved, the basic assumption is to take the DOS of normal metal to be continuous states with density $N(0)$ including energies below, as well as above Fermi level. Within this model, called as the semiconductor model, illustrated in *Figure 1.3*, tunnelling current from metal 1 to metal 2 can be written as,

$$I_{1\to 2} = A \int_{-\infty}^{\infty} |T|^2 N_1(E) f(E) N_2(E + eV)[1 - f(E + eV)]dE \quad (1.13)$$

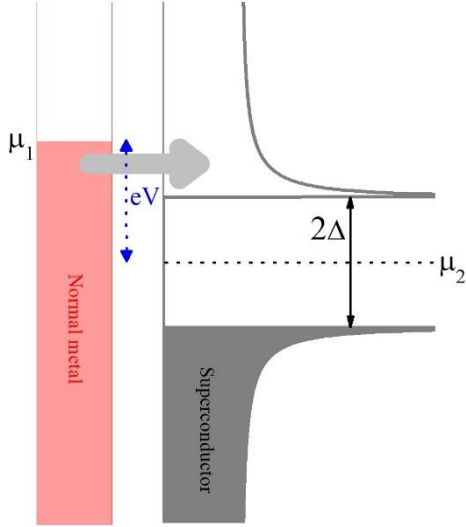

*Figure 1.3:* Schematic of tunnelling between a superconductor and a normal metal, which are filled up to their Fermi levels $\mu_2$ and $\mu_1$ respectively. Here tunnelling will occur if the chemical potential of these two materials are separated by $|\mu_1 - \mu_2| = eV > \Delta$. If this criterion is satisfied electrons will tunnel from filled states to empty states.

Subtracting the reverse current, the net current is:

$$I = A|T|^2 \int_{-\infty}^{\infty} N_1(E) N_2(E + eV)[f(E) - f(E + eV)]dE \quad (1.14)$$

### 1.2.5.1      *Normal to normal (N-N)*

If both of the materials are normal metals, then

$$I_{nn} = A|T|^2 N_1(0) N_2(0) \int_{-\infty}^{\infty} [f(E) - f(E + eV)]dE = A|T|^2 N_1(0) N_2(0) eV \equiv G_{nn} V \quad (1.15)$$

### 1.2.5.2      *Normal to superconductor (N-S)*

If one of the materials is superconducting whose DOS is given by $N_{2s}(E)$, the tunnelling equation becomes,

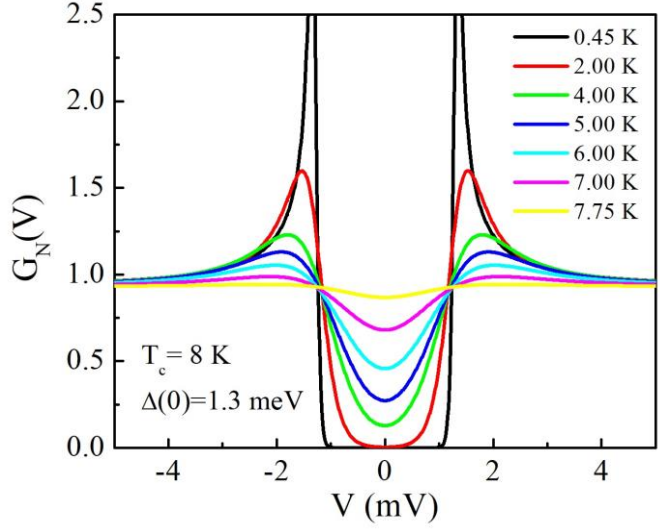

*Figure 1.4:* Simulated differential conductance curves for a normal-superconductor tunnelling for different temperatures, assuming $\Delta(0) = 1.3 \ meV$ and $T_c = 8 \ K$. The $G(V)$ are normalized by dividing the curves by $G(V = 5 \ mV)$ values. Here, $\frac{\Delta(0)}{k_B T_c} \sim 1.88$ which is close to the BCS value of 1.76. The thermal broadening of the tunnelling curves, especially of the coherence peaks are prominent.

$$I_{ns} = A|\boldsymbol{T}|^2 N_1(0) \int_{-\infty}^{\infty} N_{2s}(E)[f(E) - f(E + eV)]dE$$
$$= \frac{G_{nn}}{e} \int_{-\infty}^{\infty} \frac{N_{2s}(E)}{N_2(0)} [f(E) - f(E + eV)]dE \qquad (1.16)$$

$$G_{ns} = \frac{dI_{ns}}{dV} = G_{nn} \int_{-\infty}^{\infty} \frac{N_{2s}(E)}{N_2(0)} \left[ -\frac{\partial f(E + eV)}{\partial (eV)} \right] dE \qquad (1.17)$$

The magnitude of the tunnelling current is independent of sign of $V$, because hole and electron excitations have equal energies. For $T > 0$, tunnelling can occur in region below $eV = |\Delta|$ giving rise to an exponential tail in current and differential conductance. This variation of $G_{ns}(eV)$ is used to extract value of $\Delta$ from experimental data. Characteristics of normal-superconductor tunnelling, the differential conductance are simulated for different temperatures, shown in *Figure 1.4*.

### 1.2.6 Variation of $\lambda_L$

In clean limit, temperature variation of penetration depth ($\lambda_L$) is given by the BCS nature of variation given by,

$$\frac{\lambda_L^{-2}(T)}{\lambda_L^{-2}(0)} = \left[ 1 - 2 \int_{\Delta}^{\infty} \left( -\frac{\partial f}{\partial E} \right) \frac{E}{(E^2 - \Delta^2)^{\frac{1}{2}}} dE \right] \qquad (1.18)$$

In dirty limit, on the other hand, the temperature variation of $\lambda_L$ becomes,

$$\frac{\lambda_L^{-2}(T)}{\lambda_L^{-2}(0)} = \frac{\Delta(T)}{\Delta(0)} \tanh \frac{\beta \Delta(T)}{2} \qquad (1.19)$$

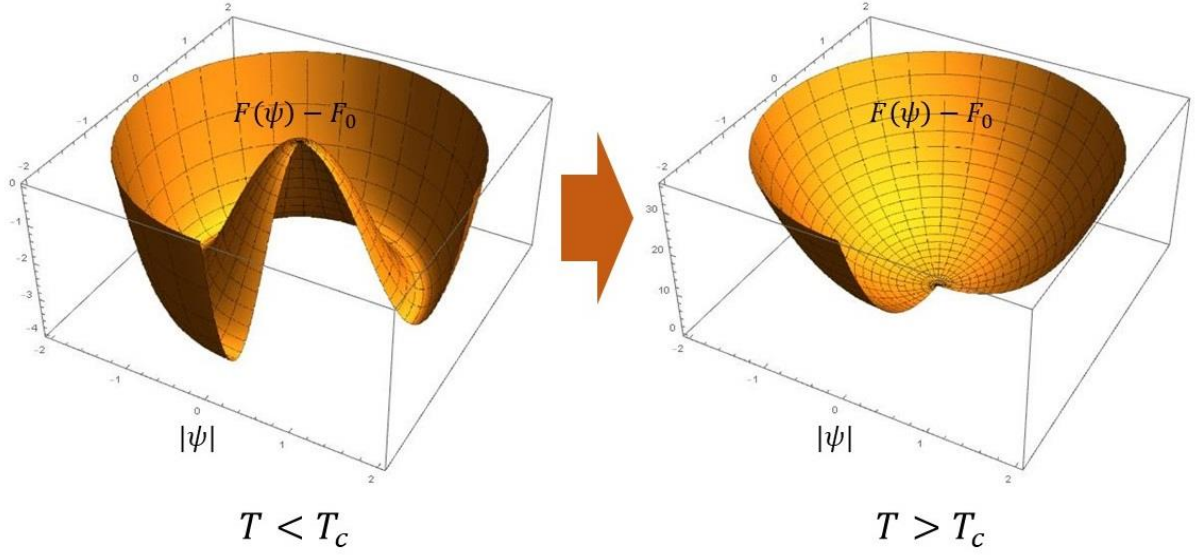

*Figure 1.5:* 3D plot of variation of GL free energy ($\alpha|\psi|^2 + \frac{\beta}{2}|\psi|^4$ part) without the magnetic field contribution as a function of $|\psi|e^{i\phi}$ for $T < T_c$ to $T > T_c$. The radial part signifies $|\psi|$ while the angular part represents $\phi$.

Now it is interesting to note that the penetration depth in thin films (of thickness $t$) changes significantly, as was shown by Pearl[16], to be, $\Lambda \approx \frac{2\lambda_L^2}{t} \gg \lambda_L$. The utilization of this Pearl penetration depth is that the supercurrent around a vortex falls off at $r \sim \Lambda$ instead of $r \sim \lambda_L$ for films of thickness $t < \lambda_L$.

## 1.3 Ginzburg Landau (*GL*) theory basics

In situations where application of the microscopic theory becomes difficult, much reliance is placed on the more macroscopic Ginzburg-Landau theory[3]. Although this theory came into being from a phenomenological postulation, Gor'kov showed[17] that GL theory can be derived from the BCS theory.

### 1.3.1 Order parameter and GL free energy

The triumph of GL theory is the introduction of a pseudowavefunction, $\psi(\vec{r}) = |\psi|e^{i\phi}$ as a complex order parameter. Local density of superelectrons $n_s(\vec{r})$ is represented as $|\psi(\vec{r})|^2$. The theory was developed by assuming an expansion of free energy in powers of $|\psi|^2$ and $|\nabla\psi|^2$ as follows:

$$f = f_{n0} + \alpha|\psi|^2 + \frac{\beta}{2}|\psi|^4 + \frac{1}{2m^*}\left|\left(\frac{\hbar}{i}\nabla - \frac{e^*}{c}\vec{A}\right)\psi\right|^2 + \frac{h^2}{8\pi} \qquad (1.20)$$

This free energy function go from a parabola-like shape at $T > T_c$ to a Mexican hat-like shape at $T < T_c$ as is shown in *Figure 1.5*.

### 1.3.2 GL equations

The variational treatment of the GL free energy gives rise to the celebrated GL equations:

$$\alpha\psi + \beta|\psi|^2\psi + \frac{1}{2m^*}\left(\frac{\hbar}{i}\vec{\nabla} - \frac{e^*}{c}\vec{A}\right)^2\psi = 0 \qquad (1.21)$$

$$\vec{J} = \frac{e^*}{m^*}|\psi|^2\left(\hbar\vec{\nabla}\phi(r) - \frac{e^*}{c}\vec{A}\right) = e^*|\psi|^2\vec{v}_s \qquad (1.22)$$

Where, $\vec{v}_s = \hbar\vec{\nabla}\phi(r) - \frac{e^*}{c}\vec{A}$ is the velocity of superelectrons. Assuming $f = \frac{\psi}{\psi_\infty}$, where $\psi_\infty^2 = -\frac{\alpha}{\beta} > 0$, in 1D the first GL equation becomes:

$$\frac{\hbar^2}{2m^*|\alpha|}\frac{d^2 f}{dx^2} + f - f^3 = 0 \qquad (1.23)$$

### 1.3.3 GL coherence length and GL parameter

We define, the Ginzburg Landau coherence length as, $\xi_{GL}^2(T) = \frac{\hbar^2}{2m^*|\alpha|} = \frac{\Phi_0}{2\sqrt{2}H_c(T)\lambda_L(T)}$, over which characteristic lengthscale a small disturbance of $\psi$ from $\psi_\infty$ will decay. ($\Phi_0 = hc/2e$)

We also define GL parameter, $\kappa = \lambda_L(T)/\xi_{GL}(T)$, as the ratio of two characteristics lengths. In classic superconductors in clean limit $\kappa \ll 1$, but in dirty superconductors or in high-$T_c$ superconductors, $\kappa$ is greater than 1.

#### *1.3.3.1  Type I and Type II superconductor classification*

Based on the value of surface energy in intermediate states in strong magnetic fields, superconductors are subdivided into two groups, type I and type II. The exact crossover from positive to negative surface energy in presence of magnetic field occurs for $\kappa = 1/\sqrt{2}$. This shows that for $\kappa < 1/\sqrt{2}$, named as type I superconductors would have laminar intermediate state in $H \sim H_c$. However, for $\kappa > 1/\sqrt{2}$, Abrikosov showed[18] that the negative surface energy causes the flux bearing normal regions to subdivide until a quantum limit is reached in which each quantum of flux $\Phi_0 = hc/2e$ passes through the sample as a distinct flux tube. These flux tubes form a regular array and have $\psi \to 0$ at the centres of each of these regions. Unlike type I superconductors, the intermediate mixed state in type II superconductors span over a substantial field range, lower critical field, $H_{c1}$ and upper critical field, $H_{c2}$.

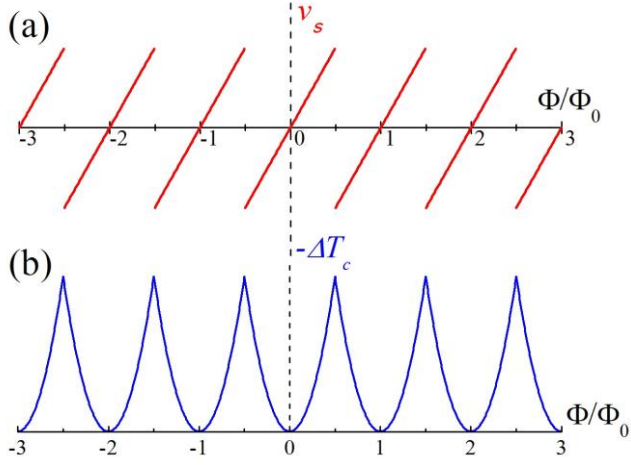

*Figure 1.6:* (a) Variation of velocity of superelectrons, $v_s$ as a function of filling factor, $\Phi/\Phi_0$. For integer filling $v_s = 0$, which means that at $\Phi/\Phi_0$ equals integer, order parameter, $|\psi|^2 = \psi_\infty^2$. (b) Variation of $-\Delta T_c \propto v_s^2$ as a function of $\Phi/\Phi_0$. For integer filling $-\Delta T_c$ shows a minima, which means for integer filling $T_c$ is enhanced than its non-integer filling counterparts.

### 1.3.4 Fluxoid quantization

One important consequence of GL theory is to explain quantization of 'fluxoid', which is defined as,

$$\Phi' = \frac{c}{e^*} \oint \left( m^* \vec{v}_s + \frac{e^* \vec{A}}{c} \right) \cdot d\vec{s} = \frac{c}{e^*} \oint \vec{p} \cdot d\vec{s} = n \frac{hc}{e^*} = n\Phi_0 \qquad (1.24)$$

using Bohr-Sommerfeld quantum condition. Here, fluxoid quantum is, $\Phi_0 = \frac{hc}{2e} = 2.07 \times 10^{-15}\, Wb$. This gives a powerful tool to deal with many problems of superconductors penetrated by magnetic flux.

### 1.3.5 Little Parks experiment

An ingenious experiment which shows this quantization of fluxoid was performed by Little and Parks[19] who showed, using a thin-walled superconducting cylinder in an axial magnetic field, the shifts $\Delta T_c(H)$ in critical temperature depending on the magnetic flux enclosed by the cylinder. If $R$ is the radius of the thin-walled cylinder and $H$ is the applied field, we have, $\Phi = \pi R^2 H$ and $\Phi' = n\Phi_0$. Now one can use $\Phi_0 = \Phi + \frac{m^*c}{e^*}\oint \vec{v} \cdot d\vec{s}$ to find that, velocity of superelectrons, $v_s = \frac{\hbar}{m^* R}\left(n - \frac{\Phi}{\Phi_0}\right)$ which shows a variation as is shown in *Figure 1.6(a)*.

Since, $|\psi|^2 = \psi_\infty^2\left[1 - \left(\frac{\xi m^* v_s}{\hbar}\right)^2\right]$ and superconductor becomes normal if $|\psi|^2 = 0$, this results in a condition, $\frac{1}{\xi_{Gl}^2} = \frac{1}{R^2}\left(n - \frac{\Phi}{\Phi_0}\right)$. Taking into account the temperature variation of $\xi_{GL}$ one can show that,

$$\frac{\Delta T_c(H)}{T_c} \propto \frac{\xi_{GL}^2(0)}{R^2}\left(n - \frac{\Phi}{\Phi_0}\right)^2 \qquad (1.25)$$

This variation is shown in *Figure 1.6(b)* which shows that maximum suppression of $T_c$ occurs at $n - \frac{\Phi}{\Phi_0} = \frac{1}{2}$. If $v_s$ is zero for some $\Phi/\Phi_0$, then $|\psi|^2 = \psi_\infty^2$ and $T_c$ is enhanced. Otherwise for any nonzero $v_s$, $T_c$ is suppressed to some $T_c'$, where $T_c > T_c'$.

However in experiments a background (possibly $\propto H^2$) is observed in the $\Delta T_c(H)$ variation, which may have many sources. One of them could be experimental limitation to obtain a cylinder with very thin thickness. Any substantial width would contribute a $H^2/8\pi$ kind of term to the energy which may give rise to such quadratic background. Other sources of this effect is often assigned to misalignment between the field axis and the cylinder.

### 1.3.6 Linearized GL equations and H$_{c2}$

If the magnetic field is such that it has reduced the value of $|\psi|^2 \ll \psi_\infty^2$, we can drop the term in the GL equation with $\beta$ and the GL equation becomes linearized,

$$\left(\frac{\vec{\nabla}}{i} - \frac{2\pi \vec{A}}{\Phi_0}\right)^2 \psi = \frac{\psi}{\xi_{GL}^2(T)} \tag{1.26}$$

Now, for an infinite sample if we take the vector potential to be $A_y = Hx$ then, a field along $z$ axis is obtained and the solution of the linearized GL equation shows that, the magnetic field $H$ has a maxima, also known as upper critical field, given by,

$$H_{c2} = \frac{\Phi_0}{2\pi \xi_{GL}^2(T)} \tag{1.27}$$

### 1.3.7 Abrikosov vortex state

For a bulk sample (type II) there are infinite number of interior solutions of the linearized GL equations at $H_{c2}$ each of the form[20],

$$\psi(x, y) = \sum_{n=-\infty}^{\infty} C_n \exp(inky) \exp\left(-\frac{1}{2}\kappa^2 \left(x - \frac{nk}{\kappa^2}\right)^2\right) \tag{1.28}$$

Here, $\kappa^2 = 1/\xi_{GL}^2$ and the periodicity condition is given by, $C_{n+N} = C_n$. Furthermore, Gibbs free energy density of the system is minimized for minimum value of $\beta_A = \langle \psi^4 \rangle_{avg} / \langle \psi^2 \rangle_{avg}^2$ for different geometries. For a triangular lattice, $\beta_A$ is found to be minimum and the lattice constant is given by,

$$a_\Delta = 1.075 \times \left(\frac{\Phi_0}{B}\right)^{\frac{1}{2}} \tag{1.29}$$

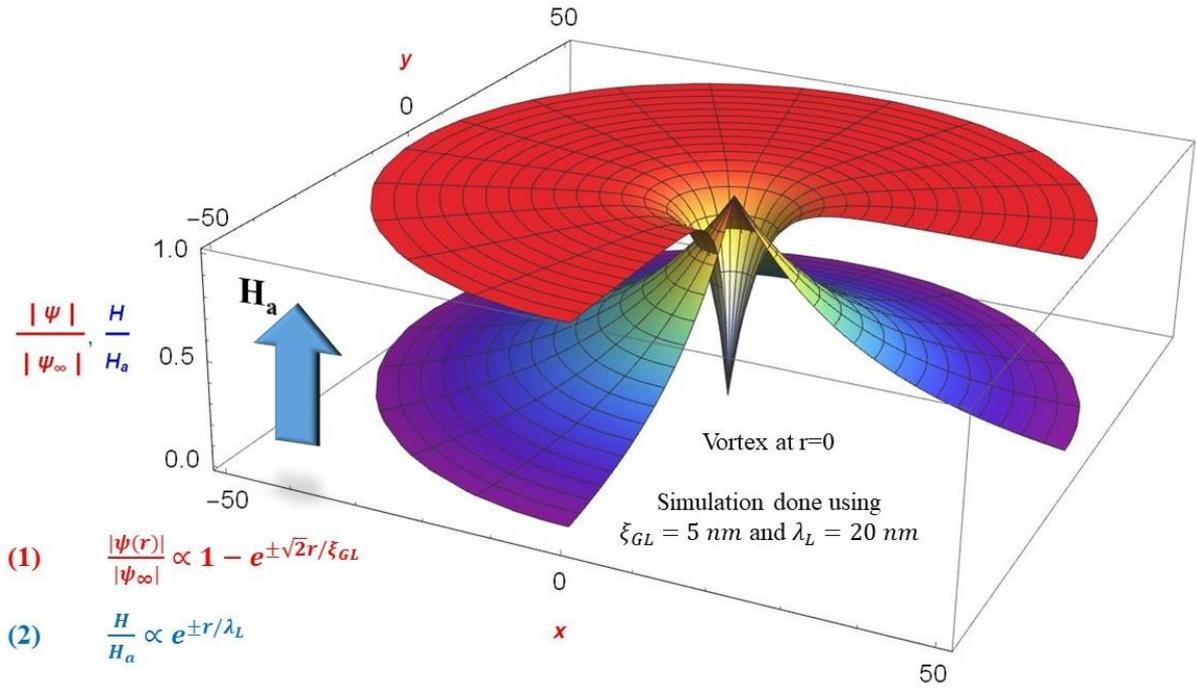

(1) $\frac{|\psi(r)|}{|\psi_\infty|} \propto 1 - e^{\pm\sqrt{2}r/\xi_{GL}}$

(2) $\frac{H}{H_a} \propto e^{\pm r/\lambda_L}$

*Figure 1.7:* 3D plot of variation of $|\psi(r)|/|\psi_\infty|$ is shown in red, which falls of from its value unity to zero at the centre of the vortex core. Similarly 3D variation of $H/H_a$ is shown in blue. The vortex is situated at $r = 0$ and the plots are simulated assuming, $\xi_{GL} = 5\ nm$ and $\lambda_L = 20\ nm$, i.e., $\kappa = \lambda_L/\xi_{GL} = 4 > 1/\sqrt{2}$.

These solutions, the periodic normal regions in the superconductor are known as *vortices* and they carry one flux quanta each and has a circulating supercurrent around the normal region. For a vortex in a material with high $\kappa$, one single vortex looks like *Figure 1.7*, where field falls off inside the superconductor with a lengthscale of $\lambda$, while magnitude of order parameter, $|\psi(r)|$ falls off inside vortex core from $\psi_\infty$ with a lengthscale of $\xi_{GL}$.[21]

## 1.4 Interaction between vortex lines

It can be shown using London equations that for $\kappa \gg 1$, magnetic field due to a vortex is given by,

$$H(r) \to \frac{\Phi_0}{2\pi\lambda^2}\left(\frac{\pi}{2}\frac{\lambda}{r}\right)^{\frac{1}{2}} e^{-\frac{r}{\lambda}} \text{ for } r \to \infty$$
$$H(r) \approx \frac{\Phi_0}{2\pi\lambda^2}\left[\ln\frac{\lambda}{r} + 0.12\right] \text{ for } \xi \ll r \ll \lambda \quad (1.30)$$

This can be used to determine the interaction energy between two vortex lines separated by a distance $r_{12}$. Free energy per unit length for the interaction between these two vortex lines is given by:

$$F_{12} = \frac{\Phi_0^2}{8\pi^2\lambda^2} K_0\left(\frac{r_{12}}{\lambda}\right) \quad (1.31)$$

Here, $K_0$ is the zeroth order Hankel function of imaginary argument. The force of interaction per unit length between these two vortex lines is given by,

$$\vec{f_2} = \vec{J_1}(\vec{r_2}) \times \frac{\widehat{\Phi}_0}{c} \qquad (1.32)$$

And for an arbitrary array of vortices, force on vortex per unit length due to a current density $J_s$ is given by, $\vec{f} = \vec{J_s} \times \widehat{\Phi}_0/c$.

## 1.5 Flux dynamics in type II superconductor

### 1.5.1 Flux flow

If one passes a current $\vec{J}$ through a superconductor, the Lorentz force on a vortex is given by, $\vec{F} = \vec{J} \times \widehat{\Phi}_0/c$. This moves a vortex from its position and this change of flux induces an electric field given by, $\vec{E} = \widehat{\Phi}_0 \times \vec{v}/c$ where $\vec{E} \parallel \vec{J}$. Now, according to Bardeen-Stephen model[22] power dissipated due to this induced voltage comes dominantly from the normal vortex cores moving through the superconductor. The resistance of the superconductor does not remain zero anymore and is characterised by *flux flow resistance*, $R_{ff} = V/I$, where $V$ is the induced voltage due to the current $I$.

Another way to view this power dissipation is to consider a viscous drag force that tends to oppose the flow of vortices due to Lorentz force. This concept is analogous to how power dissipation is visualized in normal conductor. In normal conductors the applied electric field causes the electrons to move but the drag term opposes their motion and hence in equilibrium the normal conductor has a constant current density. Similar to these electrons in normal conductors, the vortices too move with a constant velocity when driven by a constant dc current. If the viscous drag coefficient is $\eta$, then viscous force per unit length of vortex line moving with velocity $v$ is $-\eta v$. In an equilibrium state this viscous force must be countered by the force on each vortex, $J\Phi_0/c$.

Since, $\vec{E} = \vec{B} \times \vec{v}/c$ and $J\Phi_0/c = \eta v$ we get, the flux flow resistivity,

$$\rho_{ff} = \frac{E}{J} = B\frac{\Phi_0}{\eta c^2} \qquad (1.33)$$

### 1.5.2 Flux pinning

Therefore, in principle superconductivity cannot exist in presence of a drive current because of flow of vortices. But what prevents the flow of vortices even when a current is passed, is the presence of "pinning centres", which arises due to inevitable atomic vacancies or defects or impurities or boundaries even in the cleanest of superconducting materials.

In principle, when a flux line passes through a superconductor, locally a region of radius $\sim \xi_{GL}$ becomes normal metal. Hence, for a sample of thickness $t$, the free energy increment to create a vortex will be, $\sim \frac{1}{2} N(0) \Delta^2 \xi_{GL}^2 t$, where $N(0)$ is density of states at Fermi energy and $\Delta$ is BCS gap. However, if the material has a region of size of the order of $\xi_{GL}$, where superconducting order parameter is already suppressed, then it will be energetically favourable to create a vortex in that region, since passing the flux line through that region would cost less energy. These regions are called pinning centres. Now, because the vortices form a lattice in a superconductor, and the lattice is rigid (up to energetically favourable distortions by a lengthscale determined from the equilibrium between pinning energy and elastic energy of the flux tubes; described as *collective pinning theory* by Larkin and Ovchinikov[23]), if one vortex is pinned, it tends to pin other vortices near to it. Likewise, when one pinned vortex moves, it tends to move a bundle of neighbouring vortices with it. However the exact form of average pinning force depends on the size and distribution of the pinning sites as well as the rigidity of the lattice.

Passing a current is countered by these pinning centres up to the point when the Lorentz force on the vortices is larger than the pinning force. In this 'flux flow regime' when the vortices start to flow by application of a drive current naturally gives rise to flux flow resistance, which is given by,

$$R_{ff} = \frac{V}{I - I_c} \tag{1.34}$$

, where $I_c$ is the critical current (also called *depinning current*, in contrast with *depairing current* which is associated with pair breaking energyscale) often determined from linear fits to the $I - V$ characteristics in the flux flow regime. This $R_{ff}$ is the same flux flow resistance as to when no pinning was considered; pinning only moves the zero of voltage to the critical current value.

Quantification of the pinning strength comes from the average restoring force on an individual vortex (per unit length) due to the pinning potential when subjected to a small displacement from its equilibrium position, which is given by, $\alpha_L = \left(\frac{B}{4\pi}\right) \left|\frac{dH}{dx}\right| = B\Phi_0/\mu_0 \lambda_L^2$. This $\alpha_L$ is called Labusch parameter.[24]

### 1.5.3 Presence of ac perturbative magnetic field

It was shown by Campbell[25] that in presence of ac perturbative magnetic field in presence of vortices, the ac penetration depth is different from the London penetration depth, $\lambda_L$. This Campbell penetration depth is given by, $\lambda_C^2 = B^2/\mu_0 \alpha_L$. It was later found out by Brandt[26,27] that the effective penetration depth due to this ac perturbation in presence of vortices follows, $\lambda_{ac}^2 = \lambda_L^2 + \lambda_C^2$. This ac penetration depth can in principle be a complex number and can have the form,

$$\lambda_{ac}^2 = \lambda_L^2 + \left(\frac{B\Phi_0}{\mu_0}\right)(\alpha_L + i\omega\eta)^{-1} \tag{1.35}$$

In presence of ac perturbation in an array of vortices, the viscous coefficient becomes, $\eta = \Phi_0 B/\mu_0 \omega \delta^2$, where $\omega$ is frequency of the ac perturbation and $\delta$ is imaginary part of the ac penetration depth, also known as skin depth. From the expression 1.35, it is apparent that there could be two regimes of vortex response depending on the frequency of the ac perturbative magnetic field. A characteristic frequency demarcating these two regimes, is given by, $\omega_0 = \alpha_L/\eta$, called the depinning frequency, which generally is in the microwave regime. Above $\omega_0$, the vortices do not feel the restoring force on them as they fluctuate at the very bottom of the pinning potential and are dominated by the viscosity term.

### 1.5.4 Thermally Activated Flux Flow (*TAFF*)

Anderson and Kim[28,29] showed that at finite temperatures flux lines can jump from one pinning point to another jumping over the pinning barrier $U$ in response to the driving force of the current and flux-density gradient. The resulting flux creep is revealed in two ways, first it leads to measurable resistive voltages and secondly slow changes in trapped magnetic fields. While in absence of the current, the flux line has equal probability to go either side of the pinning potential but in presence of transport current or flux-density gradient, the tilted potential gives precedence of the downhill creep over uphill. The small resistance due to this thermally activated flux flow (*TAFF*) is expressed as,

$$R_{TAFF} = R_{ff} \exp\left(-\frac{U(I)}{k_B T}\right) \tag{1.36}$$

### 1.5.5 Vortex lattice melting

Apart from a current driven flux flow regime where the vortices overcome the pinning barrier, the vortices can be distinguished into solid and liquid depending on their behaviour at vanishingly small current drive. The vortex solid phase, characterised by its long range positional and orientational order, is also restrained from thermally activated flux flow in the limit $I \to 0$. It was shown[30,31,32] that for a vortex solid the pinning barrier, $U(I)$ depends on the current as, $U(I) = U_0 \times (I_c/I)^\alpha$, where $\alpha$ is a constant close to unity, such that $R_{TAFF} \to 0$ for $I \to 0$. On the other hand a vortex liquid phase, which has short range positional and orientational order can hardly resist even an infinitesimally small drive current. For a vortex liquid[33], $U$ is independent of the current, which makes $R_{TAFF}$ independent of current for $I \to 0$.

Although the existence of a true vortex liquid has been a topic of extensive research, it is widely accepted that thermal and/or quantum fluctuations of the flux lines, result in a measurable change in the vortex lattice. The most important effect of these fluctuations is possibility of melting of the vortex lattice[34,35] at temperatures well below $T_c$ or at magnetic fields well below $H_{c2}$. The resulting vortex-liquid phase may occupy a significant part of the $H - T$ parameter space[6]. In addition to this, existence of pinning in the system gives rise to various possible vortex states. For example in presence of quenched disorder the vortex lattice can become a vortex glass[32,36] or in very weakly pinned superconductor, a two-dimensional vortex lattice can melt via two continuous phase transition.

### 1.5.5.1 BKT transition

Whenever melting of a two-dimensional vortex lattice is discussed, the possibility of a BKT transition is considered. Berezinskii,[37] Kosterlitz and Thouless[38] showed for a system of spins in XY model that, a low temperature quasi-long-range order can exist in the spin system which can melt into a disordered spin system with short-range correlation due to formation of vortices of the spins. For a system of size $N$ the energy associated with a single vortex (similar to a domain wall in spin chain) can be shown to be $\sim \pi J \ln N$ while the entropy is $\sim k_B \ln N$. The Helmholtz free energy of the system is, $F = E - TS \approx (\pi J - k_B T) \ln N$. Therefore, vortices, which are the topological defects in XY model drives the system across an order-disorder transition at $T_0 = \pi J/k_B$. Hence at $T < T_0$ the spin system will be in an ordered state and at $T > T_0$ the spin system becomes disordered. This order-disorder transition is famously called the BKT transition.

### 1.5.5.2 BKTHNY melting

As an extension to Berezinskii, Kosterlitz and Thouless's work in order-disorder transition of spins in XY model mediated via topological defects, Halperin and Nelson[39] and Young[40] showed that this concept of melting can also be applied to 2-dimensional materials which have long range interaction similar to XY spin system. This BKTHNY melting shows that a solid, which is characterised by its long range positional as well as orientational order, melts into a hexatic liquid (if we are dealing with a triangular lattice, like vortex lattice) which retains only a quasi-long range orientational order. Hexatic fluid has zero shear modulus and short-range positional order. This hexatic liquid is characterised also by randomly appearing topological defects, called *dislocations* and also its directional flow along one of the principal axes. Next the dislocation pairs dissociate into *disclinations* which destroys the orientational order too, giving rise to a transition of the hexatic fluid into an isotropic liquid. This is shown in *Figure 8* as schematics.

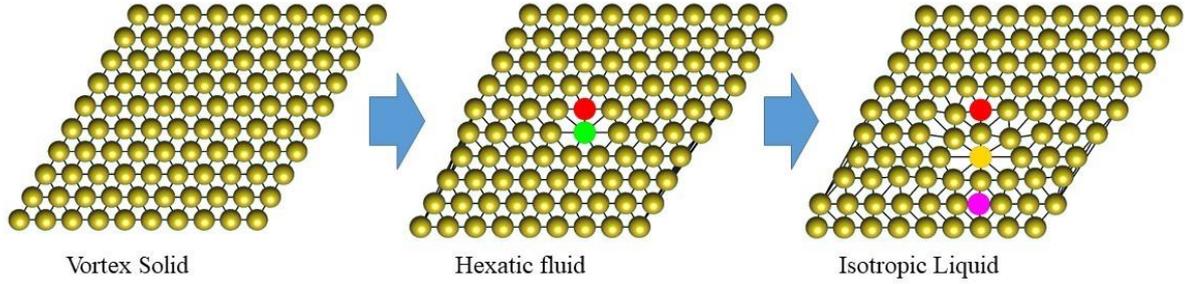

*Figure 1.8:* Schematics of a BKTHNY two-step melting is shown for a hexagonal lattice of golden balls. The vortex solid is free from any topological defect, which undergoes a continuous phase transition to hexatic fluid. The topological defects in hexatic fluid, namely *dislocation* is the pair of 5- and 7-fold, red and green balls. There could also be pairs of 4- and 8-fold, magenta and yellow balls. This hexatic fluid melts into an isotropic liquid, which develops isolated 5-, 7-, 4-, 8-fold topological defect, named as *disclinations.*

Since vortices have a long range logarithmic interaction, the vortex lattice forms one of many potential candidates for observing melting.[41,42,43,44] Other candidates include electrons on the surface of superfluid helium,[45] inert gas monolayers adsorbed on graphite, colloidal crystals[46,47,48,49] etc. According to various experimental conditions either the occurrence of the melting transition or existence of orientational order in absence of translational one has been proved. But simultaneous observation of the two features is only available in colloidal crystals. From that viewpoint, vortex lattice is a very promising candidate where the experimental conditions such as thickness of the superconductor, pinning geometry and strength of the superconductor on the vortices etc. can be controlled with extraordinary precision. However the two step BKTHNY melting in vortex lattice has been very elusive and is only observed recently.[50]

## 1.6   Background on disordered superconductivity

Superconductivity in presence of disorder in recent times has garnered renewed attention due to observation of novel phenomena in the vicinity of the critical disorder for destruction of superconductivity. Most of the works on strongly disordered superconductors are on 2D systems where at critical disorder the superconducting ground state transforms into an insulating state. This phase transition is usually referred to as the superconductor-insulator transition. The destruction of superconductivity can be thickness controlled,[51,52] magnetic-field controlled[53] or disorder controlled.[54] The quantification of disorder in NbN thin films is done using the Ioffe Regel parameter,[55] $k_F l$, where $k_F$ is the Fermi wavevector and $l$ is mean free path of the electrons. In NbN thin films, this quantity decreases with increasing disorder.

### 1.6.1 Phase stiffness of GL order parameter

The superconducting order parameter $|\psi|e^{i\phi}$ has two parts, the amplitude $|\psi|$ and the phase $\phi$. This phase part is same for all of the superconductor, i.e. superconductor is a phase coherent macroscopic state (this can be shown from BCS theory that the BCS wavefunction has a single phase). One can destroy superconductivity by simply making the amplitude part zero, otherwise, if the phase is twisted in different parts of the superconductor, the phase coherence is lost, hence destroying superconductivity. There are two routes one can attain phase incoherence in a superconductor, firstly due to quantum fluctuation and the other due to thermal fluctuations.

The quantum route of phase fluctuation involves the number-phase uncertainty relation. If the number of particles involved in the process is $N$, then accuracy in their phase can be determined using the uncertainty relation, $\Delta N \Delta \phi \geq 1$.[12] Consider that the superconductor gets divided into regions with linear dimension of the order of $\xi_{GL}$, which are coupled to each other. If fluctuation in $N$ between these neighbouring regions causes charge pile-up at some places in the superconductor, then the phase fluctuation will start to dominate. However in 3D superconductors in clean limit, this fluctuation is screened very efficiently making the phase fluctuation insignificant. But if the disorder is increased or one goes to 2D limit, the charge screening goes down and as a result the phase fluctuation dominates.[7]

The thermal route is followed when thermal energy is enough to twist the phase of the superconductor and gives rise to destruction of the phase coherent state. The characteristic energy-scale associated with this process is called superfluid stiffness, given by,

$$J_S = \frac{\hbar^2 a n_s}{4m} \quad (1.37)$$

Here, $n_s$ is superfluid density given by $n_s = m/\mu_0 \lambda_L^2 e^2$. $a = \min(\xi_{GL}, d)$ is taken to be the minimum lengthscale of the problem. In principle, one can attain a phase incoherent state, if $k_B T \sim J_S$, and consequently destroy zero resistance state.

### 1.6.2 Superconductor to insulator-like transitions

For clean conventional superconductor, $J_S$ is found be very large compared to the other energy scale of the problem, *viz.* pairing amplitude, $\Delta$. In this case, $T_c$ is determined by $\Delta$ and $J_S$ remains insignificant. However, in presence of disorder, or low dimensionality, $J_S$ crashes down and can become comparable to $\Delta$, can be even less depending on the strength of disorder. Here $J_S$ determines $T_c$. This scenario destroys superconductivity by phase fluctuation, but the pairing amplitude (or in other words the amplitude part of order parameter, $|\psi|$) still remains finite.[56,57,58,59,60,61] This regime in $H - T$ parameter space is called the *Pseudogap regime*. The destruction of superconductivity in strong disorder can give rise to an insulating phase above a specific magnetic field, $H_{SIT}$. This phenomenon is often referred to as the superconductor-insulator transition (SIT). If there are Cooper pairs in the insulating phase it is called a *bosonic* SIT.

The other path through which one can obtain SIT is via quasiparticle fluctuation affecting the amplitude $|\psi|$ part of order parameter, which correlates to the *fermionic* SIT.[62] Here disorder-enhanced Coulomb interaction breaks Cooper pairs into fermionic states leading to superconductor-(bad) metal transition (SMT) with $\Delta \to 0$ and $T_c \to 0$. At even higher disorder metal-insulator transition (MIT) follows due to Anderson localization.[63]

# Chapter II: Details of experimental techniques and analysis

The procedure to characterize a superconducting thin film starts with sample deposition and subsequent bulk measurements and STS/M measurements.

## 2.1 Sample deposition techniques

Although there are several sample deposition technique like DC and RF magnetron sputtering, pulsed laser deposition (PLD), evaporation, molecular beam epitaxy (MBE), and several other chemical deposition techniques, I have used in this thesis DC magnetron sputtering and PLD to deposit the superconducting thin films of NbN and $a$-MoGe. NbN is an s-wave superconductor with $T_c \sim 16\ K$ in the clean limit. One can change the deposition conditions to create structural disorder, for example Nb vacancies which can reduce the $T_c$. Similarly, in its bulk limit $a$-MoGe has a $T_c \sim 8\ K$. Here we have used thickness as a parameter to control the disorder in the system.

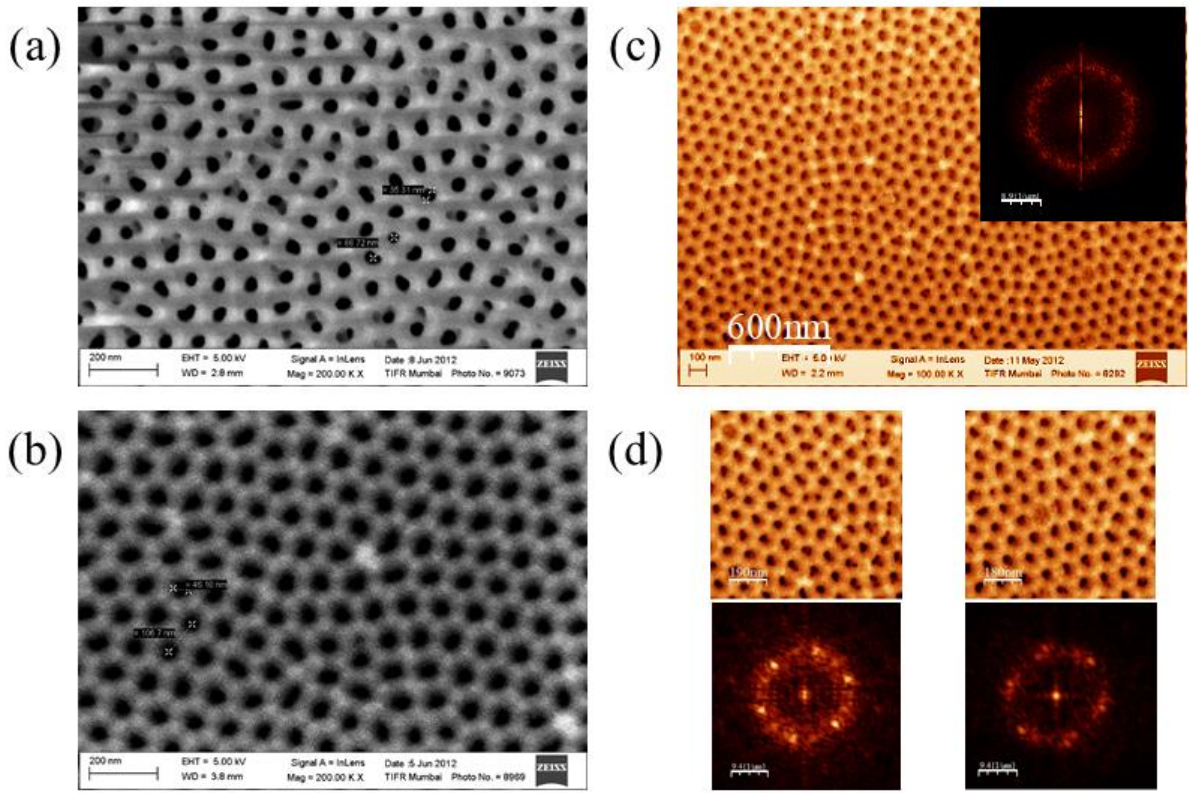

*Figure 2.1:* Anodic alumina template (AAM) (a) Scanning electron microscope (SEM) image of AAM before deposition. (b) SEM image of AAM after deposition. (c) SEM image of AAM after deposition of a larger area, FFT of which shows an isotropic ring conforming to a disordered lattice. (d) SEM of AAM after deposition of smaller areas, both of which locally show six-spots in FFT, but are disoriented with each other.

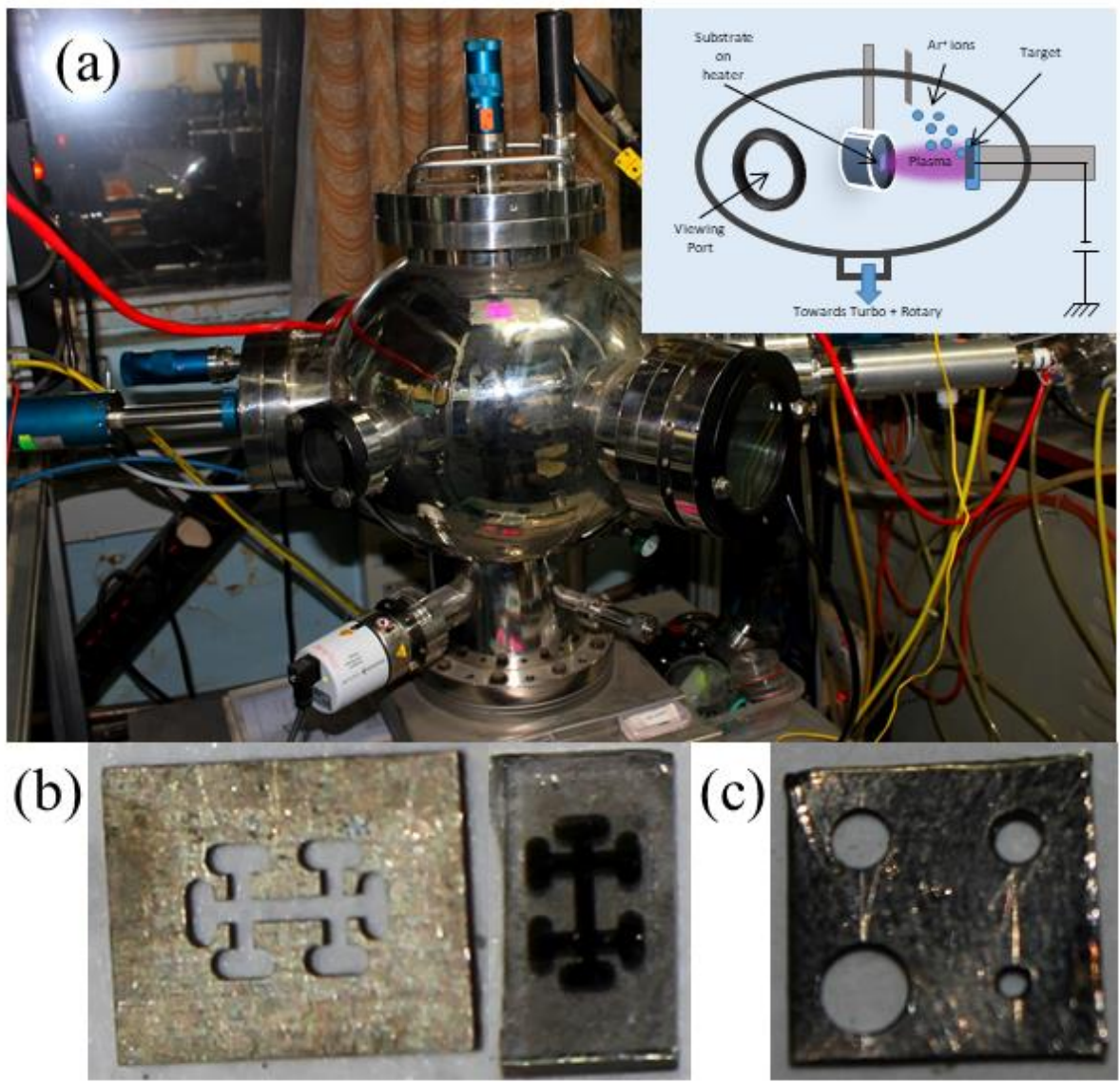

*Figure 2.2:* DC magnetron sputtering (a) Photograph of a sputtering chamber with its schematic diagram in the *inset*. The schematic is a simplified model to explain working principle. (b) (*left*) Stainless steel (SS) mask for deposition of thin film with Hall bar geometry. (*right*) NbN sample with Hall bar geometry, used for thickness and transport measurement. (c) SS mask for deposition of circular film on AAM, of diameters 1, 2, 3 and 4 mm. Among these the 3 mm diameter NbN thin film on AAM was used in Chapter 3.

### 2.1.1 Anodic alumina membrane (AAM)

For deposition of the NbN thin film for the study of *effect of periodic pinning centres* (Chapter 3), the substrate is taken to be anodic alumina membrane (AAM),[1] commercially bought from *Synkera Technologies Inc.*[2] The anodizing of the alumina creates pores which self-organize themselves in hexagonal lattice.[3,4] The AAM used in this study have pores of diameter $\sim 35\ nm$ and lattice constant $\sim 85\ nm$. After deposition of the NbN thin films, SEM image shows a pore diameter of $\sim 46\ nm$ and lattice constant $\sim 106\ nm$. SEM images in these two cases are shown in *Figure 2.1*, along with fast Fourier transform (FFT) of the lattice. The FFT shows that although locally the lattice has hexagonal order, apparent from the six spots in Fourier space, globally the orientational order is absent, visible from the ring-like structure in Fourier space. These pores in the AAM, as a whole are called antidot array.

### 2.1.2 DC magnetron sputtering

In this process the target material is bombarded with Argon ions and this transfers momentum to the target atoms, which are ablated and subsequently form a plasma. This plasma gets deposited on a substrate. There are also provisions of introducing a reactive gas which can react with the plasma to create compounds which then gets deposited on the substrate. The target is maintained at negative potential to accelerate the Ar ions towards the target. A dc power source is used in case of metallic targets, like Nb to create the flow of plasma (in contrast with rf sputtering where ac power source is used to avoid charge build-up on the target material). Neodymium magnets kept behind the target creates field lines such that the ions can be trapped and directed towards the target more efficiently. This in turn increases effective ionization and hence higher sputter yield, which makes it possible to carry out the deposition at a much lower pressure. Sputtering in general is a very popular method for commercial usages for its high reliability and control.

The dc magnetron sputtering system was used to grow NbN thin films on a substrate of AAM for the study of periodic pinning centres. The setup, shown in *Figure 2.2a* is fabricated by *Excel instruments*[5] and the target is 99.999% pure two-inch Nb-target manufactured by Kurt-Leskar. For standardization purposes the film is deposited on MgO single crystal with (100) orientation. MgO substrates are cleaned with tri-chloroethylene (TCE) via ultrasonication and vapour cleaning. In case of AAM substrates, they are only cleaned using TCE vapour. The sputtering chamber is evacuated using rotary-backed turbo to $\sim 10^{-6}\ mbar$ before deposition starts. During deposition, the substrate is maintained at a constant temperature of $600°\ C$. The dc power supply from Aplab is used to keep the sputter power in between 30-230 W, depending upon the $T_c$ requirement. The flow of Ar and N$_2$ were controlled by flow meters and were kept at 74 sccm and 11 sccm respectively after a long trial and error check for best $T_c$. During deposition the pressure inside the chamber is usually kept at $5 \times 10^{-3}\ mbar$.

For thickness calibration and possible transport measurements, a SS mask with Hall bar geometry is placed on top of the substrate, shown in *Figure 2.2b*. For the deposition of NbN films on AAM with circular geometry, specially fabricated SS mask, shown in *Figure 2.2c* with different diameters of openings was used. Among these the sample with 3 mm diameter was used for the full study.

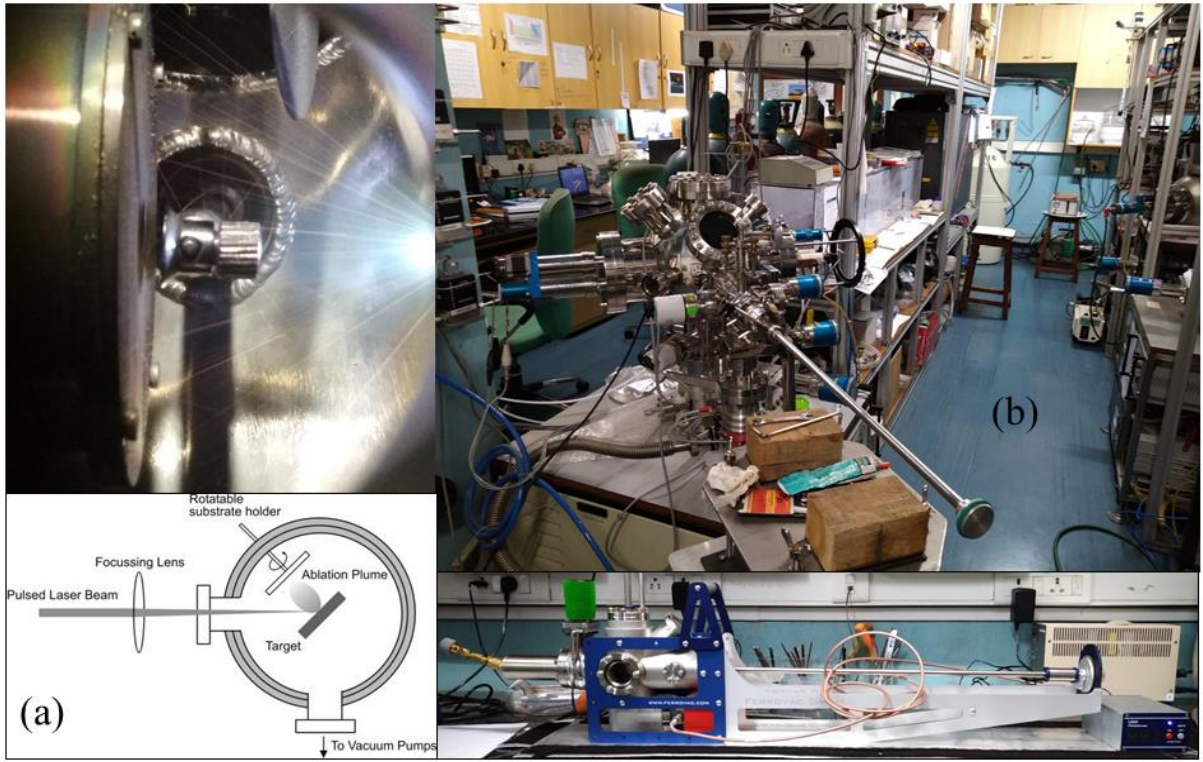

*Figure 2.3:* Pulsed laser deposition (PLD) and vacuum suitcase (a) Schematic of PLD setup; (*top*) Bright plume of *a*-MoGe ablated by laser and getting deposited on the sample holder placed in front of it. (b) Photograph of PLD setup; (*bottom*) Photograph of vacuum suitcase.

The thickness of the deposited sample is measured using Ambios XP2 Stylus profilometer, where the resolution is $\sim 10 \, nm$. Hence, films with different thicknesses ($> 10 \, nm$) are made by varying deposition time to generate a thickness vs. deposition time curve to extrapolate and find film thicknesses $< 10 \, nm$.

### 2.1.3 Pulsed laser deposition

Pulsed laser deposition is a very effective tool in vapour deposition technique which is used to grow systems with multilayers and complex stoichiometric ratio. A focused high energy laser beam is used to ablate the target material which creates the plasma which then condenses on the substrate placed in front of it, as is shown in *Figure 2.3a*. In high vacuum, the plasma plume takes a shape often described by a $\cos^n x$ form. In our depositions we have used a $248 \, nm$ KrF excimer laser in pulsed mode.

For the growth of NbN,[6,7] the chamber is evacuated to $\sim 10^{-6} \, mbar$ and the MgO (100) substrate is kept at $700° \, C$. Introduction of nitrogen is controlled by a mass flow meter and pressure reaches $\sim 5 \times 10^{-3}\ mbar$ range in presence of N$_2$. The 1 inch 99.999% pure Nb target is ablated using the laser, which is operated at constant energy mode to deliver an energy density of $\sim 200\ mJ/mm^2$ at a repetition rate of $10\ Hz$ for clean NbN samples. To deposit the disordered NbN thin films, flow rate of $N_2$ was changed, which is believed to create atomic vacancies in the NbN crystal. However, the repetition rate for this deposition was kept at $4\ Hz$ to obtain a smooth surface. AFM and SEM images of a typical NbN thin film grown using PLD are shown in *Figure 2.4a* and *Figure 2.4b*.

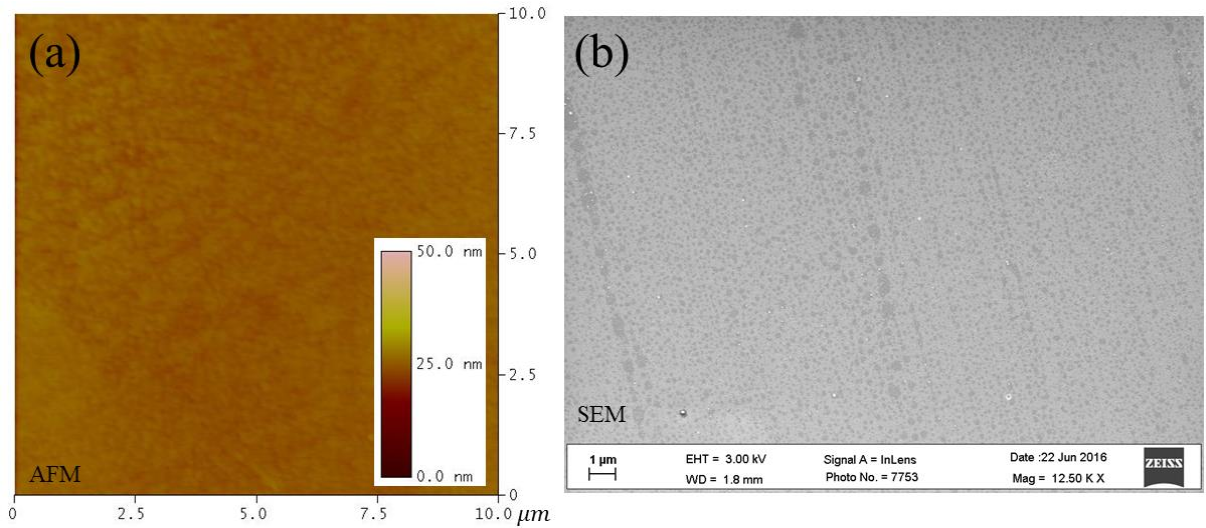

*Figure 2.4:* NbN thin film (a) Atomic force microscope (AFM) image of NbN thin film grown using PLD; the surface is very smooth and free from any particulates. (b) SEM image of the same NbN thin film.

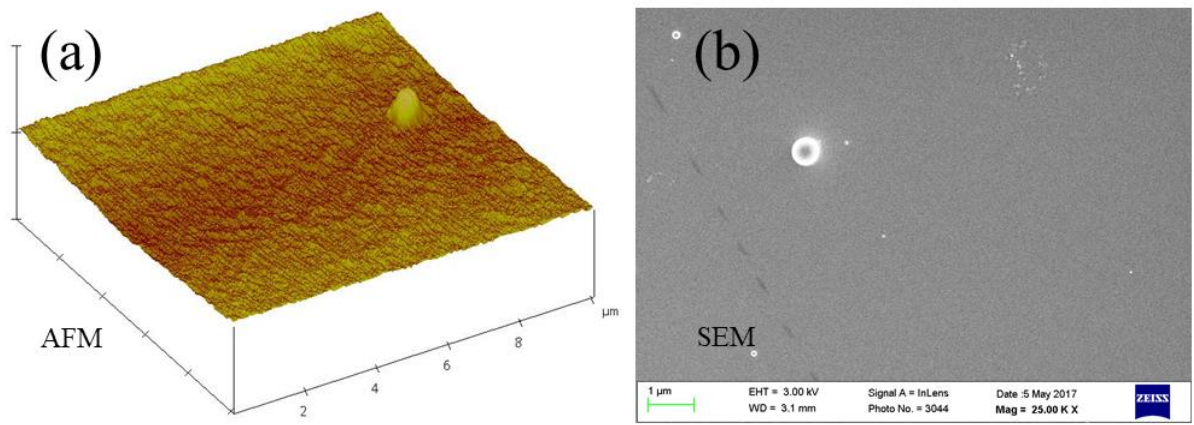

*Figure 2.5:* a-MoGe thin film (a) AFM image of a-MoGe thin film grown using PLD; the surface is very smooth except for occasional particulate. (b) SEM image of the same a-MoGe thin film, showing a particulate.

For the growth of $a$-Mo$_{70}$Ge$_{30}$,[8] surface oxidised Silicon wafer (Si/SiO$_2$) is used as substrate. The target used in the ablation is arc melted pellet of MoGe of the same stoichiometry. The target is ablated at a base pressure of $\sim 1 \times 10^{-6}\ mbar$ at room temperature and to keep the stoichiometric ratio constant, the energy density of laser is increased to 240 $mJ/mm^2$ at a repetition rate of 10 $Hz$. AFM and SEM images of the films, in *Figure 2.5a* and *Figure 5b* show a very smooth surface with rarely occurring particulates.

### 2.1.4 Sample transfer: Vacuum suitcase

For studies involving scanning tunnelling spectroscopy/microscopy (STS/S) on the thin films, the surface of the film should be very clean. We have used a vacuum suitcase manufactured by *Ferrovac GmbH*[9] which can maintain a base pressure $\sim 10^{-10}\ mbar$ inside the suitcase using a battery operated ion pump and a getter material for adsorption of nonreactive gases. We connect this vacuum suitcase to PLD cross and deposit our sample using specially built sample holder for STS/M and transfer the holder to STS/M without ever exposing the sample to atmosphere. *Figure 2.3b* shows the vacuum suitcase.

## 2.2 Bulk measurement techniques

There are two bulk measurements, low-frequency ac susceptibility and four-probe resistivity measurements which were used to conduct mutual inductance and transport measurements respectively.

### 2.2.1 Two coil mutual inductance measurement

The two coil mutual inductance measurement is an ingenious technique to measure low frequency ac susceptibility. Here, the primary quadrupolar coil has 28 turns, 14 turns on each direction and the secondary dipolar coil has a total 120 turns wound in four layers. The thin film is sandwiched between these two coils where the film faces the secondary coil as is shown *Figure 2.6a*. We send an alternating current, $I_d$ of frequency, $f = \frac{\omega}{2\pi} = 31\ kHz$ through the primary coil and use two lock-in amplifiers (Stanford Research Systems SRS 830) to record the induced voltage, $V_p$ in the secondary coil. The ac in the primary is generated by a voltage to current amplifier as is shown in *Figure 2.6b*. Mutual inductance is given by,

$$M = M' + iM'' = \frac{V_p}{\omega I_d} \qquad (2.1)$$

This two coil mutual inductance measurement is used to calculate London penetration depth, $\lambda_L$. By solving coupled Maxwell's equation and London equation for the geometry for our two coil apparatus we can in general obtain supercurrent density, $J_s^{ac}$ as a function if radial distance from the centre of the sample ($r$). The self-consistent equations used in this method are:

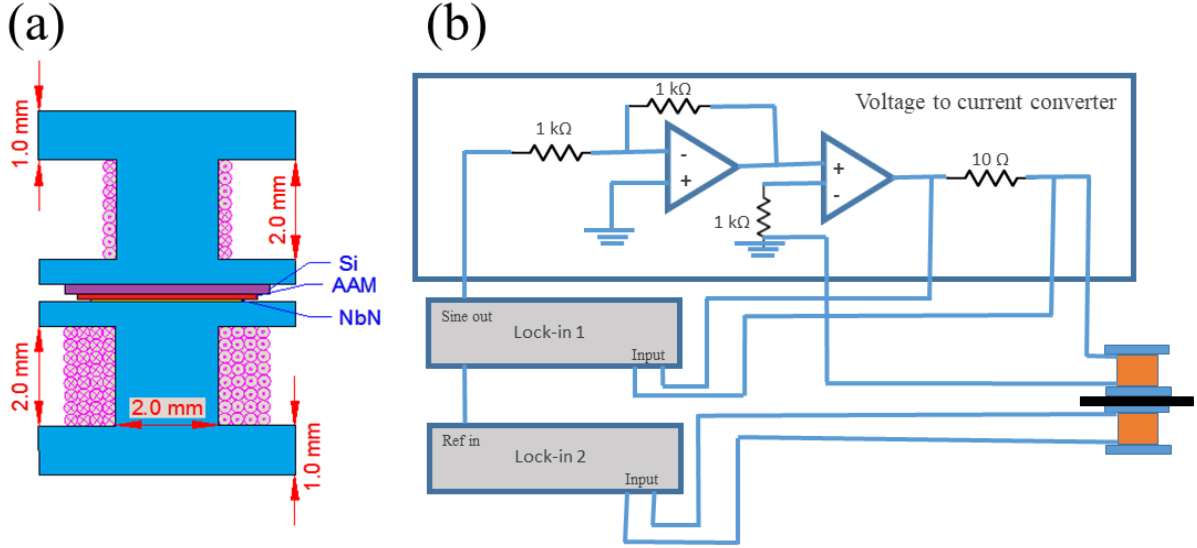

*Figure 2.6:* Two-coil mutual inductance setup (a) Schematic of the 2-coil setup; the top coil is quadrupolar and the bottom one is dipolar and the sample is sandwiched between these two. For quantitative calculation purpose the sample is placed face down in between the two coils. (b) Schematic diagram of the electronics connected with the 2-coil setup, which utilizes two lock-in amplifiers and a voltage to current converter.

$$\vec{A}_{film}(\vec{r}) = \vec{A}_{drive}(\vec{r}) + \frac{\mu_0}{4\pi} \int d^3\vec{r}' \frac{\vec{J}_s^{ac}(\vec{r}')}{|\vec{r}-\vec{r}'|} \qquad (2.2)$$

$$\vec{J}_s^{ac}(\vec{r}) = -\frac{\vec{A}_{film}(\vec{r})}{\mu_0 \lambda_L^2} \qquad (2.3)$$

Here the vector potential on the film, $\vec{A}_{film}(\vec{r})$ is a combination of the vector potential resulting from the current passing in the drive coil and the induced supercurrent density in the drive coil. These equations can be used to find theoretical value of mutual inductance, following:

$$M = \frac{1}{I_d} \sum_{j}^{n_p} \oint \vec{A}_j \cdot d\vec{l} \qquad (2.4)$$

This theoretical value of mutual inductance is compared with the experimental value to find out the unknown of the problem, $\lambda_L$. The scheme for numerically solving these equations for a film with circular geometry is identical to that established previously.[10,11,12,13]

### 2.2.2 Four probe resistivity measurement

The resistivity measurements are done using regular four probe geometry using either a resistivity insert in a He$^4$ VTI (*Figure 2.7*) or a He$^3$ resistivity VTI made by *Janis Research*,[14] which have rotatable sample mounting facility. The He$^3$ VTI can go down to 280 mK using charcoal pumping over liquid He[3]. We make four contacts for two current and two voltage probes on the Hall bar geometry, shown in *Figure 2.2b*. For NbN samples, the contacts are made using Ag-In solder which connect with four points in a PCB via copper wires. For *a*-MoGe thin films we use two component silver epoxy[15] to bind gold wires with secondary contact pads made on glass plate, which is then connected to the PCB via copper wires. For the resistivity measurements, the current was sourced from a Keithley 6220 precision current source and the voltage was measured using a Keithley nanovoltmeter.

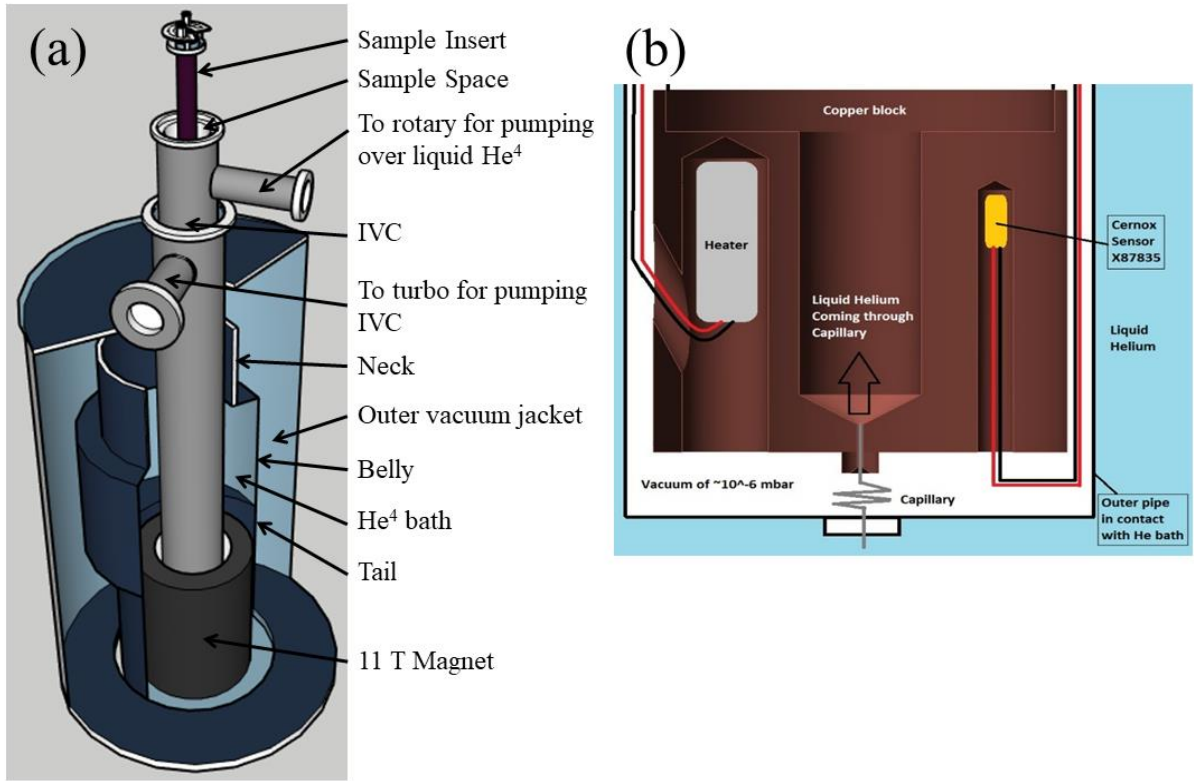

*Figure 2.7:* He[4] VTI and the 110 kOe cryostat (a) Schematic diagram of the 110 kOe cryostat along with the He[4] variable temperature insert (VTI). This VTI is specially designed in which a sample insert can be loaded. (b) Bottom part of the VTI, consisting of a cooper block which has a capillary to flow liquid helium in the sample space. The flow of liquid helium is controlled by the heater which is placed in the copper block.

### 2.2.3 Cryostat for bulk measurements

Both of the mutual inductance and resistivity measurements were done in He[4] cryostat fitted with a superconducting solenoid which can generate up to 110 kOe magnetic field. The current for the magnet is drawn from *Cryogenics* magnet power supply. The He[4] VTI is used for both mutual inductance and resistivity measurements down to 1.4 K using rotary pumping over liquid helium drawn into sample space via a SS capillary. This flow of liquid helium through the capillary is controlled by an adjacent heater. Otherwise for both apparatus temperature is monitored and controlled by PID using *Lakeshore 340* temperature controller.

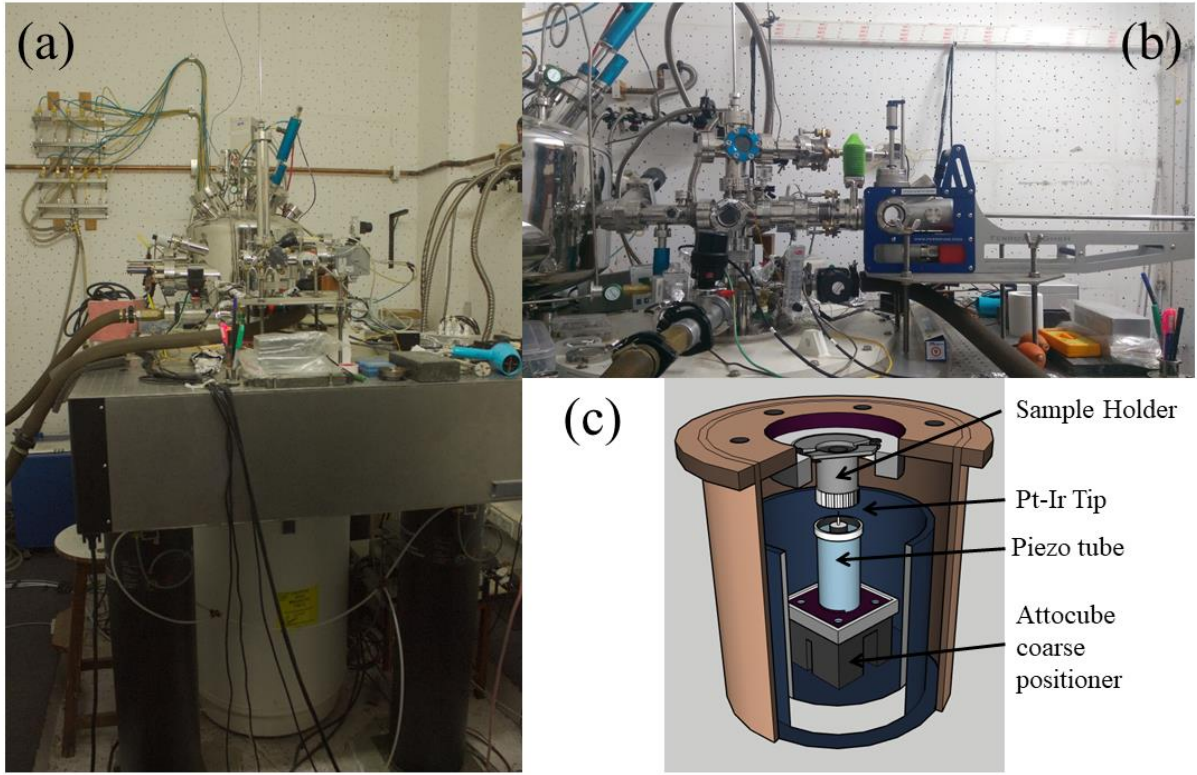

*Figure 2.8:* He³ STM (a) Photograph of the scanning tunnelling microscope (STM) cryostat, preparation chamber, vibration isolator table. (b) Photograph of the vacuum suitcase connected to the STM cross. (c) Schematics of the STM head, where the Pt-Ir tip is mounted on the piezoelectric tube, which in turn is mounted on top of attocube coarse positioner. The Molybdenum-made sample holder is facing the tip.

## 2.3  Scanning tunnelling spectroscopy/microscopy (*STS/S*)

All STS/M measurements on the NbN and *a*-MoGe thin films and single crystals of NbSe$_2$ and Nb$_x$Bi$_2$Se$_3$ are done using home-built He3 STM[16] which can operate down to 350 mK and is fitted with a superconducting magnet of 90 kOe. Attached to the STS/M cross there is a sample preparation chamber, fitted with dc magnetron sputter gun, Ar ion milling and newly installed *Ferrovac GmbH* ion pump. The controller for the STS/M is *Rev 9* from *RHK Tech*.[17] which can be used for both data acquisition and basic analysis. We use a Pt-Ir tip to scan the sample, which is mounted on a molybdenum sample holder, which is shown in *Figure 2.8c*. The tip is mounted on a piezoelectric tube, which can be controlled in $X\pm, Y\pm$ and Z directions. The piezo-tube sits on a coarse positioner, controlled by the sixth channel of *Attocube controller ANC300*,[18] operated at 30 V and 250 Hz mode utilizing the raising part of the pulse. For vibration isolation the cryostat floats on four air-cushion legs, which is further damped by piezo controlled active vibration isolator from *Newport Smart Table*.[19]

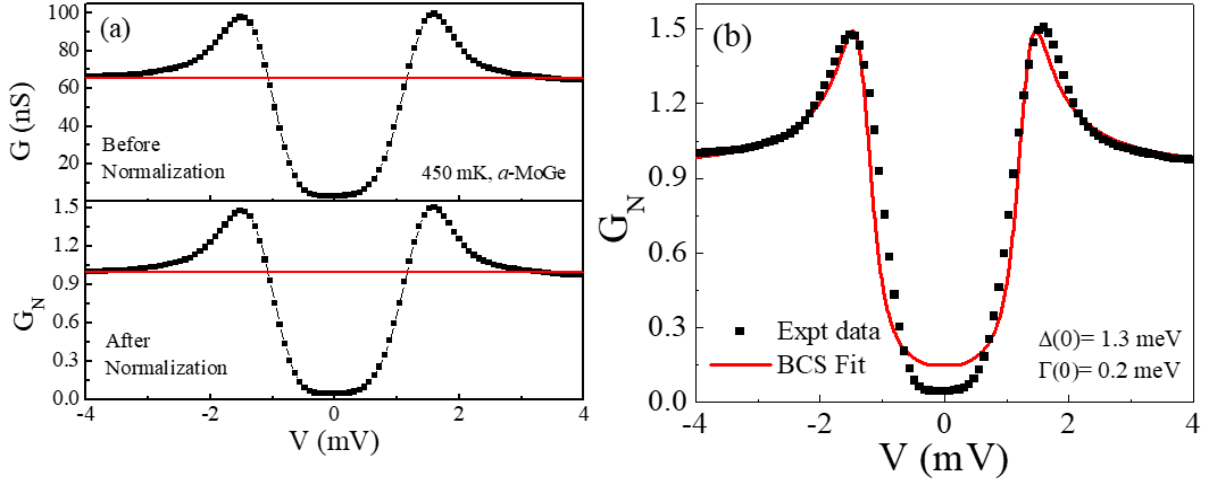

*Figure 2.9:* Bias spectroscopy (a) (*top*) The black connected squares are conductance ($G = dI/dV$) in units of nS vs. bias in mV for *a*-MoGe thin film at 450 mK. The red line represents value of $G$ at $V = 4\ mV$ with respect to which the full spectra will be normalized. This red line can also represent flat spectra as is expected for a normal metal. (*bottom*) Black connected squares represent normalized spectra $G_N(V)$ and the red line now is 1. (b) The black squares are experimentally obtained $G_N(V)$ of *a*-MoGe at 450 mK. The red line is a trial BCS fit of the spectra using $\Delta(0) = 1.3\ meV$ and $\Gamma(0) = 0.2\ meV$. Although there is a good match between the two spectra near coherence peak, the spectra near zero bias have not matched at all.

The operation of STS/M is based on the tunnelling phenomenon explained in section 1.2.5. From an instrument point of view, the tunnelling current is in pico-nano-ampere range, which is pre-amplified by a factor of $10^9$ and then using an analog to digital converter to be acquired by the computer. Depending on the set current for the measurement a feedback loop controls the distance between the tip and the surface of the sample. This *constant current* mode of operation allows scanning over a rough surface of sample. There are some specific parameters for different purposes, which are set on a trial and error basis. For example for topography imaging the bias is set at 10 mV and set current is 50 pA.

### 2.3.1 Bias spectroscopy

Apart from topography the most interesting data one can acquire using STS/M is bias spectra, i.e., differential conductance ($dI/dV$ or $G_{ns}$) as a function of bias. We add a $dV = 150\ \mu V$ and $\sim 2.677\ kHz$ alternating voltage to our bias voltage and momentarily stop the feedback loop and record the $dI$ using inbuilt lock-in amplifier. In this fashion the bias is varied from $-V$ (for example, -6 mV) to $+V$ (for example, +6 mV) and the $dI/dV$ vs. $V$ is obtained. This $G_{ns}(V)$ is further normalized by dividing the full spectra by $G_{ns}(V \gg \Delta)$ such that at high bias $G_{ns}$, as well as, $G_{nn}$ become unity. Procedure of normalization is shown in *Figure 2.9a*.

### 2.3.1.1 BCS fitting of the spectra

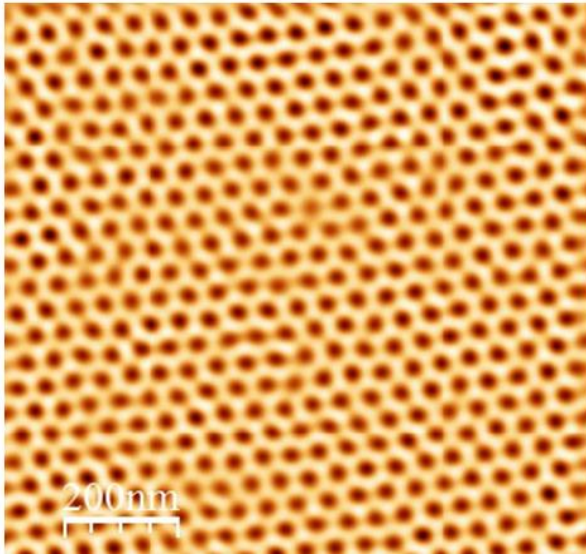

*Figure 2.10:* Vortex image in NbSe$_2$ single crystal acquired at 450 mK for 9 kOe over a area of $1~\mu m \times 1~\mu m$. The dark spots are the positions of the vortices which represent minima in conductance map taken at a fixed bias of 1.2 mV, i.e. $\Delta G(V = 1.3~mV)(nS)$. The lighter regions surrounding the vortices are the superconducting regions where the conductance value is higher due to occurrence of the coherence peaks.

From the $G_{ns}(V)$ vs $V$ data one can extract information such as BCS gap, $\Delta$ by fitting the $G_{ns}(V)$ curve using equation 1.17. However, the BCS DOS is modified slightly with an introduction of a $\Gamma$ factor to take into account non-thermal sources of broadening.[20,21] which takes the form:

$$\frac{N_s(E)}{N(0)} = \begin{cases} Re\left(\dfrac{|E| + i\Gamma}{\sqrt{(|E| + i\Gamma)^2 - \Delta^2}}\right) & (E > \Delta) \\ 0 & (E < \Delta) \end{cases} \quad (2.5)$$

*Figure 2.9b* shows such a fitting of experimental $G_{ns}(V)$ obtained from 20 nm *a*-MoGe thin film, with equation 5, which gives $\Delta = 1.3~meV$ and $\Gamma = 0.2~meV$ at 450 mK. In principle one can do the same experiment and fitting for different temperatures and from that can extract $\Delta(T)$ values which then can be fitted with the universal BCS curve.

### 2.3.2 Vortex imaging

Based on the fact that a vortex core is normal metallic, where the superconducting gap and the coherence peaks are suppressed (*Figure 2.9a*) we fix our bias at the voltage ($V_{peak}$) where coherence peak is located and record the differential conductance over a surface. This $G(V = V_{peak})$ map, shown in *Figure 2.10* portrays vortices as minima because the vortices have value of $G(V = V_{peak})$ less than the superconducting regions.

### 2.3.3 Vortex core spectroscopy

One of the most interesting applications of bias spectroscopy is vortex core spectroscopy, where the bias spectroscopy is done in a vortex core. The core which is a normal metal, can act as a potential well for the normal electrons residing inside and because of Andreev reflection of them, can form a bound state at the centre of the vortex core in a very clean superconductor. This bound state, shown in *Figure 2.11*, is called Carloi-de Gennes-Matricon bound state.[22] The energy gap of these bound states is $\epsilon_0 \sim \Delta^2(0)/E_F$ which for the current resolution of STS falls roughly at zero bias. For a dirtier superconductor this bound state vanishes because of scattering of the bound electrons by the defects and hence shows a regular metallic behaviour.

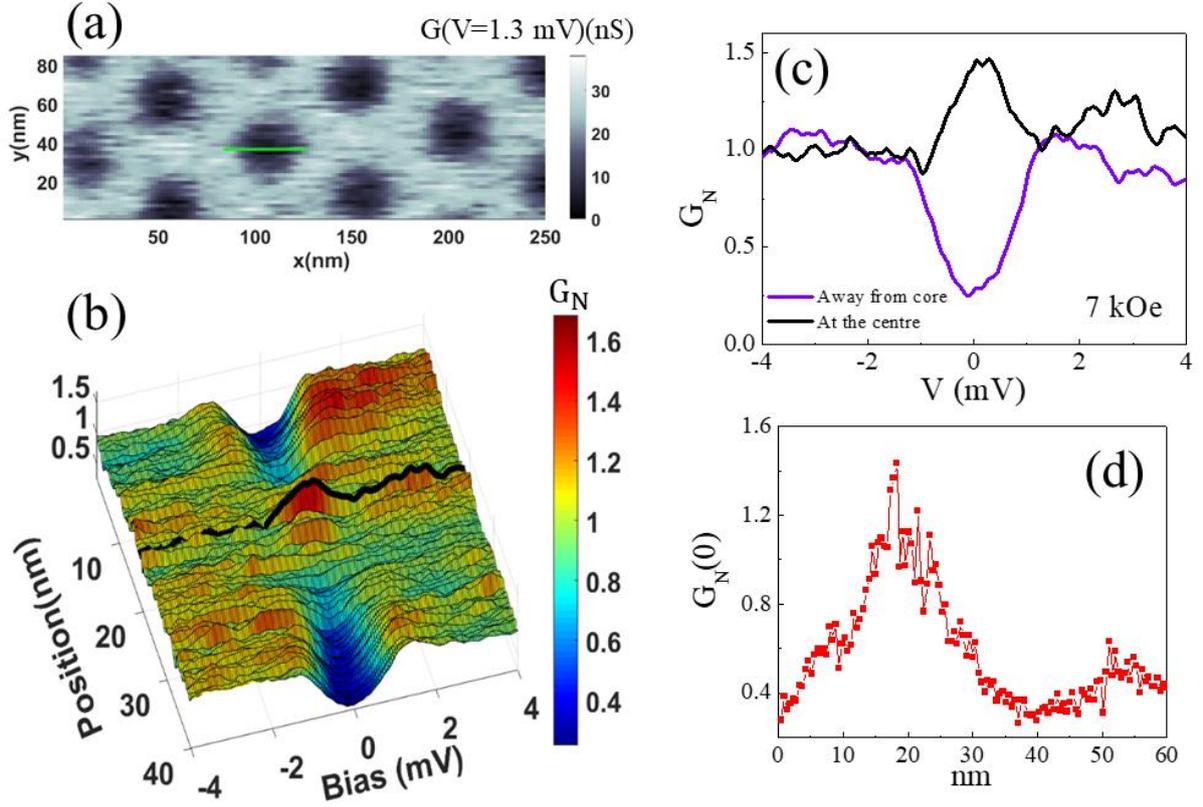

*Figure 2.11:* Caroli-de Gennes-Matricon (CdGM) bound state (a) Vortex image at 7 kOe at 450 mK for NbSe$_2$ obtained by acquiring fixed bias conductance map ($G(V = 1.3\ mV)(nS)$) (b) $G_N(V) - V$ along the green line shown in (a) shows a zero bias peak at the centre of the vortex. (c) The spectra marked black (in (b)) is the spectra at the centre of the vortex core while the purple spectra is obtained at a point far away the vortices. (d) $G_N(0)$ variation along the green line of (a) going along the centre of the vortex core.

## 2.4   Vortex Image analysis

The first step in $G(V = V_{peak})$ map or in other words the vortex image analysis are done using *WSxM 5.0* by *Nanotec Electronica*,[23] where the scan lines are removed by removing the $k_x = 0$ and $k_y = 0$ lines in the FFT.[24] The resultant 2D array of data are analysed in *MATLAB* to find out the minima in the images. These minima finding routine can incorrectly detect a noise point as a minima, which are then corrected by hand.

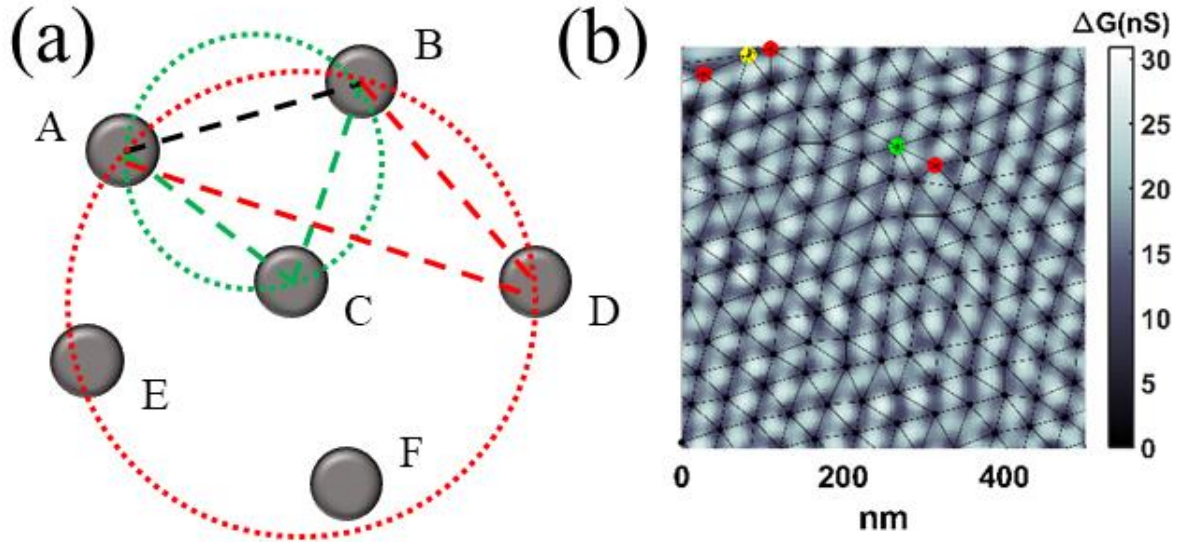

*Figure 2.12:* Delaunay triangulation (a) Lattice points A, B, C, D, E and F are being Delaunay triangulated. The triangle among A, B and C and their circumcircle are shown by the green dotted lines. The green circle does not contain any other lattice points. Hence A, B and C are designated as nearest neighbours. On the other hand the triangle among A, B and D and their circumcircle, shown in red dotted lines, contain other lattice points and hence cannot be nearest neighbours. (b) Delaunay triangulated vortex lattice of *a*-MoGe at 10 kOe at 450 mK (vortices are shown in black dots and are connected by black broken lines). The defects in this lattice are the red, green, magenta and yellow dots corresponding to 5, 7, 4 and 8-fold coordination.

### 2.4.1 Delaunay triangulation and defect finding

Post processing of the images, like finding out nearest neighbour of each vortex is done using Delaunay triangulation routine. As is shown in *Figure 2.12a*, Delaunay triangulation connects three lattice points and draw a circumcircle around the triangle. If there are no other points within this circle, then those three points are identified as the nearest neighbours of each other. This procedure is repeated for all the vortices and the number of nearest neighbours of each vortex is hence found out. Next a routine finds out vortices which have 5-, 7-, 4- or 8-fold coordination, which are identified as the topological defects of the vortex lattice. A vortex lattice with topological defects is shown in *Figure 2.12b*.

### 2.4.2 Finding trajectories of vortices

In this work the trajectories of the vortices have been studied in great detail for vortex lattice in *a*-MoGe thin film. To do that usually 12 vortex images are recorded one after another. Then from the first image one vortex is chosen and a circle of radius $r < 0.5 \times a_0$ is drawn around the vortex. Now the second image is checked to find if a vortex lies within this circle. In this way all the vortices in the parent image are tagged in a one-to-one correspondence with its daughter image, as is shown in *Figure 2.13*. This is repeated for all 12 images and hence the trajectory of each vortex is found out. However this procedure of determining the trajectory of vortices lacks perfect accuracy because of the indistinguishable character of the vortices. If, for example, a vortex jumps by of the order of or more than a lattice constant, it cannot be tagged with its original parent. Therefore, the trajectories in the solid phase are more accurate than their counterparts in the liquid phases, where vortex motion is much larger.

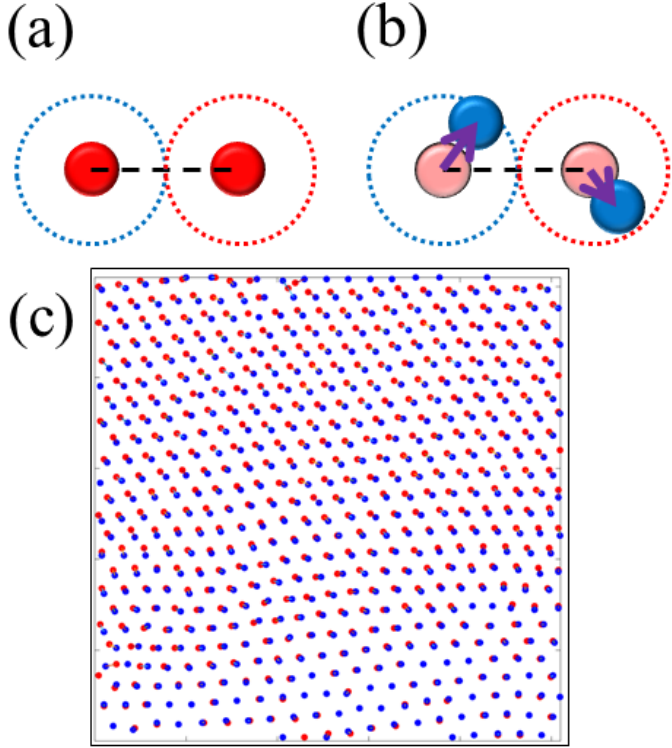

*Figure 2.13:* Vortex trajectories (a) The red balls are representations of vortices in the first frame, around which circles of $0.5 \times a_0$ is drawn ($a_0$ is lattice constant). (b) The blue balls are vortices in the next frame which fall within the circles drawn before. The parent and daughter balls are connected by arrows. (c) Red points are vortices in first frame while blue dots are vortices in second frame. Circles of $0.5 \times a_0$ would map the parent vortex to its daughter.

### 2.4.3 Calculation of metrics related to vortex lattice

A lattice is quantified by its positional and orientational order to differentiate different phases of matter, like solid, liquid, glass, hexatic fluid etc. In my study of vortex images, different metrics and quantities related to vortex lattice have been calculated.

#### 2.4.3.1    $G_K(\vec{r})$: *Positional order parameter*

The positional order parameter is defined as,

$$G_K(\vec{r}) = \langle \cos(\vec{K}.\vec{r}) \rangle \tag{2.6}$$

where $\vec{K}$ is the reciprocal lattice vector and $\vec{r}$ is distance between two lattice points in the real lattice. Since for a perfect lattice, $\cos(\vec{K}.\vec{r}) = 1$, this acts a measure of perfectness of positional order of a vortex lattice. In the limit, $r \to \infty$, for a perfect lattice $G_K(\vec{r})$ remains unity, however in real life no lattice can be perfect in that limit, hence $G_K(\vec{r} \to \infty) \to const.$

However for hexatic fluid and isotropic liquid, positional order is short range and hence, $G_K(\vec{r} \to \infty) \sim \exp(-r/\xi_K)$. For a Bragg glass, positional order is quasi-long range and falls off algebraically, $G_K(\vec{r} \to \infty) \sim 1/r^a$ ($a$ is a nonzero positive number).

In our experiments we use a *MATLAB* routine to find out $G_K(\vec{r})$ for each vortex image and then take an average over 12 consecutive images to increase statistics.

### 2.4.3.2  $G_6(\vec{r})$: Orientational order parameter

The orientational order parameter is defined as,

$$G_6(\vec{r}) = \langle \cos 6(\theta(r) - \theta(0)) \rangle \tag{2.7}$$

where, $\theta(0)$ is an arbitrarily chosen fixed axis in the vortex lattice. Since for a perfect hexagonal lattice $\cos 6\theta = 1$, in the limit $r \to \infty$, $G_6(\vec{r})$ remains unity. However for a real lattice and a Bragg glass $G_6(\vec{r} \to \infty) \to const$. For isotropic liquid, orientational order is short-range and hence $G_6(\vec{r} \to \infty) \sim \exp(-r/\xi_6)$. For hexatic fluid phase orientational order is quasi-long range and hence, $G_6(\vec{r} \to \infty) \sim 1/r^b$ ($b$ is a nonzero positive number).

Similar to $G_K(\vec{r})$, we use a *MATLAB* routine to calculate $G_6(\vec{r})$ for each vortex image and then are averaged over 12 images. We also define an averaged orientational order parameter by,

$$\Psi_6 = \frac{1}{N} \langle \sum_{k,l} e^{6i(\phi_k - \phi_l)} \rangle \tag{2.8}$$

Here, $\phi_k$ is the angle between a fixed direction in the plane of the vortex lattice and the $k$-th bond, and the sum runs over all the bonds in the lattice. For a perfect hexagonal lattice, $\Psi_6 = 1$.

### 2.4.3.3  Mean square displacement

From the trajectories of the vortices, one can calculate mean square displacement ($\langle \Delta r^2 \rangle$) of the vortex motion as a function of time. This is given by,

$$\langle \Delta r^2 \rangle(\Delta t) = \langle (r(\Delta t) - r(0))^2 \rangle = \frac{1}{N} \sum_{n=1}^{N} (r_n(\Delta t) - r_n(0))^2 \tag{2.9}$$

In order to do this, $\langle \Delta r^2 \rangle$ of vortices for each time segment, $\Delta t$ is calculated. Suppose, for calculating $\langle \Delta r^2 \rangle$ for $\Delta t = 1$, we calculate $\langle \Delta r^2 \rangle$ between image 1 and image 2, image 2 and image 3,...., image 11 and image 12, and then are averaged. This procedure we repeat for $\Delta t = 1,2,3,\ldots,11$. However, the $\langle \Delta r^2 \rangle$ for $\Delta t = 11$ can have a large inaccuracy, since for this only image 1 and image 12 can be compared in the present scenario.

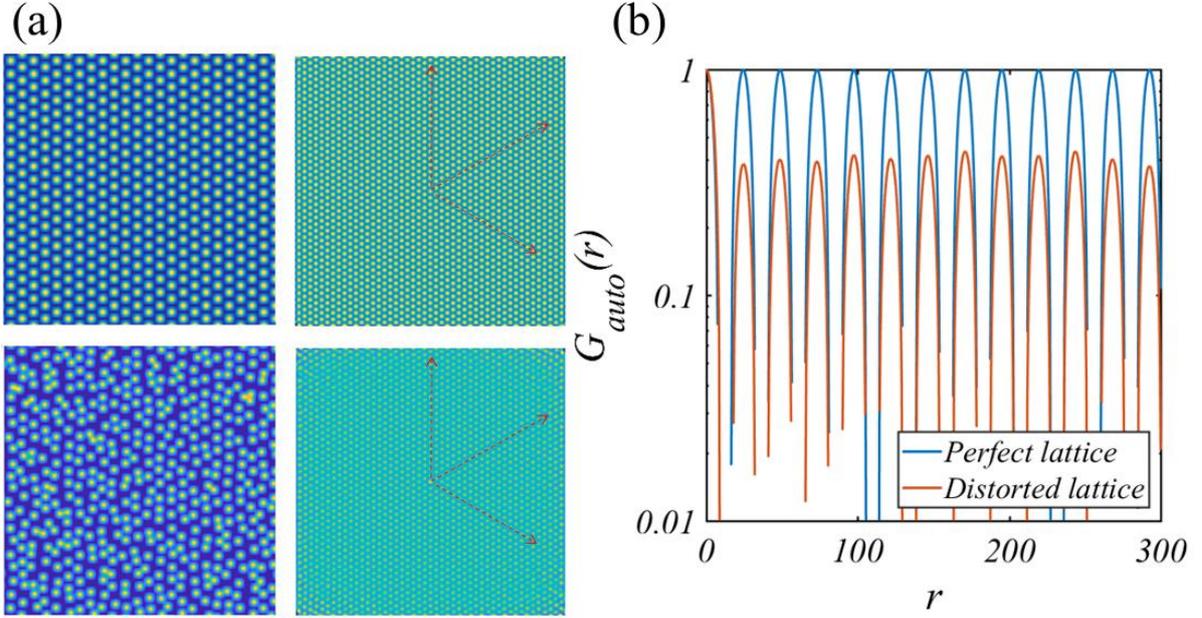

*Figure 2.14:* Autocorrelation calculation (a) (*top left*) A perfect lattice, where the yellow dots are the lattice points. (*top right*) Autocorrelation map of the perfect lattice. The red arrows represent the three principal axes. The yellow dots now represent the peaks of the autocorrelation map, which remain same even far away from centre. (*bottom left*) A distorted lattice where distortion is randomly given to each lattice point by a maximum of $0.3 \times a_0$. (*bottom right*) Autocorrelation map of the distorted lattice. Although the central peak of the map is higher, rest have decayed over length. (b) Comparison of $G_{auto}(r)$ between the perfect and the distorted lattice. For the perfect lattice $G_{auto}(r)$ remains 1 for all $r$, but for distorted lattice it falls off to a lower value.

There is a theoretical prediction[25] that a solid to hexatic to liquid transition (BKTHNY melting) can be identified from the $\langle \Delta r^2 \rangle (\Delta t)$ plots, where for large $\Delta t$, $\langle \Delta r^2 \rangle$ will be constant, while for the isotropic liquid phases, $\langle \Delta r^2 \rangle$ would increase rapidly.

### 2.4.3.4 *Autocorrelation length*

If one has to refer to structural uniformity of a lattice (or a signal in general), one looks for correlation of the lattice points with itself. This metric is called autocorrelation and in 2D it is given by,

$$G_{auto}(k,l) = \frac{1}{(N-k+1)(N-l+1)} \sum_{m=|k|}^{N} \frac{X(m,n)X(m-|k|,n-|l|)}{\sigma_X^2} \quad (2.10)$$

$$X(i,j) = I(i,j) - \langle I \rangle \quad (2.11)$$

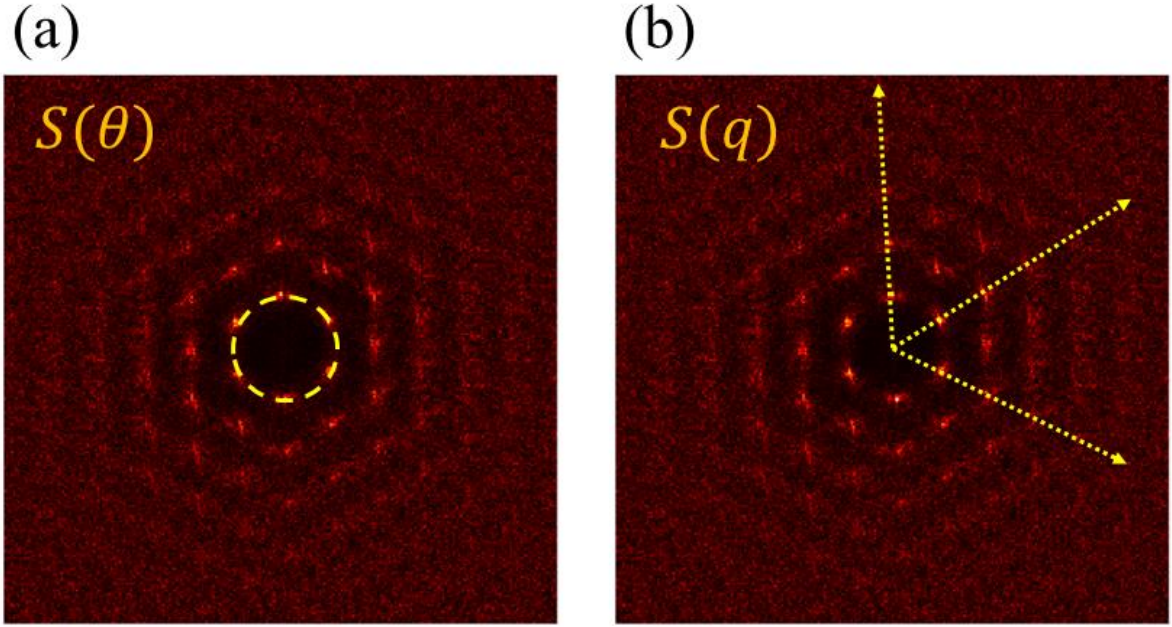

*Figure 2.15:* Structure factor (a) Definition of $S(\theta)$: In the FFT of a lattice, $S(\theta)$ is the variation of $S(q, \theta)$ as a function of $\theta$ at $q = q_{peak}^1$, the first peak position. (b) Definition of $S(q)$: $S(q, \theta)$ along three principle axes are noted, which are $\theta = 60°$ apart from each other.

Here, $I(m, n)$ refers to the intensity of each pixel of the vortex image, and $\sigma_X^2$ is the variance in $X$ and $\langle I \rangle$ is average over entire image which has a dimension of $N \times N$. For a perfect hexagonal vortex lattice, $G_{auto}(k, l)$ will also be a hexagonal lattice with unit intensity at each lattice point, as is shown in *Figure 2.14a*. However for an imperfect vortex lattice the intensity of each lattice point will decay from unity as one moves away from origin. To make sure that only the lattice points contribute to the autocorrelation map, the vortex images have been transformed into binary maps of 1 and 0, with 1 being at a position of vortex and 0 elsewhere. This $G_{auto}(k, l)$ map is used to obtain the autocorrelation function, $G_{auto}(r)$ by plotting averaged $G_{auto}(k, l)$ along three principal axes, as is shown in *Figure 2.14a*. In *Figure 2.14b*, we compare $G_{auto}(r)$ between a perfect and a random lattice. The decay length of $G_{auto}(r)$, namely autocorrelation length is calculated by fitting the function with a relation, $G_{auto}(r) \propto e^{-r/R_{auto}}$.

### 2.4.3.5 $S(\vec{q})$: *Structure factor*

It has been suggested[26,27] by T. V. Ramakrishnan and M. Yussouff that the structure factor of a lattice undergoes a definite jump during a solid to liquid transition. Hence from the FFT of the vortex images, we extract two quantities, first, radial variation of structure function at $|\vec{q}|$ =position of the first maxima, as a function of angle, $\theta$ to determine $S(\theta)$, shown in

*Figure 2.15a*. Secondly, starting from $|\vec{q}| = 0$, we map structure function vs $|\vec{q}|$ along three principal directions and average over them to determine $S(q)$, shown in *Figure 2.15b*.

# Chapter III: Dynamic vortex Mott to vortex metal transition in NbN films with periodic array of pins

In this chapter we will discuss the effect of periodic pinning centres on the vortex lattice. In section 1.3.5 it has been shown that in presence of a physical hole or void in the superconductor, the order parameter and subsequently related thermodynamic quantities show oscillating behaviour with magnetic field, due to Little Parks effect. It has been shown that[1,2] in an array of voids, or in other words, in an antidot array cancellation of circulating supercurrent around each antidot can give rise to Little-Parks-like effect. This is understood as resulting from quantum interference (QI) of supercurrent.

## 3.1 Commensurate pinning: Matching effect

Our model system NbN thin film on top of anodic alumina membrane (AAM) (discussed in section 2.1.1) is prepared using DC Magnetron sputtering. Post deposition, the lattice constant of this antidot array becomes $\sim 106\ nm$ while the diameter of the voids become $\sim 46\ nm$, giving a superconducting channel of width, $w = 60\ nm$ (*Figure 3.1a*). In this antidot array a different kind of oscillation, named matching effect takes place, which is of the dynamic quantities related to the vortex lattice (VL), arising due to periodic variation of compressibility of the VL with magnetic field. This effect, known as commensurate pinning occurs when each antidot carry equal integer multiple of flux quanta and each vortex gets strongly caged by the inter-vortex repulsion from adjacent antidots. While both of Little-Parks effect and matching effect has a period equal to first matching field, given by, $H_m = 2\Phi_0/\sqrt{3}a_0^2$ ($a_0$ being the lattice constant of the antidot array), matching effect manifests itself only in dynamic properties of the VL.

The antidot array acts as a periodic but extremely strong pinning centres and mutual repulsion of vortices can cage the vortices even in a stronger fashion, if the magnetic field, $H$ is an integer multiple of $H_m$. In this scenario, the VL mimics a Mott insulator[3,4,5] Mott insulators[6,7] are materials where each free electron is in a bound state about each atom due to Thomas-Fermi screening and do not participate in transport. However thermal energy or high electric field larger than the potential well of each atom, can in principle delocalize the bound electrons and give rise to a metallic state. The parallel between the Mott insulator and the vortex Mott state can be drawn where each vortex can be mapped to each bound electron and the antidots mimicking the potential well of the protons. This vortex Mott insulator-like state, in general, can be driven to a vortex metal-like state, if sufficient energy to delocalize the localised vortices is provided in form of temperature, magnetic field or drive current.[8] However it is interesting to ask whether the same Mott-like to metal-like transition can be achieved dynamically by shaking the vortices about the minima of their cage. Recently this dynamic transition of vortex-Mott to vortex-metal state has been observed,[9] induced by an external current.

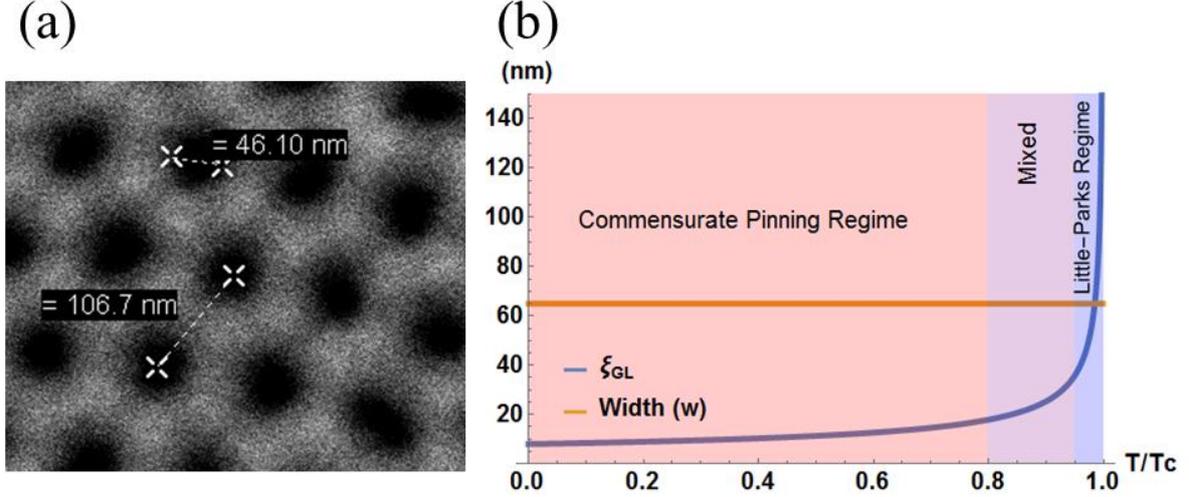

*Figure 3.1*: (a) SEM image of AAM after deposition. The diameter of the antidots is ~46 nm and the lattice constant is ~106 $nm$, which gives width of the superconducting region to be $w$~60 $nm$. (b) Variation of $\xi_{GL} \propto \sqrt{1 - T/T_c}$ in violet along with the width of the superconducting region, $w$ in orange. For $w \gg \xi_{GL}$ commensurate pinning dominates while for $w \sim \xi_{GL}$ the Little-Parks effect dominates.

### 3.1.1 Difference from Little Parks effect

The main difference between the Little-Parks effect and matching effect is in their origin and the geometry in which these two phenomenon take place. In order to observe Little-Parks-like effect in an array of antidots, the superconducting channel width ($w$) must be of the order of $\xi_{GL}$, such that $J_s$ remains uniform over the superconductor and the circulating supercurrent gets cancelled to give rise to enhancement of order parameter to its zero field value at integer filling. However, commensurate pinning relies on the inter-vortex repulsion and the pinning potential of the antidots and is always present independent of $w$ or $\xi_{GL}$. Now, if one looks at *Figure 3.1b*, the GL variation of $\xi_{GL} \propto \left(1 - \frac{T}{T_c}\right)^{1/2}$ is plotted as a function of $T/T_c$ and is compared with $w$. It is apparent from this figure that only at $T \to T_c$ the condition of Little-Parks-like effect is satisfied, however, for $T \ll T_c$, $w \gg \xi_{GL}$ and hence Little-Parks is not at all dominant at lower temperatures. It has also been reported that for NbN thin films with higher disorder (for example sample with $T_c$~4 $K$), matching effect is observed down to $T \sim 0.38\ K < 0.1 T_c$ in contrast with samples with less disorder ($T_c \sim 12\ K$), where matching effect is observed only down to $T \sim 6\ K \sim 0.5 T_c$.[10]

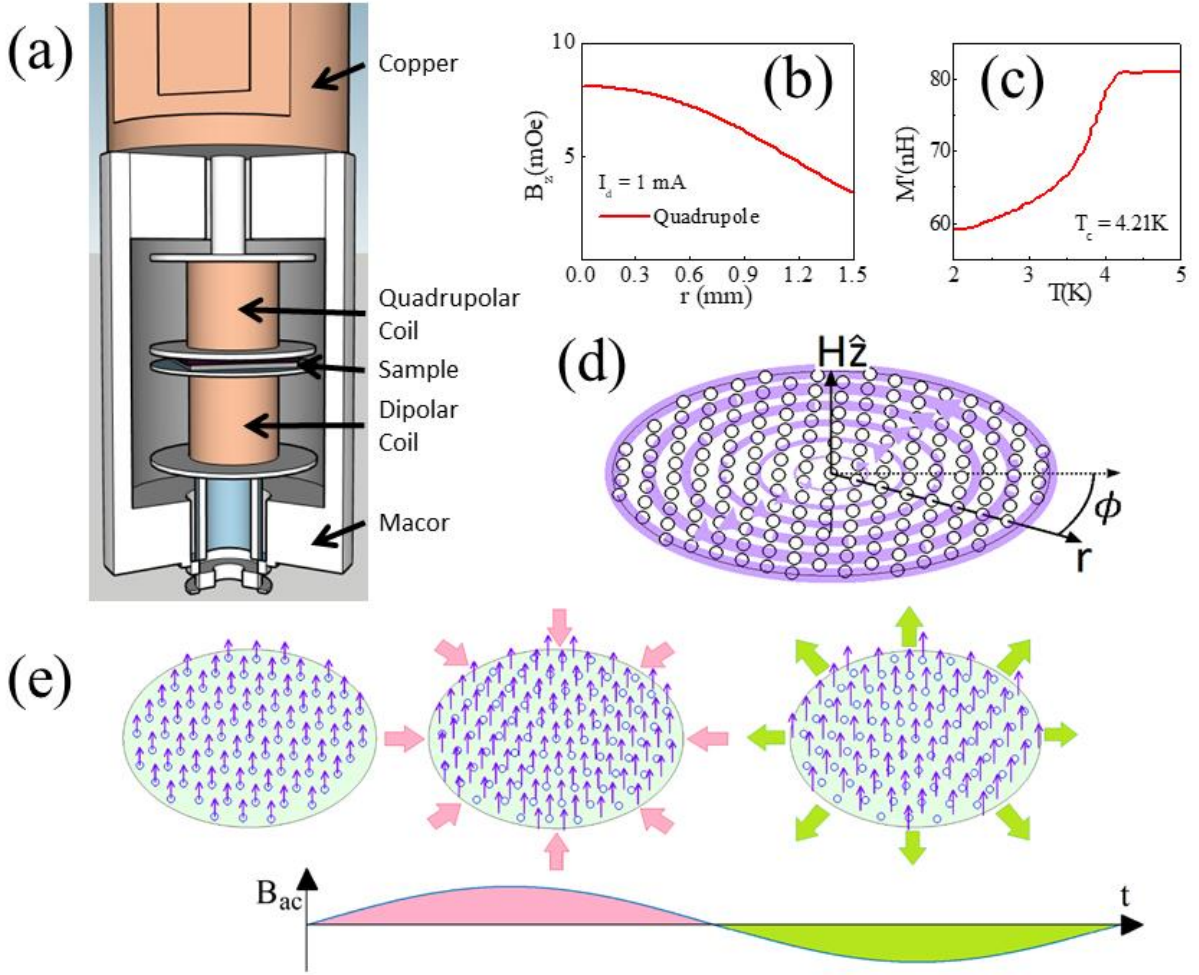

*Figure 3.2*: (a) Schematic diagram of the two-coil mutual inductance setup, where the sample is sandwiched between the quadrupolar primary and the dipolar secondary coils. The bobbins and the head of the setup is made of macor, which is connected to a copper block on which the electrical connections are made on a pcb. (b) ac magnetic field produced by the quadrupolar primary coil as a function of radius of the sample, which has a peak value of ~7 mOe/mA. (c) $M' - T$ data of the sample showing $T_c \sim 4.21\ K$. (d) Schematic diagram of the AAM, where the field is applied in the z direction. The coarse grained average current density grows linearly with radius from the centre, depicted by the width of the violet circles. (e) Schematic of the effect of the ac perturbating field on the VL. In the positive part of the ac perturbation, the VL gets compressed, while in the other cycle the VL gets relaxed.

Further confirmation that matching effect is not related to thermodynamic quantities have been proved in Ref. 2 and 8, that matching effect does not affect thermodynamic quantities such as superconducting energy gap. In contrast Little-Parks effect is reflected in all thermodynamic quantities including gap.

## 3.2 Sample details and Measurement scheme

### 3.2.1 Sample

The sample used here[11] consists of a 3-mm diameter disorder NbN thin film of thickness $t \sim 25\ nm$ grown using DC reactive magnetron sputtering using AAM as the substrate. The reason a disordered film is chosen is that in order to observe matching effect down to $T \ll T_c$, we should always maintain inter-vortex repulsion to be of the order of pinning force from the antidots. Although for $T \to T_c$, this is satisfied because of a diverging $\xi_{GL}$ for any disorder, only in strong disorder $\xi_{GL}$ is enhanced and superfluid density is reduced to make the condition viable down to low temperatures. $T_c$ of our film is $\sim 4.21\ K$ (*Figure 3.2c*) and matching effect is observable down to $\sim 1.45\ K$.

### 3.2.2 Measurement technique

Although the matching effect is customarily observed in resistance, the changes in dynamic quantities do not reflect well in resistivity measurements. Furthermore, resistivity measurement is only prominent close to $T_c$, which makes matching effect in juxtaposition with Little-Parks-like effects. Hence we rely on two-coil mutual inductance measurement setup, which can directly measure the effect on dynamic quantities such as compressibility due to matching effect. The thin film on AAM was sandwiched between the primary and the secondary coils of the two-coil setup, shown in *Figure 3.2a*, where the quadrupolar primary coil has total 28 turns, producing a peak field of ~7 mOe/mA (*Figure 3.2b*). The dipolar secondary coil has 120 turns, which picks up the induced voltage, $V''^{(\prime)} = 2\pi I_{ac} f \times M'^{(\prime\prime)}$, using lock-in amplifier measured at a frequency of 31 kHz. We have varied the amplitude of the ac excitation ($I_{ac}$) from 0.5-15 mA and measured the in phase and out of phase components of mutual inductance which represent the dissipative and inductive responses respectively. The two coil setup is placed in a He$^4$ cryostat assembly operating down to 1.4 K and fitted with a superconducting solenoid capable of generating a dc magnetic field perpendicular to the film plane, up to 90 kOe.

### 3.2.3 Measurement Scheme

To understand the quantities $M'$ and $M''$, we note that the magnetic field produced by the primary coil will generate a coarse grained averaged time varying circulating supercurrent of the form $\vec{J}_S^{ac}(\vec{r}, t) = J_S^{ac}(r)\sin(\omega t)\hat{\phi}$ in the superconducting film, shown in *Figure 3.2d*, which from symmetry considerations will be zero at the centre of the sample and will increase towards the periphery. ($J_S^{ac}$ is a function of *r* only because, in order to have a field, $\vec{B} = B_0 \hat{z}$, we should have, $\vec{A} = \hat{\phi}\frac{1}{2}rB_0$. This gives, $\vec{J}_S = -\frac{rB_0}{2\mu_0 \lambda^2}\hat{\phi}$.) However, here we have ignored the pattern of supercurrent on a shorter lengthscale around the antidots. Nonetheless, the circulating supercurrent will apply a radial oscillatory force on each vortex, per unit length of the vortices, given by,

$$\vec{F}_{ac} = \vec{J}_S^{ac} \times \hat{z}\Phi_0 = J_S^{ac}(r)\sin\omega t\ \Phi_0 \hat{r} \tag{3.1}$$

causing each vortex to oscillate about its equilibrium position at the centre of the antidot. Hence, the effect of the ac excitation is to produce an oscillatory compaction and rarefaction of the VL with the vortex at the centre in its equilibrium position, as is shown in *Figure 3.2e*. This is equivalent to a periodic change in the flux density which will induce a voltage in the secondary coil. Thus the real part of mutual inductance, $M'$ measures the compressibility of the vortex lattice.

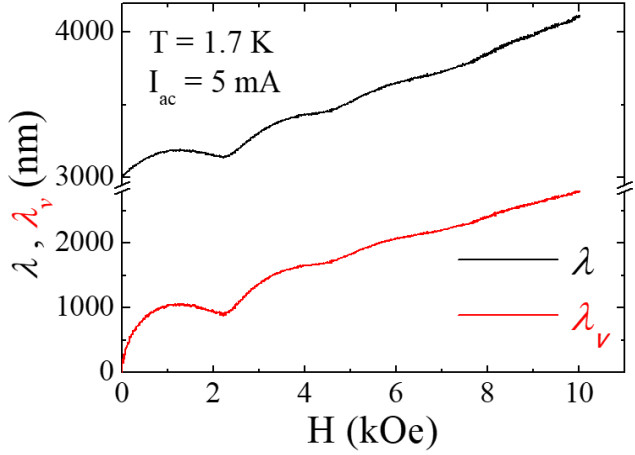

*Figure 3.3*: Variation of $\lambda$ and $\lambda_v = \sqrt{\lambda^2 - \lambda_M^2}$ as a function of magnetic field. $\lambda(0) = \lambda_M(0) \sim 3009\ nm$ is at $T = 1.7\ K$ obtained using $I_{ac} = 5\ mA$. The oscillation in $\lambda_v$ is the source of the oscillations in mutual inductance, which arises due to $\lambda_v$-s dependence on Labusch parameter, $\alpha$ and magnetic field.

### 3.2.4 Source of oscillation

In our experiment the screening response in the presence of vortices arises from two sources: The screening of the magnetic field from Meissner supercurrents and the periodic compaction and rarefaction of the vortex lattice due to the ac magnetic field. The effective penetration depth is thus given by, $\lambda^2(H) = \lambda_M^2 + \lambda_v^2(H)$, where $\lambda_M$ is from Meissner supercurrent and $\lambda_v$ is from the motion of vortices. These two penetration depths are analogous to the London penetration depth, $\lambda_L$ and the Campbell penetration depth[12], $\lambda_C$ in a bulk Type II superconductor. *Figure 3.3* shows the magnetic field variation of $\lambda(H)$ at 1.7 K measured with $I_{ac} = 5\ mA$ and is determined using the method outlined in section 2.2.1. If we assume that to the lowest order the vortex lattice is empty in zero field then, $\lambda_M \approx \lambda(0) = 3009\ nm$. On the other hand at the first matching field, $\lambda(H = H_m) \approx 3144\ nm$. This gives $\lambda_v(H = H_m) \approx 911$ nm, which is the vortex contribution to the penetration depth at this field.

Therefore the dominant response to the shielding response is always from the Meissner current, which accounts for the relatively small amplitude variation in M' over the large background.

In this context, it is also instructive to analyse the magnetic field variation of $\lambda_v$ between $H = 0$ and $H = H_m$. $\lambda_v$ which arises from this oscillatory motion of the vortex inside a potential well is given by, $\lambda_v^2(B) = \frac{\Phi_0 B}{4\pi\alpha}$, where $\alpha$ is the Labusch parameter[13] which is proportional to the restoring force on the vortex when it is moved from its equilibrium position. At very small magnetic field, vortices are far from each other and $\alpha$ is determined essentially by the field independent restoring force arising from the antidot. In this regime $\lambda_v^2(B)$ increases with increasing $B$. However, as the magnetic field is increased, very soon the vortices start experiencing the repulsive interaction from all surrounding vortices, which further confines the vortex in its equilibrium position, thereby increasing $\alpha$. Thus $\alpha$ increases monotonically with field, and attains its maximum value at the first matching field, $H_m$ where each vortex is confined in a cage formed by the vortices in neighbouring antidots. Thus, the variation of $\lambda_v^2(B)$ is governed by a combination of two effects: (i) an increase with $B$ due to the $B$ appearing in the numerator of the expression and (ii) a decrease with $B$ due to the increase in $\alpha$ appearing in the denominator. As a result $\lambda_v^2(B)$ first increases with field, goes through a maxima and somewhere midway between *0* and $H_m$ and then exhibits a minima at $H_m$ where $\alpha$ is maximum. Consequently, $M'$ mimics the same variation.

### 3.2.5 Source of dissipation

The amplitude of the imaginary part of mutual inductance, $-M''$ is related to the dissipation in the superconductor. At small ac excitations when each vortex undergoes small oscillations about its mean position, the dissipation arises due to Bardeen-Stephen loss inside the superconductor surrounding the antidot.

However, if the excitation is large enough to a force a vortex to hop from one antidot to the next, the dissipation comes predominantly from the $2\pi$ phase slip in the intermediate superconducting strip separating the two antidots. This can be understood in terms of Josephson relation[14,15]

$$\frac{\partial \Delta\phi}{\partial t} = \frac{e^*}{\hbar}\Delta V \quad (3.2)$$

The $2\pi$ phase slip gives rise to a voltage drop across the superconducting channel following this relation which in turn results in the dissipation.

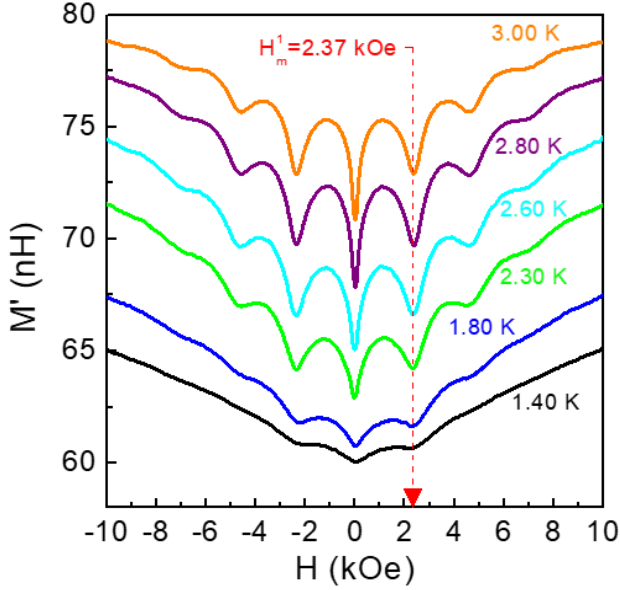

*Figure 3.4*: Variation of $M'$ as a function of $H$ ($kOe$) for different temperatures from 3 K down to 1.4 K. The first matching field is identified as $H_m^1 = 2.37\ kOe$. At the lowest temperatures only the first matching field is clearly visible.

## 3.3 Temperature variation of matching effect

*Figure 3.4* shows the $M' - H$ scans recorded at different temperatures between *3 K* and *1.45 K* using $I_{ac} \sim 5\ mA$. $M' - H$ shows a dip at multiples of matching field, $H_m = \frac{2\Phi_0}{\sqrt{3}a_0^2} = 2.37\ kOe$ consistent with experiments. The dips in $M'$ show that the VL becomes more rigid at the matching fields, consistent with the description of a vortex-Mott state. Decreasing the temperature has two effects: first, the overall magnetic screening response increases and secondly the minima in $M'$ at matching fields become less pronounced. Both of these effects can be understood from the temperature variation of the coherence length $\xi_{GL}$ and magnetic penetration depth $\lambda$ of the superconductor. Since the individual pinning force is given as, $F_p \sim \frac{\Phi_0^2}{4\pi\mu_0 \Lambda \xi_{GL}}$, where, $\Lambda = 2\lambda^2/t$ is called the Pearl's penetration depth. Assuming dirty BCS relation of variation of $\lambda$, $\frac{\lambda^{-2}(T)}{\lambda^{-2}(0)} = \frac{\Delta(T)}{\Delta(0)} \tanh\left(\frac{\Delta(T)}{2k_B T}\right)$, it is apparent that $\frac{1}{\lambda^2}$ increases by 30% if the sample is cooled from *3* to *1.45 K*. Thus at low temperatures $F_p$ increases and each vortex gets strongly pinned, which in turn increases the magnetic screening. On the other hand, amplitude of oscillation in mutual inductance is governed by inter-vortex repulsive force, $F_I \sim \frac{\Phi_0^2}{4\pi\mu_0 \Lambda a_0}$, because of which the VL becomes less compressible at matching fields when each vortex is surrounded by a vortex at neighbouring antidots. Since, $\frac{F_I}{F_p} \sim \frac{\xi_{GL}}{a_0}$, and $\xi_{GL} \propto \left(1 - \frac{T}{T_c}\right)^{-1/2}$ tells that going to 1.45 K from 3 K would reduce the ratio $F_I/F_p$ by 30% and hence suppressing the oscillations at lower temperatures. We also observe that at 1.45 K, oscillation only in the first matching field is clearly visible, because of which we concentrate primarily on the behaviour around the first matching field, $H_m^1$.

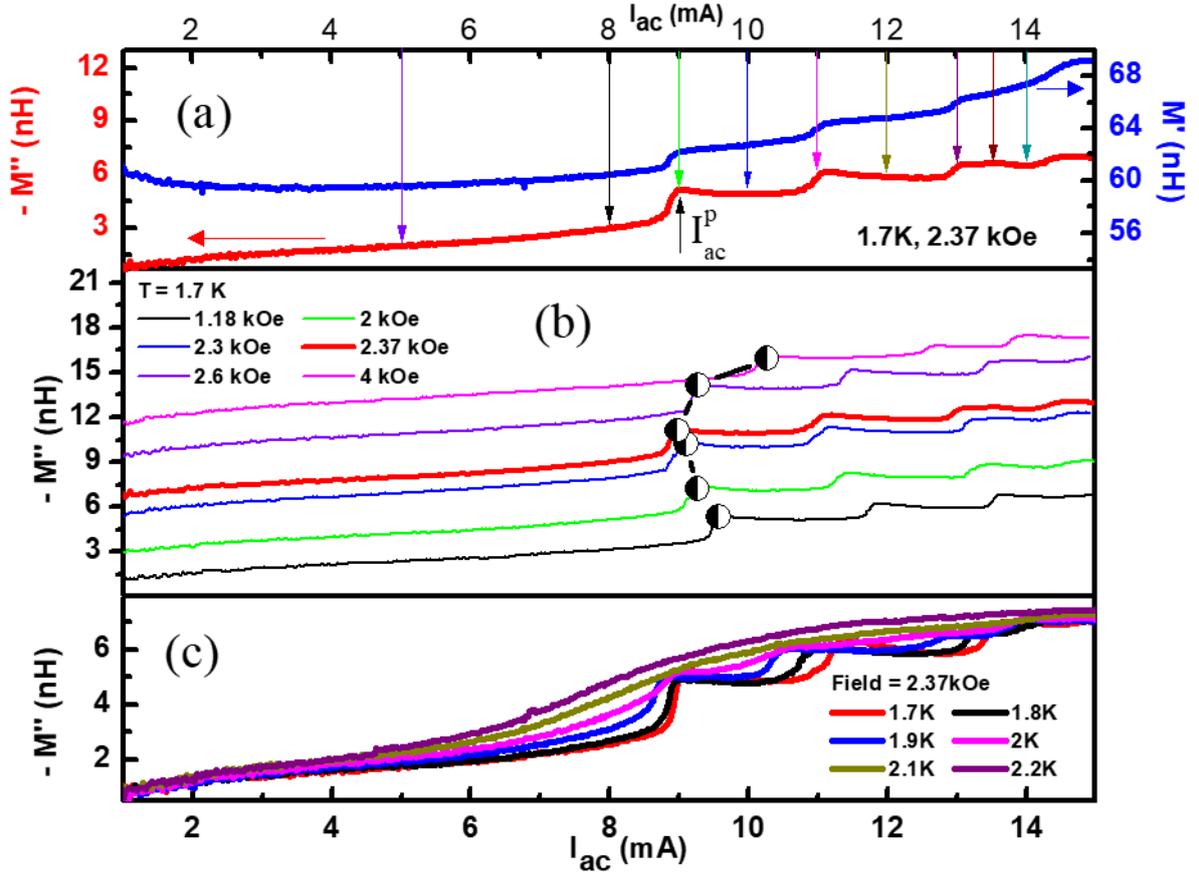

*Figure 3.5*: (a) Variation of $M'$ and $-M''$ as a function of perturbation amplitude, $I_{ac}$ at 1.7 K for the first matching field, $H_m^1 = 2.37\ kOe$. The amplitude corresponding to the first jump in $M'^{('')}$ is designated as $I_{ac}^p$ and the rest are designated as $I_{ac}^1$, $I_{ac}^2$ etc. The vertical arrows correspond to $I_{ac}$ values for which $M - H$ curves are shown in *Figure 7*. (b) Variation of $-M''$ as a function of $I_{ac}$ for different magnetic fields at 1.7 K close to the first matching field, $H_m^1$. The black-n-white circles represent the locus of the first jump, which is minimum for the first matching field. (c) Variation of $-M''$ as a function of $I_{ac}$ for different temperatures at the first matching field. Above 2 K, the jumps have vanished due to thermally activated jumps of the vortices.

## 3.4   Determination of depinning current and its dependencies

In *Figure 3.5a*, we study screening response of the vortex state as a function of the perturbation current, $I_{ac}$ at constant magnetic field. At *1.7 K*, we study variation of $M'$ and $-M''$ for $H = H_m$ and observe that at low $I_{ac}$ both the quantities vary smoothly. At $I_{ac} \sim 8.8\ mA$ on the other hand both of these quantities exhibit a sharp jump which is followed by a series of further jumps at higher excitation currents. To understand the connection between matching effect and these jumps, we study in *Figure 3.5b*, variation of $-M''$ as a function of $I_{ac}$ for different $H$ close to $H_m$ between 1.18 and 4 kOe at 1.7 K. Here we unfold the strong dependency of the value of $I_{ac}$ of first jump with magnetic field. The jump occurs for the smallest value of $I_{ac}$ when $H = H_m$ and increases when $H$ is shifted away from $H_m$. This gives a direct understanding of the effect integer filling has on dynamic quantities such as this. We further our case with the study of variation of $-M''$ with $I_{ac}$ at $H = H_m$ now for different temperatures in *Figure 3.5c*. And we observe that the steps become a smoothly increasing curve for higher temperatures, which suggest that the discrete jumps have been smeared out by thermally activated motion of vortices.

The fact that $I_{ac}^p$ is lowest for $H \approx H_m$, when the vortex lattice is the most rigid, can be physically understood from the interplay between individual pinning potential and inter-vortex interaction. Away from the matching field the antidot array either contains some empty sites for $H < H_m$ or some doubly occupied sites for $H > H_m$, which can accommodate the small alternating change in flux through a local rearrangement of vortices. In this scenario the vortex lattice gets collectively delocalized only when these local rearrangements are energetically unfavourable compared to the sliding of the entire vortex lattice within an annulus. In contrast at $H_m$ each antidot is filled with exactly one vortex such that local rearrangements trigger a collective delocalization of the entire vortex lattice, at a lower $I_{ac}^p$.

## 3.5 Simulation and calculation of the collective depinning scenario

In order to understand the series of jumps in $M'$ and $-M''$ we note that for low $I_{ac}$, the induced $J_S^{ac}(r)$ will cause small alternating compression and rarefaction of the VL. This is similar to the elastic response of a solid under oscillating stress. However the slow increment in both $M'$ and $-M''$ shows an anharmonic nature of the pining force, without which both of the quantities should have remained flat. As $I_{ac}$ is increased to a critical value, $I_{ac}^p$, for some range of $r$, $J_S^{ac}(r)$ will reach the depinning current $J_c$ and the vortices will be delocalized from the antidots. Above $I_{ac}^p$, towards the periphery of the sample vortices in a circular annulus starts to slide over the antidot array, which brings about the sharp increment in dissipation, $-M''$. This also means that the vortex lattice at the periphery has suddenly become compressible. The subsequent jumps can be understood from a similar picture. Let's assume that at $I_{ac}^p$ the vortex lattice within a circular annulus slides back and forth over one lattice constant of the antidot array in each cycle of the ac perturbation. However, as $I_{ac}$ will be increased, the circular annulus will gradually grow. Also, the amplitude of the oscillatory displacement of the vortices will grow. Thus one can conjecture that the subsequent jumps corresponding to the excitation values, $I_{ac}^1$, $I_{ac}^2$, ... gives rise to sliding of vortices over higher integer multiples of the lattice constant, *viz.* $2a, 3a, \ldots$. This can, in principle, explain the almost same gap of $I_{ac}$-value between two subsequent jumps. However, it is interesting to note that in between two successive jumps, $-M''$ slightly deceases before showing another jump. This opens up a probabilistic picture where other mechanisms of strain relaxation not involving entire delocalized VL annulus, such as sliding of a domain of vortices over another is possible. This might be the reason why the magnitude of subsequent jumps are smaller in magnitude, other than a straightforward answer of decreasing annulus-size.

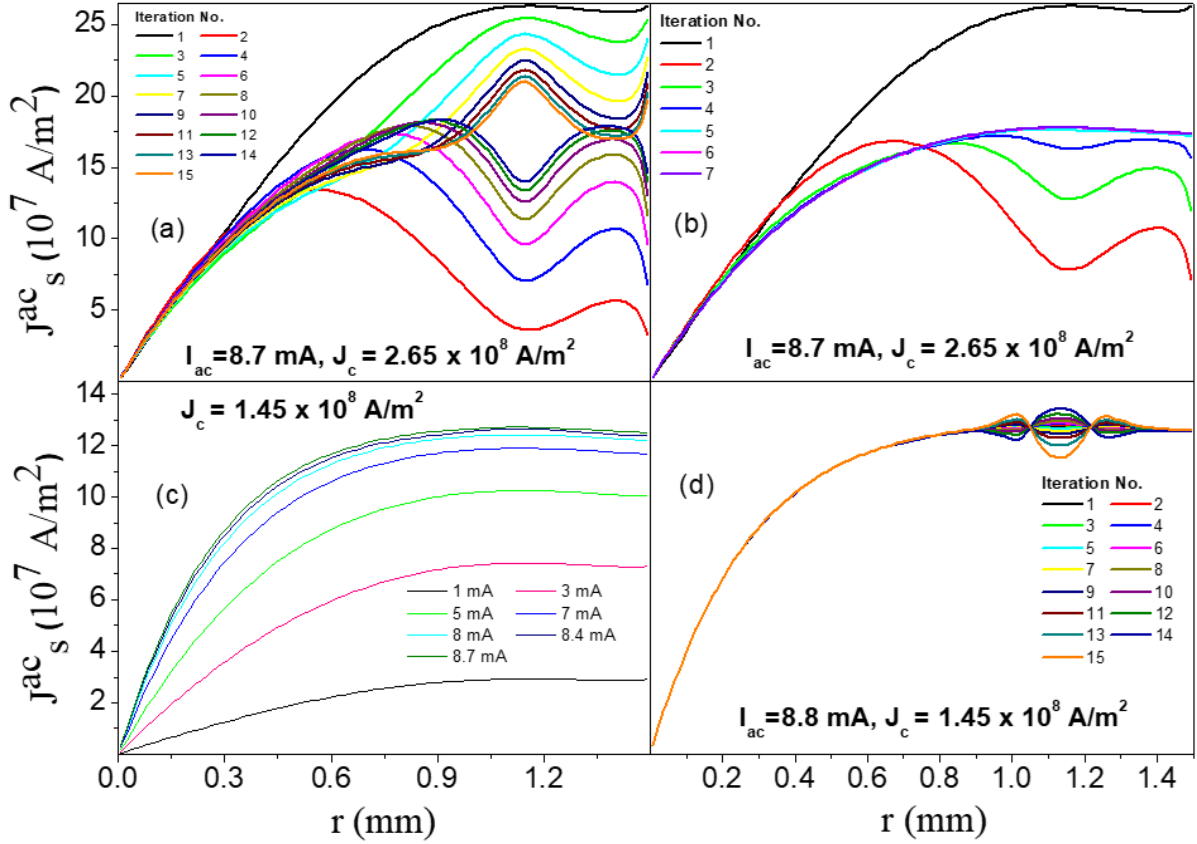

*Figure 3.6*: (a) Oscillatory convergence of $J_S^{ac}$ as a function of radial distance from the centre of the film, for $I_{ac} \sim 8.7\ mA$ with a guess of $J_c \sim 2.65 \times 10^8\ A/m^2$. Even after 15 steps the curves do not converge. (b) Faster convergence utilizing the modified iteration procedure, where the convergence is achieved in 7 steps. Values of $I_{ac}$ and $J_c$ is same as previous. (c) $J_S^{ac}$ as a function of r, for different $I_{ac}$, and the value of $J_c$ is found to be $\sim 1.45 \times 10^8\ A/m^2$. (d) $J_S^{ac}$ as a function of r for $I_{ac} = 8.8\ mA$, but the convergence is not achieved even after 15 iterations.

### 3.5.1 Simulation strategy

At this point, we direct our attention to the fact that the depinning of the vortices could either be an individual depinning or a collective one. To investigate the possibilities, we need to compare the force exerted by the supercurrent on a vortex, when $J_S^{ac} \sim J_c$, with the pinning by an antidot. The first step of this is to find out $J_c$, which can in principle be done by solving the following self-consistent equations,

$$\vec{A}_{film}(\vec{r}) = \vec{A}_{drive}(\vec{r}) + \frac{\mu_0}{4\pi}\int d^3\vec{r}'\, \frac{\vec{J}_S^{ac}(\vec{r}')}{|\vec{r}-\vec{r}'|} \quad (3.3)$$

$$\vec{J}_S^{ac}(\vec{r}) = -\frac{\vec{A}_{film}(\vec{r})}{\mu_0 \lambda_L^2} \quad (3.4)$$

as is described in section 2.2.1, following the prescriptions of Ref. [16],[17],[18],[19]. However, our system has one additional complication arising from the response of the vortex lattice in presence of ac excitation, which is captured within Campbell penetration depth, $\lambda_p$ and our measurement measures a mixture of $\lambda_L$ and $\lambda_p$. This $\lambda_p$ is a function of the local supercurrent density, $J_S^{ac}(r)$ and depinning current density, $J_c$, following,

$$\lambda_p(J_s) = \frac{\lambda_p(0)}{(1 - J_S^{ac}/J_c)^{1/4}} \quad (3.5)$$

Because of this interdependency of $\lambda_p$ and $J_S^{ac}$, we adopt an iterative procedure to obtain $J_S^{ac}$. First, using the value of $M'$ for $I_{ac} = 1\, mA \ll I_{ac}^p$, we obtain the value of $\lambda_p(0) \approx 3100\, nm$ using standard procedure. For this $I_{ac}$ value $J_S^{ac}(r) \ll J_c$ for all values of $r$ and therefore $\lambda_p$ is assumed to be constant over the entire film, i.e. $\lambda_p(J_S^{ac}) \approx \lambda_p(0)$.

To obtain the $J_S^{ac}(r)$ at higher currents, we first note that the depinning at matching field occurs at $I_{ac} \sim 8.8\, mA$; at this value $J_S^{ac}(r) \sim J_c$ on the outer circular annulus of the sample. At this excitation it should theoretically not be possible to obtain $J_S^{ac}(r)$ over the entire film since eqn. 6 would no longer remain valid. Therefore we take a value just below, i.e. $I_{ac} \sim 8.7\, mA$ and first obtain $J_S^{ac}(r)$ as a function of $r$ using constant value of $\lambda_p = \lambda_p(0)$. Now to invoke the $J_S^{ac}(r)$ dependence of $\lambda_p$, we choose a trial value for $J_c$ as $2.65 \times 10^8 A/m^2$, which is marginally larger than the maximum value of $J_S^{ac}(r)$ (typically $J_S^{ac}(r = 1.5\, mm)$) obtained from the previous constant-$\lambda_p$ simulation. This is done so as to keep the value of $\lambda_p$ obtained from eqn. 6 real in all steps during the iteration. Using this value of $J_c$ and $J_S^{ac}(r)$ we calculate $\lambda_p(r)$. Then using the obtained $\lambda_p(r)$ refine the value of $J_S^{ac}(r)$ and keep repeating this process iteratively. We observe that this iterative procedure leads to an oscillatory convergence of $J_S^{ac}(r)$ as is shown in *Figure 3.6a*. To achieve a faster convergence we therefore take the average of the values of $\lambda_p(r)$ in the last two steps of the iteration and solve to obtain the new values for $J_s^{ac}(r)$. This modified algorithm leads to a faster convergence as shown in *Figure 6b*.

We observe that, as expected, at small $r$ the final value of $J_s^{ac}(r)$ is similar to the value calculated for constant $\lambda_p$ but becomes significantly smaller at larger values of $r$. Therefore to obtain $J_c$ for our sample we again take a new trial value which is slightly higher than the maximum value of $J_s^{ac}(r)$ obtained from the last iteration. Then we run the iterative procedure again, starting from with $\lambda_p(r)$ obtained from the last step of the previous iteration, to obtain $J_s^{ac}(r)$ for this new trial $J_c$. When this step is repeated several times we observe that below $J_c \sim 1.45 \times 10^8 A/m^2$ the iterative procedure fails to converge. This signals the depinning threshold where eqn. 3.6 is no longer applicable. We observe that our iterative procedure converges till the maximum value of $J_s^{ac}(r)$ is within 20% of $J_c$.

To cross-check the consistency of the procedure we fix $J_c \sim 1.45 \times 10^8 \, A/m^2$ for which we get a stable solution for $I_{ac} \sim 8.7$ mA and run the simulation from the other end by gradually increasing the value of $I_{ac}$ from 1 mA in steps of 0.1 mA and run the iteration for each value. For each $I_{ac}$ we start the iteration using the converged value of $\lambda_p(r)$ obtained for the previous value of $I_{ac}$. This gives us $J_s^{ac}(r)$ for different $I_{ac}$ (*Figure 3.6c*). At $I_{ac} = 8.8$ mA the iterative procedure fails to converge (*Figure 6d*). We thus conclude that this is the physical value of $J_c$ in our film with an error of about 20%.

This simulation not only gives us the insight about the depinning current, but shows that the profile of $J_s^{ac}(r)$ which increases almost linearly with small $r$ and saturates for $r > 0.9 \, mm$. As $I_{ac}$ is increased, this saturation current reaches $J_c$ for some critical $I_{ac}$. This critical value $I_{ac} \approx I_{ac}^p$ is where the outer annulus of the VL for $r > 0.9 \, mm$ starts to slide over the antidot array.

### 3.5.2 Calculation of depinning energies

Based on the value of $J_c \sim 1.2 - 1.4 \times 10^8 \, A/m^2$ at the first matching field, $H_m^1$, the force exerted on a single vortex for $J_s^{ac} \sim J_c$ is $F_{ac}^c = J_c \Phi_0 t \sim 6.4 - 7.5 \times 10^{-15}$ N. Next we estimate the confining potential created by the superconductor surrounding an antidot using,[20,21]

$$E(r) = \frac{\Phi_0^2}{2\pi\mu_0\Lambda} \ln\left[\frac{2w}{\pi\xi_{GL}} \sin\left(\frac{\pi\left(r - \frac{d}{2}\right)}{w}\right)\right] \quad (3.6)$$

where, $\Lambda = \frac{2\lambda^2}{t}$ is Pearl's penetration depth and $w$ is the width of superconducting strip separating two antidots ($\sim 60\ nm$); with a constraint that the vortex is completely inside the superconductor, i.e., $r > 2\xi_{GL} + d/2$. The potential is such that the restoring force one each vortex is maximum at the edge of the strip and is minimum at the centre of the strip. Thus the maximum pinning force is given by, $F_p = \frac{\partial E}{\partial r}\big|_{r\sim(2\xi_{GL}+d/2)} \sim \frac{\Phi_0^2}{4\pi\mu_0\Lambda\xi_{GL}} = 4.41 \times 10^{-14}\ N$, (assuming $\xi_{GL}\sim 8\ nm$[22]) which is one order of magnitude larger than $F_{ac}^c$. However, possibility of thermally activated hopping can be ruled out from energetic considerations. A vortex has to cross a potential barrier to hop from one antidot to the other, which is given by, $E(r)|_{max} = \frac{\Phi_0^2}{2\pi\mu_0\Lambda}\ln\left(\frac{2w}{\pi\xi_{GL}}\right) \sim 7\ meV$. On the other hand thermal energy at *1.7 K* is $k_BT \sim 0.15\ meV \ll E(r)|_{max}$, which eliminates the possibility of thermally activated hopping. Therefore the delocalization of the vortices at $I_{ac}^p$ can only be understood in terms of a collective delocalization driven by interplay of inter-vortex interaction and localization.

The same conclusion can be obtained if one considers the upper bound of vortex displacement due to ac magnetic field. If we assume that the pinning is infinitesimally small for the time being, the effect of ac magnetic field becomes non-negligible. In this case, the vortex lattice constant is given $a_v = \left(\frac{2\Phi_0}{\sqrt{3}B}\right)^{1/2}$, where $B = B_{dc} + B_{ac}\sin\omega t$ and the role of the ac field is to introduce periodic variation of $a_v$ centred about its value for $B = B_{dc}$. Since the vortices enter and exit from the periphery of the sample, the vortex at the centre of the sample remains at rest whereas the remaining vortices undergo periodic oscillations along the radial direction whose amplitude increases as we go towards the periphery (see section 3.2.3 for a coarse variation of the force on the vortices). Thus for a sample of radius $R = 1.5\ mm$, the maximum displacement from its mean position by a vortex located close to the periphery of the sample is given by, $D_{max} = \left(\frac{R}{a_v}\right) \times \Delta a_v$, where, $\Delta a_v = \left(\frac{2\Phi_0}{\sqrt{3}B_{dc}}\right)^{1/2} - \left(\frac{2\Phi_0}{\sqrt{3}(B_{dc}+B_{ac})}\right)^{1/2} \approx a_v\left(\frac{B_{ac}}{2B_{dc}}\right)$. For the first matching field, $a_v = 107\ nm$. Therefore, for $B_{ac}\sim 62\ mOe$ corresponding to the first jump in mutual inductance, $D_{max}/a_v \sim 0.2$, which shows that $D_{max}$ is less than half the lattice constant. Therefore, the ac field cannot cause a vortex hop even for infinitesimally small pinning. In the real system the pinning strength is way too large such that $D_{max}$ will further be suppressed. This is another way to understand why the Mott to metal transition is a collective phenomenon.

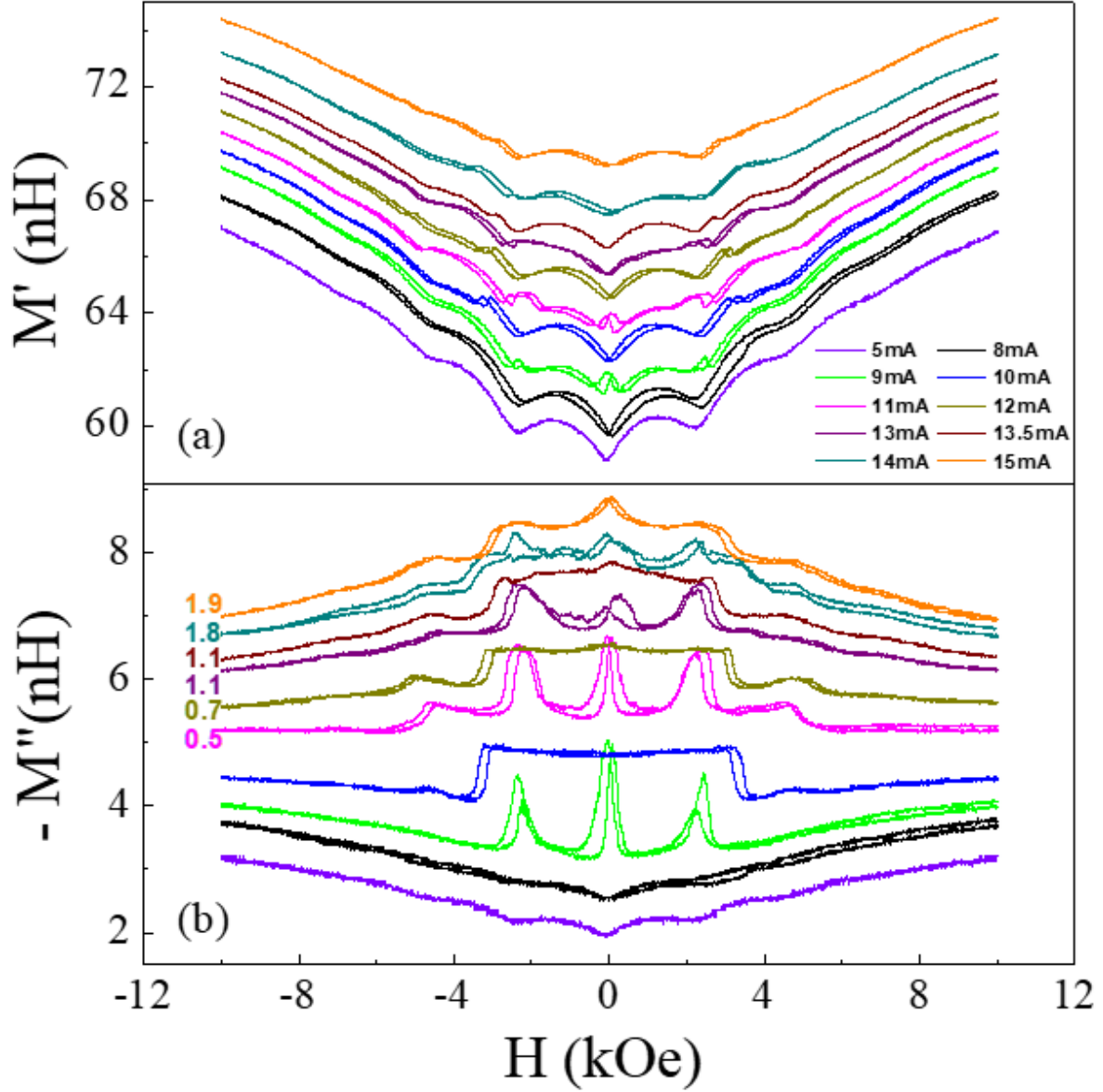

*Figure 3.7*: (a) $M' - H$ and (b) $- M'' - H$ at different $I_{ac}$ at 1.7 K. The values of $I_{ac}$ correspond to the vertical dashed arrows shown in *Figure 5a*. The data are recorded by sweeping $H$ from 10 to −10 kOe and back. A small hysteresis between positive and negative field sweeps is observed in all curves. In (b) some of the curves have been shifted upwards for clarity, by an amount mentioned next to the curve in the same colour.

## 3.6 Amplitude dependence of matching effect

To get the full picture we look at magnetic field evolution of $M'$ and $M''$ for different $I_{ac}$ in *Figure 3.7a* and *Figure 3.7b*. Here we vary $H$ from $-10\ kOe$ to $+10\ kOe$ and back at 1.7 K corresponding to $I_{ac}$ values marked by the vertical dashed arrows in *Figure 3.5a*. We observe that for $I_{ac} < I_{ac}^p \sim 8.8\ mA$, $M' - H$ curves are qualitatively similar, showing minima at the matching field showing less compressible phase at integer filling fraction. $(-M'') - H$ also shows shallow dips at the matching field signifying least vortex movement at matching fields. Both of these are consistent with the elastic regime where the vortex lattice becomes more rigid at $H_m$ and therefore each vortex oscillates with a much smaller amplitude. In this regime no vortex hop occurs and hence the only dissipation comes from Bardeen-Stephen dissipation due to oscillating vortices. This behaviour drastically changes at $I_{ac} = 9\ mA \sim I_{ac}^p$, where $M'$ develops small peaks at $H_m$, showing that the vortex lattice is more delocalized at the matching field. In $-M''$ the signature of the dissipation is more pronounced at $H = H_m$ which shows more dissipation at matching field. However as we move away from the matching field the elastic response is restored partially. As we approach $I_{ac} = 10\ mA$, which lies between $I_{ac}^p$ and $I_{ac}^1$, the vortices are delocalized even away from $H_m$ and the peaks in $M'$ shifts to higher field. At the same excitation current, $-M''$ becomes quite featureless and flat. This can be understood from the fact that even after delocalization of the vortices in the outer annulus, the inner core remains elastic, which continues to contribute to the matching effect while the sliding of the annular region adds a nearly field independent background. But in terms of dissipation the picture is inverted because $-M''$ will mostly be dominated by the sliding of the annular region, while the core region contributes little to it. The nearly constant $-M''$ shows that at this excitation the vortex lattice in the outer annular region remains delocalized over the entire field range between $+H_m$ and $-H_m$. We observe that this qualitative behaviour repeats between $I_{ac}^1$ and $I_{ac}^2$, $I_{ac}^2$ and $I_{ac}^3$, and so on though the matching effect in $M'$ weakens as a larger fraction of the vortex lattice gets delocalized.

One interesting observation is the small difference between +ve sweep and –ve sweep of fields, similar to hysteresis. This can in principle have two origins, one of which is unrelated to the physics of the problem, but might be instrument-related, where the positive and negative sweeps keep a different reminiscent field which shifts the true zero to $\pm \delta H$ which gives the discrepancy between the two sweeps. To eliminate this effect, the field ramping has been very slow (0.003 kOe/s) which in principle precludes this scenario. However the other scenario is related to the hysteresis in the system itself which manifests itself as different fractionation of the vortex lattice in different initial conditions. But this conjecture needs to be investigated more, and if possible using microscopic techniques.

## 3.7    Zero field anomaly

In the previous section the evolution of $M'$ and $M''$ at zero field shows an evolution which is similar to that at matching fields. In *Figure 3.8* we compare $-M''$ as a function of $I_{ac}$ at zero field and at the first matching field, $H_m \sim 2.37\ kOe$. The interesting fact is that both curves at $H = 0$ and $H = H_m$ show similar jumps. Besides, the $H = 0$ data show additional smaller jump which is absent is $H = H_m$. The most likely explanation of this observation is that the $H = 0$ state is not empty but is populated with vortex and antivortex pairs in the antidot array, which are expected to spontaneously form in a 2D superconductor. These vortex-antivortex pairs can also get delocalized under ac excitation, but its dynamics will also be governed by annihilation of vortex-antivortex. Although this is not captured in our model, the existence of vortex-antivortex pairs in zero field in an antidot array has also been conjectured in an earlier study[23] studying transport current driven depinning of the vortex lattice. Here zero field state showed similar behaviour as that of at matching fields.

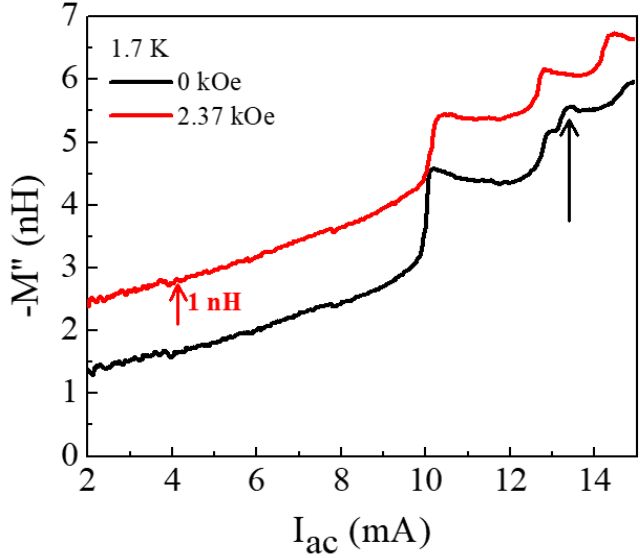

*Figure 3.8*: Variation of $-M''$ as a function of $I_{ac}$ ($mA$) for 0 kOe and first matching field $H_m^1 = 2.37\ kOe$ (shifted by $1\ nH$ upwards for better visibility) at 1.7 K. Surprisingly, both of them show jumps in $-M''$, while zero field data shows an additional feature marked by the black arrow.

## 3.8 Difference of depinning current from Little Parks-like effect

In this study our observation is that the current at which the vortex lattice dynamically gets delocalized, $I_{ac}^p$ is lowest at the matching field and hence $J_c$ also has the same behaviour. This is in contrast with variation of critical current very close to $T_c$, where critical current shows a maxima at the matching field.[2,24,25] This difference can be traced back to the difference between matching effect and Little-Parks-like quantum interference effects described in section 3.1.1. When the width (*w*) of the superconducting strip between two antidots is less than $\xi_{GL}$, quantum interference cancels the supercurrent around each loop and hence superconducting order parameter is enhanced to its zero field value, which is reflected in increment in $T_c$ and critical current at matching field. This effect is even dominating close to $T_c$ since $\xi_{GL}$ diverges at these temperatures. In contrast in the limit of our experiment, $\xi_{GL} \ll w$ such that the quantum interference effects play insignificant role. Hence, the delocalization current in our case is only because of matching effect and dominated by the variation of the dynamic quantities such as compressibility. In the light of this, it is natural to have a minimum delocalization current if it is driven by dynamic collective depinning.

## 3.9 Discussion

To summarise, we have shown that a Mott-like state of the vortex lattice can be realized in superconducting films with an antidot array, where each vortex is localized in an antidot and the state is characterized by its least compressibility. This Mott-like state can be transformed with application of ac magnetic field perturbation into a metal-like state where each vortex is delocalized. The fact that this transition occurs well below the conventional depinning threshold is a proof that the transition is a collective effect where strength of individual pinning and inter-vortex interaction play the controlling roles. From this standpoint our observation is on similar grounds compared to Ref. 9 where vortex Mott insulator-to-metal transition is induced by a transport current. However, the measurement scheme in our measurement being different, we can directly extract compressibility and dissipation, while the latter is analogous to the differential resistance measured in Ref. 9. Nonetheless, further investigations are required to understand whether our observation represents a true dynamic phase transition in light of scaling of the data which was adopted in previous reports. We also need to understand why the characteristic jumps in $M$ disappear well below $T_c$. Further study using microscopic tools like STS/M will also be helpful to understand this physics.

# Chapter IV: Magnetic field evolution of the superconducting state in strongly disordered NbN thin films

The vortex lattice in presence of strong random pinning is a completely different arena compared to clean BCS-type conventional superconductors. Disorder in its strongest limit plays a crucial role to induce spatial inhomogeneity of order parameter and creates preformed domains or puddles of superconducting regions separated by regions where superconductivity is suppressed. This occurs because disorder brings down phase stiffness and the resultant puddles carry different phases. In moderate disorder limit these puddles can cooperate with each other via Josephson tunnelling to give rise to zero resistance state. However, in this chapter we'll see that this zero resistance state breaks down with magnetic field, as magnetic field starts to play the role of disorder. The central theme of this chapter is to show an alternate route of destroying superconductivity with magnetic field, in contrary to the clean limit case where with increasing magnetic field vortex density increases and at $H = H_{c2}$ the vortices start to overlap with each other to go into the normal state.

## 4.1 Sample details

The samples used in this study are NbN thin films grown using PLD and have been transferred to the STS/M using the vacuum suitcase. The sample deposition technique has been discussed in section 2.1.3. We have used two samples, one[1] with $T_c \sim 9\ K$ and another[2] with $T_c \sim 1.65\ K$. The first one is moderately disordered while the second one is strongly disordered. However from a Ginzburg-Landau perspective both of these samples are in the dirty limit, such that $\xi_{BCS} = \frac{\hbar v_F}{\pi \Delta} \gg l$. Here $v_F$ is the Fermi velocity and $l$ is the mean free path. The transport measurements were performed after the STS/M measurements were done.

### 4.1.1 Quantification of strength of disorder

It has been extensively shown[3] in case of NbN thin films that disorder in the system can be quantified by the dimensionless quantity, $k_F l$, where $k_F$ is Fermi wave-vector. On the basis of $k_F l$ values varying from $k_F l \sim 10$ to $k_F l \sim 0.42$, disorder in NbN samples can be tuned over a wide range. As the disorder is increased the superconductor progressively passes through three regimes: Regime I ($10 \gtrsim k_F l \gtrsim 3$) where superconducting energy gap in zero field vanishes exactly at the same temperature where resistance appears; Regime II ($3 \gtrsim k_F l \gtrsim 1$) where superconducting gap in zero field vanishes at $T^*$ and resistance appears at $T_c$ and $T^* > T_c$. The region between these two temperatures where superconducting order parameter is finite but zero resistance state is lost is called the pseudogap state. Regime III ($k_F l \lesssim 1$), where superconductivity is completely suppressed down to 300 mK. The NbN samples used here have $k_F l \sim 4$ for $T_c \sim 9~K$ and $k_F l \sim 1.5$ for $T_c \sim 1.65~K$.[4]

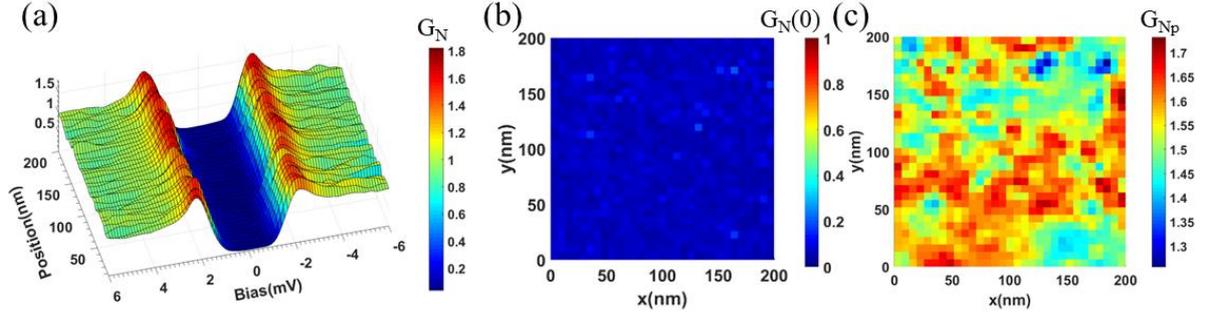

*Figure 4.1*: (a) Representative $G_N(V)$ vs. $V$ spectra along a 200 nm line at 350 mK. (b) $G_N(0)$ map over a 200 $nm \times 200~nm$ area, which shows uniform values of $G_N(0)$ over the space. (c) $G_{Np}$ map over the same 200 $nm \times 200~nm$ area showing puddle-like structures of high and low values of $G_{Np}$.

## 4.2 Magnetic field induced emergent granularity

In this section we shall explore effect of magnetic field, which makes the superconducting state inhomogeneous, in a manner similar to what disorder of much larger strength alone would have done. The experimental observation is backed by numerical simulations which shows that flux tubes enter a disordered superconductor at locations where disorder has partially suppressed superconducting correlations. Thus disorder creates a network of weak links where vortices enter and suppresses superconducting order parameter even more. The resultant superconducting state persists in regions of tens of nanometres in size, where superconducting order parameter is still finite separated by chains of vortices where order parameter is suppressed. Consequently the system exhibits a field induced pseudogap that progressively widens as the magnetic field is increased.

### 4.2.1 Zero field superconducting state

We start with measuring tunnelling conductance spectra using STS/M ($G(V)$ vs $V$) over a 200 $nm \times 200~nm$ area at 450 mK on the sample with $T_c \sim 9~K$. The tunnelling conductance here shows uniform energy gap and presence of coherence peak at the gap edge over the entire area. In *Figure 4.1a* we show such representative spectra of a 200 $nm$ line. Now to look at the zero bias conductance, $G_N(0)$ and coherence peak height, $G_{Np}$ variation, first we normalize the tunnelling spectra using: $G_N(V) = G(V)/G(V = 3.5~mV)$ (as shown in section 2.3.1). In *Figure 4.1b* we show $G_N(0)$ map and its histogram (in *Figure 4.4e*), which shows that $G_N(0)$ has a very narrow distribution sharply peaked at 0.05, consistent with a fully gapped superconducting state. However, $G_{Np}$ map shown in *Figure 4.1c* shows a large spatial variation. It has been theoretically shown using quantum Monte Carlo simulations that in a disordered superconductor $G_{Np}$ provides a measure of the local superconducting order parameter.[5] It is visible in the map that $G_{Np}$ varies smoothly from 1.1 to 1.8 forming an inhomogeneous structure varying tens of nanometres in size. However the presence of a prominent coherence peak at all points signify a finite order parameter.

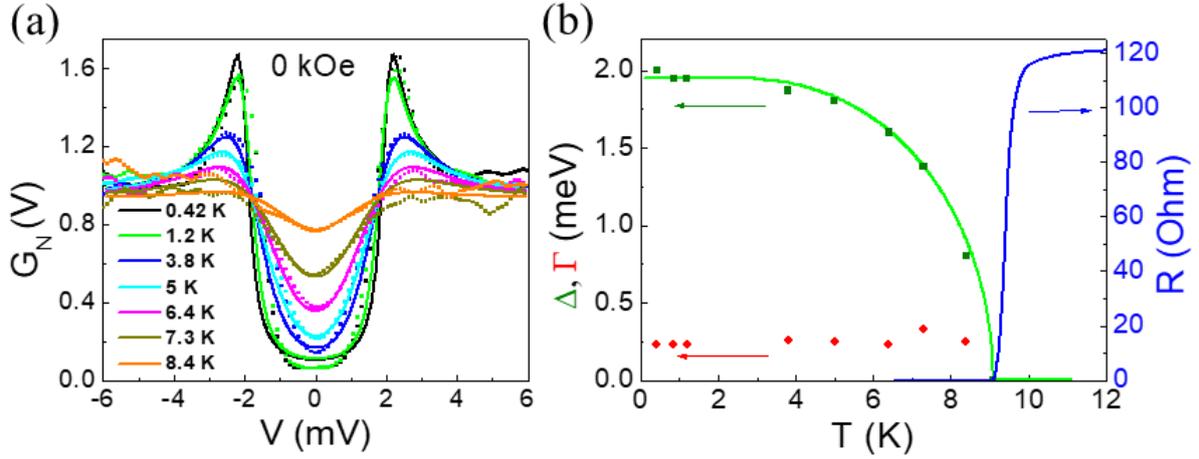

*Figure 4.2*: (a) Fitting of $G_N(V)$ vs. $V$ spectra with tunnelling equation for different temperatures, where Δ and Γ plays the roles of fitting parameters. (b) Variation of Δ (green dots) and Γ (red dots) as a function of temperature, along with BCS fit of $\Delta(T) - T$ shown in green line. Resistance variation as a function of temperature is shown in blue line. Both $T_c$ and $T^*$ are same here, $9\,K$.

To check the nature of the superconducting state in detail, we focus on temperature variation of BCS gap Δ. To obtain Δ, we fit our $G_N(V)$ vs. $V$ curves following the procedure discussed in section 2.3.1.1, as is shown in *Figure 4.2a*. The fitting incorporates a phenomenological broadening parameter Γ to take into account non-thermal sources of broadening in the DOS. Although Γ varies between 0.23 and 0.33 and this relatively large value of Γ signifies the presence of regions where coherence peak is suppressed Next we compare the variation of $\Delta(T)$ vs. $T$ with the universal BCS curve in *Figure 4.2b*. We also compare temperature variation of resistance and observe that the gap vanishing temperature, $T^*$ and temperature at which zero resistance state is destroyed, $T_c$ are same.

Similar measurements done on the $T_c \sim 1.65\,K$ sample shows that even at zero field the $G_N(0)$ maps show fragmented structure of regions of high and low values of $G_N(0)$. The histogram of the $G_N(0)$ map at zero field also shows a broad distribution around $\sim 0.45$. This is shown in *Figure 4.14*.

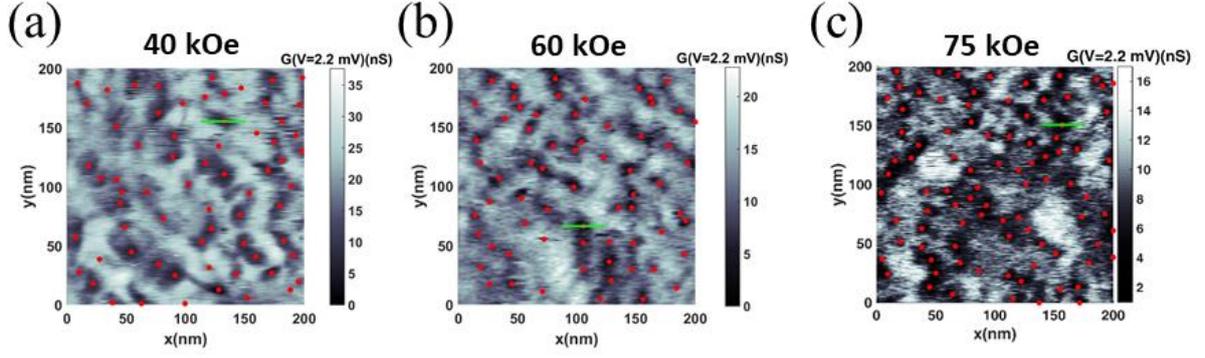

*Figure 4.3*: (a)-(c) Differential conductance map ($\Delta G(nS)$) for bias $V = 2.2\ mV$ for magnetic fields 40, 60 and 75 kOe at 350 mK. The red dots show the positions of the minima of $\Delta G$, or the positions of the vortices.

### 4.2.2 Emergence of granular superconducting state in magnetic field

When magnetic field is applied, a type-II superconductor develops vortices for $H > H_{c1}$. As is discussed in section 2.3.2 that the vortices can be identified in STS/M measurements from the recording the conductance map where bias is fixed at coherence peak, $V = 2.2\ mV$. The $\Delta G(V = 2.2\ mV)$ maps shown in *Figure 4.3* study evolution of the vortices at 350 mK as a function of magnetic field for the $T_c \sim 9\ K$ sample, where positions of the vortices, *viz.* minima in the maps, are shown as red dots. It is interesting to note that with increasing magnetic field, the maximum value of conductance even far from the vortex, decreases. This is a direct result of the orbital current, which for an extreme type-II superconductor extends well beyond the vortex core, since, $\lambda_L \gg \xi_{GL}$. Nonetheless, the conductance maps were acquired at the same area for different magnetic fields at zero field cooled (ZFC) state.

We observe that at 40 kOe, the vortices enter the superconductor forming laminar structure separating regions where $G(V = 2.2\ mV)$ is high. As field is increased, further entry of vortices progressively widens the regions with suppressed coherence peaks, and the laminar structure becomes denser. However, one discrepancy should be noted here, which is the number of vortices (minima of conductance map or number of red dots in general) does not follow the general rule of quantization of fluxoid. Number of vortices for 60 and 75 kOe multiplied by $\Phi_0$ does not account for the entire flux trapped in that specific region. This discrepancy arises either because of our inability to detect two closely spaced vortices, or a multiple quanta-vortex or because the route of vortex creation might be different which precludes the possibility of fluxoid quantization.

Since the manifestation of vortices locally suppress the superconducting order, the inhomogeneous distribution of vortices produce boundaries of suppressed superconductivity which separate the superconducting puddles. To visualize the superconducting state in more details, we show in *Figure 4.4*, data from tunnelling conductance spectra over the same 200 $nm$ × 200 $nm$ area. *Figure 4a-d* show the $G_N(0)$ maps at different fields up to 75 kOe corresponding to the same area as in *Figure 4.3a-c*. We observe that with increase in field the superconducting state develops large inhomogeneity, forming regions where $G_N(0)$ is large and regions where $G_N(0)$ is small. This is also reflected in the distribution of $G_N(0)$, which with increase in field develops large tails shown in *Figure 4.4e-h*. *Figures 4.4i-k* show the coherence peak height maps corresponding to the same fields. We observe an inverse correlation of the $G_{Np}$ maps with the $G_N(0)$ maps, implying that in regions where $G_N(0)$ is large, the coherence peak is suppressed. The anticorrelation is also apparent from the two-dimensional histogram of $G_N(0)$ and $G_{Np}$ which shows a negative slope over a large scatter, which suggests that the anticorrelation is not perfect. We quantify the anticorrelation using the cross correlator,

$$I = \frac{1}{n} \sum_{i,j} \frac{\left(G_N^{i,j}(0) - \langle G_N(0) \rangle\right)\left(G_{Np}^{i,j} - \langle G_{Np} \rangle\right)}{\sigma_0 \sigma_p} \qquad (4.1)$$

where, $\sigma_0$ and $\sigma_p$ are the standard deviations in the values of $G_N(0)$ and $G_{Np}$, respectively; $i, j$ refer to the pixel index of the image; and $n$ is the total number of pixels. We obtain $I \sim -0.15\ to -0.2$ where $I = -1$ implies perfect anticorrelation. This is qualitatively similar to earlier observation in strongly disordered NbN samples in zero field. The weak anticorrelation suggests that $G_N(0)$ is probably not governed by the local superconducting order parameter alone. As expected, the vortices are preferentially located in the regions where $G_N(0)$ is high, i.e., superconducting order parameter is suppressed.

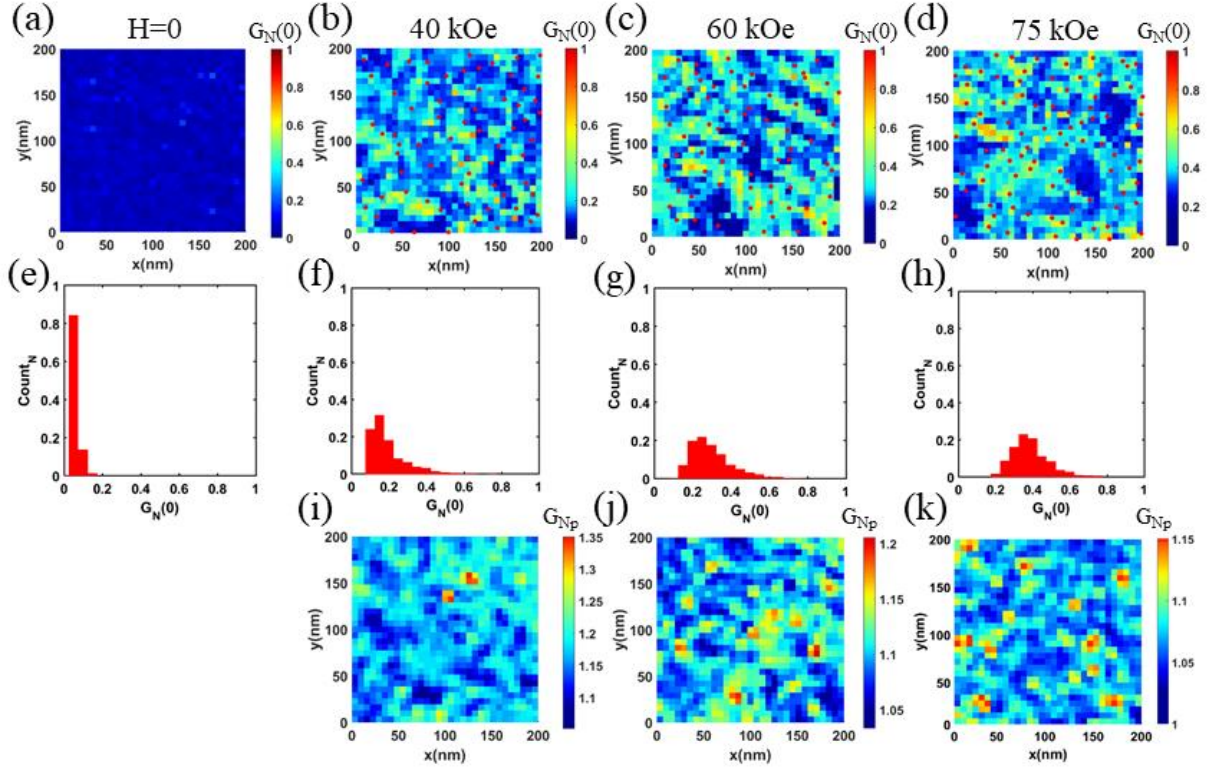

*Figure 4.4*: (a)-(d) $G_N(0)$ maps over a 200 $nm$ × 200 $nm$ area at 350 mK for 0, 40, 60 and 75 kOe. The red dots are determined in *Figure 3 (a)-(c)* as the positions of the vortices. The red dots are only positioned on regions of higher values of $G_N(0)$. (e)-(f) Histograms of $G_N(0)$ maps obtained from the images above. The histograms broaden monotonously with field. (i)-(k) $G_{Np}$ maps over the same area as (b)-(d). The regions with high values of $G_N(0)$ can be associated with low values of $G_{Np}$.

In case of the sample with $T_c \sim 1.65\ K$, the granularity is present even in zero field, which gets enhanced with increasing magnetic field, shown in *Figure 4.14*. The width of the histograms, shown in *Figure 4.14c*, remains almost same from 0 kOe to 60 kOe and shows a continuous increment above 60 kOe.

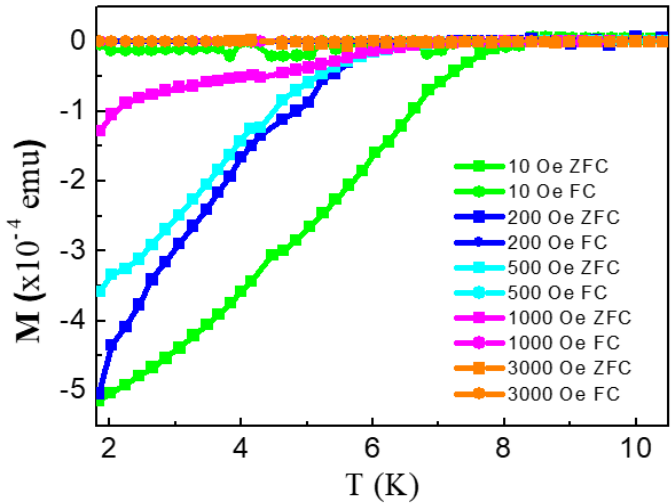

*Figure 4.5*: Comparison of $\mathcal{M} - T$ measurements between zero field cooled (ZFC) and field cooled (FC) states in different magnetic fields. Up to 1000 Oe, the FC and ZFC states show a significant difference, however from 3000 Oe and above the difference is absent.

### *4.2.2.1    FC and ZFC states*

Since the vortices enter the thin film through the periphery in general, for strong pinning it is possible that full flux penetration of the sample is not happening and because of this in 60 and 75 kOe, the number of vortices might not portray a true fluxoid quantization. However, this scenario can be checked with $\mathcal{M} - T$ measurement in the ZFC and FC state for different magnetic fields using a SQUID magnetometer ($\mathcal{M}$ is magnetization). The ZFC state is created by applying the magnetic field after cooling the sample to the base temperature (1.8 K) in zero field. The FC state is created by applying the field at 15 K and cooling the sample to the base temperature in the magnetic field. The $\mathcal{M} - T$ measurements are carried while warming up the sample from this initial ZFC or FC state. In *Figure 4.5* it is shown that, at 10 Oe the ZFC curve shows pronounced diamagnetic response whereas the FC curve is flat and very close to zero as expected for a strongly pinned type-II superconductor. However, as the magnetic field is increased the difference between the FC and ZFC curves progressively decreases, and at 3 kOe the two become indistinguishable. Thus beyond this field the flux completely enters the ZFC state and the role of flux pinning on the entry of flux in the ZFC state is negligible.

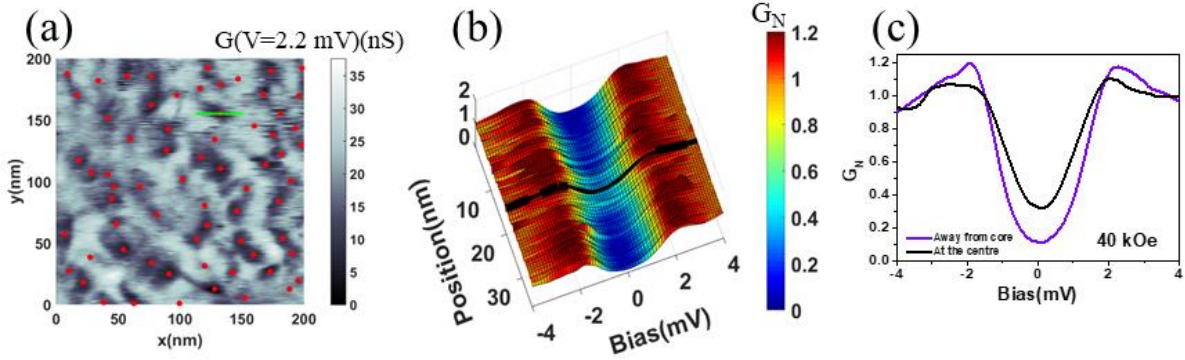

*Figure 4.6*: (a) $\Delta G(V = 2.2\ mV)$ map for 40 kOe at 350 mK (exactly same as *Figure 3a*). The green line across one red dot (minima in $\Delta G$) is studied in (b) as $G_N(V)$ vs. $V$ spectra. The spectra at the centre of the red dot (vortex) is shown by the black line. (c) The spectra at the centre of the vortex (black line) is compared with the spectra away from the core (violet line).

### 4.2.3  Nature of vortex core

In section 2.3.3 it is shown how the vortex core look like for a clean NbSe$_2$ sample. The vortex core in such clean systems contain the Caroli-de Gennes-Matricon bound state.[6,7] However in dirtier system often the bound state is not found due to scattering of the normal electrons inside the core and the tunnelling spectra becomes flat, $G_N(V)\sim 1$. In NbN, on the other hand we observe in *Figure 4.6b*, a line cut across a vortex line shows a soft gap surviving inside the core. *Figure 6c* compares a spectra at the centre of the core and away from the vortex line and one can see that the coherence peak is suppressed inside the vortex line, even though the pairing amplitude remains finite. This scenario is similar for both the $T_c \sim 9\ K$ and $T_c \sim 1.65\ K$ samples.

This prompts us to think about a different mechanism of vortex creation in these materials. In regular materials such as NbSe$_2$, when a flux line enters the superconductor, supercurrent revolves around the core starting from $r \sim \xi_{GL}$. This size of the vortex core can be naively derived from the fact that if the kinetic energy of the circulating Cooper pairs reaches the binding energy of the pair, the pairing breaks down, pushing pairing amplitude to zero at the vortex core. However, in the disordered systems, the superfluid stiffness, $J_S \sim \Delta$ and flux lines pass through regions where order parameter is weak. And in the process it doesn't always need to go through the pair breaking mechanism, rather the phase can be twisted locally to destroy local superconductivity. This would allow pairing amplitude to be finite inside the core.

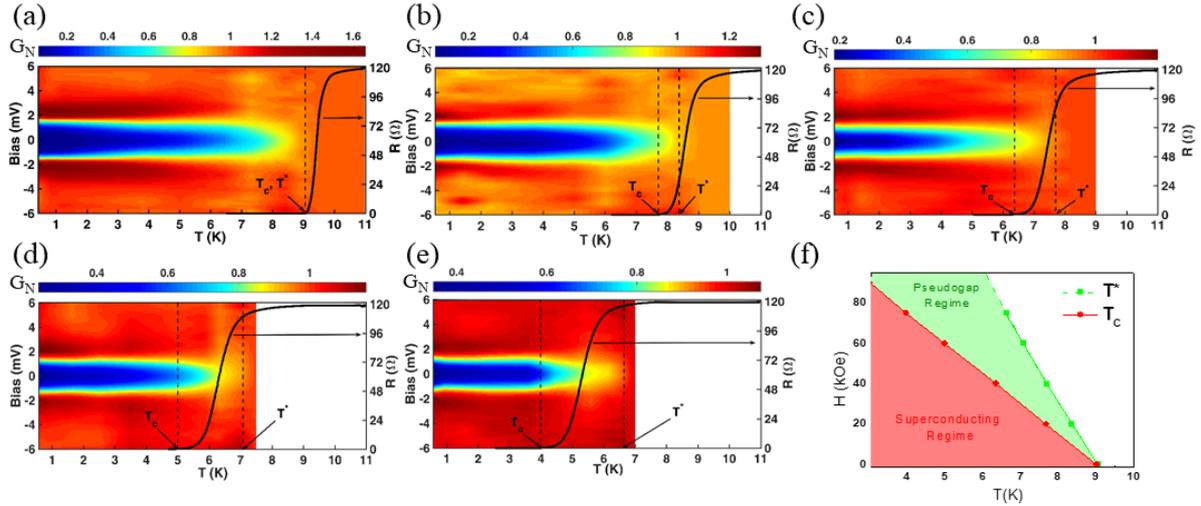

*Figure 4.7*: (a)-(e) The colour maps show $G_N(V)$ vs. $V$ as a function of temperature for different magnetic fields, 0, 20, 40, 60 and 75 kOe. $R - T$ data are superposed on the gap variation data. The temperatures $T_c$ and $T^*$ are shown by the vertical dashed lines. For 0 kOe, $T_c = T^*$. (f) $T_c$ and $T^*$ lines are plotted in the $H - T$ parameter space, where the pseudogap regime progressively widens in presence of magnetic field.

### 4.2.4 Field induced pseudogap state

In this section we investigate the temperature evolution of superconducting state in presence of magnetic field. For this we take averaged $G_N(V) - V$ spectra over a 200 $nm$ × 200 $nm$ area for different temperatures at different magnetic fields and plot them along with corresponding temperature variation of resistance in *Figure 4.7a-e*. Here we define the temperature where $G_N(0)$ is 95% of its normal state value (~1) as $T^*$ or the gap vanishing temperature and $T_c$ as the temperature where resistance is 0.05% of its normal state value. We observe that although at zero field $T^* = T_c$, with increasing magnetic field $T^* > T_c$. Plotting $T_c(H)$ and $T^*$ in the *H-T* parameter space in *Figure 4.7f*, we observe that the pseudogap phase progressively widens as magnetic field is increased.

It is however a valid question to ask if the observed pseudogap is caused by a local distribution of superconducting and non-superconducting regions which in average shows a gap. To rule out this possibility, we have separately tracked temperature dependence of the tunnelling spectra in small areas inside and outside a vortex patch in *Figure 4.8b-c*. From this we plot variation of $G_N(0)$ for both of these regions in *Figure 8d* and observe that at both locations $G_N(0) \to 1$ at the same temperature.

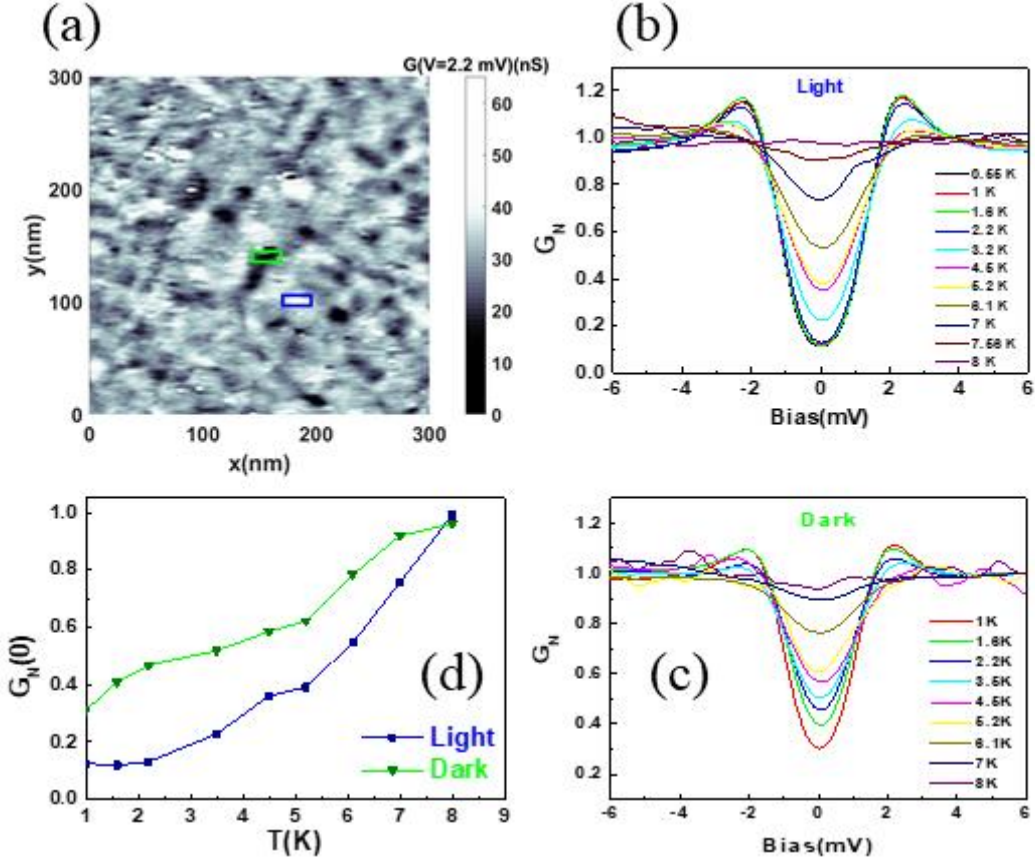

*Figure 4.8*: (a) $\Delta G(V = 2.2\ mV)$ map over $300\ nm \times 300\ nm$ area at 40 kOe at 350 mK, where the green box surrounds a laminar vortex and the blue box surrounds a superconducting region. (b)-(c) Evolution of bias spectra in the area corresponding to the blue (green) box, i.e. superconducting region (laminar vortex) for different temperatures. (d) Variation of $G_N(0)$ values as a function of temperature for the superconducting and laminar vortex region. In both cases $G_N(0) \to 1$ at the same temperature.

The definition of $T_c$ is verified from comparing in *Figure 4.9*, the $R - T$ data with $M - T$ data acquired from two-coil mutual inductance measurements, where in the latter $T_c$ is defined as the temperature where shielding response starts to appear due to appearance of a global superconductivity. Hence we prove that magnetic field induced granularity gives rise to a pseudogap phase similar to disorder alone.

### 4.2.5 Comparison with numerical simulation

Numerical simulation was carried out[1] by *Anurag Banerjee* and *Amit Ghosal, IISER, Kolkata*, following an attractive Hubbard Hamiltonian,[8,9] which in presence of disorder and magnetic field has the form:

$$H = -t \sum_{<i,j>,\sigma} e^{i\phi_{ij}} c_{i\sigma}^{+} c_{j\sigma} - |U| \sum_{i} \hat{n}_{i\uparrow}\hat{n}_{i\downarrow} + \sum_{i,\sigma}(V_i - \mu)\hat{n}_{i\sigma} \quad (4.2)$$

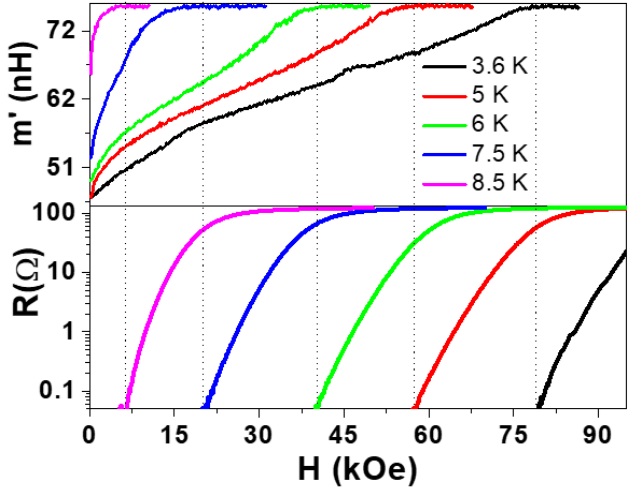

*Figure 4.9*: Comparison of real part of mutual inductance, $m' - H$ data with $R - H$ data for different temperatures. The criterion of $H_{c2}$ (or equivalently $T_c$) from $R - H$ (or $R - T$) is determined as the field (or temperature) where $R$ is 0.05% of its normal state value. This criterion matches with $m' - H$ data, where $m'$ starts to fall off due to circulating shielding current.

where, $c_{i\sigma}^+$ ($c_{j\sigma}$) is the creation (annihilation) operator of an electron with spin $\sigma$ at site $i$ of a two-dimensional square lattice; $\hat{n}_{i_\sigma} = c_{i\sigma}^+ c_{i\sigma}$ is the occupation number operator of site $i$ with spin $\sigma$; and the phases $\phi_{ij} = \frac{\pi}{\Phi_0} \int_i^j \vec{A}.\vec{dl}$ are the Peierls factor of an applied orbital magnetic field. In the simulation Landau gauge, $\vec{A} = Bx\hat{y}$ was used. The attraction $U = 1.2t$ induces s-wave superconductivity in the system and the large value of $U$ (beyond the weak coupling limit) gives the coherence length $\xi \sim 5 - 6$ lattice spacing in presence of disorder, which is necessary to work on a small $36 \times 36$ two-dimensional grid available at the time of simulation. Chemical potential $\mu$ fixes the average density, $\rho = \frac{1}{N} \sum_i \hat{n}_i$ and $\rho$ is fixed at 0.875 (although $\rho$ equals to 1 gives strongest superconductivity, it also gives rise to locally ordered charge density wave even in the presence of disorder, which can act as a competing order against superconductivity; hence $\rho = 0.875 \lesssim 1$ is used). Disorder at each site is given by $V_i$ which is chosen as an independent random variable from a uniformly distribution between $-V$ and $+V$. Value of $V$ is chosen to be $V = 0.5\,t$, such that qualitatively the histogram of $D_p$ (peak height of the single particle DOS, $D(E)$) matches with the histogram of experimentally acquired $G_{Np}$. In *Figure 4.10c*, normalized distribution of $G_{Np}$ at 450 mK, defined as $\tilde{G}_{Np} = \frac{G_{Np} - G_{Np}^{min}}{G_{Np}^{max} - G_{Np}^{min}}$ is compared with normalized distribution of $D_p$, defined as $\tilde{D}_p = \frac{D_p - D_p^{min}}{D_p^{max} - D_p^{min}}$.

Next we study magnetic field evolution of the superconducting state in *Figure 4.11*. In *Figure 4.11a-c*, we show spatial variation of the phase of the superconducting order parameter $\phi$ for number of flux quanta, $n = 2, 4, 6$, while the colour scale of the arrows represent the pairing amplitude. The positions of the vortices are the points around which phase twists and amplitude of order parameter goes to a minima. The spatial variation of $D(0)$ and $D_p$ shown in *Figure 4.11d-i*, qualitatively capture all the broad features observed in our experiment. The histograms of the $D(0)$ maps shown in *Figure 4.11j-l*, show that the distribution progressively widens with increasing *n* and form long tails in agreement with the experiments. We also note in the inset of *Figure 4.11k*, that the soft gap inside the vortex core is also observed similar to the experiments. The survival of the soft gap in the vortex core, in our opinion, is strongly tied to the phase fluctuations of the order parameter due to the inhomogeneous background that depletes the superfluid stiffness in the core regions, but keeps the pairing amplitude finite.

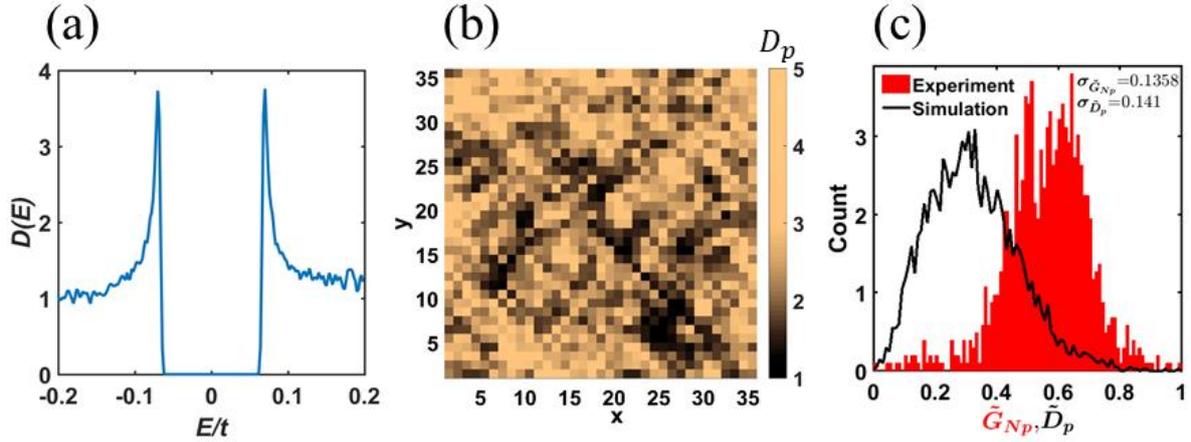

*Figure 4.10*: (a) DOS obtained from simulation for zero field, where the sharp coherence peaks and well-formed gap is clearly visible. (b) Coherence peak height ($D_p$) map for zero field over a $36 \times 36$ lattice. The inhomogeneity is qualitatively similar to the experimental map of coherence peak height. (c) Comparison of histograms of coherence peak heights obtained from simulation (black curve) and experiment (red bars). Both of these having standard deviation roughly equal qualitatively establishes a similar strength of disorder.

## 4.3   Superconductor to insulator-like transition

In presence of very strong disorder, application of magnetic field often results in a phenomenon, popularly known as *superconductor to insulator transition*, where on application of a magnetic field, a very strongly disordered superconductor transforms into an insulator, characterized by a diverging resistance as $T \to 0$ and also shows a magnetoresistance peak, with in extreme cases, sheet resistance as high as few Giga-Ohms.[10,11,12,13,14] Even when the effect is not as dramatic the basic feature of a peak in resistance with magnetic field, is observed for a wide range of disorder close to the superconductor to insulator transition.[15,16] Several theoretical models try to explain this superconductor to insulator-like transition, for example, percolation based model invoking Coulomb blockade,[17,18] charge-vortex duality,[19,20] dirty boson model,[21,22,23,24,25] etc. However, the percolation based model is till date the most successful one to explain both disorder and magnetic field induced superconductor-to-insulator-like transition.

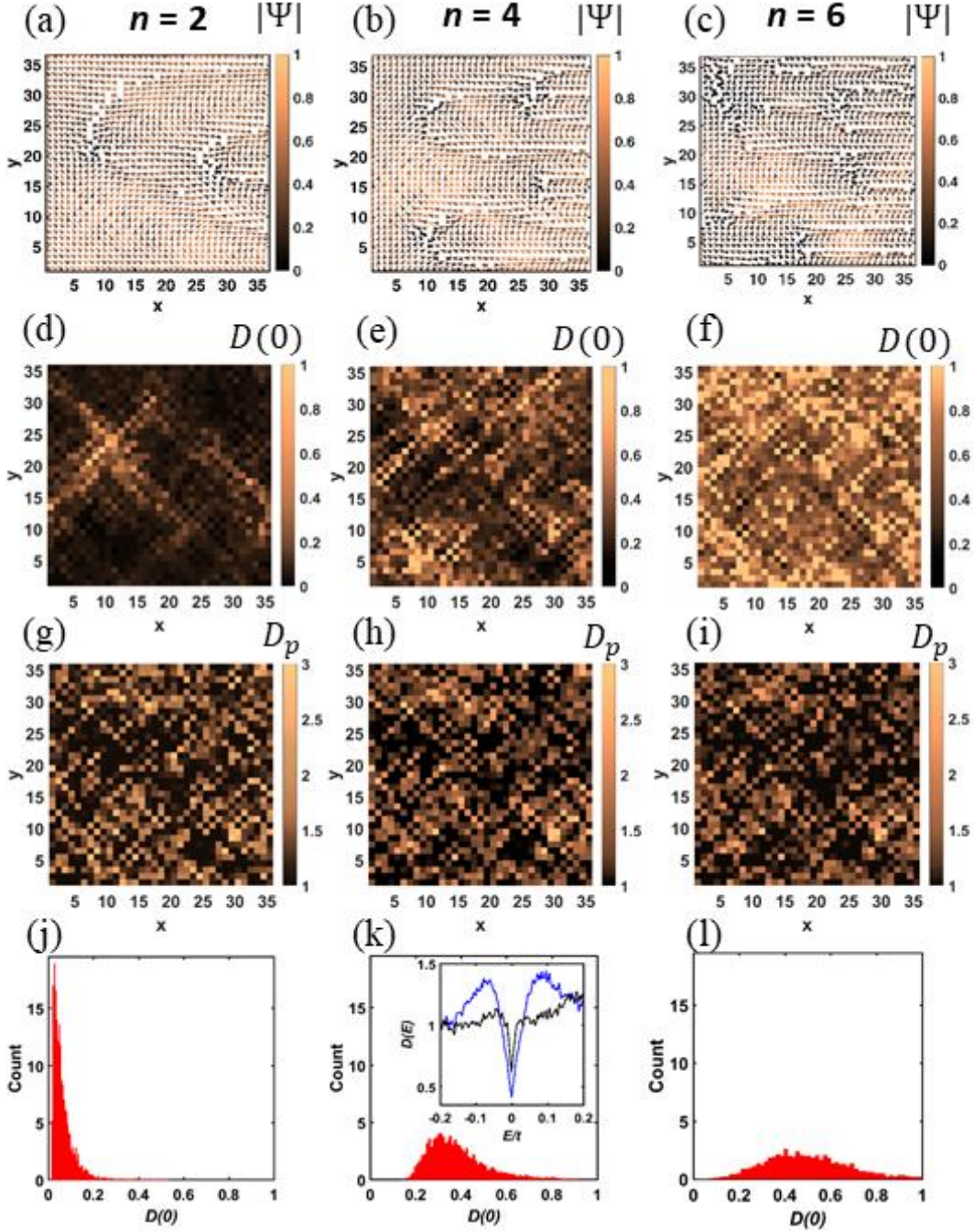

*Figure 4.11*: (a)-(c) The arrows represent phase ($\phi$) of local order parameter ($|\Psi|e^{i\phi}$), while the colour bar indicates the amplitude part ($|\Psi|$) of local order parameter for number of flux entering in the system 2, 4 and 6 respectively. The phase twists around regions of low order parameter amplitude, are the regions where vortices enter the system. (d)-(f) $D(0)$ maps for different fields. (g)-(i) $D_p$ maps for different fields. $D(0)$ and $D_p$ maps show weak anti-correlation. (j)-(l) Histograms of $D(0)$ for different magnetic fields, which progressively widens with increasing magnetic field. *Inset* of *k* shows DOS inside and outside a vortex patch for $n = 4$. In the vortex core the soft gap is still present.

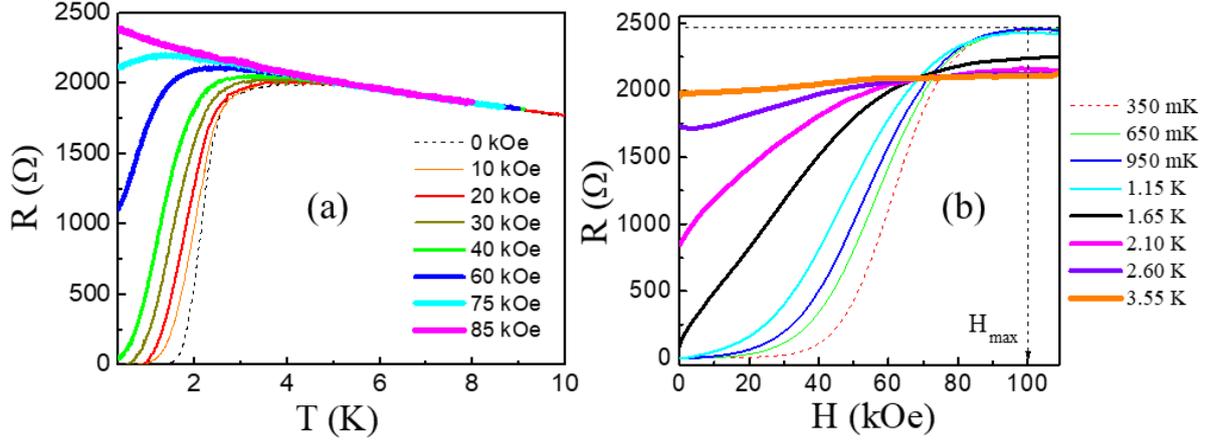

*Figure 4.12*: (a) $R - T$ data for different magnetic fields for the $T_c \sim 1.65\ K$ NbN thin film. There is a superconducting transition for fields less than 40 kOe. For 85 kOe, on the other hand $dR/dT < 0$ down to 300 mK, implying an insulator-like state. (b) $R - H$ data for different temperatures. The isothermal magnetoresistance curves cross each other within 74-84 kOe. The magnetoresistance curve for 350 mK shows a weak peak at about $H_{max} \sim 100\ kOe$.

### 4.3.1 Magnetoresistance measurements

We study here the variation of sheet resistance of the $T_c \sim 1.65\ K$ sample, as a function of temperature for different magnetic fields in *Figure 4.12a*. We observe here that with increasing magnetic field, $T_c$ gets suppressed as expected and 40 kOe onwards we observe no superconducting transition. The behaviour of the 85 kOe $R - T$ curve shows $dR/dT < 0$ down to 300 mK, similar to an insulator-like behaviour. Here we define $T_c$ as the temperature where the sheet resistance falls below 0.05% of its normal state value. In *Figure 4.12b*, we show variation of sheet resistance as a function of magnetic field for different temperatures. As is customary for superconductor to insulator-like transition, the isothermal $R - H$ curves cross each other. But instead of a single crossover field, here we observe a magnetic field range of 74 kOe to 84 kOe for the crossover. We also observe here for temperatures below 1.65 K, the $R-H$ curves have reached a maxima at around 100kOe, denoted as $H_{max}$, above which the curves start to decrease slightly. We call this maxima the 'magnetoresistance peak'.

Generally the magnetoresistance data presented in *Fig. 4.12b* is further used for scaling analysis[26] to find out critical exponent of the transition. But close inspection of our data shows that there isn't one single crossover field because of a large temperature range. This has been previously reported in moderately disordered InO$_x$ films[27] that with increasing temperature the crossover field decreases. This precludes us from the possibility of scaling of the data around a single crossover field unlike previous celebrated reports.[28,29,30,31]

### 4.3.2 Altshuler-Aronov background

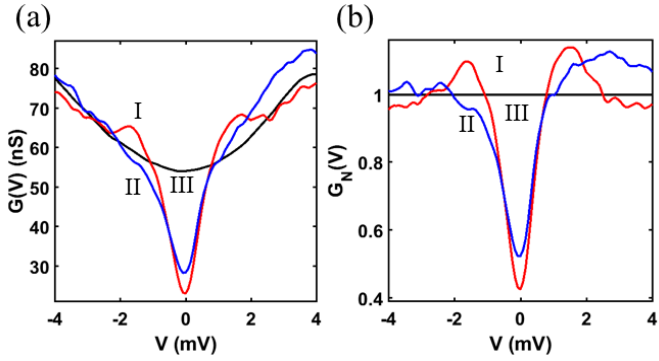

*Figure 4.13*: (a) Spectra I (II) is bias spectra at 350 mK for zero field at a point where coherence peak and gap are (not) suppressed. Spectra III is taken at 10 K, and is similar for both of the points. (b) Normalized spectra I (II) obtained by dividing spectra I (II) in (a) by spectra III in (a).

Highly disordered samples in general has a temperature independent V-shaped background in the spectra extending up to high bias, arising from Altshuler-Aronov type electron-electron interactions.[32,33] Unlike clean NbN case where the spectrum is fully gapped, the spectra in strong disordered samples such as this, show a substantial $G_N(0)$ (ZBC) value. Furthermore, the height of coherence peaks differ significantly over the space. For example, *Fig. 4.13a* shows a spectrum (Spectrum-I) with significant coherence peak, a spectrum (Spectrum-II) without that obtained at a different point and a V-shaped spectrum (Spectra-III) taken at 10 K at zero field which is similar for all points and is used to normalize all spectra to remove the V-shaped background. *Fig. 4.13b* shows the normalized spectra obtained by dividing all spectra by spectrum-III of *Fig. 4.13a*.

### 4.3.3 Omnipresence of pseudogap phase

Analysis of the tunnelling spectra over a $200\ nm \times 200\ nm$ area shows that granularity in $G_N(0)$ is present starting from zero field, and gets enhanced with increasing magnetic field (*Figure 4.14a-c*). This is similar in nature with the moderately disordered sample in application of magnetic field. However, temperature evolution of the superconducting gap for different magnetic fields, shown in *Figure 4.15b*, gives us a very significant insight about the system. Similar to the moderately disordered sample, here too we plot $T^*$ (temperature where, $G_N(0) \sim 0.95$) and $T_c$ (temperature where, $R = 0.05 R_N$) in the $H - T$ parameter space in *Figure 4.15c*. It is interesting to note that unlike the moderately disordered sample, the $T^*(H)$ line remains almost constant, giving rise to an omnipresent pseudogap phase in all magnetic fields.

### 4.3.4 Smooth transition of bias spectra

In *Figure 4.15a*, we plot average spectra for different magnetic fields, and we observe no sudden change across the crossover fields, which is generally claimed to be a *quantum critical point*. This triggers the idea that in a superconductor to insulator-like transition, both superconducting and insulator-like states are intrinsically similar characterized only by different volume fraction of superconducting regions and average size of superconducting puddles.

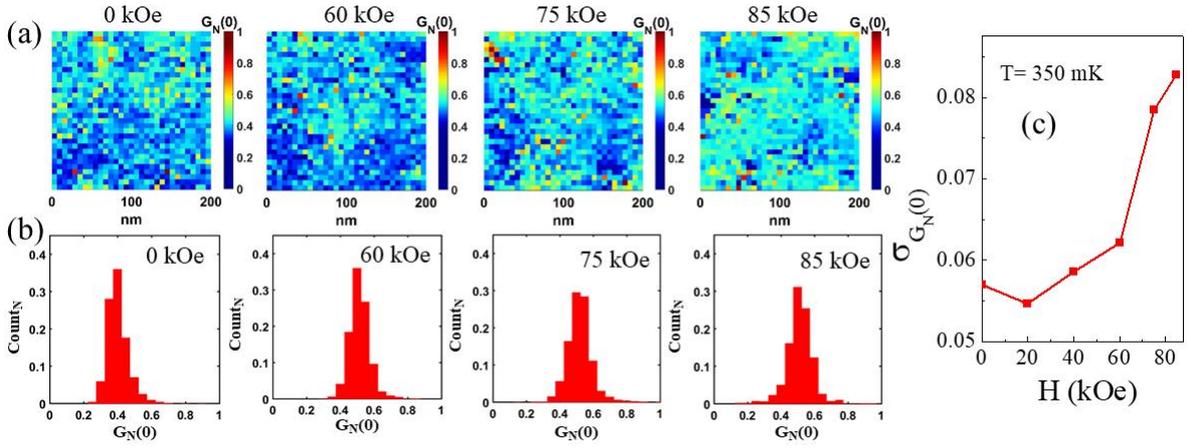

*Figure 4.14*: (a)-(b) $G_N(0)$ maps and their histograms for different magnetic fields at 350 mK. The histograms start to broaden slightly above 60 kOe. (b) Standard deviation of the histograms as a function of magnetic field, which starts to increase above 60 kOe.

### 4.4 Simulation using random resistor network

Till now the data gathered suggests a superconductor to insulator-like transition which is driven by the vanishing superconducting volume fraction which consequently gives rise to smaller islands of superconducting puddles. At higher magnetic fields, the appearance of high resistance values is dominated by vanishing of Josephson tunnelling amongst these puddles because of Coulomb blockade. However, this scenario deals with the concept of charging energy of the puddles, which is hard to capture in a back-of-the envelope simulation which I'll present in the next subsections. Hence we try to capture in the following subsections only the initial increment of magnetoresistance as the volume fraction of the superconducting regions decrease.

#### 4.4.1 Conversion of STS images to binary maps

The simulation directly involves the evolution of the zero bias conductance maps in magnetic field and tries to map them with the diminishing superconducting volume fraction concept. We take the $G_N(0)$ maps shown in *Fig. 4.14* and apply a thumb rule: points with $G_N(0)$ values lower than the average value are superconducting points, whereas points with $G_N(0)$ values higher than the average value are the insulating points. After several trial and error, we set this specific value of $G_N(0)$ to be 0.52 to match with the experimental variation of magnetoresistance at the lowest temperature. Representative resultant binary maps are shown in *Figure 4.16a-b*.

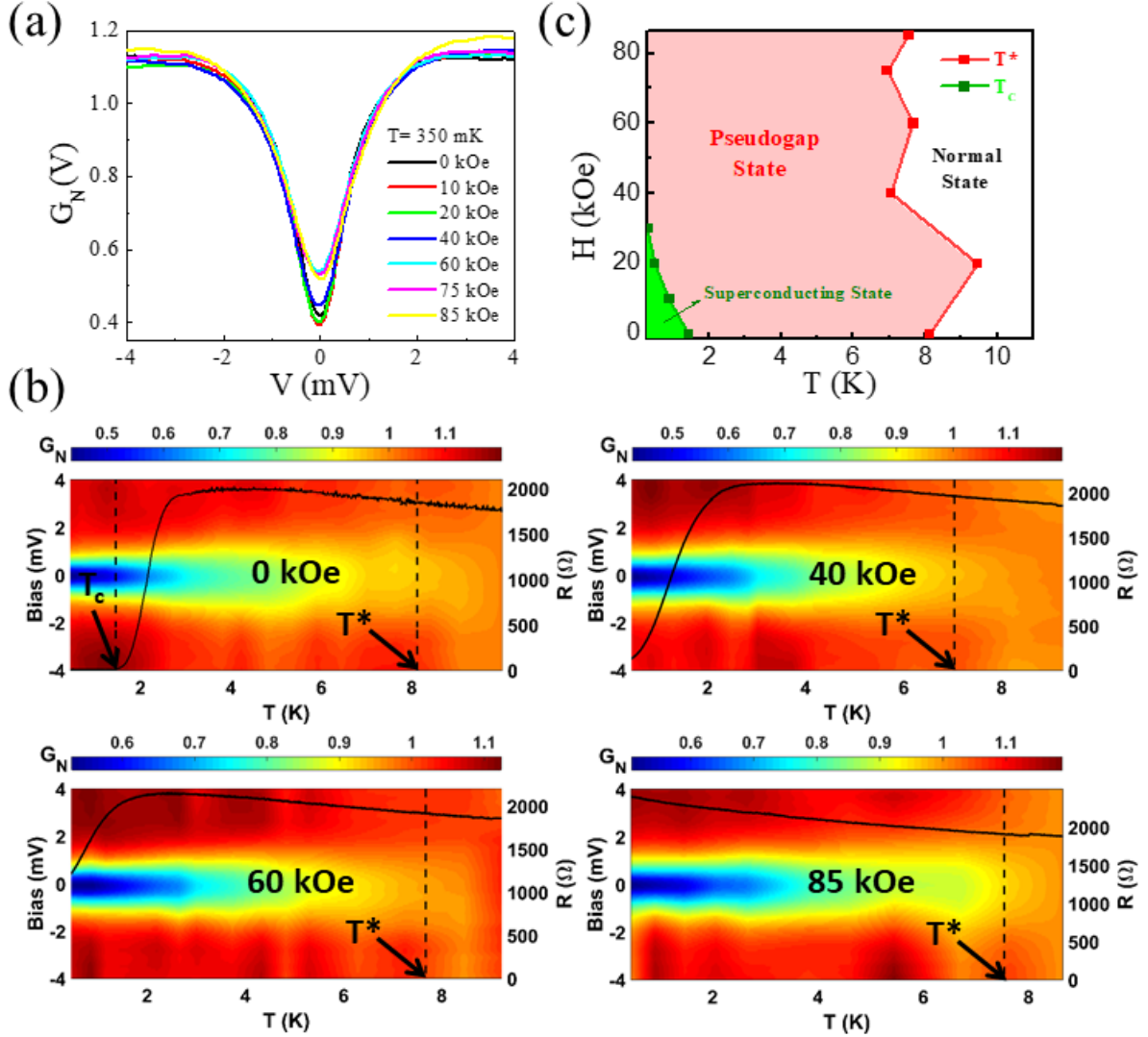

*Figure 4.15*: (a) Average $G_N(V) - V$ for different magnetic fields at 350 mK. (b) $G_N(V) - V$ as a function of temperature for different magnetic fields. $T_c$ and $T^*$ differ even starting from 0 kOe, while $T_c$ only exists below 40 kOe. (c) Variation of $T_c$ and $T^*$ as a function of magnetic field in the $H - T$ phase diagram. It shows the omnipresence of the pseudogap phase.

### 4.4.2 Simulation strategy

The binary maps are used as a precursor to the random resistor network. To assign the resistances we follow the rules similar to Ref. [34]. We assign three types of resistance values for three types of junctions:

(i) Superconductor-Superconductor junctions: $R_1 = R_0 \times T$, where $R_0$ is taken as *0.1* and *T= 0.35 K*.

(ii) Superconductor-Insulator junctions: $R_2 = R_0 \exp\left(\frac{E_c}{k_B T}\right)$, where $E_c$ is the charging energy of the superconducting islands, taken to be *4*. To avoid complexity and additional parameters $E_c$ is taken to be same for all size of islands.

(iii) Insulator-Insulator junctions: $R_3 = R_N \exp\left(\frac{|\epsilon_i|+|\epsilon_j|+|\epsilon_i-\epsilon_j|}{k_B T}\right)$, where $R_N$ is taken to be $5R_0$ and $\epsilon_i$ is the energy of the $i$-th site measured from the chemical potential, taken from a uniform distribution $\left[-\frac{W}{2}, \frac{W}{2}\right]$, where $W$ is taken to be *0.4*. *T* is *0.35 K* as usual.

In all cases mentioned above, $k$ is taken as 1 and $R_0$ is chosen such that it gives the best match with the experimentally obtained $R - H$. By choice, we have taken $R_1 \ll (R_2, R_3)$, but nonzero, to avoid numerical runaways. By using the above prescription, we obtain a resistor network as is shown in *Fig. 4.16c*. To solve the resistor network problem and find out the equivalent resistance value of the network, we utilize Kirchhoff's law in the following form[35]: Resistance between any two nodes ($k$ and $l$) of the resistor network,

$$R_{kl}^{equivalent} = \frac{|A^{(kl)}|}{|A^{(l)}|} \quad (4.3)$$

$$A_{ij} = \begin{cases} \sum_{k \neq i} \frac{1}{R_{ik}} & \text{for } i = j \\ -\frac{1}{R_{ij}} & \text{for } i \neq j \end{cases} \quad (4.4)$$

where $R_{ij}$ is value of resistance between $i$-th and $j$-th nodes in the network. Furthermore, $|A^{(l)}|$ is the determinant of $A$ with the $l$-th row and column removed, and $|A^{(kl)}|$ is the determinant of $A$ with $l$-th and $k$-th rows and columns removed.

The equivalent resistance of the resistor network is calculated from two nodes lying on the opposite edges. All possible combinations are averaged over to get the final value of the resistance, which is plotted in *Fig. 4.16d*. Though the model is oversimplified, it still captures very well the key feature of the data, which is the increase in the resistance of the system as the superconductor fragments into smaller islands.

### 4.4.3 Limitation of the simulation

In the simulation we have assumed that the charging energy is same for all different sizes of superconducting puddles, which is not at all practical. Since, charging energy is inversely proportional to the size of the puddle, one should in principle, find out from the $G_N(0)$ maps domains of superconducting points and then assign appropriate values of $E_c$. This charging energy plays the most important role when the size of the superconducting puddle shrinks so much that *Coulomb blockade* does not let any tunnelling between two superconducting puddles take place. However, the simulation sketched above, tries to describe the increasing behaviour of the $R - H$ curve in terms of the $G_N(0)$ maps, and always remain $H < H_{max}$.

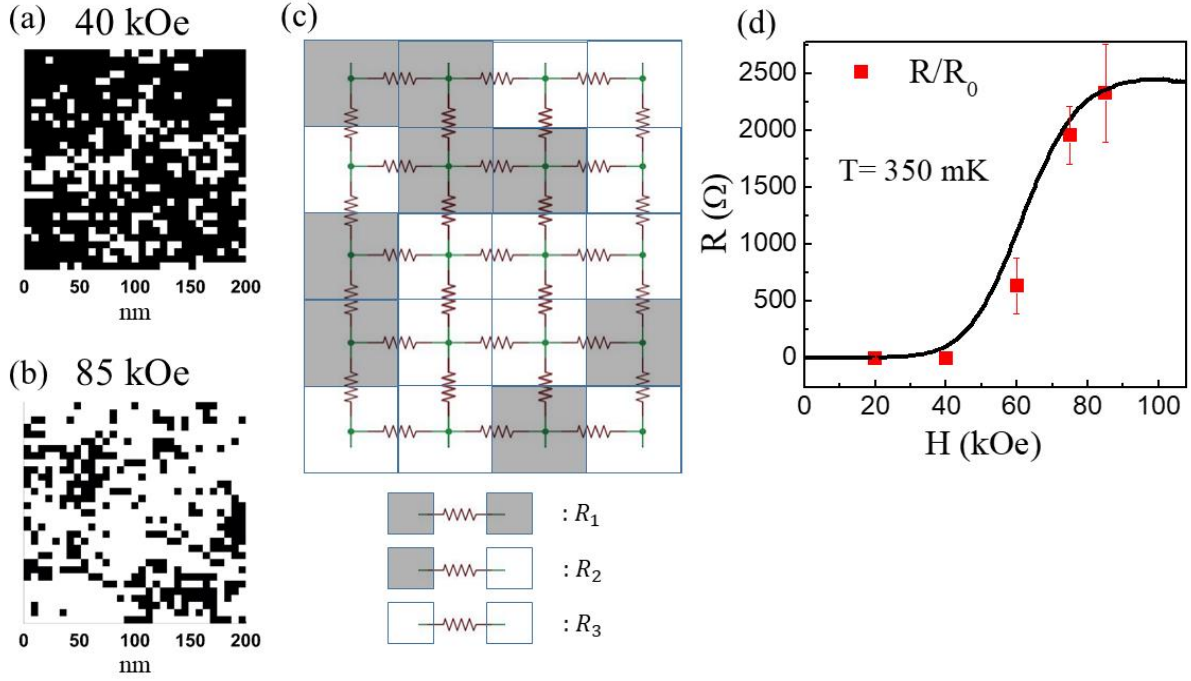

*Figure 4.16*: (a)-(b) Binary maps of $G_N(0)$ for 40 and 85 kOe obtained from $G_N(0)$ maps at 350 mK. (c) Schematic of the random resistor network, where two black points give resistance $R_1$, black-white points give $R_2$ and white-white points give $R_3$. (d) Simulated variation of averaged resistance in arbitrary units of $R_0$ (red dots) as a function of magnetic field is compared with $R - H$ data at 350 mK.

### 4.4.4 Physical picture and discussion

The emerging physical picture from the experiments and the simulations is that the random disorder potential makes the superconducting order parameter spatially inhomogeneous even in the absence of magnetic field. As a result the flux tubes enter the system through locations where the local amplitude of the order parameter $|\Psi|$ is low, which naturally causes the spatial vortex distribution aperiodic. This in turn kills the superconducting correlations by introducing an emergent granularity, which gives rise to the pseudogap regime where Cooper pairs continue to survive even when the zero resistance state is lost due to phase fluctuations between superconducting puddles.

In presence of stronger disorder, the picture remains similar, apart from additional features which can be explained in the light of the back-of-the-envelope simulation and an extension of the magnetic field induced emergent granularity picture. The primary increment of the $R-H$ curves can be understood from the decreasing volume fraction of superconducting regions with increasing magnetic field, due to emergent granularity. Although zero resistance state can be achieved in between some puddles via Josephson tunnelling, which comes at the cost of a charging energy. Due to this, there are two channels for transport, one is via Josephson tunnelling through the superconducting islands, giving a resistance $R_s$ due to the charging energy, and another one via normal patches, giving a resistance $R_I$. Below the magnetoresistance peak $R_s < R_I$ and the tunnelling path through the superconductors is preferred while the increment of resistance is well captured by the decreasing volume fraction of superconducting islands. As magnetic field is increased the decreasing size of the superconducting islands give rise to an increasing charging energy, which in turn increases $R_s$. At $H = H_{max}$, the two channels are equivalent which is where the magnetoresistance peak occurs. For $H > H_{max}$, the superconducting islands acquire a size where transport through the superconducting islands is energetically unfavourable due to Coulomb blockade and transport occurs through the normal channels. Therefore as the magnetic field is increased the decrease in the size of superconducting islands results in increase in the number of normal channels thereby decreasing the resistance. Whether this picture would fundamentally change for samples with $k_F l \ll 1$ is at the moment an open question.

# Chapter V: Two step melting of the vortex lattice in a very weakly pinned *a*-MoGe thin film

In three-dimensions, vortices act like cylindrical tube which traverse the thickness of the sample and while doing so it maps the pinning centres across sample. Since passing through a pinning centre minimizes the energy to be paid, the flux tubes can bend in order to accommodate a pinning centre. This elastic energy competes with the condensation energy saved by passing through a pinning centre and in turn gives a 3D vortex lattice which can have topological defects similar in nature to two-dimensional melting.[1,2] However, not being in a 2D limit, the vortex lattice does not follow a true BKTHNY transition (discussed in section 1.5.5.2), neither it shows a hexatic fluid phase during the two-step melting. The vortex state in the 3D materials studded with topological defects, behaves like a hexatic glass due to its inability to flow. In this hexatic glass state, the topological defects get frozen in space.

However, if one goes to the 2D limit (where thickness, $t \sim \xi_{GL}$), the vortices behave effectively like point objects and the effect of pinning on the vortices is largely reduced. Furthermore, being in a 2D space, the vortices now can undergo a true BKTHNY transition, also existence of a hexatic fluid phase is possible. Hence, it is very interesting to study the effect of dimensionality on vortex lattice, which deals directly with dynamics of vortices and its melting transitions. There are reports[3] on different effects of pinning on 3D and 2D systems, giving rise to different vortex phase diagrams owing to different dimensions.

## 5.1  Melting of 2-dimensional vortex lattice: Observation of hexatic fluid

In this chapter we shall primarily explore the melting of 2D vortex lattice.[4] Following the predictions of the BKTHNY theory,[5,6,7,8] the 2D vortex solid (VS) undergoes a continuous phase transition to a hexatic vortex fluid (HVF) phase, where randomly generated topological defects, *dislocations* appear in the vortex lattice. This HVF phase, characterized by its flow of vortices along the principal axes of the lattice, retains its orientational order. At a second continuous transition this HVF phase loses its orientational order altogether and transforms into an isotropic vortex liquid phase (IVL).

### 5.1.1 Sample

The sample used to study the melting of the 2D vortex lattice is *a*-MoGe thin film of thickness, $t \sim 21\ nm$ and $T_c \sim 7\ K$. The thin films are grown using PLD and are transferred *in-situ* using the vacuum suitcase to the STS/M, without ever exposing it to air. For transport measurements, Hall-bar patterned sample of same $t$ ($\lesssim \pm 10\%$) and $T_c$ ($\lesssim \pm 10\%$) is used. Bias spectroscopy of the sample for different temperatures gives spectra as is shown in *Figure 5.1b*, from which one can determine the temperature variation of superconducting gap, $\Delta(T)$, shown in *Figure 5.1a*. From this it is visible that the sample is a BCS type superconductor. The BCS nature is also obtained from variation of $1/\lambda_L^2$ as a function of temperature in zero field, shown in *Figure 5.1c*, obtained from the two coil $M - T$ measurement following the technique discussed in section 2.2.1 and also in Ref [9]. $1/\lambda_L^2$ follows the BCS nature, $\frac{\lambda_L^{-2}(T)}{\lambda_L^{-2}(0)} = \frac{\Delta(T)}{\Delta(0)} \tanh \frac{\Delta(T)}{k_B T}$. Both of these measurements give superconducting gap, $\Delta(0) \sim 1.3\ meV$. The mutual inductance measurement gives value of $\lambda_L(0) \sim 534\ nm$. Also, from temperature variation of $H_{c2}$ close to $T_c$, obtained from $R - T$ measurements (shown later) shows[10] a $H_{c2}(0) \sim 0.69 \times \left(\frac{dH_{c2}}{dT}\right)_{T=T_c} \sim 180\ kOe$, which gives, $\xi_{GL} \sim 4.3\ nm$.

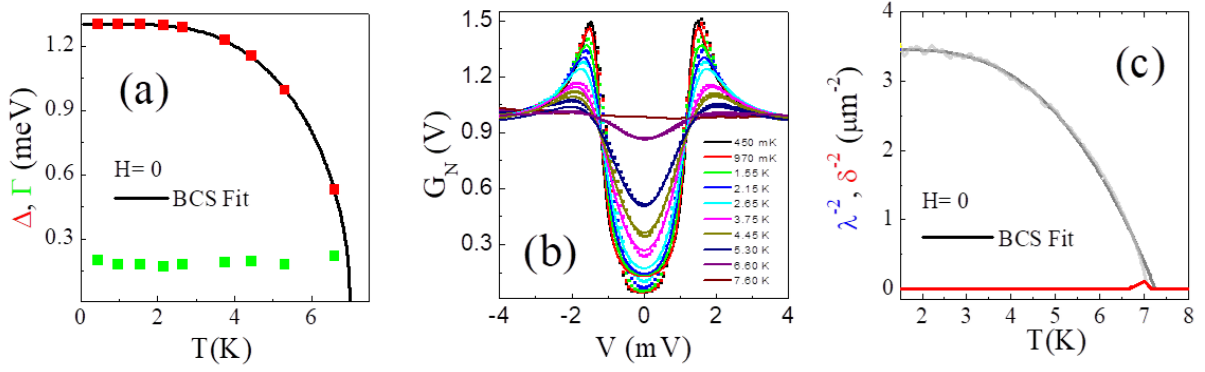

*Figure 5.1*: (a) Red squares show variation of Δ as a function of temperature, while green squares show variation of Γ. The black line is BCS fit to Δ − T. The Δ and Γ values are obtained by fitting $G_N(V) - V$ spectra for different temperatures as is shown in *(b)*. (c) Grey (red) line shows variation of $1/\lambda_L^2$ ($1/\delta^2$) as a function of temperature. The black line superposed on the grey line is the BCS fit to $1/\lambda_L^2$ data.

### 5.1.2 Advantages of *a*-MoGe thin film

However, the observation of the two-step BKTHNY melting and the true HVF phase is not easy, since, apart from going into the 2D limit, another hurdle to cross is the pinning strength of the superconductor and effects from the superconducting lattice. Strong pinning in the form of quenched disorder can drive the vortex lattice into a vortex glass, which therefore melts into a vortex liquid phase. On the other hand lattice orientation of the superconductor affects the orientation of the vortex lattice.[11] In order overcome these two effects an amorphous superconducting material, *a*-MoGe was chosen. Being amorphous there is no orientational coupling between the superconducting material and the vortex lattice. Furthermore, amorphous materials, such as Bi,[12,13] MoSi,[14] *a*-MoGe[15] are well-known for their low pinning strength, *a*-MoGe is not an exception.

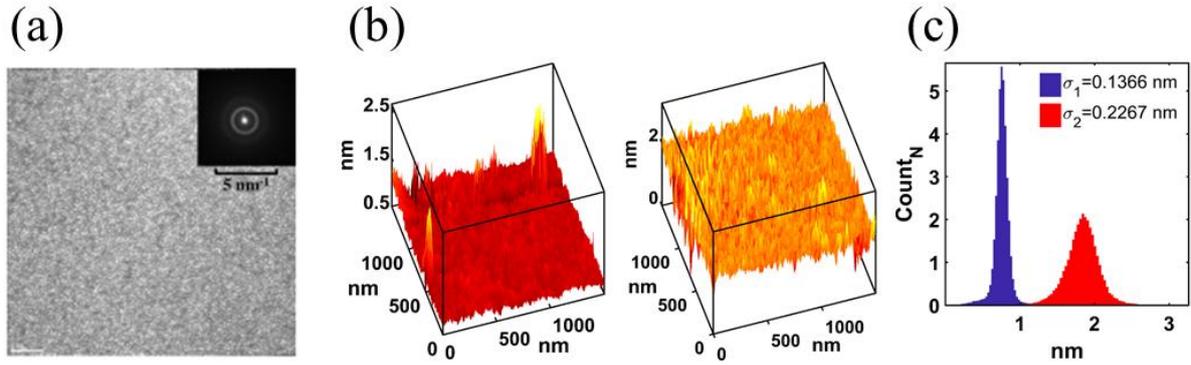

*Figure 5.2*: (a) High resolution SEM data along with ring-like electron diffraction pattern obtained from TEM. Both shows amorphous nature of the sample. (b) Two topographic images obtained from STM showing $1.9\ \mu m \times 1.9\ \mu m$ area of the film. The left area shows two particulates of tens of nanometres in size. (c) Histograms of the topographies shown in (b), where blue (red) corresponds to the left (right) image. The standard deviation of the histograms (equivalent to *rms* roughness) is $0.1366 - 0.2267\ nm$.

### 5.1.2.1    *Characterization of topography*

To check the amorphous nature of the *a*-MoGe thin film, we have done high resolution SEM which shows no lattice formation in the thin film, shown in *Figure 5.2a*. Electron-diffraction using high resolution TEM also has showed ring pattern, shown in *Figure 5.2a inset*, signifying the amorphous nature. We have also used STM to acquire topography image of the surface with *rms* roughness $\sim 0.1 - 0.14\ nm$, shown in *Figure 5.2b* and it does not show any regular lattice-like structure whatsoever. However, sometimes there exists one or two particulates of tens of nanometres in size.

### 5.1.2.2    *Characterization of pinning strength*

To have a detailed characterization of the pinning strength we have performed two independent sets of experiments. The first one is to check difference between the field cooled (FC) and zero field cooled (ZFC) states utilizing two-coil mutual inductance measurement,[16] shown in *Figure 5.3* (FC state was prepared by cooling the sample in field and then doing $M - T$, while for ZFC field was applied after reaching lowest temperature and then doing $M - T$). Since in the presence of pinning the vortex lattice can support a number of metastable states with different degrees of order and FC and ZFC states represent the limiting cases, with increasing pinning both of these states start to show significant differences. Here we observe that there is no difference between FC and ZFC, performed in 500 Oe, 1 kOe and 3 kOe, and by induction we can safely claim that for higher magnetic fields there is no chance for a difference between the two states. Since all our STS/M and transport measurements deal with magnetic fields in the regime $H \gtrsim 1\ kOe$, we always have a very low pinning in our system.

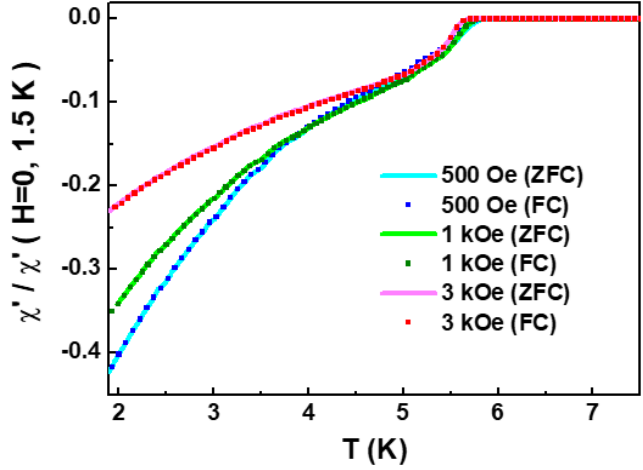

*Figure 5.3*: Comparison between FC and ZFC states for different magnetic fields. $\chi'/\chi'(H=0, 1.5\,K)$ as a function of temperature shows that for all magnetic fields FC and ZFC states match with each other.

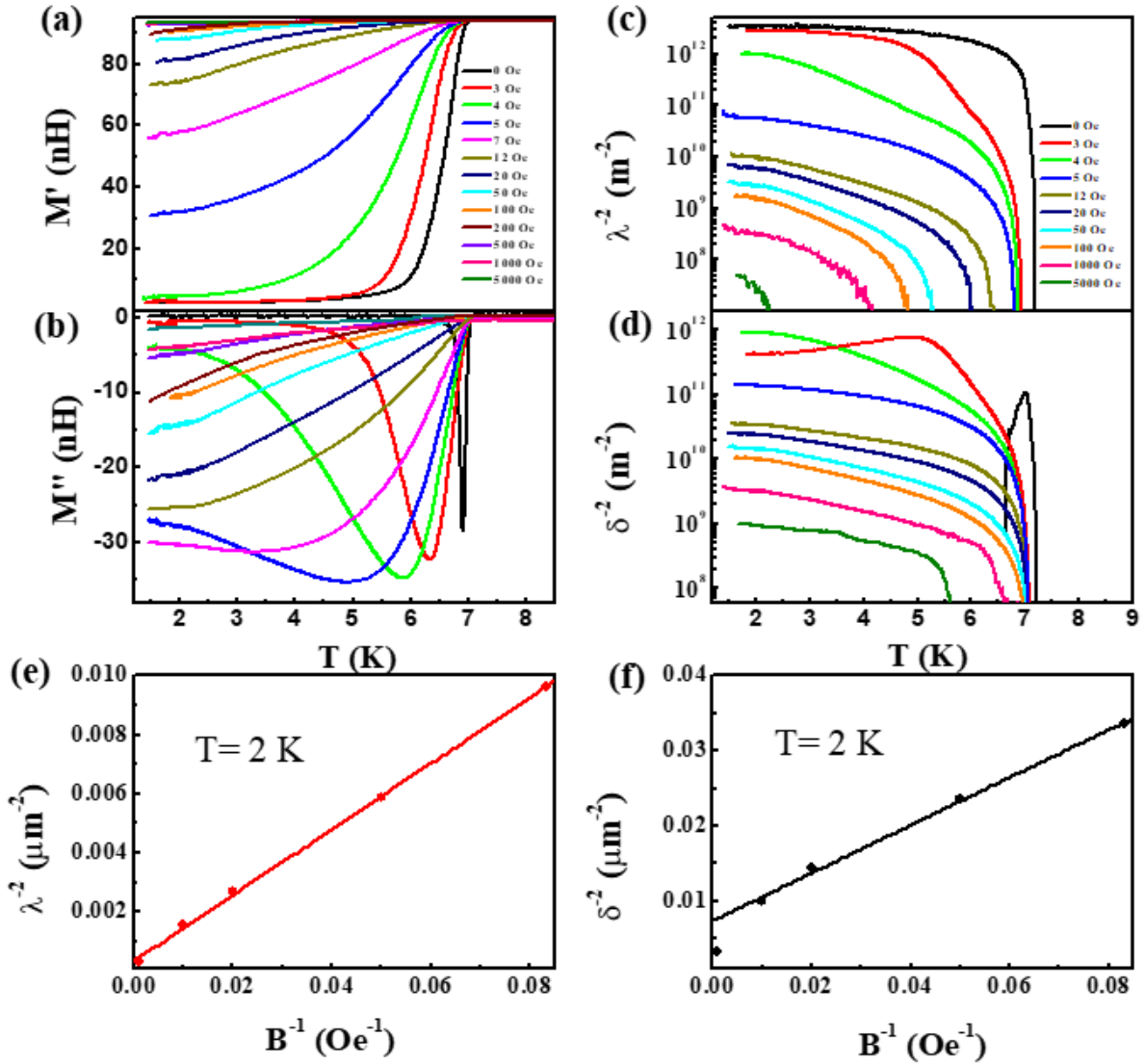

*Figure 5.4*: (a)-(b) Real ($M'$) and imaginary ($M''$) parts of mutual inductance as a function of magnetic field for different magnetic fields. (c)-(d) Variation of $\lambda^{-2}$ and $\delta^{-2}$ as a function of temperature for different magnetic fields obtained using (a)-(b). (e)-(f) Variation of $\lambda^{-2}$ and $\delta^{-2}$ at 2 K as a function of $B^{-1}$. The straight lines are fits to the data points.

Further quantification of the pinning strength is obtained from two-coil mutual inductance measurement for different magnetic fields. Shown in *Figure 5.4a-b* are the real ($M'$) and imaginary ($M''$) parts of mutual inductance as a function of temperature for magnetic fields ranging between 0-5000 Oe. From this one can obtain the variation of $1/\lambda_\omega^2$ and $1/\delta^2$ as a function of temperature, shown in *Figure 5.4c-d*, following the methods discussed earlier. Since, for a weakly pinned vortex solid, the low frequency penetration depth is given by $\lambda_\omega^2 = \lambda_L^2 + \lambda_v^2 = \lambda_L^2 + \left(\frac{B\Phi_0}{\mu_0}\right)(\alpha_L + i\omega\eta)^{-1}$, where $\alpha_L$ is the Labusch parameter (average restoring force on an individual vortex per unit length) (see section 1.5.2) and $\eta$ is the Bardeen-Stephen viscous drag per unit length (see section 1.5.1) and $\lambda_v$ is vortex contribution to penetration depth, namely Campbell penetration depth. Since, $1/\lambda_L^2 = 1/\lambda_\omega^2(H = 0)$ at 2 K is two orders of magnitude larger than $1/\lambda_\omega^2(H = 5\ kOe)$, and assuming $\lambda_L$ remains roughly field independent, the change in $1/\lambda_\omega^2$ can be attributed to dominantly the change of $\lambda_v^2$ and we can ignore the $\lambda_L^2$ term altogether. Therefore we obtain, $\lambda^{-2} = \lambda_v^{-2} = \frac{\mu_0 \alpha_L}{B\Phi_0}$ and $\delta^{-2} = \frac{\mu_0 \omega \eta}{B\Phi_0}$. In *Figure 5.4e-f* we plot $\lambda^{-2}$ and $\delta^{-2}$ as a function of $1/B$ for 2 K in the field range between 12 Oe to 1 kOe, and from the linear slope of these two curves we obtain, $\alpha_L \sim 0.026\ N/m^2$ and $\eta \sim 7.15 \times 10^{-7}\ N.s/m^2$. It is interesting to note that these values of $\alpha_L$ is very small compared to values observed for Nb [17] or YBa$_2$CuO$_7$ [18] films, while $\eta$ values are comparable. Consequently, the depinning frequency is found to be $\omega_0 = \frac{\alpha_L}{\eta} \sim 35\ kHz$, which is about 6 orders of magnitude less than corresponding values in both Nb and YBCO of comparable thickness. This also proves that the *a*-MoGe thin film used in this study is in the very weak pinning limit.

### 5.1.3 Real space evolution of vortex lattice in magnetic field

At first we acquire vortex lattice images by STS/M (section 2.3.2) by recording differential conductance map at fixed bias, $G(V = 1.3\ mV)$, staying at a fixed temperature of 2 K and varying magnetic field. The vortex images shown in *Figure 5.5*, where the vortices are represented by the black dots and are connected by black line via Delaunay triangulation, gives an idea of how vortex lattice evolves with magnetic field: In *Figure 5.5a*, the vortex lattice at 1 kOe does not have any topological defect in the visible area, representative of the vortex solid (VS) phase. FFT of the vortex image shows prominent six spots, which grouped with the absence of topological defects, is a proof of having long-range positional and orientational order. In *Figure 5.5b-e*, i.e., from 3 kOe onwards, topological defects start to appear in the vortex lattice, designated by the red, green, magenta and yellow dots which correspond to 5-, 7-, 4- and 8-fold coordination. These defect appear in a pair, like 5-7, 4-8 and are called a dislocation, which are indication of a hexatic phase, which has lost its positional order owing to the appearance of topological defects. Simultaneously we inspect the FFT of the vortex images, and observe that the six-spots in the Fourier space exists up to ∼70 $kOe$, in *Figure 5.5e*, however the spots are broadened and diffused. Nonetheless, the six-fold symmetry proves the existence of orientational order. At 80 kOe, in *Figure 5.5f*, a noticeable change is observed in FFT, where the six-spots turn into a ring, indicating existence of an isotropic vortex liquid (IVL) phase.

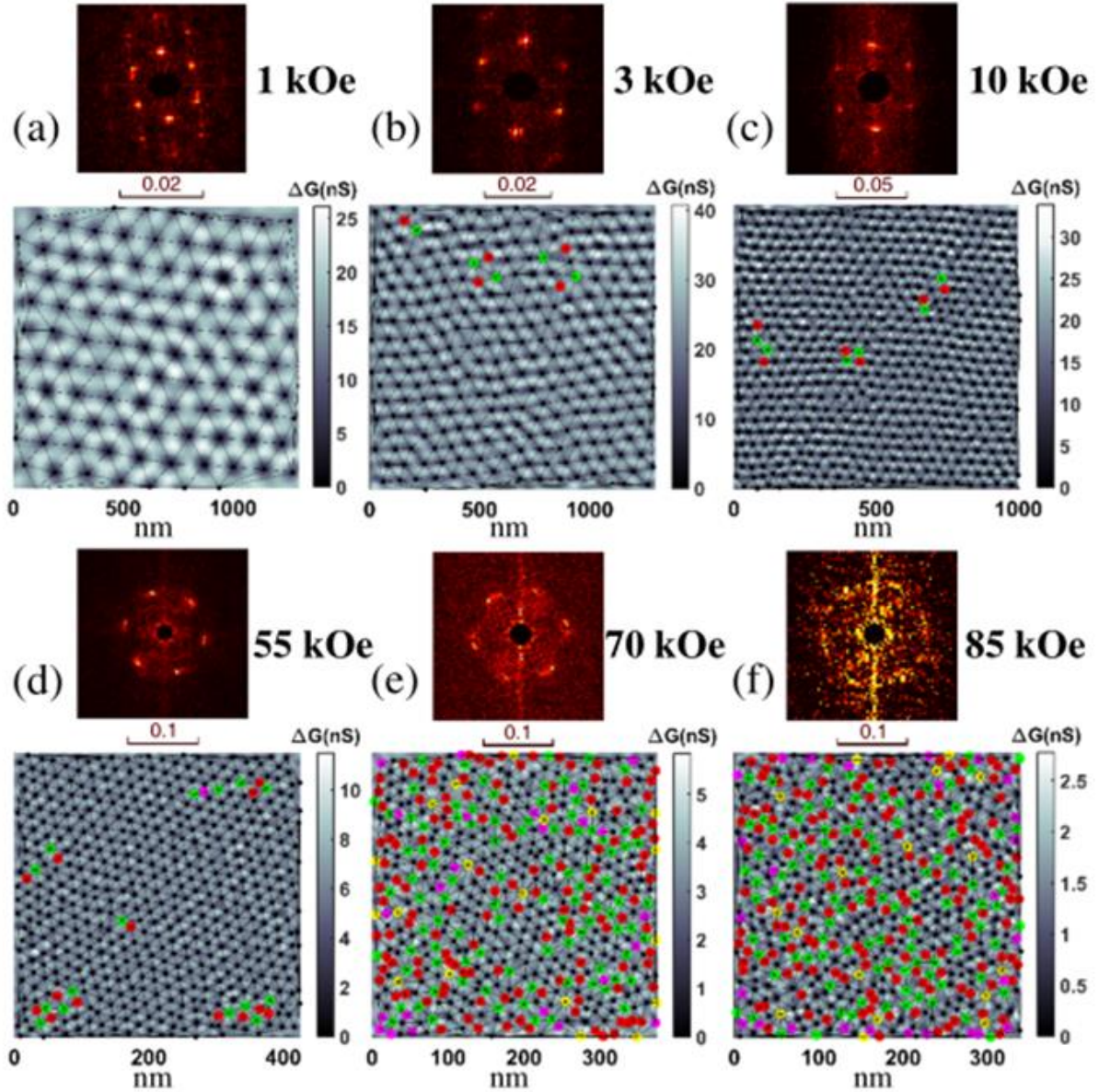

*Figure 5.5*: (a)-(f) Vortex image ($\Delta G$ at fixed bias, $V = 1.3\ mV$) at 1, 3, 10, 55, 70 and 85 kOe at 2 K. The black dots are the vortices which are connected by black line to show the Delaunay triangulation. The topological defects are the red, green, magenta and yellow points corresponding to 5, 7, 4 and 8-fold coordination. On top of each image is their FFT, scale of which is in $nm^{-1}$. FFT shows six spots up to 70 kOe, but in 85 kOe, FFT shows ring.

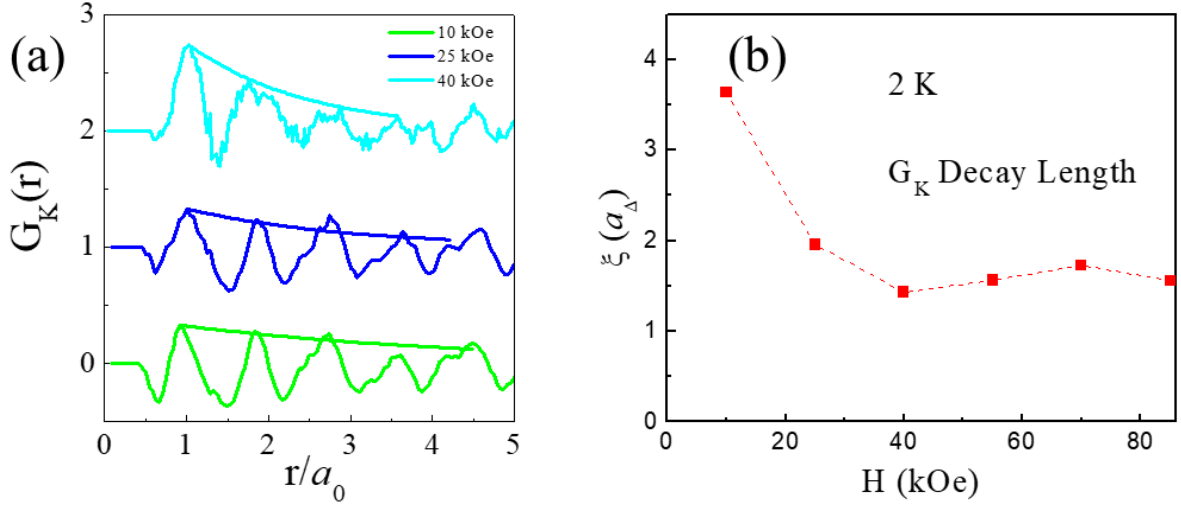

*Figure 5.6*: (a) $G_K(r)$ vs. $r/a_0$ for different magnetic fields at 2 K. The fits to the curves of the form $\exp(-r/\xi_K)$ are in the same colours. (b) Variation of $\xi_K$ as a function of $H$ at 2 K, which falls off to $\lesssim 2a_\Delta$ for $H \geq 40\ kOe$.

We also study same evolution of vortex lattice with magnetic field at 450 mK and observe here that the VS exists up to $\sim 4\ kOe$ and from $\sim 6 - 70\ kOe$, we observe existence of hexatic phase and in $\sim 85\ kOe$ we observe the IVL phase. However, in both 2 K and 450 mK, we still do not know the nature of the hexatic phase, be it a glass or a liquid, which we shall uncover in the next section.

### 5.1.3.1  *Spatial variation of $G_{\vec{K}}(\vec{r})$*

Apart from the appearance of topological defects, a quantitative proof of having a VS to hexatic phase transition can in principle be obtained from the variation of positional order parameter, $G_{\vec{K}}(\vec{r})$ for different magnetic fields. However, for the lowest magnetic fields, such as 1-5 kOe, the vortex image which can be acquired by the STS/M is not very large and contains roughly $\lesssim 50 - 100$ vortices, which is a very small number in order to calculate $G_{\vec{K}}(\vec{r})$ within any sensible confidence. This problem is only rectified for $H \geq 10\ kOe$, from which field onwards the vortex images contain $\simeq 500$ vortices. In spite of this limitation, in *Figure 5.6a*, we present $G_{\vec{K}}(\vec{r})$ vs $|\vec{r}|/a_0$ for different magnetic fields at 2 K fitted with a $\sim e^{-r/\xi_K}$ form to determine positional order correlation length, $\xi_K$. In *Figure 5.6b* we plot variation of $\xi_K$ as a function of magnetic field and observe that $\xi_K \sim 4 \times$ lattice constant for $H = 10\ kOe$. Beyond $25\ kOe$, $\xi_K$ falls below twice lattice constant and we can claim that positional order is completely destroyed. However, the mismatch between the appearance of the topological defects and disappearance of positional order comes from the small size of the vortex images for low magnetic fields, which is why $G_{\vec{K}}(\vec{r})$ here is not a very reliable metric to determine the transition. We shall show later on that the first transition is well determined from transport measurements which matches with the vortex images with extraordinary accuracy.

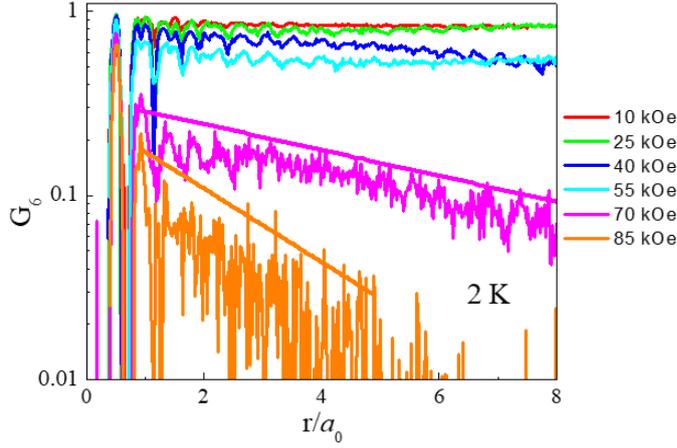

*Figure 5.7*: $G_6$ as a function of $r/a_0$ for different magnetic fields at 2 K. For 70 and 85 kOe the variation shows exponential decay of $G_6(r)$, implying an IVL phase.

### 5.1.3.2    *Spatial variation of $G_6(\vec{r})$*

The hexatic phase to IVL phase transition can be identified from the variation of orientational order parameter, $G_6(\vec{r})$ for different magnetic fields, apart from the change in FFT from six-Fourier spots to ring. Since this transition occurs at a higher magnetic field $H \simeq 70 - 85\ kOe$, the size of the vortex images are good enough to conclude sensible remarks. In *Figure 5.7*, we plot $G_6(\vec{r})$ vs $|\vec{r}|/a_0$ for different magnetic fields at 2 K, and observe that for $H < 70\ kOe$, $G_6(\vec{r})$ remains finite up to large $|\vec{r}|$, and the slow decay of $G_6(\vec{r})$ in the hexatic phase is consistent with a quasi-long range orientational order. However for 70 and 85 kOe, $G_6(\vec{r})$ falls off exponentially and an exponential fitting of the form $\sim e^{-r/\xi_6}$ results in orientational order correlation length, $\xi_6 \sim 6.2 a_0$ and $2.2 a_0$ for 70 and 85 kOe respectively. Although for 70 kOe, $G_6(\vec{r})$ falls off exponentially, the FFT shows much diffused six spots. The reason for this can be found in the large value of $\xi_6$, which is of almost the same size as that of the image. However, it can be safely claimed that close to $\sim 70\ kOe$, the hexatic to IVL transition takes place.

### 5.1.4   Thermally generated random dislocations

In this section we shall study the nature of the hexatic phase. In order to do this, we have studied temporal evolution of the vortex lattice at the same temperature (maximum drift of temperature $\pm 50\ mK$) and same magnetic field (magnet in *persistent* mode) over the same area. At first we have acquired three consecutive vortex images very slowly, each of which takes about 1 hour 30 minutes to complete. These three vortex images for different magnetic fields, shown in *Figure 5.8a-d* show temporal evolution of vortex lattice for 3, 10, 25 and 55 kOe at 2 K, all of these in the hexatic regime. Here we observe that the dislocations appear and disappear randomly over the space over time, which is a direct proof of the presence of a hexatic vortex fluid (HVF) phase.

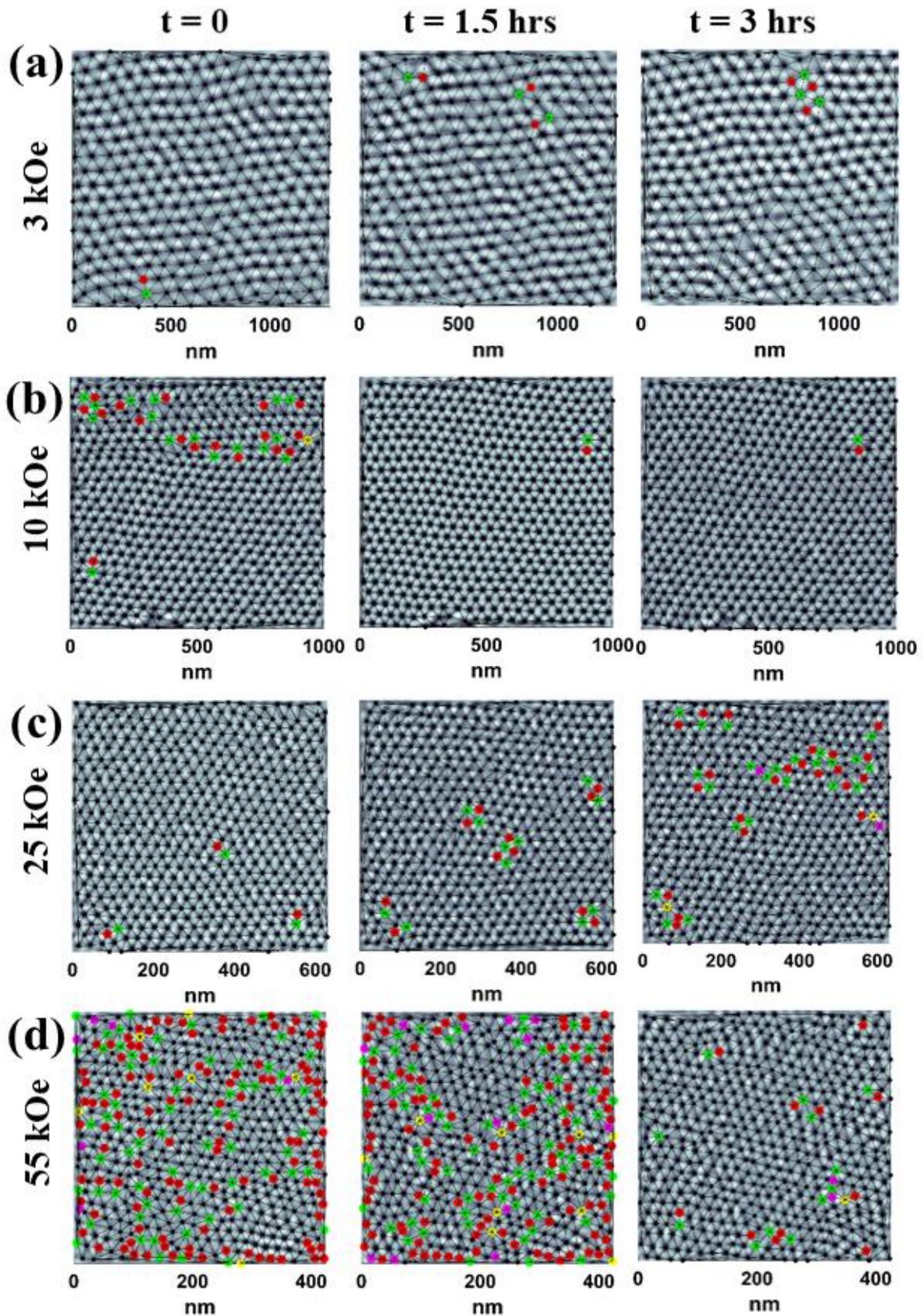

*Figure 5.8*: (a)-(d) Temporal evolution of vortex lattice for different magnetic fields in the HVF regime. Each vortex image takes 1.5 hours to complete. The colour codes of the VL are same as *Figure 5.5*. For all four magnetic fields the dislocations move in space over time.

Since, in a BKT-type melting the free energy is related to the number of topological defects involved, or in other words to the entropy of the system, as long as net entropy remains same, the location where topological defects appear or disappear does not matter. Since, we are dealing with a finite size vortex image, within our frame of view the number of topological defects might not look the same. However the randomness in their appearance and disappearance would prove the HVF phase.

The source of the dislocation is also debatable, since they can be generated by either global stress due to the system being trapped in a metastable state or local stress due to thermally generated flux creeps. Since the pinning of the system in question, is very weak there is little chance of being in a metastable state. However as an added precaution we have applied magnetic field pulses of 30 $Oe/kOe$ (for example 300 $Oe$ for 10 $kOe$) to the VL to bring the system down to its most stable state. This process of applying field pulses to the VL has been very effective to transform a FC state into a ZFC state, even in 3D and relatively strongly pinned system such as VL in NbSe$_2$ single crystal. Hence it can safely be claimed that the source of the randomly appearing and disappearing dislocations is from local stress due to thermally generated vortex creeps.

### 5.1.5 Thermally generated vortex creep: directionality

Another prominent proof in favour of a HVF phase is obtained from studying the trajectories of the vortex motion created by thermally generated vortex creep. In order to do that we have acquired 12 consecutive vortex images at the same area and temperature for each magnetic field. Each of these vortex images takes about 15 minutes to complete. Each vortex in a single time frame is then mapped to its daughter vortex in the next time frame (see section 2.4.2) which gives a total of 11 arrows for each vortex corresponding to its trajectory. In *Figure 5.9* and *Figure 5.10*, trajectories of the vortices in three vortex phases are shown for 2 K and 450 mK respectively.

In the VS phase (1 kOe for 2 K and 2.5 and 4 kOe for 450 mK) the vortices undergo small meandering motion about their mean positions, similar in nature to particles in a cage. In the HVF phase on the other hand (3-55 kOe for 2 K and 6-55 kOe for 450 mK) the trajectories of the vortices on a longer timescale are directionally concomitant with the principal axes of the VL, although over smaller timescale the jittering motion of the vortices still exist. This directionality is only disturbed in the vicinity of the topological defects. In the IVL phase (70-85 kOe for 2 K and 85 kOe for 450 mK) the vortices move randomly without any directional preference, consistent with the description of the two-step BKTHNY melting.

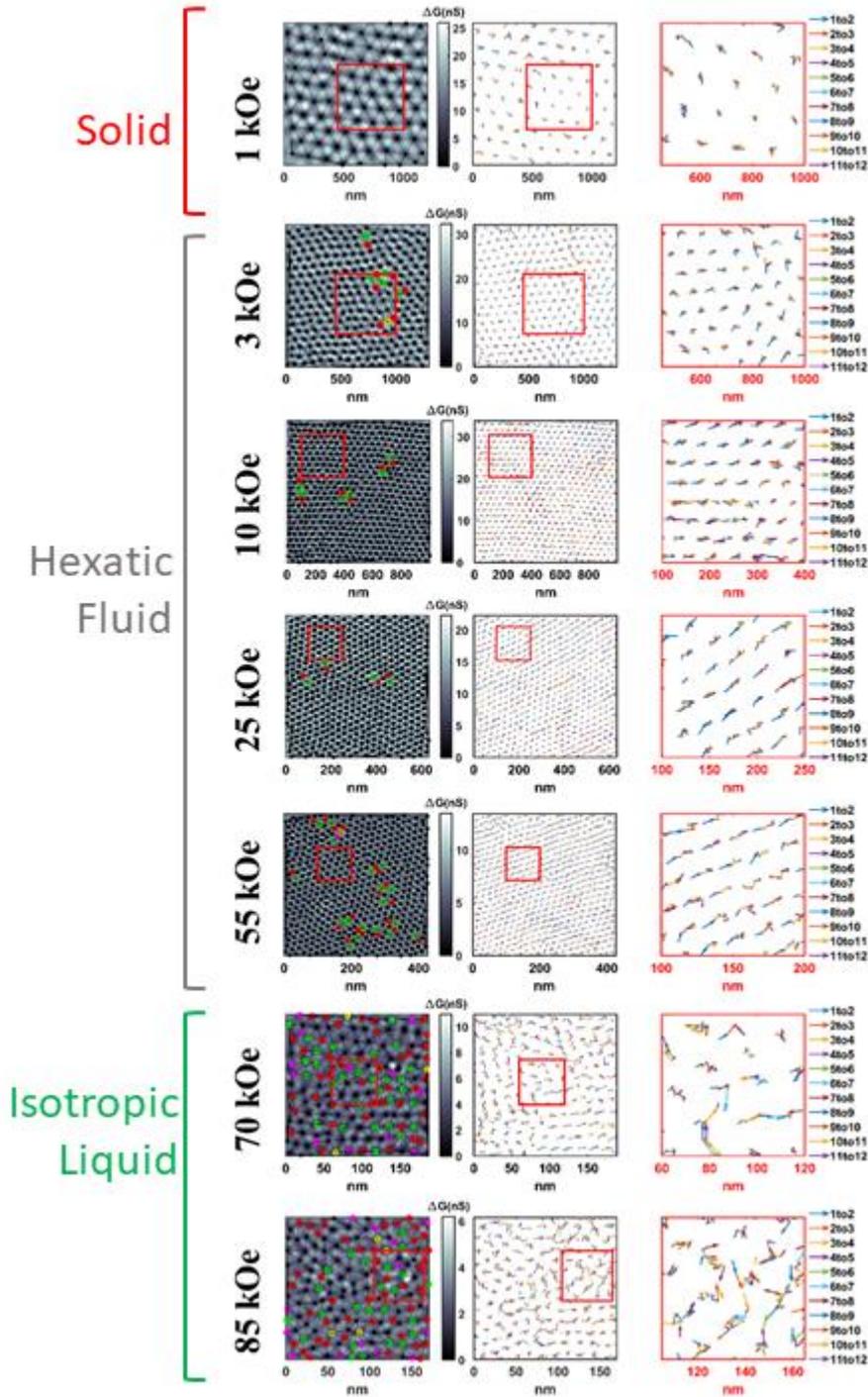

*Figure 5.9*: The left panel shows first vortex image of 12 consecutive images taken 15 minutes apart at 2 K. The colour codes are same as *Figure 5.5*. The middle panel shows arrows connecting 12 consecutive vortex images. The right panel shows expanded view of the red box of left and middle panel. The VS, HVF and IVL phases are marked.

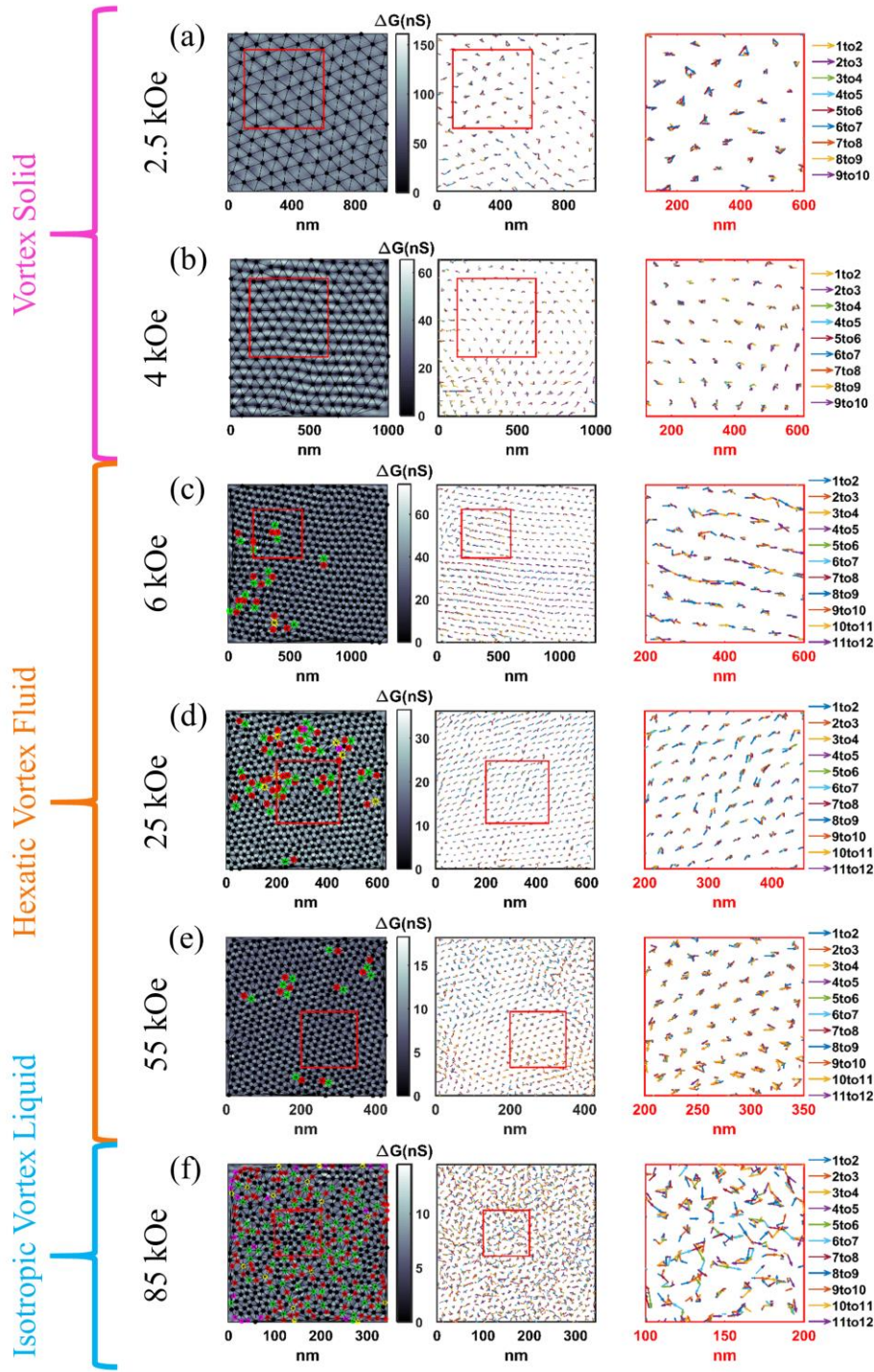

*Figure 5.10*: The left panel shows first vortex image of 12 consecutive images taken 15 minutes apart at 450 mK. The colour codes are same as *Figure 5.5*. The middle panel shows arrows connecting 12 consecutive vortex images. The right panel shows expanded view of the red box of left and middle panel. The VS, HVF and IVL phases are marked.

It is however interesting to note that for the IVL phase if the vortices are undergoing large jumps of the order of a lattice constant, the method we have adhered to determine the trajectories of the vortices, fails and issues a wrong parent-daughter mapping. Furthermore, a detailed analysis of true randomness should be performed by comparing trajectories of one region of vortices with a different region. This should in principle cement the claim of a 'perfect' liquid phase.

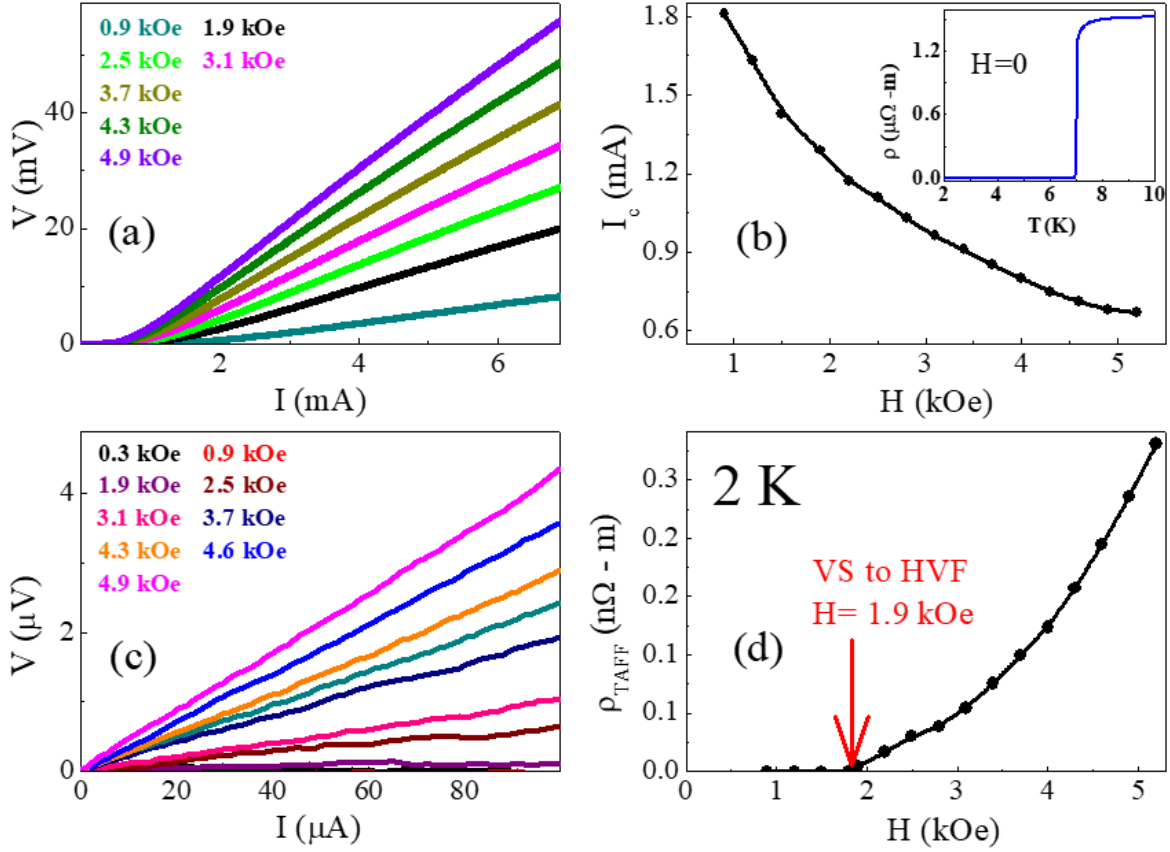

*Figure 5.11*: (a) $V-I$ characteristics for different magnetic fields in the range 0-7 mA at 2 K. The linear region of the curves are fitted with the flux flow equation to obtain $I_c$ as a function of $H$ shown in (b) (*inset*) $\rho - T$ at zero field shows $T_c \sim 7\ K$. (c) $V-I$ curves for different fields in the range $0 - 100\ \mu A \ll I_c$. This shows $V \sim 0$ below 1.9 kOe and above 1.9 kOe it shows positive finite slope. (d) $\rho_{TAFF} - H$ obtained from fitting the linear parts of $V-I$ curves of (c), which shows that the VS to HVF transition occurs at 1.9 kOe at 2 K.

### 5.1.6 Signatures in transport measurements

The last and one of the most important proofs of the two-step melting is obtained from transport measurements performed on similar samples.[19] Since for current $I > I_c$, there is no difference between a VS and vortex liquid, because both of them go into a flux flow regime, where, flux flow resistance, $R_{FF} = V/(I - I_c)$ arises due to Lorenz force on the vortices exceed the pinning force. As it is already been discussed in section 1.5.4, a VS can be differentiated from a vortex liquid (HVF or IVL) only by studying the thermally activated flux flow (*TAFF*) resistivity, given by,

$$R_{TAFF} = \frac{V}{I} = R_{FF} \exp\left(-\frac{U(I)}{k_B T}\right) \quad (5.1)$$

in the limit $I \ll I_c$. Seminal works on vortex melting has shown that the effective pinning potential $U(I)$ strongly depends on the nature of the vortex lattice.[20] For example, for a vortex solid,[21,22] $U(I) \propto \left(\frac{I_c}{I}\right)^\alpha$ and For $I \to 0$ this pinning potential diverges and hence $R_{TAFF} \to 0$ in this limit. However for a vortex liquid[23] $U$ is independent of the drive current and as a result, $V \propto I$. To check these two limiting cases, we study the $I - V$ characteristics at 2 K in the range $\sim 0 - 6\ mA$ for different magnetic fields in *Figure 5.11a*. Consequently we fit the linear regime of the curves with the flux flow nature and obtain $I_c$ for different magnetic fields shown in *Figure 5.11b*, which is $\sim 0.6 - 1.8\ mA$ for magnetic fields up to 5 kOe. Hence we acquire $I - V$ data in the limit $I \sim 0 - 100\ \mu A \ll I_c$ and observe in *Figure 5.11c-d* that in this limit $\rho_{TAFF}\ (= R_{TAFF} \times A/l)$, or $V$ remains zero for magnetic fields up to 1.9 kOe and above it a finite $\rho_{TAFF}$ starts to appear, i.e., $V = I \times R_{TAFF}$, signifying a VS to HVF transition at 1.9 kOe for 2 K.

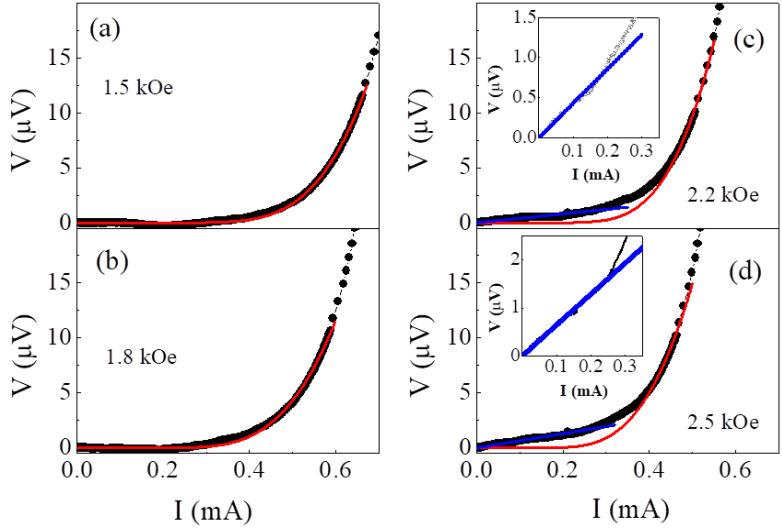

*Figure 5.12*: (a)-(b) Fitting the $V - I$ data (black dots) with exponential form (red line) of *eqn. 1*, for 1.5 and 1.8 kOe at 2 K. (c)-(d) The $V - I$ data (black dots) for 2.2 and 2.5 kOe doesn't fit (red line) with *eqn. 1*, whereas the *insets* show that the data fits well with the linear nature (blue line) in the low current regime. This determines that for 2 K the VS to HVF transition occurs at 1.9 kOe.

The $I - V$ curves can also be tested whether they fit with the form, $V = IR_{FF} \exp\left(-\left(\frac{I_c}{I}\right)^\alpha \cdot \frac{1}{k_B T}\right)$. In *Figure 5.12a-d* we try to fit $I - V$ curves using this form for magnetic fields, 1.5, 1.8, 2.2 and 2.5 kOe and observe that the fits are pretty good for 1.5 and 1.8 kOe using $\alpha \sim 1$, while the $I - V$ curves for $H > 1.9\ kOe$ deviate a lot from the form mentioned above. Rather a linear fit to the latter two $I - V$s describe better the nature of these two curves in the low current regime. This also proves that at 2 K, the VS to HVF transition occurs at 1.9 kOe. For 450 mK, this transition point is found to be $\sim 5\ kOe$. This is in complete agreement with the STS/M images, which show for 2 K VS state in 1 kOe and HVF from 3 kOe; similarly for 450 mK VS state up to 4 kOe and HVF phase from 6 kOe onwards.

The other transition, i.e., HVF to IVL can also be determined if one plots the variation of the linear slope of the $I - V$ curves for $I \leq 100\ \mu A$, $\rho_{lin}$ (equivalent to $\rho_{TAFF}$ in lower fields, since above $H \gtrsim 80\ kOe$, $I_c$ becomes small and $\rho_{lin}$ and $\rho_{TAFF}$ cannot be unambiguously differentiated) as a function of magnetic fields, shown in *Figure 5.13a*. This shows a sudden increase in $\rho_{lin}$ above 70 kOe at 2 K, signifying the HVF to IVL transition. Since in HVF phase the vortices are constrained to move only along the principle axes and in IVL this constraint is lifted, the sudden increment of $\rho_{lin}$ is completely understandable. This also matches with STS/M images, which showed 70 kOe to be the HVF to IVL transition field at 2 K. In 450 mK this transition takes place at $\sim 74 \, kOe$.

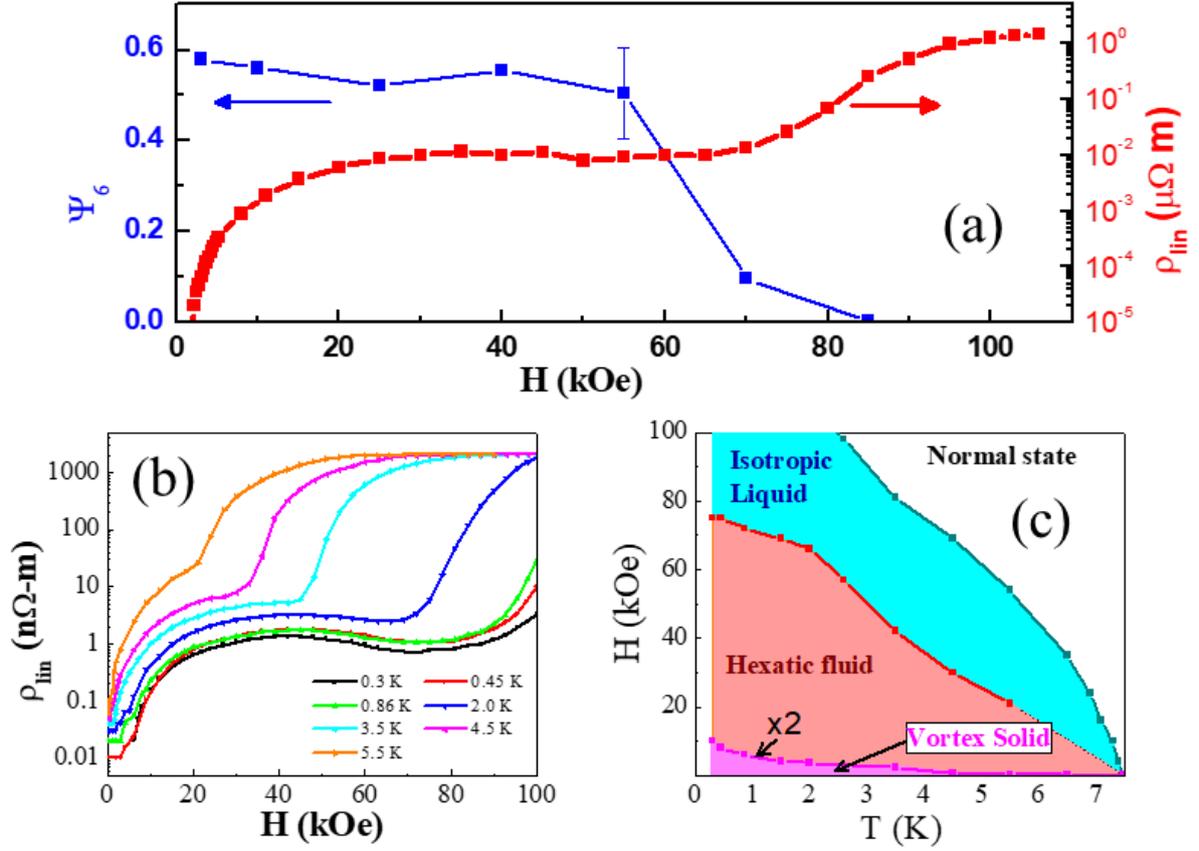

*Figure 5.13*: (a) Variation of $\Psi_6$ as a function of magnetic field (blue squares) show that $\Psi_6$ falls off to zero within error-bar (shown on the 55 kOe data) at about 70 kOe at 2 K. This is compared with the variation of $\rho_{lin}$ (in red squares) which start to increase rapidly above about 70 kOe 2 K. (b) Variation of $\rho_{lin}$ as a function of magnetic field for different temperatures, which is used to construct the $H - T$ phase diagram in (c). The VS-HVF boundary is magnified by 2 to increase clarity.

### 5.1.7 Phase diagram and discussion

In *Figure 5.13a*, along with the variation of $\rho_{lin} - H$, we also show variation of averaged orientational order parameter ($\Psi_6$), defined as,

$$\Psi_6 = \frac{1}{N} \langle \sum_{k,l} e^{6i(\phi_k - \phi_l)} \rangle \tag{5.2}$$

Here, $\phi_k$ is the angle between a fixed direction in the plane of the vortex lattice and the $k$-th bond, and the sum runs over all the bonds in the lattice. For a perfect hexagonal lattice, $\Psi_6 = 1$. We observe that $\Psi_6$ falls off to about zero within error-bar at $H \approx 70\ kOe$. The finite yet very low, non-zero value of $\Psi_6$ at $H = 70\ kOe$ has the same origin as that of the relatively large value of orientational correlation length at this field (discussed in section 5.1.3.2).

So far we have observed a tremendous consistency between STS/M images and transport measurements. On the basis of which, we construct the phase diagram of the VL from the $\rho_{lin} - H$ variation for different temperatures, shown in *Figure 5.13b* and noting down three magnetic fields, $H_m^1$, the VS to HVF transition field, $H_m^2$, the HVF to IVL transition field and $H_{c2}$, the upper critical field. These three set of points are plotted in the $H - T$ parameter space in *Figure 5.13c*. The VS to HVF phase boundary is multiplied by 2 to make the VS phase visible, which is otherwise restricted in a very narrow part of the phase diagram. The large extent of the HVF and IVL phases down to 300 mK is a very intriguing part which possibly can have its roots in quantum fluctuations of the vortices.

It is interesting to note that the $\rho_{lin} - H$ variation for temperatures, $T \lesssim 2\ K$, there exists a broad minima before the HVF-IVL transition. This phenomenon, related to *peak effect* in an order-disorder transition, can be explained very well[24] within the framework of Larkin-Ovchinikov collective pinning scenario,[25] where the role of elastic moduli of the ordered state is played by *Frank constant* of the HVF phase.[26]

Further investigation is needed to determine to what extent this phase diagram could be generic, for example, in thin crystalline superconductors, such as monolayer[27] NbSe$_2$ or for the quasi-2D VL in layered high-T$_c$ cuprates.[28] We would also like to note that, although we have used the magnetic field (or, alternatively, the density of vortices) as the tuning parameter, one would also expect to observe the two-step melting as a function of temperature. However, it might be more difficult to observe the transition as a function of temperature in imaging experiments because the contrast in STS images becomes poor at elevated temperatures.

## 5.2  Variation of structure function

The HVF to IVL transition can also be verified from the variation of structure function of the VL (see section 2.4.3.5). However for this measurement a different sample was used which had $T_c$ and thickness within $\pm 10\%$ error. In *Figure 5.14* we show evolution of 2D FFT-s of the vortex images for different magnetic fields taken in smaller steps of fields close to the

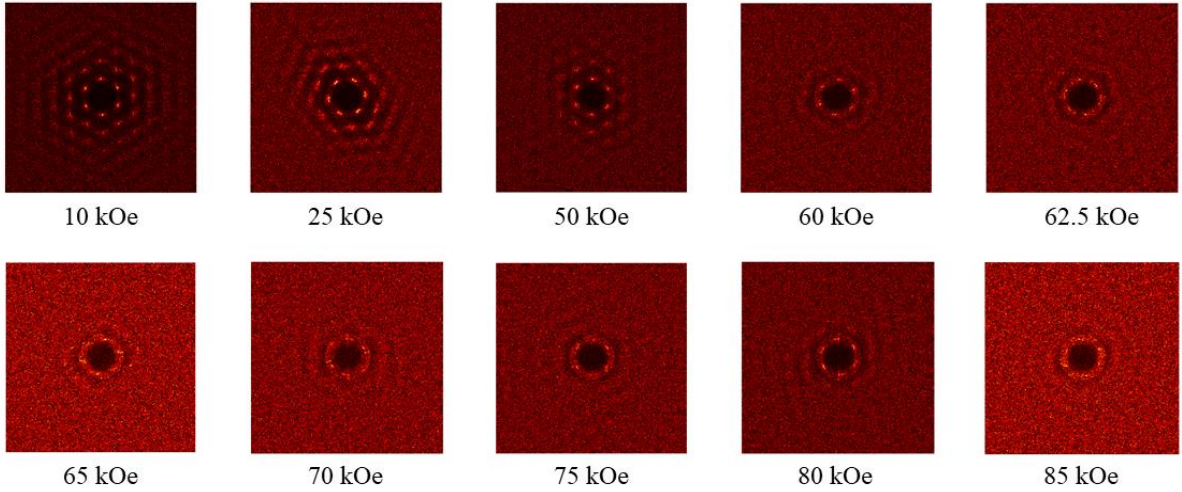

*Figure 5.14*: Evolution of FFTs of vortex images as a function of magnetic field at 2 K close to HVF-IVL transition.

HVF to IVL transition. At first we calculate averaged orientational order parameter, $\Psi_6$ ($\equiv \frac{1}{N}\sum_{i=1}^{N} G_6(r_i)$) as a function of magnetic field (shown in *inset of Figure 5.15a*) and find that the HVF to IVL transition occurs at $H \sim 65\ kOe$ here (since $\Psi_6 = 1$ for perfect orientational order; for HVF phase $\Psi_6$ is finite yet less than unity because of its quasi-long-range orientational order). Now in *Figure 5.15a* we show $S(q)$ for different magnetic fields, where $S(q)$ is determined from the structure function by averaging the FFT along the three principle axes of the FFT (shown in *inset of Figure 5.15b*). (For statistics this has been calculated for three VL at each field and averaged.) In *Figure 5.15b* we plot the height of the first peak of $S(q)$ obtained from *Figure 5.15a*, as a function of magnetic field. The values of $S_q^{peak}$ falls off from about 5-7 in the hexatic phase to about 2-4 in the IVL phase. Similarly we study in *Figure 5.15c*, $S(\theta)$ for different magnetic fields, determined from the angular variation of structure function at a radius equal to the position of the first peak (shown in *inset of Figure 5.15d*). The peak heights of $S(\theta)$ are averaged for each field and then over three vortex images and are plotted as a function of magnetic field in *Figure 5.15d*. Although not normalized similar to $S(q)$ using the limiting value $S(q \to \infty) \to 1$, $S(\theta)$ shows a jump from a higher value in the HVF phase to a lower saturating value in IVL at about 65 kOe. However, the noisy jumps in the variation of $S(q)$ and $S(\theta)$ are possibly occurring because of a low number of vortex images are used in the study. Nonetheless the HVF to IVL phase transition can be observed using the variation of the structure factor.

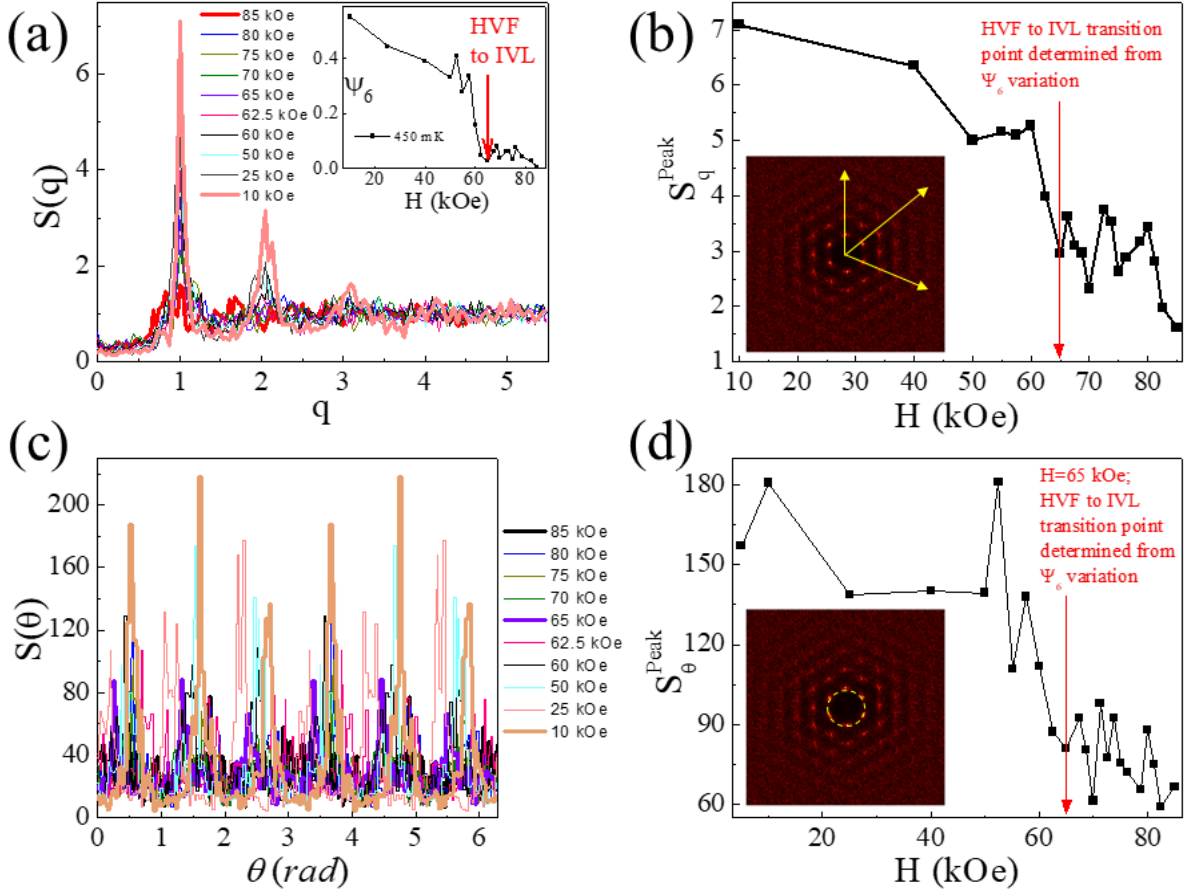

*Figure 5.15*: (a) $S(q)$ vs. $q$ at 2 K for different magnetic fields, which for 10 kOe shows the highest peaks and for 85 kOe hardly shows any peak. (*inset*) Variation of $\Psi_6$ as a function of magnetic field, which falls off to zero within error-bar at about 65 kOe, signifying the HVF to IVL transition. (b) Variation of the first peak amplitude of $S(q)$ as a function of magnetic field, where the red arrow points the HVF to IVL transition. (*inset*) Representative FFT of vortex image at 10 kOe, where the yellow arrows show the three principle axes. (c) $S(\theta)$ vs. $\theta$ at 2 K for different magnetic fields, where the 10 kOe and 85 kOe data are in bold. (d) Variation of average peak heights of $S(\theta)$ as a function of magnetic field, along with the red arrow marking the HVF-IVL transition. (*inset*) Representative FFT of vortex image at 10 kOe, where the yellow dotted circle has radius equal to $q$, the position of first peak in $S(q)$.

## 5.3 Discrepancy between transport and STS/M

Till now the STS/M and the transport measurements have shown tremendous consistency with each other. However a close inspection reveals a major discrepancy between these two. From the temporal dynamics of the vortices, we can determine mean square displacement for each field as a function of time (section 2.4.3.3), shown in *Figure 5.16*. Let us assume that for 85 kOe (in the IVL phase), the vortices undergo Brownian motion, which can be related to diffusion constant, $D$ by,[29]

$$\langle r^2 \rangle = 4Dt \tag{5.3}$$

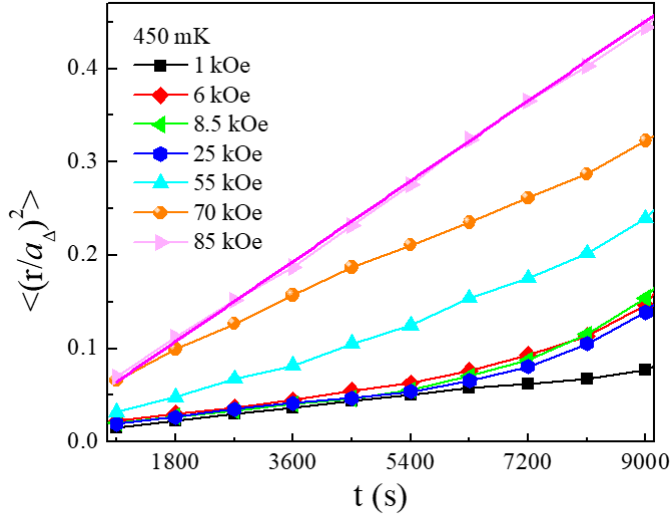

*Figure 5.16*: Mean square displacements in units of lattice constants for different magnetic fields at 2 K, as a function of time. The 85 kOe data is linearly fitted.

Hence, slope of the $\langle r^2 \rangle$ vs. $t$ curve gives $D = 3.23632 \times 10^{-21} m^2/s$. Assuming the motion is thermal in origin, the mobility of the vortices is given by, $\mu = D/k_B T$. This in turn gives effective viscosity of the vortices to be $\eta = \frac{1}{\mu d} = \frac{k_B T}{D d} = 1.37063 \times 10^5 \ kg \ m^{-1} s^{-1}$. This again can be related to flux flow resistivity given by, $\rho_{FF} = B \times \frac{\Phi_0}{\eta} = 1.28371 \times 10^{-19} \, \Omega.m$, which is well below the sensitivity of our measurements.

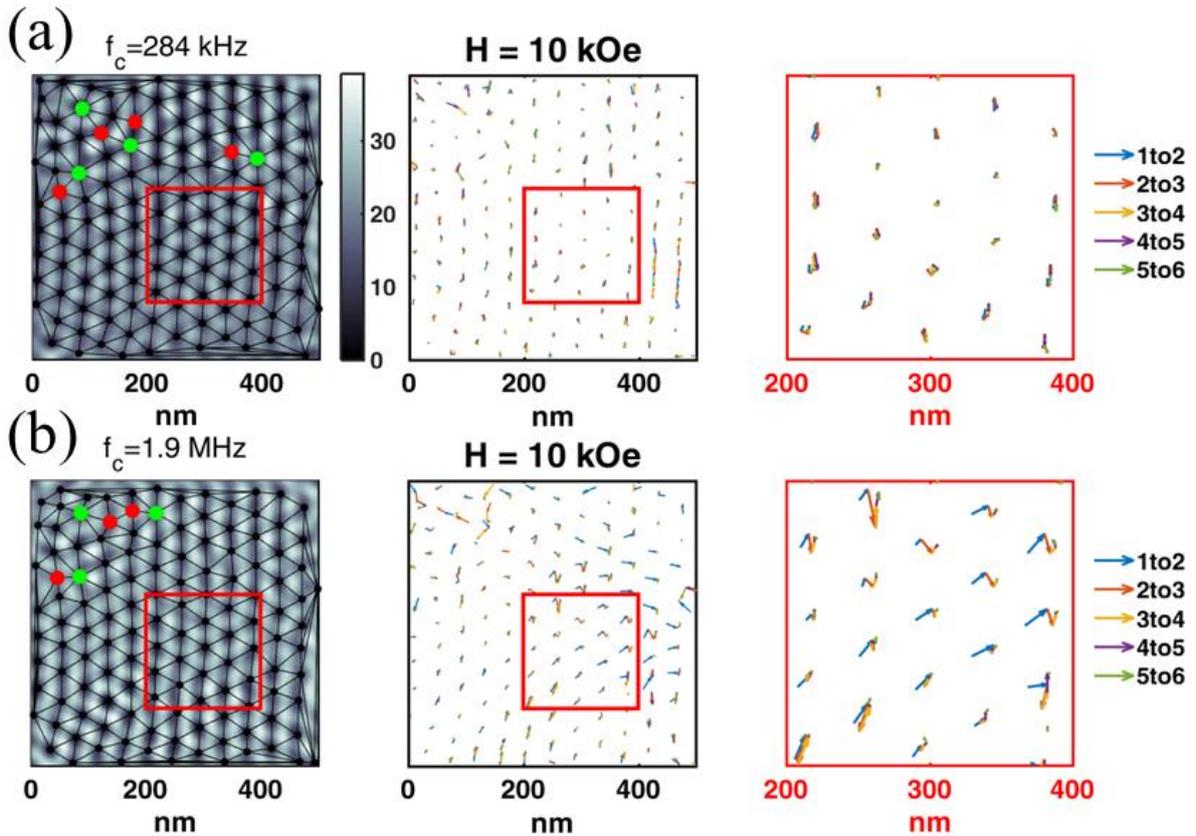

*Figure 5.17*: (a)-(b) Vortex movement at 10 kOe at 450 mK with filters of cut-off 284 kHz and 1.9 MHz. Left panel shows first image of 6 consecutive vortex images with topological defects, colour codes of which is same as *Figure 5.5*. Middle panel is arrows connecting the vortices, showing trajectories of them. Right panel is expanded view of the red box in the middle panel.

However, the estimation of vortex diffusion constant from Bardeen-Stephen-like calculation in a pinning-less scenario shows the diffusion constant to be $10^{-7} - 10^{-8} \, m^2/s$ using equation 1.33 in the thesis. This difference in the value of diffusion constant of 13-14 orders of magnitude can be somewhat bridged by taking into account a scenario where pinning is present. Since the Anderson-Kim flux-creep shows that net creep velocity in presence of a pinning potential, $F_0$ is given by, $v \sim v_0 e^{-F_0/k_B T} \times (driving\ term)$, where $v_0$ is creep velocity without any pinning, $k_B$ is Boltzmann constant and $T$ is temperature. The driving term arises due to tilt in the pinning potential due to a drive current. The net creep velocity will be unobservably small unless the drive term is huge, which is not the case in STS, since we do not apply any drive here. Therefore the diffusion constant obtained from a pinning-less Bardeen-Stephen-like estimate would be exponentially suppressed in presence of a pinning barrier. However whether this exponential factor can give a 13-14 orders of magnitude difference is a question to be explored further.

Though this kind of analysis from the vortex images is not completely free from our inability to determine large flux jumps due to indistinguishable nature of the vortices, it however gives an order of magnitude estimation of the resistivity that one should expect in the HVF phase. If a chunk of vortices slides by a large amount between acquiring two subsequent images, we cannot distinguish between individual vortices which can in principle move by more than one lattice constant. This limits our mean square displacement estimate to be less than one lattice constant squared, always. This in turn reduces the estimate of the diffusion constant.

From the transport data, the thermally activated flux flow (TAFF) resistivity (for $I \ll I_c$) turns out to be in the $\rho_{TAFF}^{Transport} \sim 0.01 - 10 \, n\Omega.m$ range. The flux-flow resistivity, on the other hand estimated from the vortex images is about $\rho_{FF}^{STM} \sim 10^{-10} \, n\Omega.m$. If we calculate the TAFF resistivity from this, we would obtain, $\rho_{TAFF}^{STM} \sim \rho_{FF}^{STM} \times \exp(-U/k_B T)$, which will reduce the order of magnitude of the estimate further. Since we do not know the exact description of the pinning potential, $U$, estimate of the exact order of magnitude of $\rho_{TAFF}^{STM}$ is not possible. However, the origin of this discrepancy[30] between $\rho_{TAFF}^{Transport}$ and $\rho_{TAFF}^{STM}$ is related to the extreme sensitivity of the HVF phase to external electromagnetic radiation.

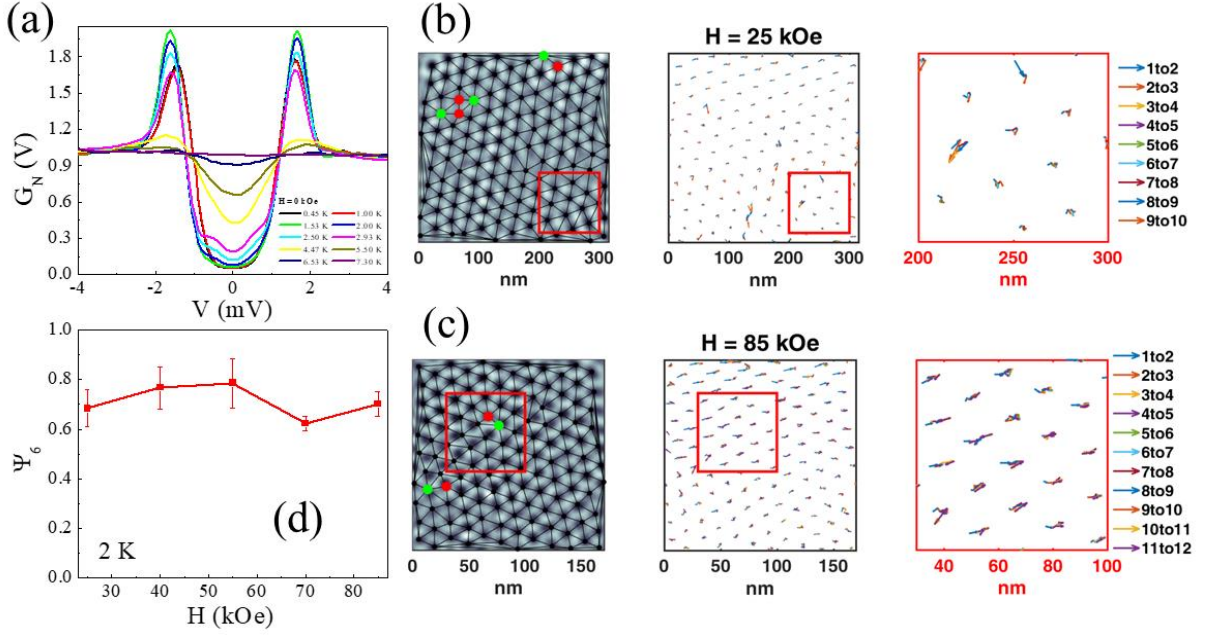

*Figure 5.18*: (a) $G_N(V) - V$ for different temperatures at zero field, shows $T_c \sim 7.3\ K$. (b)-(c) Vortex movement at 25 kOe and 85 kOe at 2 K. Left panel shows first image of 10-12 consecutive vortex images with topological defects, colour codes of which is same as *Figure 5.5*. Middle panel is arrows connecting the vortices, showing trajectories of them. Right panel is expanded view of the red box in the middle panel. (d) Variation of $\Psi_6$ as a function of magnetic field at 2 K, showing orientational order persisting even up to 85 kOe.

In Ref. 30, we have used a RC filter of cut-off frequency 340 kHz and have found that the thermally activated flux flow resistivity ($\rho_{TAFF}^{Transport}$) is vanishingly small in the HVF phase. This suggests that the HVF phase doesn't destroy the zero resistance state unless exposed to the external perturbation. We only get back the original value of $\rho_{TAFF}^{Transport} \sim 0.01 - 10\ n\Omega.m$ if we use filter cut-off frequency of 10 MHz or higher. However, in both cases where we use a filter of cut-off frequency of 340 kHz and where we use a filter of cut-off frequency 10 MHz, the flux-flow resistivity match exactly and both of them match with the Bardeen-Stephen form: $\rho_{FF}^{Transport} = \rho_N^{Transport} \times B/B_{c2}$. However there is no point in trying to match these values with $\rho_{FF}^{STM}$, since in STM we do not apply any drive current and the vortex motion is completely controlled by thermally generated flux hops. Furthermore, in the flux flow regime the pinning scenario does not matter since in this regime we are dealing with $I > I_c$. So, the immeasurably small $\rho_{TAFF}^{STM}$ could be comparable with $\rho_{TAFF}^{Transport}$ in presence of proper filtering and the discrepancy is addressed.

The effect of the external perturbation can be qualitativeley modelled in the following way. If each vortex is within a potential well given by $U$, then the external perturbation will shake the vortices and will reduce the potential by an amount $\Delta U$ by providing a kinetic energy to the vortices; hence the effective potential each vortex feels becomes, $U' = U - \Delta U$. The thermally activated flux flow (TAFF) resistivity is given by, $\rho_{TAFF} \sim \rho_{FF} \exp(-U/k_B T)$. This in presence of the external perturbation becomes, $\rho'_{TAFF} = \rho_{TAFF} \exp(\Delta U/k_B T)$ which can again be exponentially large. Hence the observation of a finite TAFF resistivity (in the range of $0.01 - 10\, n\Omega.m$) in the HVF phase is possible only in the presence of the external perturbation. Otherwise the TAFF resistivity is expected to be below our measurable limit and can be indeed compared with the value obtained from the vortex images. Such simplistic models like shaking of the vortices by the external electromagnetic signal (in the range of few hundreds of kHz) giving rise to behaviour similar to *shear thinning*[31] can in principle explain this extremely sensitive nature of the HVF phase.

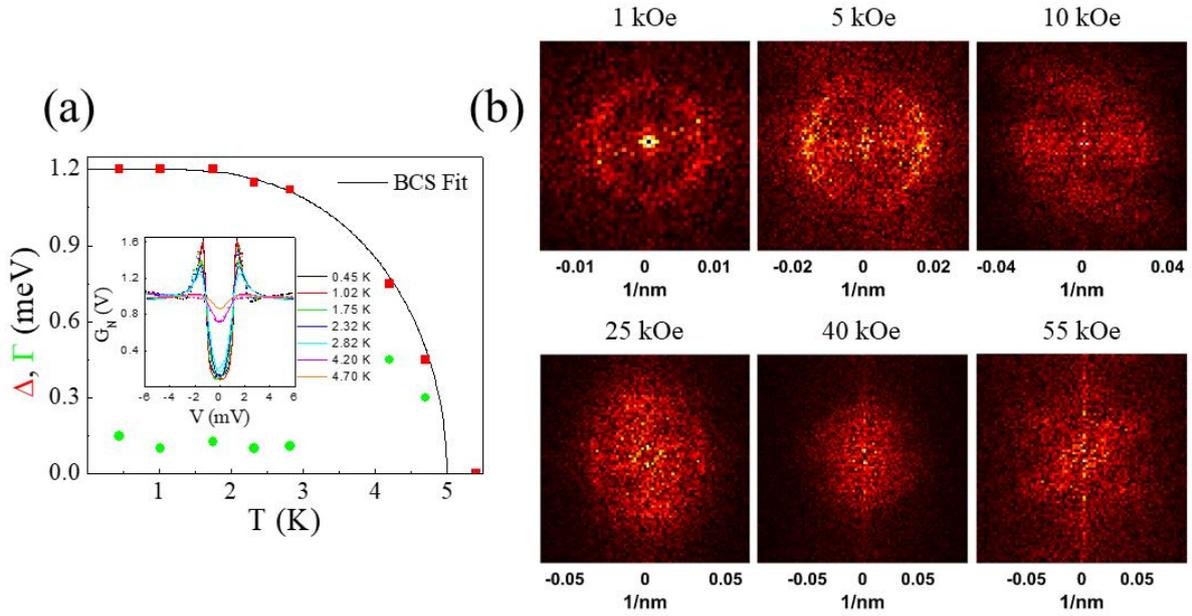

*Figure 5.19*: (a) Variation of Δ (Γ) as a function of temperature shown in red (green) squares, along with BCS fit of $\Delta - T$. (*inset*) $G_N(V) - V$ for different temperatures at zero field, shows $T_c \sim 5\,K$, with BCS fits to the spectra. (b) FFT of vortex images for different magnetic fields at 450 mK, shows ring-like feature in FFT up to 40 kOe, above which it becomes very diffused.

However a thorough theoretical analysis is required to identify the precise microscopic mechanism. Furthermore, the STS/M images and the vortex dynamics observed therein have been considered as undisturbed by the external radiation. This is explainable, since the vortex images have been gathered for 10 kOe at 450 mK utilizing one 284 kHz filter and one 1 MHz filter. Their comparison, as is shown in *Figure 5.17a-b*, shows that the strict filtering has not changed the vortex motions.

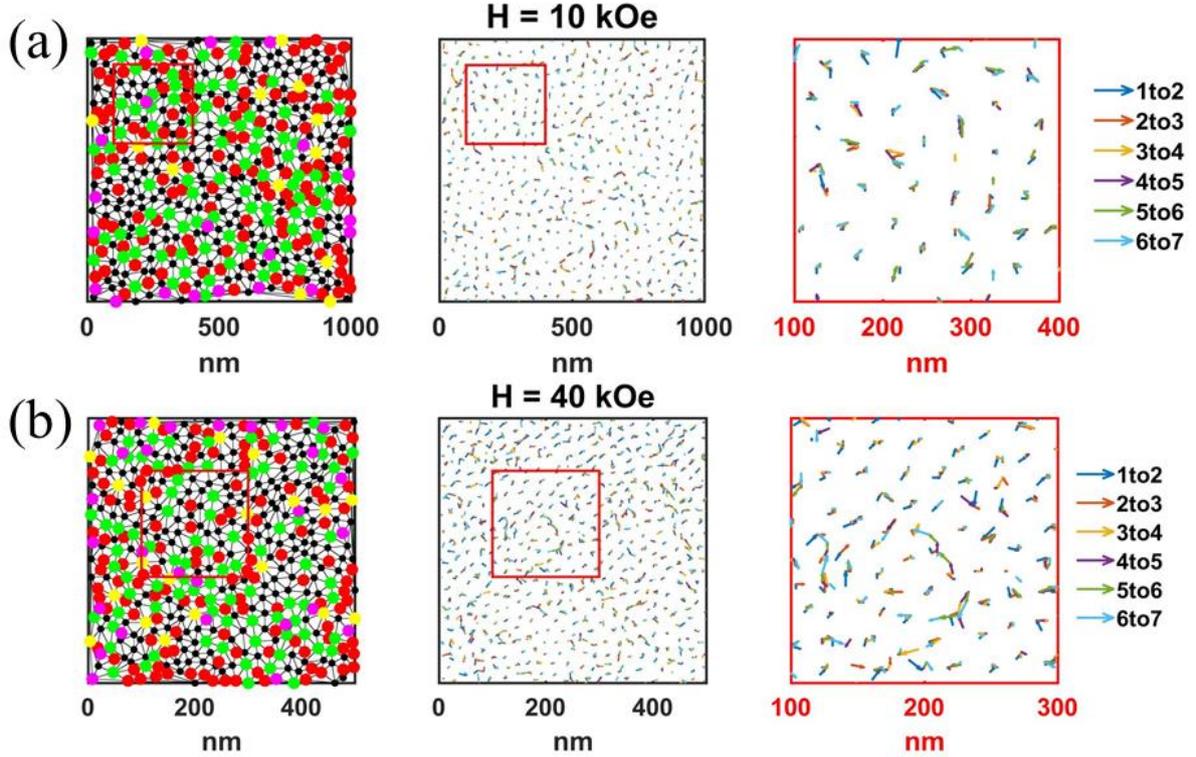

*Figure 5.20*: (a)-(b) Vortex movement at 10 kOe and 40 kOe at 450 mK. Left panel shows first image of 7 consecutive vortex images with topological defects, colour codes of which is same as *Figure 5.5*. Middle panel is arrows connecting the vortices, showing trajectories of them. Right panel is expanded view of the red box in the middle panel. Although in 10 kOe the movements are very small except for occasional large jumps, in 40 kOe the amplitude of motion is larger. But in both cases there is no directionality of motion.

### 5.4 Thickness dependence of vortex lattice melting

In this section we shall study the effect of dimensionality on vortex lattice melting by tuning the thickness of the *a*-MoGe thin films. We study here a $\sim 40\ nm$ thick sample, which has a $T_c \sim 7.3\ K$ (Bias spectroscopy at different temperatures is shown in *Figure 5.18a*) and another $\sim 5\ nm$ sample with $T_c \sim 5\ K$. The first one, although thicker than the one studied before (thickness$\sim 21\ nm$), is still within 2D limit from the fact that typical bending length of vortices is in micron range which is much larger than the thickest thin film here. On the other hand the vortex lattice in the 5 $nm$ sample is in an extreme 2D limit.

### 5.4.1 Omnipresent hexatic phase

In *Figure 5.18b-d*, we show temporal dynamics of vortex lattice in 25, 55 and 85 kOe at 2 K, along with their FFT. All the FFT-s for fields up to 85 kOe, show six clear spots. The temporal dynamics of the vortices on the other hand always show random appearance and disappearance of topological defects, as well as, directionality of the vortex flow. Although the existence of a VS at low fields is not explored, the HVF phase seems to be omnipresent all over our magnetic field window. The variation of $\Psi_6$ at 2 K as a function of field, in *Figure 5.18e*, reinstates this omnipresence of HVF phase, since $\Psi_6 \approx 0.6$ within our magnetic field window.

### 5.4.2 Vortex glass to vortex liquid melting in extreme 2D limit

Next we explore the ultrathin $\sim 5\ nm$ thin film, which shows BCS-type gap variation with $\Delta(0) \sim 1.2\ meV$ (shown in *Figure 5.19a*). However, at 450 mK, starting from 1 kOe we observe ring-structure in FFT (shown in *Figure 5.19b*) of vortex images and the ring persists up to $\sim 40\ kOe$, beyond which it becomes completely diffused. In all magnetic fields, $\Psi_K$ ($\equiv \frac{1}{N}\sum_{i=1}^{N} G_{\vec{K}}(\vec{r}_i)$) and $\Psi_6$ are zero within error bar, signifying an amorphous glassy phase starting from lowest magnetic field. The glassiness of the phase is revealed in *Figure 5.20a-b*, where we compare vortex motion between 10 and 40 kOe. In 10 kOe, the vortex motion is very low except for few occasional flux jumps, although the VL is filled with large number of topological defects. At 40 kOe, on the other hand the vortex motion is much higher and completely random, indicative of a vortex liquid. Since at 450 mK, superconducting gap persists at least up to 85 kOe, the fuzziness in vortex images for $H \gtrsim 40\ kOe$ can be attributed to the advent of a vortex liquid phase. It is widely believed[22,32] that in presence of quenched random disorder the vortex lattice becomes a vortex glass, which can melt into a liquid with temperature or magnetic field acting as the tuning parameters. In this vortex glass phase vortices are frozen into a random configuration determined by the details of the pinning centres and hence they are not free to move giving rise to zero ohmic-resistivity. The vortex glass to vortex liquid transition is also found in equivalent transport measurements, where $\log V - \log I$ shows a curvature change[33] around 35 kOe, signifying the vortex glass to vortex liquid transition, quite close to the STS/M report.

In summary the observation of a BKTHNY two step melting of the 2D vortex lattice is proven beyond any doubt and the presence of a true hexatic vortex fluid phase is also established. Although apparently there is a perfect match between the STS/M and transport data, close inspection shows that the vortex motions are greatly affected by the existence of an external electromagnetic perturbation. The external electromagnetic perturbation amplifies the transport signal and helps to detect the phase transitions. In presence of disorder, however, the hexatic phase completely vanishes and gives rise to a vortex glass which in higher magnetic field melts into a vortex liquid.

# Chapter VI: Possibility of vortex fluctuation in weakly pinned *a*-MoGe thin film

In the previous chapter we have established the phase diagram of melting transition of a 2D vortex lattice down to 300 mK.[1] However, the fluid phases are observed to exist down to the lowest temperature, as well as, over a large region in the $H-T$ parameter space. This is quite contradictory to the common rationale because in a thermally activated transition it is expected to have the liquid phases frozen as $T \to 0$.[2] It is interesting to note that at elevated temperatures a VS can be driven into an HVF and then an HVF phase into an IVL phase due to thermally generated topological defects. At lower temperatures on the other hand the advent of these defects can have either thermal or quantum origin. Nonetheless in the following sections I'll present data relating to vortex core of *a*-MoGe and will try to extract as much information from it, as possible.

## 6.1 Vortex core spectroscopy

The first set of data is of vortex core spectroscopy in the 20 nm thick *a*-MoGe thin film which we have explored in the previous chapter. In STS measurements the tunnelling conductance, $G(V) = \frac{dI}{dV}\big|_V$ between the tip and the sample gives the local tunnelling density of states (LTDOS) at energy level $eV$ with respect to Fermi energy. Here we capture $G(V)$ vs $V$ spectroscopic map at 450 mK over a small area containing 6-8 vortices. Representative vortex image at 5 and 65 kOe are shown in *Figure 6.1a-b*.

### 6.1.1 Presence of soft gap at the centre of vortex core

To study the nature of the vortex core, in *Figure 6.1c-d* we show the normalized bias spectra $G_N(V)$ vs $V$ (where $G_N(V) = G(V)/G(V = 4\,mV \gg \Delta/e)$, where $\Delta \sim 1.3\,meV$) along the line passing through the centre of a vortex core at these two representative fields. Here we observe that the superconducting gap starts to fill while approaching the centre of the vortex core. However at the centre of the vortex core a soft gap in the LTDOS continues to exist, i.e., $G_N(0) < 1$. In *Figure 6.1e-f* we show variation of $G_N(0)$ along three lines passing through the vortex centre. These three lines are at $45°$ angle with each other and show that the core is mostly symmetric in these directions. This presence of soft-gap at the centres of the vortex cores is contrary to the regular vortex cores of for example, NbSe$_2$, where in the cleanest limit the core accommodates a low energy fermionic bound state (known as Caroli-de Gennes-Matricon (CdGM) bound state).[3,4] However in slightly disordered NbSe$_2$, the core loses this bound state due to scattering and a normal metal-like flat spectra is seen at the centre of the vortex core.[5]

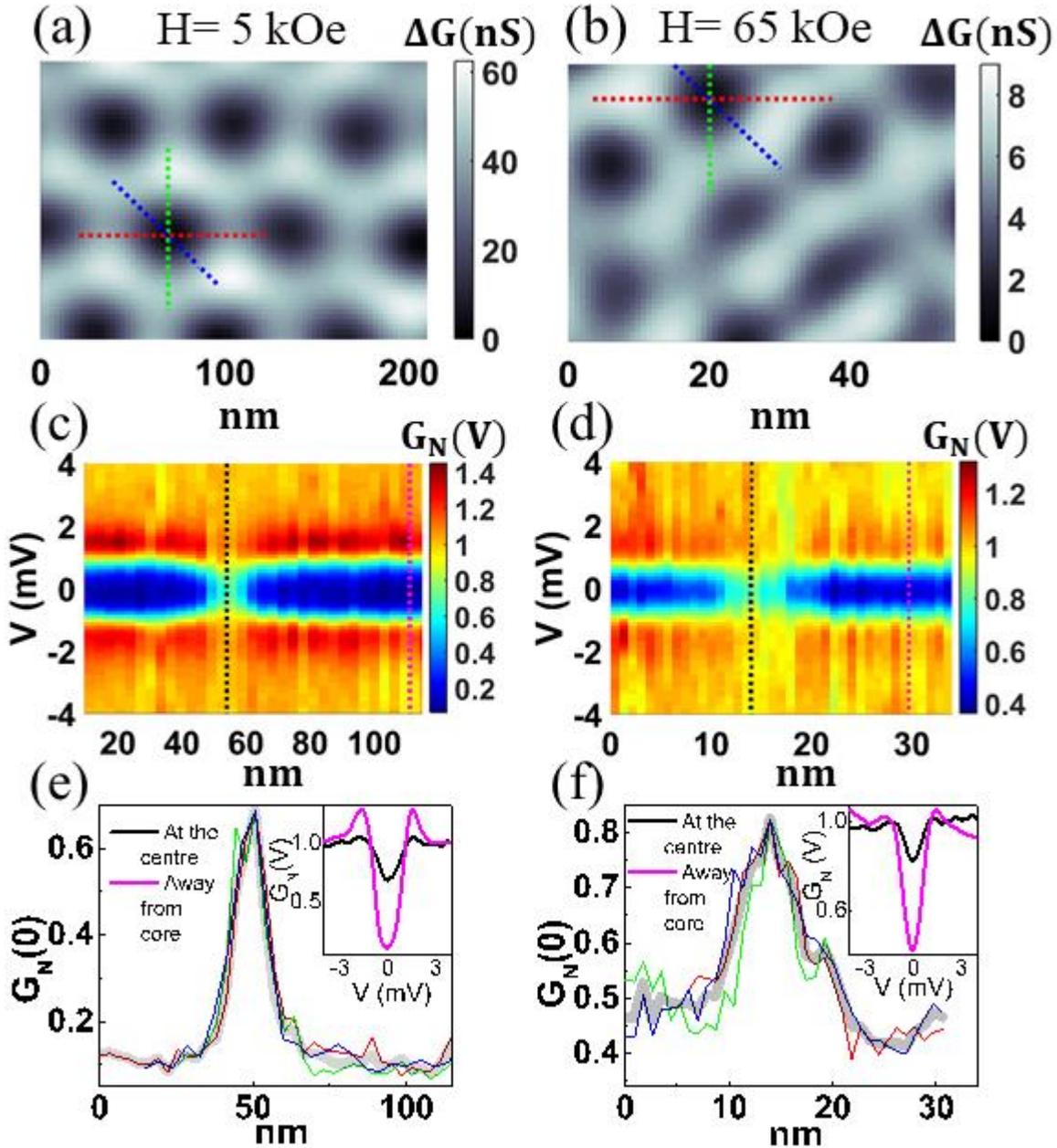

*Figure 6.1*: (a)-(b) Vortex images at 5 kOe and 65 kOe at 450 mK, showing the $\Delta G$ maps at fixed bias $V = 1.3\ mV$. The red dotted lines are shown in (c)-(d) as $G_N(V) - V$ map across the length of the vortex core. Spectra long the black (magenta) dotted lines are shown in *inset* of (e) and (f). (e)-(f) $G_N(0)$ line cuts along the vortex core, along the red, green and blue dotted lines in (a)-(b) are shown along with the average $G_N(0)$ variation in transparent grey.

A soft gap at the centre of the vortex has already been mentioned in unconventional high-$T_c$ cuprate superconductors[6,7] and in strongly disordered NbN thin films[8,9,10] (see section 4.2.3). This has been attributed mostly to the presence of phase incoherent Cooper pairs inside the vortex core. But this is true if and only if the superfluid stiffness of the material ($J_s$) is low and is of the order of Δ. This is exactly the case for disordered NbN thin films.[11] However, *a*-MoGe is a conventional superconductor following closely the BCS theory, where $T_c(0)$ and gap vanishing temperature, $T^*(0)$ are exactly same. Furthermore, using the value of $\lambda(0) \sim 534\ nm$, we get $J_s \sim 100\ K \gg \Delta/k_B$, which tells us that the possibility of phase fluctuation is rare in *a*-MoGe. This instructs us to look for other origins.

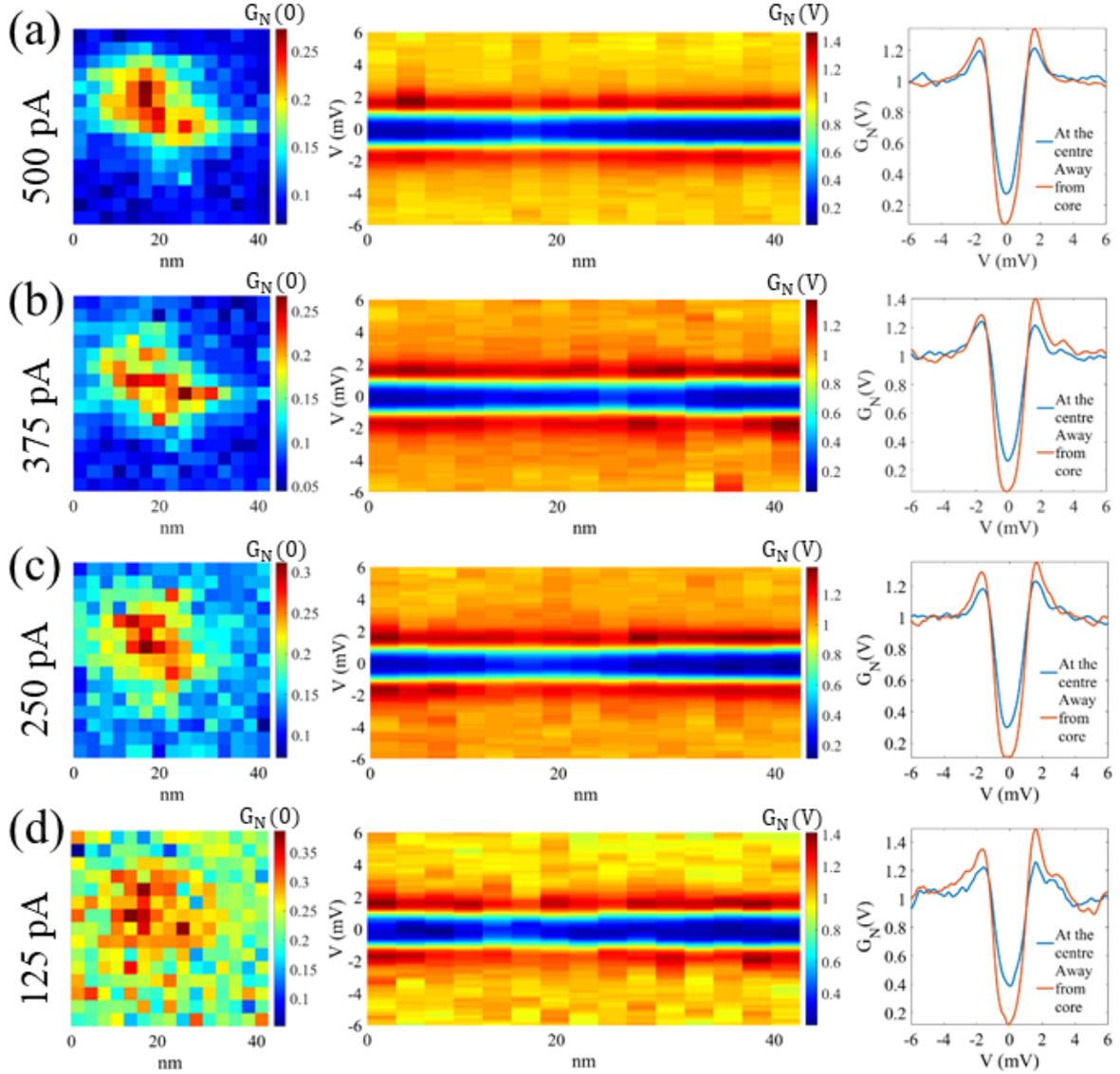

*Figure 6.2*: (a)-(d) *left* panel shows $G_N(0)$ maps at 450 mK at 10 kOe for a single vortex for set current ranging between 500 pA and 125 pA. *Middle* panel shows a $G_N(V) - V$ map along the vortex core for corresponding set currents. Blue (orange) curve in *right* panel shows spectra at the centre of the vortex (away from core).

### 6.1.1.1    *Effect of changing set current and cut-off filter*

We study the effect of set current on the vortex core, since a higher value of set current can create local heating and as a result can give rise to core states which are not in equilibrium. We inspect in *Figure 6.2a* $G_N(0)$ map of the same vortex at different set current ranging from

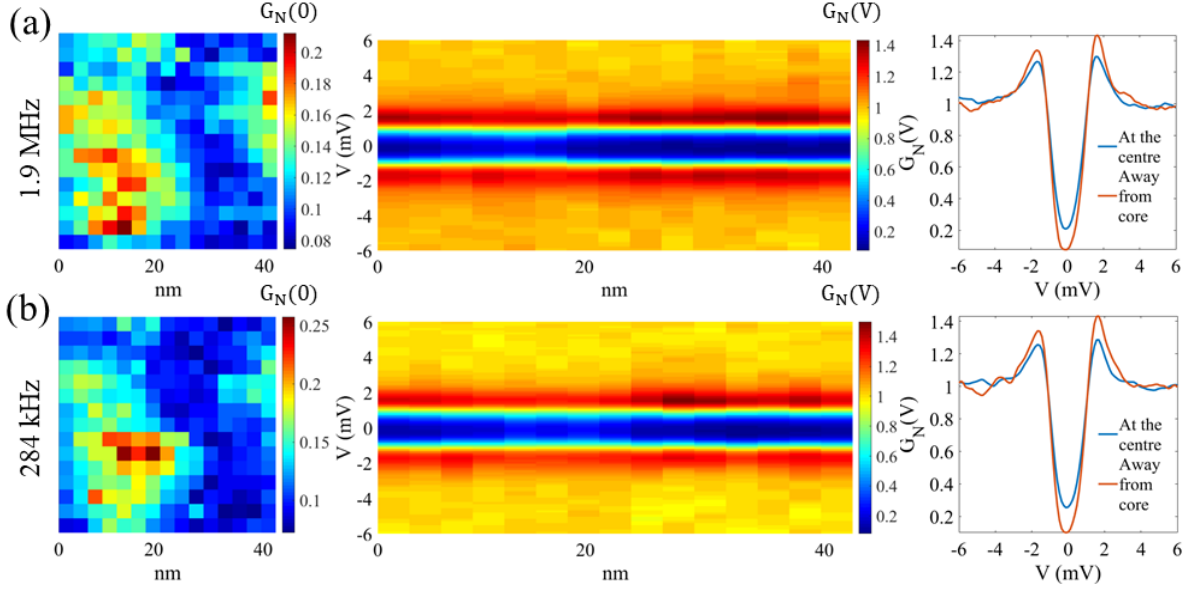

*Figure 6.3*: (a)-(b) *left* panel shows $G_N(0)$ maps at 450 mK at 10 kOe for a single vortex with filters of cut off frequencies 1.9 MHz and 284 kHz respectively. *Middle* panel shows a $G_N(V) - V$ map along the vortex core for corresponding set currents. Blue (orange) curve in *right* panel shows spectra at the centre of the vortex (away from core).

500 pA down to 125 pA at 10 kOe at 450 mK, and observe that apart from slight decrease in contrast, the vortex core looks similar. In *Figure 6.2b*, we show normalized bias spectra along a line through the vortex core and in *Figure 6.2c*, we compare a spectra at the centre of the vortex core with a spectra away from the core and observe that they all look similar and the soft gap persists for all of these different set currents.

We also study the effect of changing the low pass filter (similar to section 5.1.8) and compare two cases, one with 1 MHz cut-off filter and another with 284 kHz cut-off filter, in *Figure 6.3a-b*. Here too, there is no difference between the two cases and both of them show similar variation of bias spectra along the vortex core. This signifies that the external electromagnetic perturbation does not have any effect on the soft gap inside the vortex core.

### 6.1.2 Variation of ZBC as a function of magnetic field

In *Figure 6.4*, we pot $G_N(0)$ at the centre of the vortex (designated as $G_N(0)(Center)$) and in between two vortices (designated as $G_N(0)(Away)$) as a function of magnetic field, $H$. Here each point is obtained by taking the average value of $G_N(0)$ of 6 vortex centres, where the values for the 6 vortices differ from each other by $\lesssim 10\%$. We observe that $G_N(0)$ shows an overall increasing trend with magnetic field except for two anomalous dips around 5 kOe and 70 kOe. But most interestingly the vortex core always show a soft gap at the centre of the core.

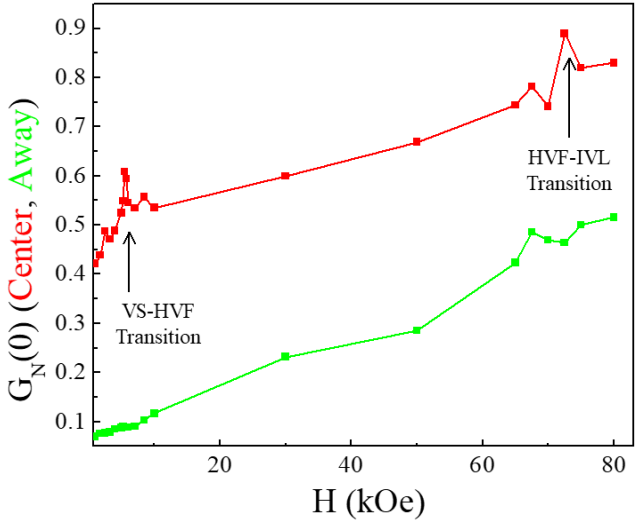

*Figure 6.4*: Variation of $G_N(0)$ as a function of magnetic field at 450 mK, where the red (green) connected points show $G_N(0)$ values at the centre (away from) of a vortx core. The value is obtained by averaging over 6-7 vortices for each field. The transition fields of VS-HVF and HVF-IVL are marked by the arrows.

## 6.2 Conjecture and simulation: fast vibration of vortex core

The observation of an omnipresent soft gap at the vortex centre over the whole magnetic field range and its unresponsiveness to changing set current and using different filters cannot be explained by phase fluctuation due to two completely different energy-scales. However it can be easily understood if we assume that the vortex core spatially fluctuates rapidly about its mean position. Since STS is a slow measurement, a rapid fluctuation of the vortices will get integrated out showing only the average tunnelling conductance at a particular location. This fluctuation, in principle, is superposed on the slow diffusive motion in the HVF and the IVL states. The qualitative effect of this kind of fluctuation is that the tunnelling conductance close to the centre of the vortex would have contribution from both the vortex core and the superconducting regions around it.

This simplistic model tells us that this kind of fluctuation of the vortices would give rise to significant Bardeen-Stephen loss if the flux lines enter the superconducting regions to a distance of the order of the lattice constant. Therefore this conjecture can, in principle, be extended to connect it to the large range to which the fluid phases exist. Although this kind of conjecture can have its faults and limitations, it is however interesting to investigate this model in order to extract as much information as possible from the data.

### 6.2.1 Simulation strategy

In order to support the conjecture in a more quantitative manner, the exact solution for the LTDOS in presence of the vortices is required which can only be obtained by solving the Usadel equations.[12] However this is very difficult and therefore here we adopt a pure phenomenological approach.

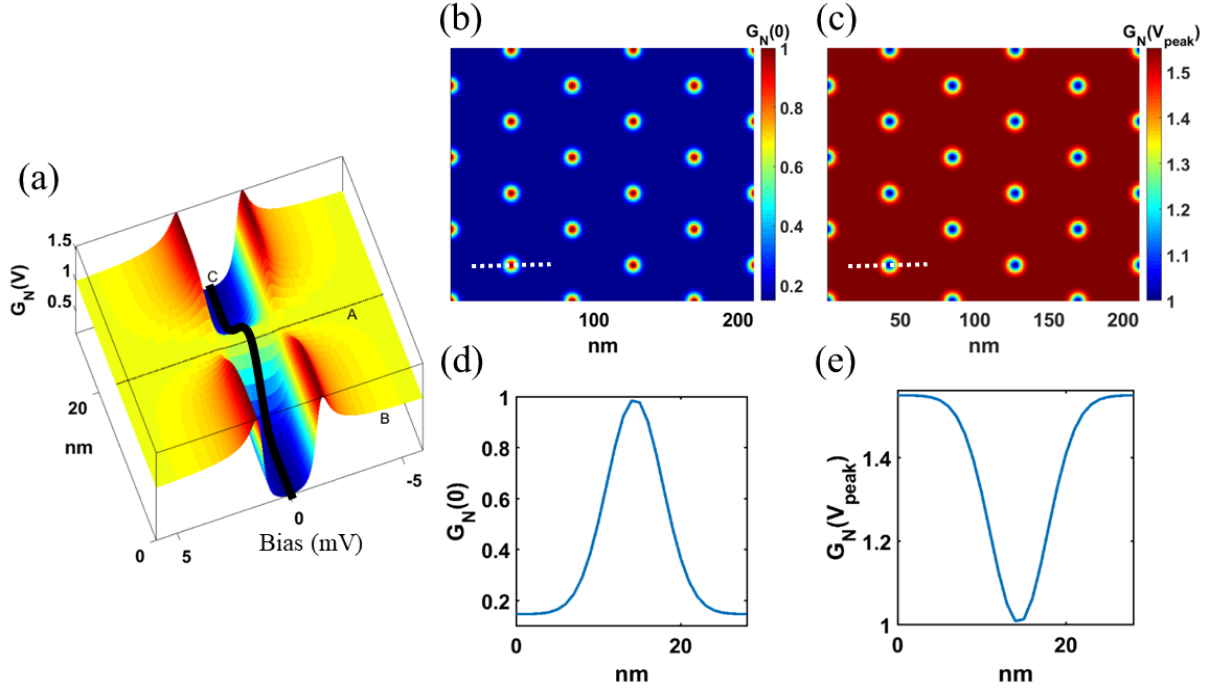

*Figure 6.5*: (a) Simulated $G_N(V) - V$ variation along a single vortex core. Line A corresponds to flat spectra at the centre of the vortex core, while B is the spectra of BCS nature, situated away from the core. Black line C gives the $G_N(0)$ profile along a vortex core, which is of Gaussian nature. (b)-(c) Simulated $G_N(0)$ and $G_N(V_{peak})$ maps of a vortex lattice, where the white dotted lines are shown in (d) and (e). In (d) and (e) it is visible that the values of $G_N(0)$ and $G_N(V_{peak})$ are 1 at the centre of the vortex core.

### 6.2.1.1   Simulating a single vortex

At first we simulate a single vortex. We assume that at the core of an isolated vortex $G_N(V) = 1$ and far away from the core, $G_N(V) = G_N^{BCS+\Gamma}(T,V)$ given by superconducting DOS of BCS theory. $G_N^{BCS+\Gamma}(T,V)$ is obtained by fitting the experimental zero field $G_N(V)$ vs $V$ data at temperature $T$ using $\Delta$ and $\Gamma$ as fitting parameters. Now an interpolation is done between these two spectra using an empirical Gaussian weight factor, $f(r) = \exp(-r^2/2\sigma^2)$, such that,

$$G_N(V, r) = f(r) + \big(1 - f(r)\big) G_N^{BCS+\Gamma}(T, V) \tag{6.1}$$

The width of the Gaussian factor is taken such that the $fwhm \sim 2 \times \xi_{GL}$, where $\xi_{GL}$ is experimentally obtained from $H_{c2}$. This gives, $\sigma \sim 0.849 \xi_{GL}$. Line cut across such a single vortex is shown in *Figure 5a*.

### 6.2.1.2   Simulating a vortex lattice

Next to simulate a vortex lattice we assume that the resultant normalized conductance,

$$\tilde{G}_N(V,r) = \frac{\sum_i G_N(V, r - r_i)}{[\sum_i G_N(V = 0, r - r_i)]_{max}} \quad (6.2)$$

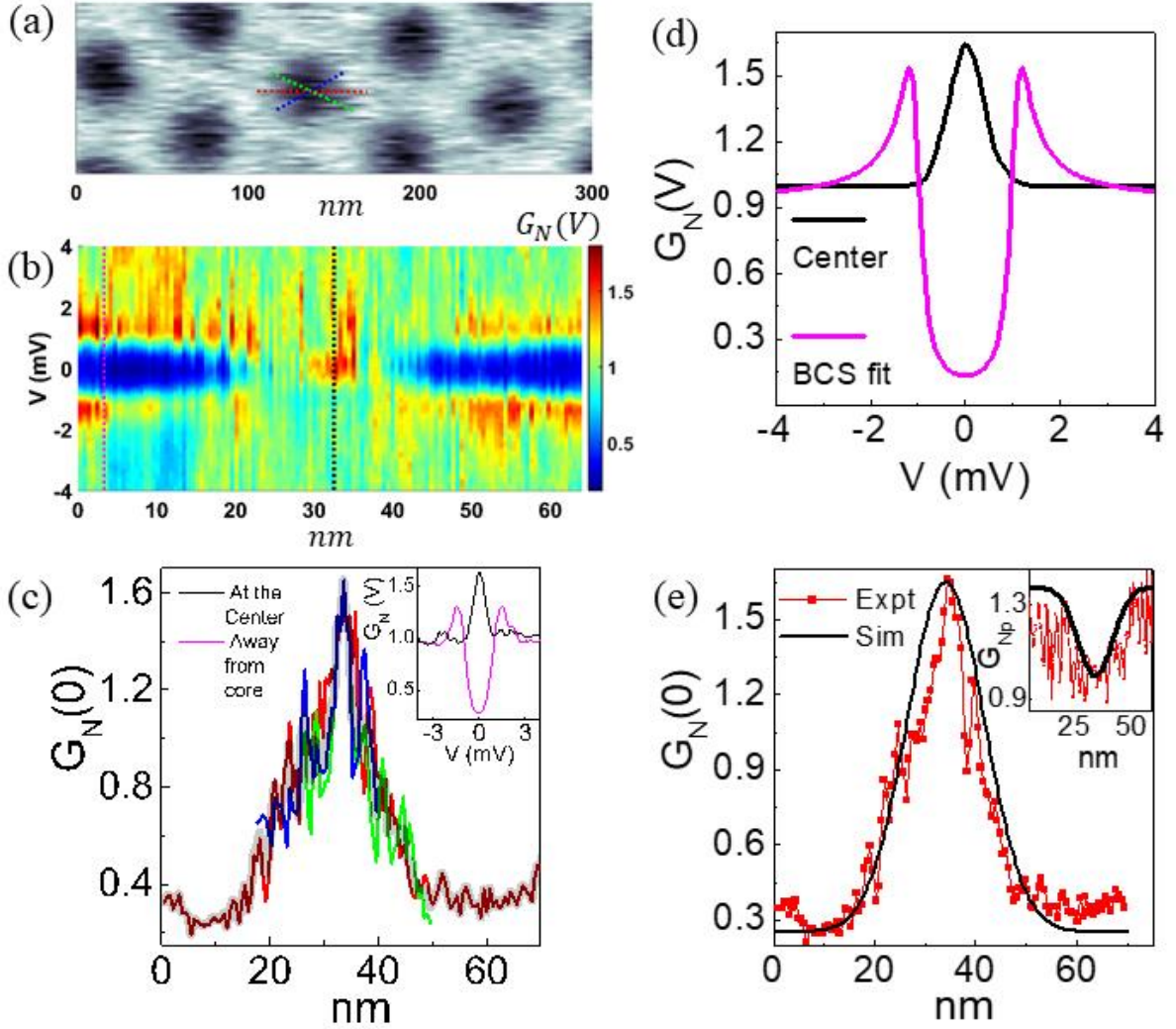

*Figure 6.6*: Simulating vortex core in NbSe$_2$ for 5 kOe at 450 mK (a) Vortex image, observed by recording $\Delta G(V = 1.3\ meV)$ across the area, showing three lines across one vortex. (b) Spectroscopic image across the red line of (a) with black dotted line denoting spectra at the centre of the vortex core and pink dotted line away from the centre. These two spectra are shown in *inset* of (c). (c) Three $G_N(0)$ line-cuts across vortex core marked in (a). The transparent black line-cut is the average of the three. (d) Black spectra corresponds to smoothened version of the spectra at the centre of the core and pink line is BCS fit to a spectra which is away from the vortex core. (e) Simulated $\tilde{G}_N(0)$ in black line is compared with averaged $G_N(0)$ line-cut in red obtained from. (*inset*) Simulated $\tilde{G}_N(V = 1.3\ meV)$ in black line is compared with averaged $G_N(V = 1.3\ meV)$ from actual data.

is a linear superposition of the conductance from all vortices, where $r_i$ is the position of the $i$-th vortex and the sum over $i$ runs over all the vortices. The normalizing factor ensures that at the centre of each vortex, $\tilde{G}_N(V = 0) = 1$. $G_N(0)$ and $G_{Np}$ maps of such a resultant vortex lattice at 10 kOe is shown in *Figure 6.5b-c*. The line cuts of $G_N(0)$ and $G_{Np}$ along one of the vortices is shown in *Figure 6.5d-e*. This part of the simulation mimics a vortex lattice similar to a slightly disordered NbSe₂ for example.

### *6.2.1.3* *Validation of the interpolating strategy*

Although the assumptions of a flat spectra at the centre of the vortex core and a BCS-like spectra far away from the core is well justified, the assumption of a Gaussian nature of the weight factor needs to be justified. Although in general the order parameter is expected to follow a Lorentzian shape towards the centre of the vortex core,[13] the difference between a Lorentzian and a Gaussian starts to grow away from $fwhm$. To check the validity of the Gaussian factor we apply it on a conventional system, the vortex state in clean 2H-NbSe₂ single crystal. *Figure 6.6a* shows the VL image on an NbSe₂ single crystal at 5 kOe at 450 mK, where the bias voltage is fixed at 1.2 mV, close to the coherence peak. In *Figure 6.6b*, we show the $G_N(V)$ vs $V$ spectra along a line passing through the centre of the vortex. Here the core of the vortex shows a zero bias conductance peak (CdGM bound state), i.e. $G_N(V = 0) > 1$. On the other hand spectra obtained inside the superconducting regions show regular BCS-like nature, partially broadened by the circulating supercurrent around the vortex core (*inset of Figure 6.6c*). The experimental variation of $G_N(0)$ along three lines passing through the centre of the vortex core along with their average is shown in *Figure 6.6c*.

As previously mentioned, far away from the vortex core $G_N^{BCS}(V)$ is found from fitting experimental tunnelling conductance spectra with BCS DOS using Δ and Γ as fitting parameters. On the other hand, for the spectra at the centre of the vortex, $G_N^{centre}(V)$ is obtained directly from experimental spectra at the vortex core. Using $\xi_{GL}^{NbSe_2} \sim 8.9\ nm$ (obtained from $H_{c2} \sim 42\ kOe$), we construct a single vortex situated at $r$, using,

$$G_N(V, r) = f(r) G_N^{centre}(V) + \big(1 - f(r)\big) G_N^{BCS}(V) \tag{6.3}$$

Using this as the building block we construct the VL by linear superposition of the conductance values from all vortices using,

$$\tilde{G}_N(V, r) = 1.6 \times \frac{\sum_i G_N(V, r - r_i)}{[\sum_i G_N(V = 0, r - r_i)]_{max}} \tag{6.4}$$

where, $r_i$ is the position of the $i$-th vortex. This construct ensures that at the centre of the vortex core the simulated value matches with the experimental one. In *Figure 6.6e* we compare the experimental variation of $G_N(0)$ with simulated $\tilde{G}_N(0)$ along a line passing through a vortex core. The *inset of Figure 6.6e* shows comparison between variation of $G_{Np} (\equiv G_N(V = 1.2\ mV))$ and $\tilde{G}_N(V = 1.2\ mV)$. This good agreement between these two cases shows the validity of our phenomenological approach to simulate the conductance map.

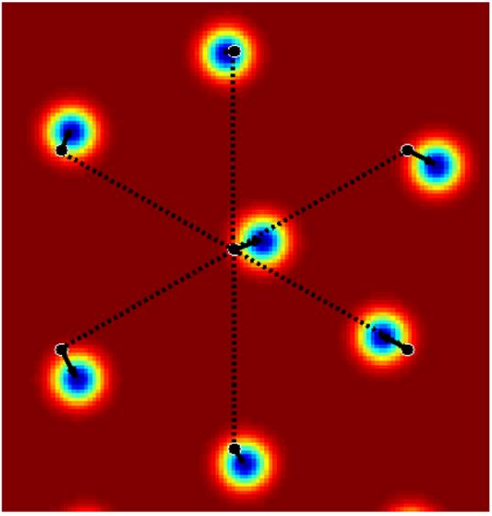

*Figure 6.7*: Simulated $G_N(V_{peak})$ map where each vortex (blue dots) is displaced from their actual position (black dots) by a random vector $\delta \vec{r}$. The maximum amplitude of the random vector is chosen such that $|\delta \vec{r}| \leq \alpha < a_\Delta$.

#### *6.2.1.4  Simulation of random displacement*

The random fluctuation of the VL can be thought of as average over an ensemble of randomly displaced vortices from their mean positions, where each element of the ensemble represent a time-snap of the fluctuating VL. To simulate this, we calculate $\tilde{G}_N(V, r)$ for 200 realizations of a distorted hexagonal lattice with lattice constant $a$. Each lattice point is displaced by a random vector, $\delta \vec{r}_i$ with the imposed constraint $|\delta \vec{r}_i| \leq \alpha < a$ and compute the average $\langle \tilde{G}_N(V, r) \rangle_\alpha$, where $\alpha$ is the maximum amplitude of fluctuations. An example of such random displacement is shown in *Figure 6.7*, where each arrow represents $\delta \vec{r}_i$.

### 6.2.2  Extraction of relative fluctuation amplitude

*Figure 6.8a-b* show representative $G_N(0)$ maps simulated for a VL at 10 kOe with and without incorporating the fluctuation of the vortices. *Figure 6.8c-d* show corresponding conductance maps for $V = 1.45\ mV$ close to the coherence peaks. It is interesting to note that the size of the vortex core in presence of fluctuation has evidently increased from its counterpart without fluctuation. Now to fit the experimental data we take $\alpha$ and $\Gamma$ as fitting parameters and constrain $\xi$ between 3.9-4.5 nm, which is within 10% of $\xi_{GL} \sim 4.3\ nm$ (obtained from $H_{c2}$). *Figure 6.9a-h* show the line cuts of $G_N(0, r)$ along with $\langle \tilde{G}_N(V = 0, r) \rangle_\alpha$ for fields ranging from 1 kOe to 80 kOe. Above 10 kOe, we have to use a larger $\Gamma$, compared to its zero field value to account for the additional broadening from the orbital current around the vortices. The variation of $\Gamma$ as a function of magnetic field is shown in *Figure 6.9i*. It should be noted here that the effect of changing $\alpha$ is to inversely change the peak height of the $\langle \tilde{G}_N(V = 0, r) \rangle_\alpha$

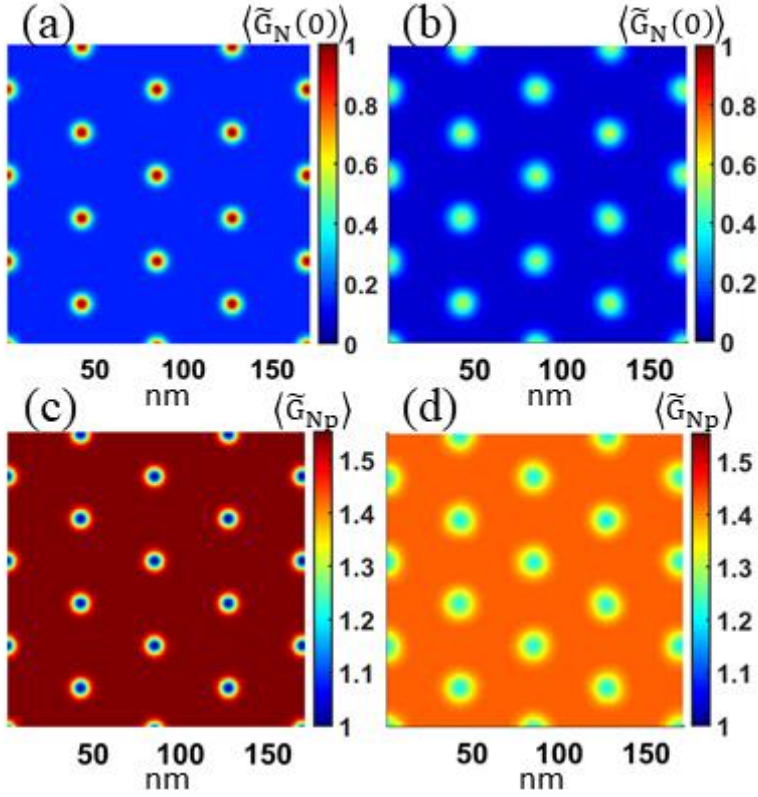

*Figure 6.8*: (a)-(b) Comparison between simulated $\langle \tilde{G}_N(0) \rangle$ maps for 10 kOe with and without fluctuation averaging. (c)-(d) $\langle \tilde{G}_N(V_{peak}) \rangle$ maps for 10 kOe with and without fluctuation averaging. In both cases the scales have been kept same to bring clarity to the difference.

line-cut across a vortex, i.e. increasing the value of $\alpha$ decreases $\langle \tilde{G}_N(V = 0, r) \rangle_\alpha^{peak}$ and vice-versa. $\alpha$ also determines the resultant width of the $\langle \tilde{G}_N(V = 0, r) \rangle_\alpha$ line-cut because larger $\alpha$ gives rise to larger mixing between the normal and the superconducting regions. On the other hand the effect of $\Gamma$ is to change the base level of $\langle \tilde{G}_N(V = 0, r) \rangle_\alpha$ far away from the vortex, i.e. increasing $\Gamma$ would increase $\langle \tilde{G}_N(V = 0, r \gg r_{core}) \rangle_\alpha$ and vice-versa.

However, for $V = 1.45\ mV$, the same set of parameters does not reproduce the experimental variation of $G_N(V = 1.45\ mV)$. Although the qualitative nature of the variation is captured in the simulated curve, the conductance value is overestimated by 15-20%. Example of such comparison between $\langle \tilde{G}_N(V = 1.45\ mV, r) \rangle_\alpha$ and $G_N(V = 1.45\ mV, r)$ for three magnetic fields is shown in *Figure 6.10a-c*. This mismatch between the simulated and experimental variation of $G_N(V = 1.45\ mV)$ is most likely because of our inability to account for the effect of orbital current on the coherence peaks, which the phenomenological parameter $\Gamma$ does not take care of.

Nonetheless, the very good fitting of the variation of $\langle \tilde{G}_N(V = 0, r) \rangle_\alpha$ with $G_N(V = 0, r)$ is used to extract the fitting parameter, $\alpha$ for different magnetic fields, and is plotted in *Figure 6.11* as relative fluctuation amplitude, $\alpha/a$ as a function of magnetic field.

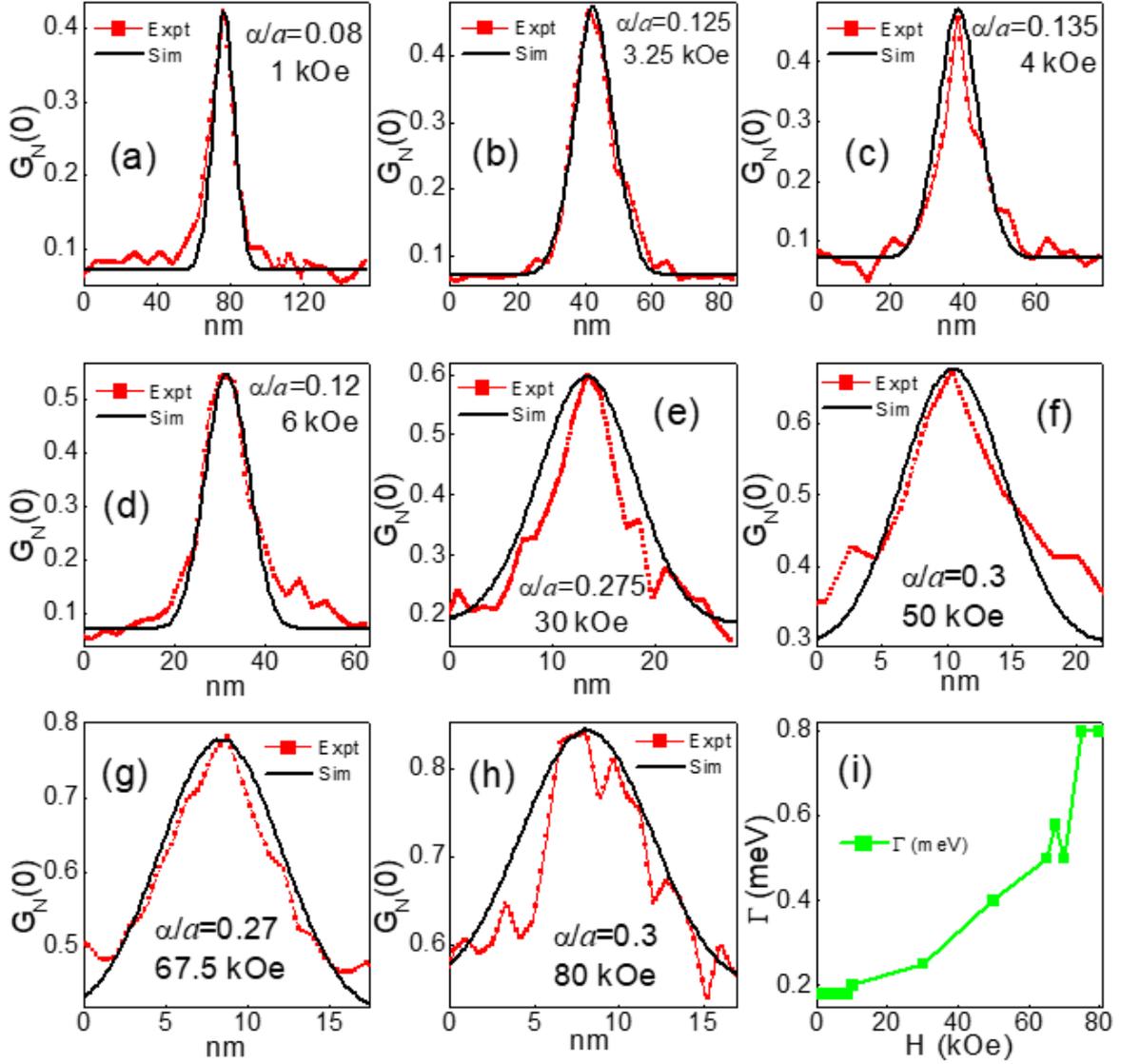

*Figure 6.9*: $G_N(0)$ variation for different magnetic fields are shown in (a)-(h) which are compared with simulated line-cuts of $\langle \tilde{G}_N(0,r) \rangle_\alpha$. Values of $\alpha/a$ for each of the fields are mentioned. (i) Variation of phenomenological BCS fitting parameter, $\Gamma$ $(meV)$ as a function of magnetic field.

Here $\alpha$ is obtained by fitting $G_N(0,r)$ profile and further averaging over 6 vortices at each magnetic field. $\alpha/a$ shows an overall monotonous increase with $H$ except for two anomalies close to 5 kOe and 70 kOe.

## 6.3   Calculation of thermal and quantum fluctuation

To understand the physical origin of this fluctuation we note that the interaction energy between two vortices, separated by a distance $R$, is given by,

$$E_{12} = \frac{\Phi_0^2 d}{2\pi\mu_0\lambda^2} K_0\left(\frac{R}{\lambda}\right) \qquad (6.5)$$

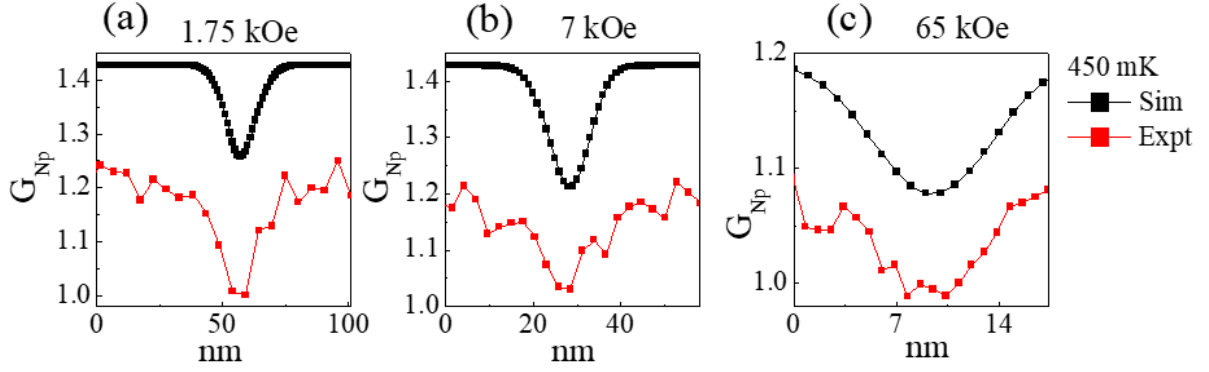

*Figure 6.10*: $G_{Np}$ variation for different magnetic fields are shown in (a)-(c) which are compared with simulated line-cuts of $\langle \tilde{G}_N(V_{peak}, r)\rangle_\alpha$. Although the nature of the curves are similar, the values differ by about 15-20%.

This for $\xi \ll r \ll \lambda$ takes the form, $E_{12} \approx \frac{\Phi_0^2 d}{2\pi\mu_0\lambda^2}\ln\frac{\lambda}{R}$ and gives the force $F_v = -\frac{\Phi_0^2 d}{2\pi\mu_0\lambda^2}\left(\frac{1}{R}\right)$ on each other (here $d$ is the film thickness). Thus each vortex is confined in a potential well formed by the repulsion of its surrounding vortices. Since the vortex motion is overdamped because of a large enough viscous drag,[14] we assume that each vortex oscillates individually in this potential well. If the oscillation amplitude $\alpha \ll a$, the potential can be approximated as a harmonic potential, since for a linear chain of vortices,

$$|\vec{F}| \sim \frac{1}{R} = \left(\frac{1}{a-\alpha} - \frac{1}{a+\alpha}\right) + \left(\frac{1}{2a-\alpha} - \frac{1}{2a+\alpha}\right) + \cdots = 2\alpha\sum_{n=1}^{\infty}\frac{1}{n^2 a^2 - \alpha^2} \sim \frac{2\alpha}{a^2}\frac{\pi^2}{6} \quad (6.6)$$

for $\alpha \to 0$. Therefore, $|\vec{F}| \propto \alpha$ mimics a harmonic oscillator with effective spring constant given by, $K = \frac{\Phi_0^2 \pi d}{6\mu_0\lambda^2}\left(\frac{1}{a^2}\right) \equiv \frac{K_0}{a^2}$. This picture of small oscillation remains valid even in the vortex fluid phases as long as the diffusive motion of the vortices is lower than the fast fluctuation. The total energy of each vortex in this potential well is given by $m_v\omega^2\alpha^2$, where $m_v$ is the vortex mass and $\omega^2 = K/m_v$.

If the vortex fluctuation is quantum in nature, then to a zeroth order approximation, the energy of oscillation will be supplied by the zero point motion of the vortices, i.e.

$$m_v\omega^2\alpha^2 \sim \hbar\omega \Rightarrow \frac{\alpha}{a} \sim \frac{\hbar^{\frac{1}{2}}}{(m_v K_0)^{\frac{1}{4}}}\left(\frac{1}{a}\right)^{\frac{1}{2}} \quad (6.7)$$

Since, $a \propto H^{-1/2}$, we get, for quantum origin of fluctuation,

$$\frac{\alpha}{a} \propto H^{1/4} \tag{6.8}$$

On the other hand, if the origin of the oscillation is thermal, then,

$$m_v \omega^2 \alpha^2 \sim k_B T \Rightarrow \frac{\alpha}{a} \sim \left(\frac{k_B T}{K_0}\right)^{\frac{1}{2}} \tag{6.9}$$

This shows that for a thermal fluctuation the relative fluctuation amplitude is independent of magnetic field.

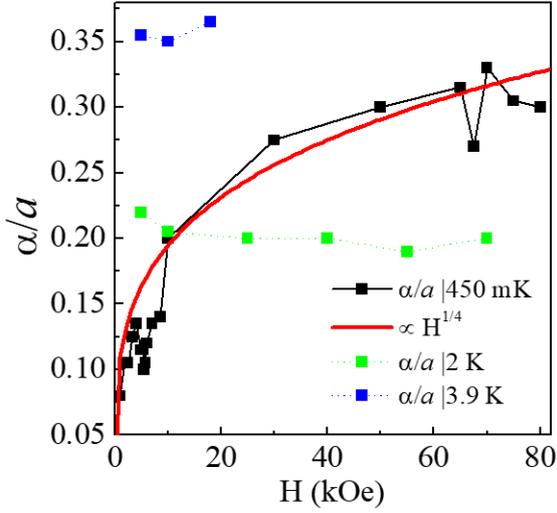

*Figure 6.11*: Variation of $\alpha/a$ for 450 mK in black dots is fitted with $H^{1/4}$ nature in red. For 2 K and 3.9 K, the green and blue dots respectively remain almost flat over the field range in which vortices are recorded.

### 6.3.1 Possibility of quantum fluctuation

In *Figure 6.11*, we plot on top of the $\alpha/a$ variation with magnetic field at 450 mK, the result of the back-of-the-envelope calculation, $\alpha/a \propto H^{1/4}$ which captures well the increasing trend. This nature of the variation of the relative fluctuation amplitude incites the idea of a quantum nature of the oscillation. The cusp-like anomalies observed near 5 kOe and 70 kOe appear very close to the VS to HVF and HVF to IVL boundaries respectively. The origin of these anomalies most likely is the anharmonicity of the confining potential close to phase boundaries, though a detailed analysis will require further theoretical investigations. It is interesting to note that the first anomaly appears when $\alpha/a \sim 0.14$, which is consistent with the Lindemann criterion for melting. However, this analysis and the model is very simplistic and a better theoretical understanding is required which can capture the observed phenomenon.

The quantum nature of the vortex fluctuation can be contradictory to the BKTHNY two step melting of the 2D vortex lattice. Since, the imaginary time dimension of the quantum fluctuation can be thought of a third dimension for the VL, the system no longer remains 2D.

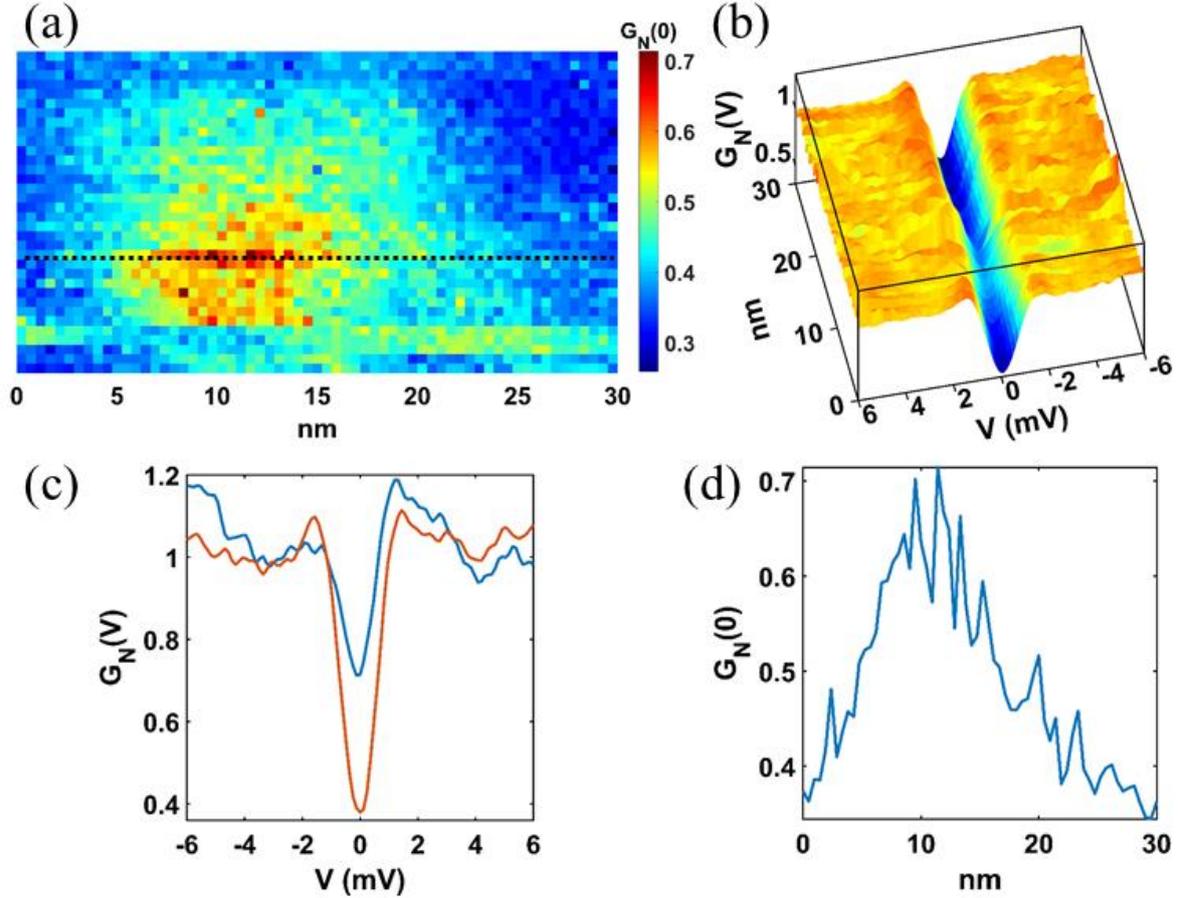

*Figure 6.12*: (a) $G_N(0)$ variation around a vortex at 40 kOe at 450 mK. (b) $G_N(V) - V$ map along the black dotted line of (a). (c) Spectra at the centre of (away from) the vortex core is shown in blue (red) curve. (d) Variation of $G_N(0)$ along the vortex core, which shows $G_N(0)$ at the centre of the core to be ~0.65.

Therefore, the very foundation on which the claim of a true BKTHNY melting has been made becomes fragile. However, a two-step melting is conjectured[15] to be possible even in the presence of quantum fluctuation, as long as the interaction remains $\sim 1/r^n$ (for $n = 1,..,3$). How this picture changes in presence of logarithmic interactions in vortices is yet unknown.

### 6.3.2 Possibility of quantum to thermal crossover

In *Figure 6.11*, we also show $\alpha/a$ variation as a function of magnetic field at 2 K and 3.9 K. In both of the cases we observe that the relative fluctuation amplitude remains almost flat over the magnetic field range. This invokes the idea of a thermal fluctuation at elevated temperatures. It is however very interesting to note that a quantum to thermal crossover is possible in the intermediate temperatures, which needs more theoretical and experimental attention.

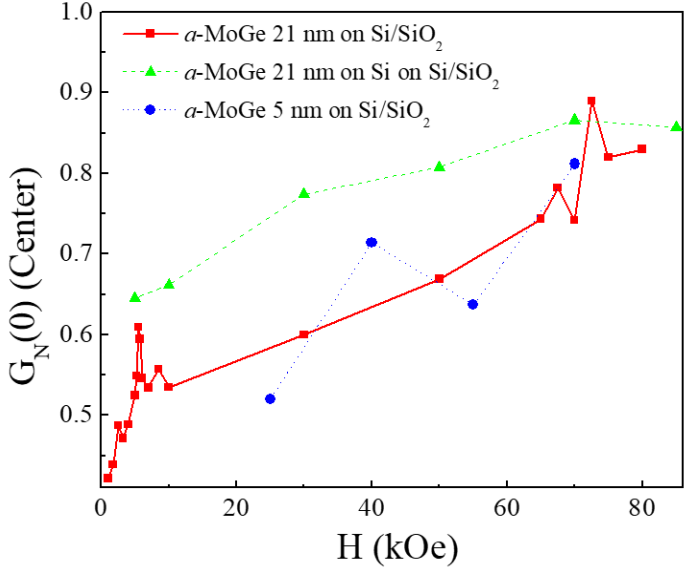

*Figure 6.13*: Variation of $G_N(0)$ at the centre of the vortex core at 450 mK as a function of magnetic field. The red connected squares are data from the 21 nm *a*-MoGe thin film grown on Si/SiO$_2$ substrate. The green triangles are data from 21 nm *a*-MoGe thin film grown on Silicon over-layer on top of Si/SiO$_2$ substrate. The blue circles are data from 5 nm *a*-MoGe thin film grown on Si/SiO$_2$ substrate.

## 6.4 Effect of stronger pinning

The fluctuation picture, whether it is quantum or thermal in nature, is valid here only because the pinning strength of the *a*-MoGe thin film is very weak. Therefore to check the random fluctuation picture in the following sections we explore the effect of stronger pinning on the vortex core. The pinning is made strong by the following paths, first by going to a reduced thickness ($d \sim 5\ nm$) where it has been established that the hexatic phase vanishes and a vortex glass state is established due to disorder. Secondly the thin film is deposited on a rougher substrate, where the Si/SiO$_2$ substrate is coated first with Si overlayer.

### 6.4.1 Effect of disorder

At first we study the vortex core in an ultrathin $\sim 5\ nm$ *a*-MoGe thin film. *Figure 6.12a* shows $G_N(0)$ map of a single vortex at 40 kOe at 450 mK and *Figure 6.12b* shows a line-cut of $G_N(V) - V$ along a line passing through the vortex core. A spectra at the centre of the core is compared with a spectra away from the vortex in *Figure 6.12c* and the $G_N(0)$ line cut across the vortex core is shown in *Figure 6.12d*. The vortex core even in the ultrathin limit shows the soft gap at the vortex core. And the variation of $G_N(0)$ values as a function of magnetic field is plotted in *Figure 6.13*, where it is compared with the same for the 21 nm sample. It is clearly visible that the $G_N(0)$ values have not changed much due to increasing disorder, even though the VL phase has become a vortex glass. Hence it can be understood from here that the disorder strength which can arrest the diffusive motion of the hexatic phase, is not enough to arrest the random fluctuation of the vortices. However, a completely different underlying reason is possible which is not related to the fluctuation picture and can explain these observed effects. But that theoretical model is yet unknown to us.

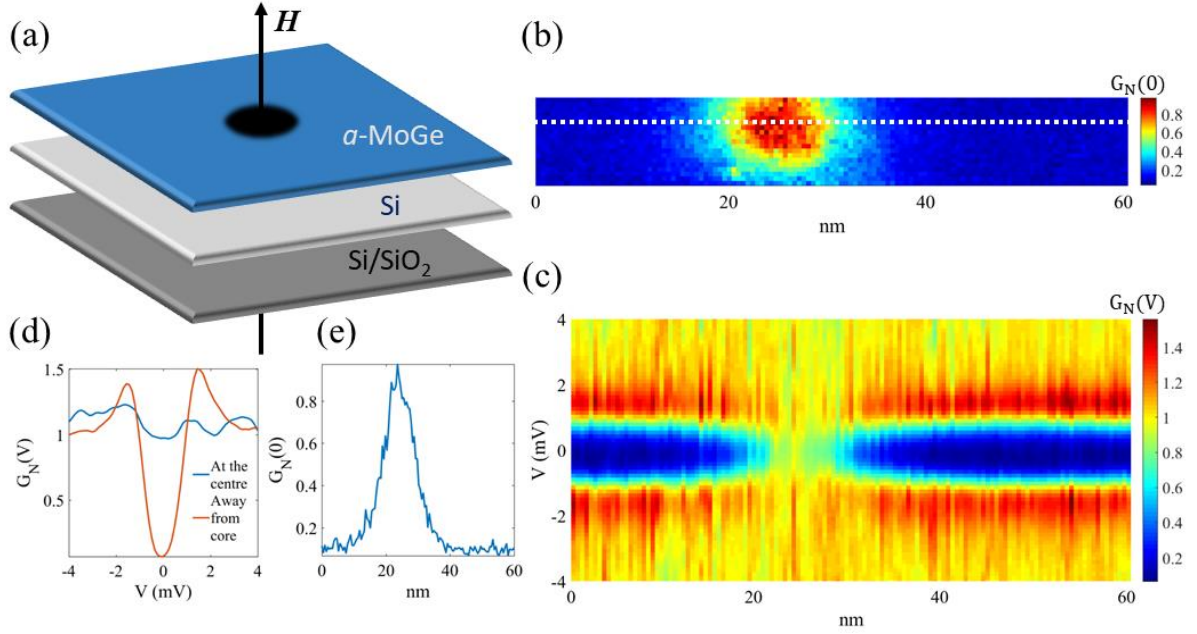

*Figure 6.14*: (a) Schematic diagram of the *a*-MoGe thin film grown on top of a Si layer which in turn is grown on Si/SiO$_2$ substrate. The black circle represents a vortex core. (b) $G_N(0)$ map at 10 kOe at 450 mK. The vortex is the red circular region. (c) $G_N(V) - V$ map along the white dotted line across the vortex core. (d) The blue (red) spectra corresponds to spectra at the centre of (away from) the vortex core. (e) Variation of $G_N(0)$ along the white dotted line of (b).

### 6.4.2 Effect of increased roughness

To explore the effect of increasing pinning strength even more, we have made the substrate rougher by depositing a thin layer ($\sim 2\ nm$ thick) of Si on the substrate before deposition of the 21 nm *a*-MoGe thin film (schematic diagram of this is shown in *Figure 6.14a*). We quantify the roughness from STM topographic images, where it is observed that the *rms* roughness of the heterostructures is $\sim 0.5\ nm$ compared to the *rms* roughness of the regular thin film $\sim 0.1\ nm$. We study the effect of this increasing roughness in vortex core spectroscopy. In *Figure 6.14b*, we plot $G_N(0)$ map of a single vortex at 10 kOe at 450 mK, where a line of $G_N(V) - V$ passing through the vortex core is plotted in *Figure 6.14c*. A spectra at the centre of the vortex core is compared with that far away from the core in *Figure 6.14d* and the variation of $G_N(0)$ along the vortex core is shown in *Figure 6.14e*. The magnetic field variation of $G_N(0)$ at the centre of the vortex core is plotted in *Figure 6.13*, which shows significant increase in the $G_N(0)$ values at the vortex core in the heterostructure compared to regular thin films of 21 nm and 5 nm. This observation again provokes the idea of the fluctuation picture, where stronger pinning can arrest the vortex fluctuation.

In summary this chapter discusses the unnatural behaviour of the vortex core in *a*-MoGe thin film and the presence of the soft gap inside the vortex core is viewed as a result of a random fluctuation of the flux lines. Although the quantum or thermal mature of this fluctuation is not known with surety, observations such as suppression of the soft gap due to increasing pinning supports the picture of the fluctuation of vortices. It is however an open question of how to tackle this phenomenon from a more detailed theoretical viewpoint.

# Chapter VII: Conclusion and outlook

## 7.1 Summary

In the above chapters we have discussed the effects of periodic pinning centres, extreme disorder and dimensionality on the nature and dynamics of vortex lattice. Although there are several other factors, such as drive current, fluctuations in temperatures close to and above $T_c$, electromagnetic noise, shape of the Fermi surface etc. which affect the stability and geometry of the vortex lattice.

Firstly, the effect of periodic pinning in vortex lattice resulted in the study of dynamic Mott-like to metal-like transition, which otherwise would have been energetically inaccessible in a real Mott insulator. Although this effect of periodic pinning has been addressed from a bulk property perspective, it would be interesting to ask what happens microscopically.

Secondly, the effect of disorder in NbN thin films has shown the emergence of granularity in order parameter with magnetic field. This in turn has helped us to explain why a strongly disordered superconductor goes into an insulator-like phase with magnetic field. Here the flux lines penetrate as laminar vortices which increases the granularity which diminishes the volume fraction of the superconducting regions. It would be interesting to test this picture in extremely disordered samples which is on the insulating side of the SIT. However, other fundamental problems like studying the ergodicity of the superconducting phase in presence of strong disorder can be addressed, where one can study the evolution of order parameter landscape in different thermal cycles of the same thin film.

The effect of going to a 2-dimensional system in presence of very weak pinning has been explored in great detail, which shows existence of BKTHNY-type two-step melting of the vortex lattice and existence of the hexatic fluid phase. This is a very great achievement from a perspective of statistical mechanics, where the hexatic fluid phase has played the role of *holy grail* for a long time.[1] However, the observation of a quantum fluctuation of vortices can be debatable, because the fluctuation in itself adds a new dimension (imaginary time) and in 3D BKTHNY melting is not possible. However, it has been conjectured[2] by Spivak and Kivelson that a two-step melting is possible even driven by quantum fluctuation as long as the interaction goes as $1/r^n$ (for $n = 1,..,3$). But how much this conjecture changes for a logarithmic interaction (as is the case with vortices) is unknown at the moment.

From a practical point of view of using superconductors to make powerful superconducting magnets, where one can send very high current without dissipation, is one of the key goals. To make that possible the critical current of the material must be very high, which can only be obtained if the pinning strength of it is high yet the superconductivity is not destroyed by the disorder. This needs extensive tuning of the material parameters with prior knowledge of the effects of different agents. Another possibility of practical application is based on the novel phenomena of SIT, where based on tuning parameters such as disorder and/or magnetic field, a superconductor can be transformed into an insulator, or in some cases a bad metal. This in principle can be used to make magnetic field controlled gates in electronic circuits.

## 7.2  Future problems

This leads us to discuss the future roadmap of the problems addressed in the previous chapters. It is a hope that these problems will be taken up and be solved very soon.

### 7.2.1  STS/M study of matching effect

Real space imaging can provide key pieces to some of the additional puzzles of the dynamic Mott-to-metal transition, for example the step-like jumps in mutual inductance even in zero magnetic field. Although the general consensus is that this is due to vortex-antivortex pair creation in zero field and consequent dynamic transition of them. This puzzle, as well as, fundamental questions like geometry of the vortex lattice at filling factor equals to half integer can be addressed using real space imaging.

Application of the periodic pinning landscape can possibly be found in the budding field of quantum computation, where vortex cores of topological superconductors are predicted to contain Majorana quasiparticles.[3,4] Deposition of topological superconducting thin films on top of a periodic pinning landscape can be utilized to manipulate the exotic states. This has recently been performed to study the braiding of the Majorana modes and subsequently creating logic gates.[5]

### 7.2.2  Universality of emergent granularity picture

Further questions like the universality of the emergent inhomogeneity in presence of magnetic field can be tested in disordered samples of other materials, for example the apparent absence of this emergent inhomogeneity in a-MoGe thin films in contrast with thin films of NbN or InO$_x$.

### 7.2.3 Possibility of vortex dynamics governed SIT

It has been suggested in many works that the superconducting state in high disorder can be tuned into an insulator by applying magnetic field due to formation of vortex superfluid. The thin films of a-MoGe being a nice platform to study the vortex dynamics can be used to study thinner samples where sheet resistance is of the order of quantum of resistance. We already have preliminary data in 2 nm thick thin films, which shows the superconductor to insulator-like transition, as well as, a single crossover point of the magnetoresistance curves. However, the preliminary STM data shows existence of vortex lattice in the insulator-like phase also, which might be an indication of the SIT governed by vortex dynamics and might act as a proof of charge-vortex duality.[6]

### 7.2.4 Explanation of bad-metallic states

However, unsolved problems like existence of the HVF and IVL phases down to lowest temperatures can provide possible explanation of the 'bad metal'[7] states often reported in literature in 2D materials and in presence of magnetic field.[8,9] The quantum fluctuation conjectured in this thesis for the vortices in a-MoGe thin films could be the possible reason behind the existence of this bad-metallic states. Although this claim needs further detailed study of the vortex states using both real space imaging and transport measurements.

# Appendix A: Study of superconductivity in $Nb_xBi_2Se_3$

$Bi_2Se_3$ is a topological insulator,[1] which is often doped to study different topological phases, such as topological superconductivity,[2,3,4,5,6] time-reversal and rotation symmetry breaking superconductivity,[7] Weyl superconductivity[8] etc. Although the existence of a topological superconductor is yet mostly in the imaginative phase, $Nb_xBi_2Se_3$ is a potential candidate for the exotic phase. Even though there are several reports of superconductivity in this material, almost all of them comes mostly from bulk measurements, except for Ref [9] which has done extensive microscopic measurements recently. However, vortex phase in $Nb_xBi_2Se_3$ is not yet observed. Vortex state is interesting in a topological superconductor, since they are predicted to accommodate Majorana quasiparticles at the vortex cores. In this study we have also studied microscopically the superconducting state in $Nb_xBi_2Se_3$ in great detail. However we too have not observed the vortex state in this material yet.

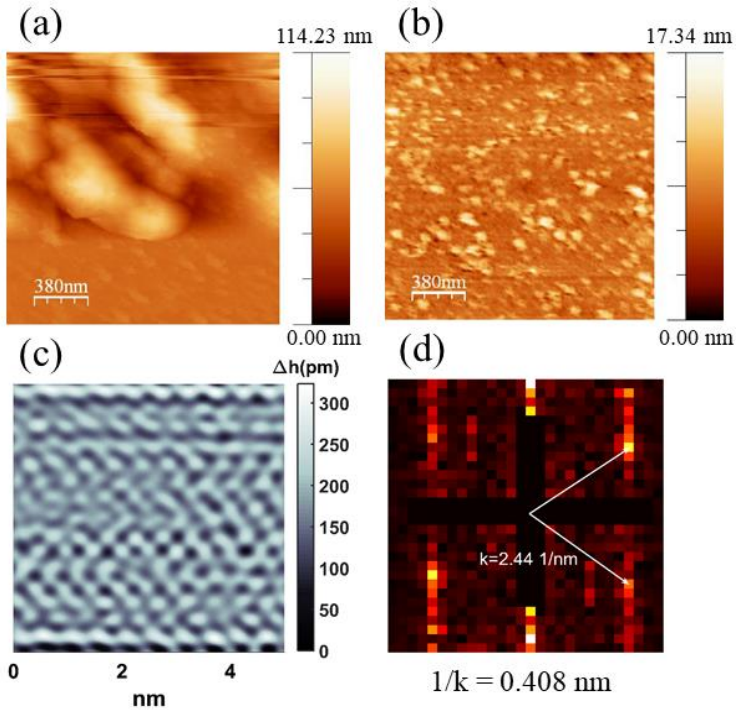

*Figure A1*: (a)-(b) Topographic image of the surface of the sample, where the first one shows large granules, the second one shows small particulates. (c) Atomic resolution image of a $5\ nm \times 5\ nm$ area. (d) FFT of the image in (c), showing faint six spots, corresponding to hexagonal symmetry of the terminating surface. Lattice constant is determined from $1/k = 4.08$ Å.

## A1.1 Topography from STS/M

As a preliminary study we do scanning tunnelling microscopy imaging of the surface topography. The surface shows many granules of several hundreds of nanometres in size (*Figure A1a*). Some area show an overall flat surface will small particulates of size tens of nanometres (*Figure A1b*). We try to acquire atomic resolution image on a granule by taking topography image (*Figure A1c*) at 200 pA set current over a $5\ nm \times 5\ nm$ area at 450 mK and from the FFT of the area, we found a triangular lattice with lattice constant to be around ~4.08 Å, which matches quite well with previous reports of crystal structure of Bi$_2$Se$_3$.

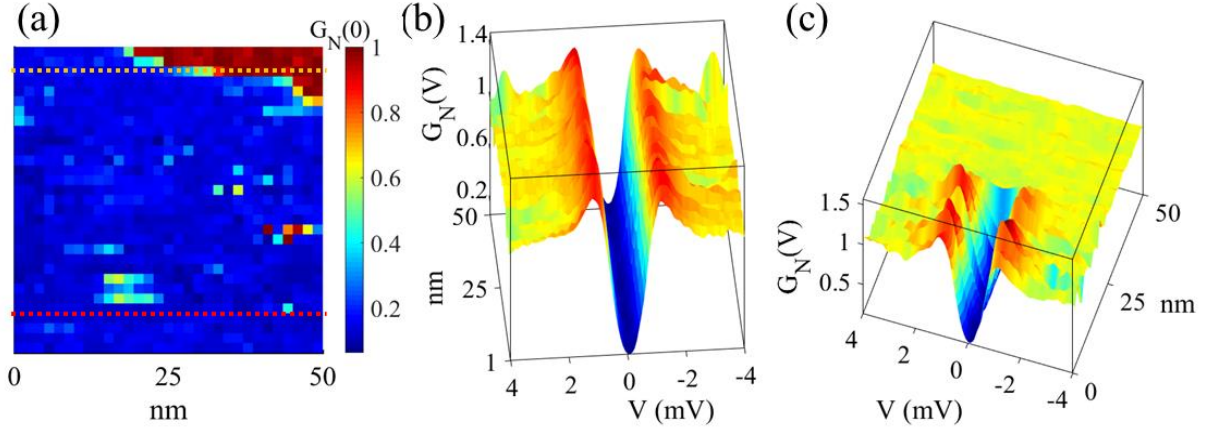

*Figure A2*: (a) $G_N(0)$ map over a 50 $nm$ × 50 $nm$ area, showing a uniform superconducting region, except for the non-superconducting region at the top-right corner. $G_N(V) - V$ variation along the red dotted line is shown in (b), and along the yellow dotted line is shown in (c).

### A1.2 Bias spectroscopy and its area dependence

Next we perform bias spectroscopy over different areas, all of which are about 50 $nm$ × 50 $nm$ in size. We observe uniform superconducting gap over these regions with prominent coherence peaks but zero bias conductance ($G_N(0)$) only going down to 0.2, instead of zero. $G_N(0)$ map obtained from bias spectroscopy over one such area is presented in *Figure A2a*, accompanied by two line cuts in *Figure A2b-c*. The right corner of the $50\ nm \times 50\ nm$ area shows a non-superconducting particulate.

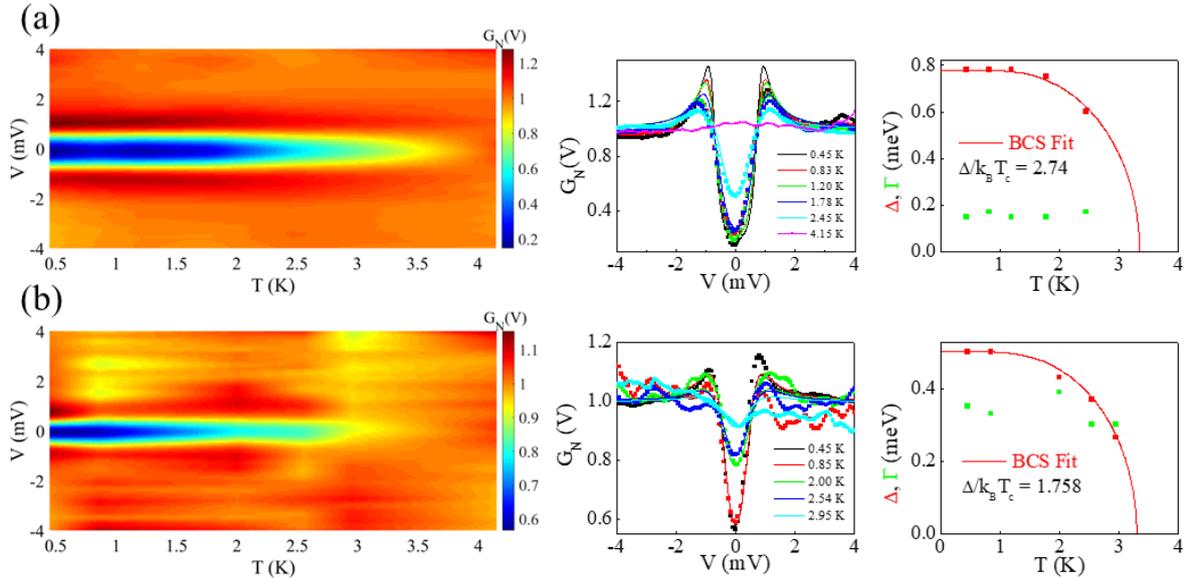

*Figure A3*: (a)-(b) *Left* panel shows $G_N(V) - V$ variation with respect to temperature for area I and II respectively. *Middle* panel shows $G_N(V) - V$ spectra for different temperatures which are fitted with BCS curves to extract Δ and Γ, shown in the *right* panel. *Right* panel shows red (green) dots as the values of Δ (Γ) as a function of temperature along with the BCS fit (red line). For both cases values of $\Delta/k_B T_c$ is quoted.

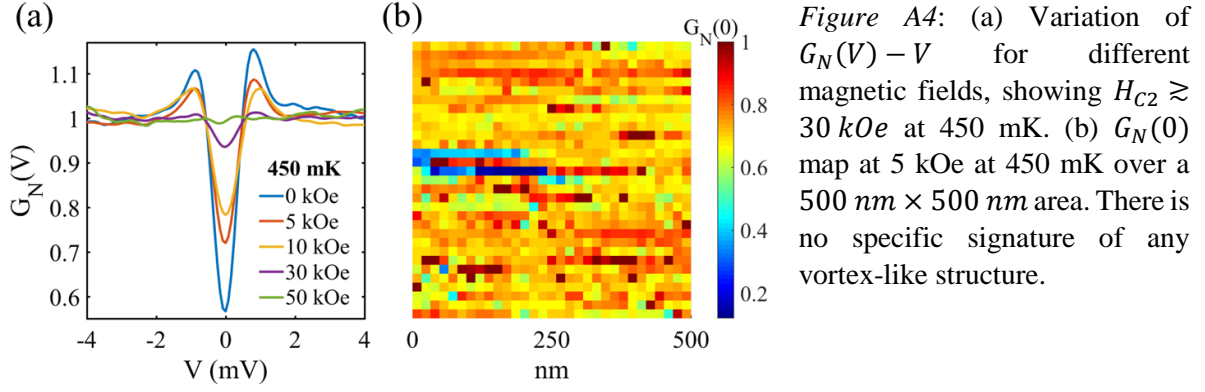

*Figure A4*: (a) Variation of $G_N(V) - V$ for different magnetic fields, showing $H_{C2} \gtrsim 30\ kOe$ at 450 mK. (b) $G_N(0)$ map at 5 kOe at 450 mK over a $500\ nm \times 500\ nm$ area. There is no specific signature of any vortex-like structure.

From the bias spectroscopy of different areas we notice that there are two types of areas. In both of these areas we have studied temperature variation of averaged bias spectroscopy. The first area shows that gap vanishes at $\sim 3.3\ K$ and $\Delta(0) \sim 0.78\ meV$ and the variation of $\Delta(T)$ (*Figure A3a*) follows BCS curve with $\frac{\Delta(0)}{k_B T_c} \sim 2.74$ which is much larger than BCS value of 1.76. The second area similarly shows that gap vanishes at $\sim 3.3\ K$ but the BCS fitting of the spectra (*Figure A3b*) needs a large value of $\Gamma$ and gives $\Delta(0) \sim 0.5\ meV$. This gives a $\frac{\Delta(0)}{k_B T_c} \sim 1.758$ close to the BCS value.

However these two types of area demands close inspection of superconductivity as a function of magnetic field. Both of the areas show that gap vanishes at about $H \gtrsim 30\ kOe$, which is previously reported as the $H_{c2}$ of the material.

### A1.3 Absence of vortex-like structures

We have also acquired conductance maps for $G(V \simeq 1\ mV)$ for both of the areas reported above for magnetic fields ranging between 5-30 kOe, but no vortex like structure is observed in the conductance maps. In *Figure A4*, we show $G_N(0)$ map over a $500\ nm \times 500\ nm$ area in 5 kOe field, which except for few noisy points, does not show any vortex like structure. This is very puzzling and needs a lot more attention.

However, this is a very basic study of superconductivity in NbxBi2Se3, which gives values of $T_c$ and $H_{c2}$ in the right ballpark, matching with the previous bulk measurements. Checking topological nature of superconductivity in this material needs more detailed work, which is not done in this thesis.